\documentclass[11pt]{article}

\usepackage[T1]{fontenc}
\usepackage[utf8]{inputenc}
\usepackage[letterpaper,margin=1in]{geometry}
\usepackage{lmodern}
\usepackage{microtype}
\usepackage{setspace}
\usepackage{amsmath,amssymb}
\usepackage{graphicx}
\usepackage{booktabs}
\usepackage{tabularx}
\usepackage{makecell}
\usepackage{threeparttable}
\usepackage{threeparttablex}
\usepackage{multirow}
\usepackage{longtable}
\usepackage{rotating}
\usepackage{pdflscape}
\usepackage{array}
\usepackage{ragged2e}
\usepackage[table]{xcolor}
\usepackage{float}
\usepackage{placeins}
\usepackage{chngcntr}
\usepackage{etoolbox}
\usepackage{adjustbox}
\usepackage{enumitem}
\usepackage{tikz}
\usetikzlibrary{arrows.meta,calc,decorations.pathreplacing,patterns,positioning,shapes.geometric}
\usepackage[font=normalsize,labelfont=bf,labelsep=colon,justification=raggedright,singlelinecheck=false]{caption}
\usepackage{natbib}
\setcitestyle{authoryear,round}
\usepackage{titlesec}
\newcommand{\periodafter}[1]{#1.}
\titleformat{\subsubsection}[runin]
  {\normalfont\bfseries}{\thesubsubsection}{1em}{\periodafter}
\IfFileExists{xurl.sty}{\usepackage{xurl}}{\usepackage{url}}
\definecolor{LovelyLink}{RGB}{52,92,150}
\definecolor{LovelyCite}{RGB}{72,130,100}
\definecolor{LovelyURL}{RGB}{70,115,180}
\usepackage[
  colorlinks=true,
  linkcolor=LovelyLink,
  citecolor=LovelyCite,
  urlcolor=LovelyURL,
  filecolor=LovelyURL,
  breaklinks=true,
  hypertexnames=false
]{hyperref}

\hypersetup{
  pdftitle={Does Training on Future Data Pay? Look-Ahead Bias in Forecasting with Pretrained Models},
  pdfauthor={Li Chen; Yunlong Chen; Difang Huang; Bo Zhang}
}

\graphicspath{{figures/}}

\setlist{noitemsep}

\DeclareCaptionFont{tabletitle}{\fontsize{11}{13.2}\selectfont}
\newcommand{\exhibitnotes}[1]{%
  \begin{minipage}[t]{\linewidth}
  \singlespacing\fontsize{9}{11}\selectfont\justifying
  \setlength{\parindent}{0pt}\textit{Notes:} #1\par
  \end{minipage}%
}
\AtBeginEnvironment{table}{\singlespacing}
\AtBeginEnvironment{figure}{\singlespacing}
\AtBeginEnvironment{sidewaysfigure}{\singlespacing}
\AtBeginEnvironment{longtable}{\singlespacing\setlength{\LTcapwidth}{\linewidth}}
\AtBeginEnvironment{landscape}{\singlespacing}

\begin{document}

\hypersetup{pageanchor=false}
\onehalfspacing
\title{%
{\Large\bfseries Does Training on Future Data Pay?\\
Look-Ahead Bias in Forecasting with Pretrained Models}%
\thanks{All authors contributed equally to this paper.
We thank participants at various seminars for helpful
comments and suggestions.
Difang Huang acknowledges financial support from
the National Natural Science Foundation of China
(Grant Nos.\ 72503232, 72622025, and 72574227).
Bo Zhang is partially supported by
the National Key R\&D Program of China
(Grant No.\ 2022YFA1008000)
and the National Natural Science Foundation of China
(Grant Nos.\ 12471268, 12001517, and 72091212).
All remaining errors are our own.}%
}
\author{
  {\bfseries Haiqiang Chen} \\
  Shenzhen University \\
  {\bfseries Li Chen} \\
  Xiamen University \\
  {\bfseries Yunlong Chen} \\
  University of Science and Technology of China \\
  {\bfseries Difang Huang} \\
  Chinese Academy of Sciences \\
  {\bfseries Bo Zhang} \\
  University of Science and Technology of China
}

\date{\today}
\maketitle
\thispagestyle{empty}

\clearpage
\doublespacing
\hypersetup{pageanchor=true}
\pagenumbering{arabic}
\setcounter{page}{1}

\begin{center}
  {\Large\bfseries Does Training on Future Data Pay? \\Look-Ahead Bias in Forecasting with Pretrained Models}
\end{center}

\begin{center}
  {\bfseries Abstract}
\end{center}

We examine whether post-origin training information inflates the measured accuracy and economic value of financial forecasts. We evaluate five sets of financial time-series foundation models, each comprising independently trained annual vintages under U.S., global, and factor-augmented training environments, across 14 equity markets and four forecast horizons. Rolling comparisons vary the annual vintage for a fixed forecast; fixed-vintage comparisons hold the vintage fixed as target windows move across its training cutoff. Each alternative forecast is paired with an origin-aligned point-in-time (PIT) benchmark using identical numerical histories and inference protocols. In the U.S.-trained reference environment, post-origin vintages materially revise informative PIT forecasts but generally reduce accuracy in both designs. Pooled rolling comparisons yield higher mean squared forecast errors in 18 of 20 U.S. model-set–horizon combinations. The origin-crossing update also performs worse on average than an equally long pre-origin update. Under a common constrained allocation rule using one-month forecasts, median exposed-minus-PIT differences in annualized certainty-equivalent returns are -1.77 percentage points in the United States and -2.14 points internationally. Global and factor-augmented training produce more mixed predictive effects. An exact squared-error decomposition shows that revisions improve accuracy when their error-correcting benefit exceeds their mean squared magnitude; under U.S. training, alignment with PIT errors generally falls short of this requirement. Temporal exposure therefore establishes an information-set violation, not sufficient evidence of inflated predictive accuracy or investor value.

\medskip
\textit{Keywords:} Look-ahead bias; Return prediction; Time-series foundation models; Large language models; Information contamination.

\medskip
\textit{JEL classification:} G11, G12, G14, C53, C58.

\clearpage
\doublespacing

\section{Introduction}\label{section-1-introduction-ux4e2dux6587ux521dux7a3f}

Historical backtests are foundational to empirical finance because they allow researchers to identify return predictability and assess whether that predictability could have supported profitable investment strategies in real time. We use pretrained models to generate financial forecasts for a sequence of historical dates. This retrospective application raises a distinct information-timing concern: even if each forecast uses only return data available at its historical forecast date, the model itself may have been trained on observations from later periods. Consider, for example, a January 2010 forecast based on returns observed through December 2009 but generated by a model trained on data through 2015. Although the input history respects the historical forecast date, the model’s learned parameters incorporate observations that would not have been available at that date. This issue is especially relevant when pretrained models are used retrospectively to reconstruct earlier forecasts \citep{sarkar_vafa_2024,he_lv_manela_wu_2025}. The distinction is becoming more important as machine learning expands the set of forecasting methods available to researchers and investors \citep{gu_kelly_xiu_2020,kelly_xiu_2023} and incorporates economic structure into asset-pricing models \citep{fan_ke_liao_neuhierl_2022,chen_pelger_zhu_2024}. Credible historical backtests must therefore account for two information cutoffs: one governing the data used to form each forecast and the other governing the data used to train the model. We examine how training data that extend beyond a historical forecast date affect measured predictive accuracy and investment performance, and whether such data create the appearance of investment opportunities that could not have been identified in real time.

Training on post-forecast observations, however, does not necessarily improve backtested performance. Evidence from financial language models is mixed: positive look-ahead premia have been documented \citep{yan_tang_gao_jiang_lu_2026}, while other findings indicate that temporally misaligned pretraining can encode future information without systematically improving classification or portfolio performance \citep{rahimikia_drinkall_2024}. Later training data can alter the relationships a model learns and, consequently, the forecasts it produces from the same return history. Relationships learned from later periods may be less useful for predicting returns in earlier market environments. In our reference U.S. training environment, later model vintages materially revise informative return forecasts but generally reduce their accuracy and portfolio value. Global and factor-augmented training yield more mixed effects. We call training coverage that extends beyond the historical forecast date temporal exposure. The existence of such exposure must be distinguished from its effects on forecast accuracy and investment performance.

We investigate this distinction using a collection of annually dated financial time-series foundation models (TSFMs) developed by \citet{rahimikia_ni_wang_2025}. These Transformer-based models are designed to learn temporal regularities shared across collections of numerical time series and to apply the resulting parameterization to new forecasting tasks. Recent theory analyzes the approximation and generalization properties of this model class \citep{wu_he_cao_fan_liu_2025}. The released collection includes five variants: \textit{Chronos Tiny}, \textit{Chronos Mini}, \textit{Chronos Small}, \textit{TimesFM 8M}, and \textit{TimesFM 20M}. Within each variant, separate models are trained independently on expanding financial datasets ending at successive year-end cutoffs. We refer to each independently trained, cutoff-specific model as an annual vintage. The annual vintages are constructed under three training-data configurations: U.S. firm returns, global firm returns, and global firm returns augmented with the JKP factor series of \citet{jensen_kelly_pedersen_2023}. We refer to these as the \textit{US}, \textit{Global}, and \textit{Augmented} training environments, respectively. We use the released parameters directly, without fine-tuning or otherwise updating them for our target markets, to forecast the U.S. equity premium and excess returns in 13 non-U.S. equity markets at horizons of one, three, six, and twelve months.

We use these annual vintages to construct matched comparisons of alternative forecasting rules. Within a given model variant and training environment, every comparison keeps the numerical return history, forecast target, and forecast construction unchanged. For a forecast target beginning in calendar year $t$, the origin-aligned point-in-time (PIT) benchmark uses the annual vintage trained on data through December of year $T-1$. Its training sample therefore contains no observations from after the historical forecast date.\footnote{The PIT designation refers only to the timing of the model’s training data. The model architecture and released parameters were not available at the historical forecast date.} The rolling design replaces this benchmark with the vintage trained through year $T-2$ or with vintages trained through years $T$, $T+1$, and $T+2$. The adjacent vintages allow us to compare two successive one-year updates: moving from $T-2$ to $T-1$ adds only pre-origin training data, whereas moving from $T-1$ to $T$ extends the training period beyond the forecast origin. Pairwise comparisons use the same target months, and pooled estimates use the months shared by the PIT forecast and all three post-origin vintages.

As a complement to these forecast-by-forecast comparisons, the fixed-vintage design follows the same fitted forecasting rule through historical time. We hold each of the 2000, 2009, and 2023 vintages fixed and apply it as forecast origins and target windows move across its training cutoff. When the training period extends through the entire target window, the resulting forecast has retrospective exposure. When the target window straddles the cutoff, the forecast represents a boundary crossing. Once the target window begins after the cutoff, the vintage is origin-aligned during the first post-cutoff year and becomes stale thereafter. At each forecast origin, we compare the fixed-vintage forecast with the corresponding PIT forecast for the same historical task. The rolling and fixed-vintage designs thus provide complementary views of how forecast performance changes with the position of the training cutoff. Because every annual vintage is trained independently, however, these comparisons capture the net consequence of replacing one fitted forecasting rule with another, not the isolated effect of adding later observations to the training sample.

A meaningful economic interpretation of vintage effects requires an informative origin-aligned benchmark. Otherwise, an unfavorable vintage effect could merely indicate that one uninformative forecast performs worse than another, rather than that temporal exposure erodes genuine predictive value. We therefore begin by evaluating whether the PIT forecasts outperform the expanding historical average, a strictly real-time benchmark, before examining how their performance changes when the annual vintage is replaced. The evidence is particularly strong for the \textit{Chronos} variants. Using all available U.S. observations under the \textit{US} training environment, the PIT forecast from \textit{Chronos Tiny} reduces mean squared forecast error (MSFE) relative to the historical-average benchmark by 3.45\% at the one-month horizon and 15.12\% at the six-month horizon. The PIT \textit{Chronos} forecasts likewise outperform the historical-average benchmark internationally: at the twelve-month horizon, their MSFE reductions in the equal-market aggregate range from 10.60\% to 12.91\% across the three variants. Having established the predictive relevance of the origin-aligned benchmark, we examine how replacing the annual vintage changes its forecasts. Holding the numerical history fixed, alternative vintages alter forecast levels, signs, and predictive distributions. For one-month U.S. forecasts from \textit{TimesFM 20M}, the first origin-crossing update—from vintage $T-1$ to vintage $T$—changes the predicted return sign in 27.9\% of matched months. Replacing the annual vintage therefore changes the economic content of the forecast rather than merely rescaling an otherwise unchanged signal.
 
The changes in forecasts caused by replacing the annual vintage generally do not improve predictive accuracy. Under the \textit{US} training environment, later vintages often continue to outperform the historical average, but they typically perform worse than the origin-aligned PIT vintage. For each model and forecast horizon, we compare the PIT (vintage $T-1$) MSFE with the average MSFE of vintages $T$, $T+1$, and $T+2$, using only target months for which all four forecasts are available. This comparison favors PIT in 18 of the 20 U.S. model--horizon combinations. Across these 20 combinations, the median MSFE advantage of PIT is 5.94\% of historical-average MSFE. Internationally, the median effect across the 13 markets favors PIT in all 20 model--horizon combinations. To distinguish the effect of crossing the forecast origin from that of ordinary annual updating, we compare two adjacent one-year updates. After normalizing the MSFE differences by historical-average MSFE, moving from vintage $T-2$ to vintage $T-1$, which adds only pre-origin training data, reduces average U.S. forecast loss by 0.72 percentage points. In contrast, moving from vintage $T-1$ to vintage $T$, which first extends the training period beyond the forecast origin, increases average forecast loss by 5.65 percentage points. The fixed-vintage analysis provides consistent evidence: in 12 of the 20 U.S. model--horizon combinations, PIT has lower MSFE in all three comparisons—the pooled rolling comparison and the retrospective comparisons with the fixed 2009 and 2023 vintages. Not a single combination has lower MSFE for the alternative vintages than for PIT in all three comparisons.

We next examine whether substituting later-vintage forecasts for PIT forecasts improves portfolio performance. We apply the same constrained mean--variance allocation rule to one-month forecasts from both sources. The PIT strategy forms its position using the origin-aligned forecast, whereas the later-vintage strategy first averages the forecasts from vintages $T$, $T+1$, and $T+2$. Both strategies are evaluated over the same investment months. Without deducting transaction costs, the median later-vintage-minus-PIT difference in annualized certainty-equivalent return is -1.77 percentage points across the five U.S. models. Across the 65 combinations of 13 international markets and five models, the median difference is -2.14 percentage points, and the later-vintage strategy underperforms the PIT strategy in 56 of those combinations. The portfolio evidence therefore mirrors the predictive-accuracy results: under the constrained allocation rule, later-vintage forecasts deliver lower median economic value in both U.S. and international markets.


An exact forecast-error decomposition clarifies why substantial forecast revisions need not improve accuracy. It separates the contribution of alignment with errors in the PIT forecast from a penalty associated with the size of the revision. Predictive gains require sufficient error correction to offset this penalty: revisions must be appropriately directed and sized, rather than simply large. In the reference U.S.-training environment, revisions are substantial, but their alignment with PIT errors is generally too weak to produce a net improvement. This accounting links parameter-induced forecast changes to their realized predictive consequences without requiring us to identify which training patterns the model retains. It also shows why a visible response to parameter replacement is not, by itself, evidence of better prediction.

Training content changes this balance. For the U.S. target, Global training shifts the median pooled rolling effect from \(-5.94\) to \(+1.37\) historical-average MSFE percentage points, where positive values indicate improvement relative to PIT. Factor augmentation brings the median back to \(-3.64\) points. For non-U.S. targets, the US environment provides an indirect-only exposure setting: the target market is absent from training, but post-origin U.S. returns can carry information relevant for non-U.S. returns \citep{rapach_strauss_zhou_2013}. Global training adds both target-market and broader cross-market information, shifting the effects toward a more balanced distribution of gains and losses. These comparisons show that excluding the target market does not eliminate the parameter-time problem and that temporal overlap alone does not determine its performance consequences. They also preserve an important distinction between baseline quality and incremental effects: a more favorable later-minus-PIT comparison need not imply a stronger origin-aligned forecast. US training supplies the strongest PIT benchmark in most model--horizon cells.

Our paper contributes to three strands of literature. The matched-vintage design contributes to research on historical information integrity in financial forecasting. \citet{harvey_leybourne_sollis_taylor_2021} develop real-time procedures for detecting regimes of U.S. equity-premium predictability. More broadly, real-time data research establishes that information available to a later researcher can differ from information available at the decision date \citep{croushore_stark_2001,koenig_dolmas_piger_2003,ghysels_horan_moench_2018}. \citet{bowles_reed_ringgenberg_thornock_2024} show that anomaly returns concentrate soon after the underlying information becomes public. \citet{li_rossi_yan_zheng_2025} show how feature engineering shapes the real-time investment performance of machine-learning strategies. We operationalize this constraint for reusable pretrained rules by pairing dated parameter states on identical return histories. The design separately evaluates the predictive relevance of an origin-aligned model and the incremental consequence of deploying another vintage. \citet{li_li_lyu_yu_2025} study how predictive information can improve on the historical-average benchmark. We use this benchmark to establish predictive content, then evaluate alternative vintages relative to the origin-aligned forecast.

Our paper also complements research on look-ahead effects in pretrained models \citep{gao_jiang_yan_2025,engelberg_manela_mullins_vulicevic_2025,zhang_stadie_2026}. Chronologically restricted language models provide a way to study the role of training dates in historical applications \citep{rahimikia_drinkall_2024,he_lv_manela_wu_2025,kelly_malamud_schwab_xu_2026}. Existing evidence includes positive look-ahead premia \citep{yan_tang_gao_jiang_lu_2026} as well as settings in which contextual knowledge offsets the benefits of future information \citep{glasserman_lin_2024}. For numerical foundation models, temporally overlapping related series can compromise evaluation even without direct inclusion of the target series \citep{meyer_kaltenpoth_zalipski_mueller_2025}. We extend this evaluation to matched numerical forecasts, forecast errors, and portfolio decisions across markets and training scopes. Whereas \citet{rahimikia_ni_wang_2025} develop the finance-native forecasting technology, we use its dated states to measure the historical predictive and investor consequences of rule substitution.

Finally, our paper connects parameter timing to the decisions forecasts support, relating to broader work on how information affects financial decision making \citep{huang_xiong_yang_2022,gao_ru_yang_2022}. Statistical predictability and investment value need not receive the same assessment \citep{kandel_stambaugh_1996,granger_pesaran_2000,marquering_verbeek_2004}. Research on learning and optimization likewise emphasizes evaluation against downstream objectives \citep{ban_elkaroui_lim_2018,elmachtoub_grigas_2022,fan_guo_yu_2024} and economic restrictions \citep{avramov_cheng_metzker_2023}. \citet{blanchet_hernandez_nguyen_pelger_zhang_2022} study the trade-off between look-ahead bias and estimation variance in imputing missing returns for portfolio allocation. We follow the same parameter substitution through forecast revision, squared loss, and constrained asset allocation. The loss accounting establishes whether a revision corrects the original forecast error; the portfolio comparison measures whether the resulting positions improve realized investor utility. Neither forecast movement nor predictive accuracy alone answers both questions. 

The practical implication is not that future-trained models are harmless or that later vintages should always be avoided. A forecast that violates the training-data constraint does not become point-in-time because its measured performance effect is negative. Rather, validation must document input dates and training cutoffs separately and compare alternative rules on matched historical tasks. Training dates establish temporal exposure. Forecast revisions, realized errors, and portfolio outcomes establish its predictive and economic consequences.

The remainder of the paper is organized as follows. Section~\ref{pretraining-historical-information-sets-and-temporal-exposure} formalizes the information constraints, and Section~\ref{markets-annual-tsfm-states-and-matched-design} presents the data and empirical design. Sections~\ref{point-in-time-predictive-relevance-and-forecast-revision} and~\ref{the-predictive-and-investor-value-of-parameter-time-exposure} examine forecast revisions, predictive performance, and investor value. Section~\ref{training-information-and-revision-to-value-accounting} studies training content and the loss accounting. Section~\ref{conclusion} concludes.


\section{Setup}\label{pretraining-historical-information-sets-and-temporal-exposure}

\subsection{Pretrained Financial Forecasting Rules}\label{pretraining-and-reusable-parameter-states}

Pretraining allows the training sample to continue influencing subsequent forecasts through the parameters after training ends. A pretrained model first estimates reusable statistical structure from a large dataset and then applies the trained parameters to new inputs or tasks. Models that reuse the same parameter state across multiple tasks or data series are commonly called foundation models \citep{bommasani_etal_2021}. Both the LLMs and TSFMs considered in this paper follow this arrangement. In financial research, the former are often used to extract predictive signals from news, corporate disclosures, and other text, while the latter generate numerical forecasts from returns, prices, or other time series.\footnote{Financial text applications include return forecasting and specialized text-learning methods \citep{lopez_lira_tang_2026,chen_kelly_xiu_2026,zhou_fan_xue_2024}.} The parameters therefore carry information from the period covered by pretraining into subsequent financial applications.

Training determines the forecasting rule; inference applies that rule to the current input. During training, the model updates its parameters based on the sample, absorbing statistical relationships in the data. During inference, it holds the resulting parameters fixed and maps new inputs into model outputs. The full training sample is not fed into the model again at inference, but its influence on forecasts persists because the parameters retain the patterns learned from that sample. A current forecast therefore reflects both the input the model sees now and the data on which the parameters that process it were trained.

Let \(\tau\) denote the cutoff date of the pretraining sample. The relationship between training and forecasting is

\[
\widehat y_{t,h}(\tau)
=
f_{\theta_\tau}\!\left(X_{\leq t};h\right),
\qquad
\theta_\tau
=
\arg\min_{\theta}
\mathcal L\!\left(\theta;\mathcal D_{\leq \tau}\right).
\tag{1}
\]

Here, \(X_{\leq t}\) is the historical input available at forecast origin \(t\) and actually supplied to the model, \(h\) is the forecast horizon, \(\mathcal D_{\leq\tau}\) is the pretraining sample through \(\tau\), and \(\mathcal L\) is the training loss used to estimate the parameters. The parameter state \(\theta_\tau\) is learned from this sample, and \(f_{\theta_\tau}\) is the forecasting rule implemented by those parameters. The forecast target \(y_{t,h}\) comprises outcomes over the future \(h\) periods after \(t\), and \(\widehat y_{t,h}(\tau)\) is the forecast obtained from the parameter state trained on data through \(\tau\). Pairing the same \(X_{\leq t}\) with different \(\theta_\tau\) can produce different outputs. Pretraining data thus continue to influence subsequent forecasts through the parameters after training ends.

\subsection{Text Signals and Numerical Return Forecasts}\label{from-llms-to-tsfms}

Within this pretraining framework, LLMs and TSFMs work with text and numerical time series, respectively. A large language model (LLM) first divides news, announcements, or reports into tokens, the discrete units of text it can process, and then learns relationships among words, contexts, and events from a large corpus. A time series foundation model (TSFM) learns dynamic structure across many numerical series and applies that structure to new forecasting tasks involving returns, prices, or other series \citep{woo_etal_2024,wu_he_cao_fan_liu_2025}. In financial applications, LLMs typically convert unstructured text into language outputs or numerical signals, whereas TSFMs directly generate forecasts of future numerical values from a numerical history. The two model classes differ in input format and in how their outputs are used, but both produce current outputs jointly from current inputs and pretrained parameters. Whenever pretrained parameters are reused for historical tasks, training information can therefore enter forecasts through those parameters, creating a temporal mismatch when the parameters are trained on data extending beyond the historical forecast origin.

Studying this mismatch requires dated training samples and repeatable forecasts under alternative parameter states. The annual financial TSFMs released by \citet{rahimikia_ni_wang_2025} provide this setting. Each annual state has a specified training-information cutoff, allowing the same historical forecast to be regenerated with another annual parameter state while holding the forecast unit, numerical history, inference rule, and output construction fixed. The resulting matched comparison measures the change in the financial forecast associated with replacing the trained forecasting rule.

This design is particularly clear in the TSFM setting. Financial forecasts from LLMs typically also depend on the prompt, retrieved context, sampling rule, and the mapping from language outputs to numerical forecasts or investment decisions. Even without prompts, representation-based designs usually include intermediate steps such as hidden-state aggregation, downstream classification, and signal aggregation. These steps jointly determine the final forecast, so changes in parameter information reach forecast outcomes through more intermediate stages. By contrast, the TSFMs we use generate numerical forecasts directly from numerical histories. Chronos discretizes scaled continuous numerical observations into a finite set of categories and autoregressively generates a probability distribution over future paths; TimesFM organizes adjacent numerical observations into continuous patches and directly generates future numerical values \citep{ansari_etal_2024,das_kong_sen_zhou_2024}. The two architectures use different numerical representations but share the same controlled parameter-replacement design.

\subsection{Historical Information Constraints}\label{the-information-set-of-a-historical-forecast}

Historical information integrity requires the forecast inputs and parameter-training data to respect the decision date. \(X_{\leq t}\) constitutes the input information set, recording the observations supplied at forecast origin \(t\); \(\mathcal D_{\leq\tau}\) constitutes the parameter information set, identifying the observations used to train the forecasting rule. The former determines what the model receives when forming the forecast, and the latter determines the data from which its parameters were learned. We call these requirements input-time integrity and parameter-time integrity, respectively.\footnote{The broader leakage literature defines the problem in terms of information unavailable for the prediction task \citep{kapoor_narayanan_2023}.} Our origin-aligned point-in-time (PIT) benchmark applies both requirements to the data under the annual matching convention.

When the historical input ends at \(t\) and the deployed parameter state is aligned with that date under our annual matching rule, \(\widehat y_{t,h}(\tau)\) is an origin-aligned point-in-time (PIT) forecast. If \(X_{\leq t}\) remains unchanged but the state is replaced by one whose training coverage extends beyond \(t\), so that \(\tau>t\), the forecast has temporal exposure. The two forecasts process the same historical series and address the same future target; their parameters differ in the periods covered by their training data.

Temporal exposure is a relationship between the model's training cutoff and the forecast origin. The same \(\theta_\tau\) is exposed when used for an earlier forecast and satisfies the training-data constraint at a later origin. We use this relationship to pair an origin-aligned PIT state with a state trained through a later date for the same historical forecast. The comparison fixes the numerical history and forecasting task and replaces the parameter state implementing the forecasting rule.

\subsection{Direct and Indirect Information Exposure}\label{direct-and-indirect-temporal-exposure}

We define post-origin information as data that become available only after forecast origin \(t\) but are included in parameter training. Such information constitutes temporal exposure only when the training coverage of the parameter state extends beyond the forecast origin, so that \(\tau>t\). The model has direct exposure when its training sample includes data from the target market after \(t\), and indirect exposure when information from other related markets, securities, financial factors, or common economic conditions enters training over the same period. Both channels can coexist. When the target market itself is excluded from training and only the latter information enters, we call this indirect-only exposure.

Indirect exposure arises because related financial series reflect common economic conditions contemporaneously.\footnote{International return predictability and the role of U.S. information are studied by \citet{rapach_strauss_zhou_2013}.} Financial crises, policy changes, and global risk conditions can affect different markets at the same time. Excluding the target market from training therefore does not ensure that model parameters are fully separated from information about that market during the test period. \citet{meyer_kaltenpoth_zalipski_mueller_2025} illustrate this channel using temporally overlapping correlated series: even when a model never observes the target series itself, related training series may carry information about the target period through common drivers. This mechanism shows how information can enter the parameters through related series.

\begin{figure}[p]
\centering
\includegraphics[width=0.95\textwidth]{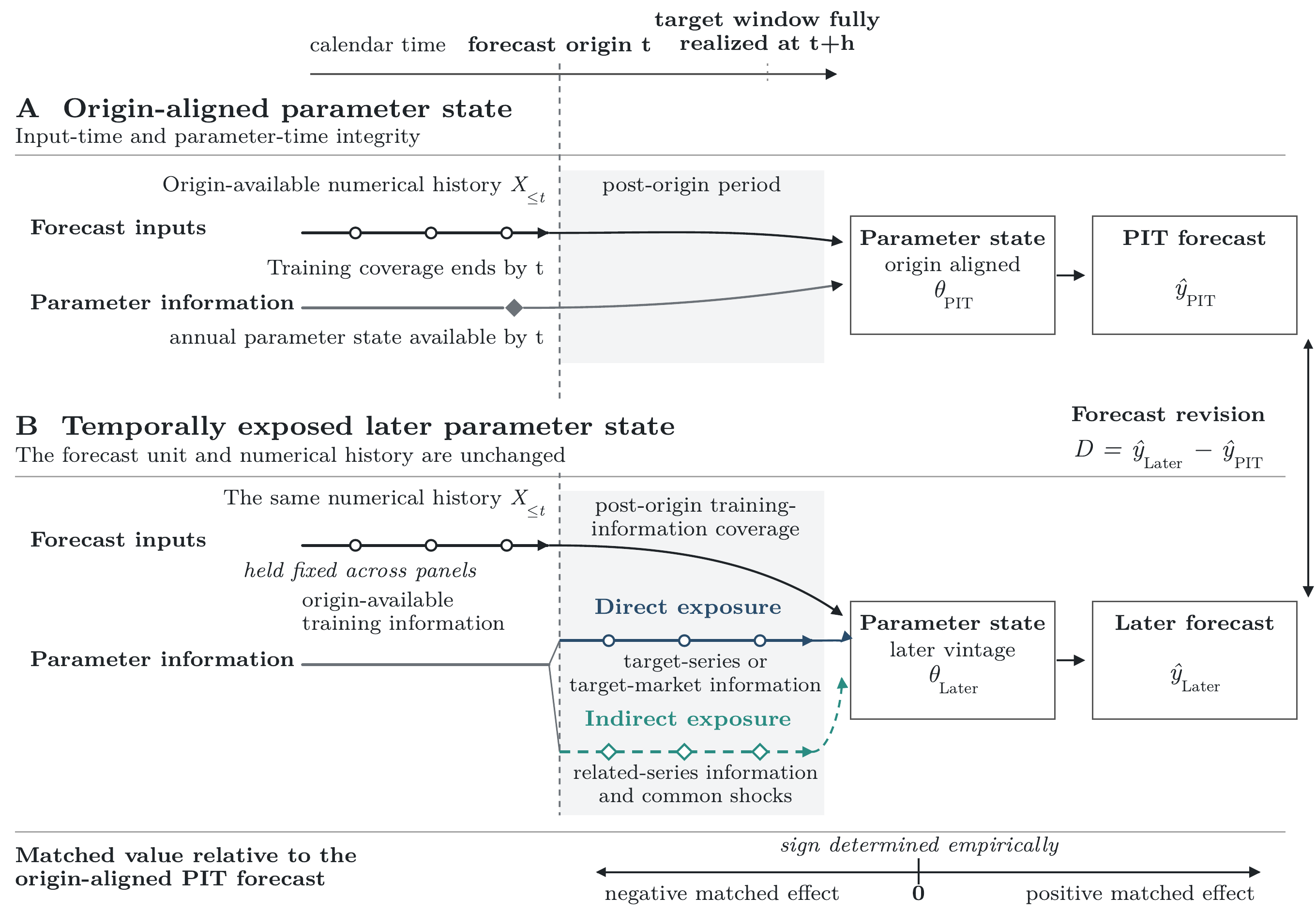}
\caption[Parameter Information and Temporal Exposure]{%
  \textbf{Parameter information and temporal exposure}.
  Both panels describe the same historical forecast, holding the target market, forecast origin $t$, horizon $h$, and numerical history $X_{\leq t}$ fixed. Panel A uses an origin-aligned point-in-time (PIT) annual state with training coverage ending no later than $t$. Panel B substitutes a later state with training coverage beyond $t$. Direct exposure refers to post-origin training information from the target market; indirect exposure refers to post-origin information from related markets, securities, factors, or common economic conditions. Forecast revision is the later-state forecast minus the PIT forecast.
}
\label{fig:temporal-exposure}
\end{figure}

\bigskip\centerline{[Insert Figure~\ref{fig:temporal-exposure} about here]}\bigskip

Figure~\ref{fig:temporal-exposure} places parameter time and exposure channels within the same historical forecast. Panel A uses an origin-aligned PIT state, with the numerical history ending at forecast origin \(t\) and training-information coverage ending no later than \(t\). Panel B holds the forecast target, horizon, and \(X_{\leq t}\) fixed but replaces the PIT state with a later state whose training coverage extends beyond \(t\). The direct and indirect paths represent post-origin information entering the parameters from the target market and related economic series, respectively. The difference between the outputs of the two states constitutes the forecast revision. The figure shows how temporal exposure reaches model output.

The evaluation follows the information from the forecasting rule to the investor's portfolio. Training dates establish temporal exposure; paired outputs measure forecast revision; realized returns determine predictive value; and the positions generated by those forecasts determine investor value. Evidence from language models similarly distinguishes memorization of training data from its exploitation in downstream tasks \citep{magar_schwartz_2022}. 

\section{Data and Empirical Design}\label{markets-annual-tsfm-states-and-matched-design}

Annual TSFM states allow the information embedded in a forecasting rule to vary along two dimensions. The annual-state year determines the rule's position in time relative to the forecast origin, while the training-information environment determines the markets and common financial series from which it can learn. Separating timing from content allows us to regenerate the same historical forecast using parameter states from different dates, without conflating when information enters the parameters with where it comes from.

\subsection{Market Returns and Forecast Horizons}\label{forecast-targets-and-point-in-time-histories}

The forecast targets are the U.S. equity premium and monthly excess returns for 13 non-U.S. equity markets.\footnote{The non-U.S. markets are Austria, Belgium, Canada, Denmark, Finland, France, Germany, Italy, the Netherlands, New Zealand, Spain, Switzerland, and the United Kingdom.} The U.S. target is the CRSP value-weighted market return less the risk-free rate in the Goyal--Welch data \citep{welch_goyal_2008}. All non-U.S. targets are local-currency excess market returns, constructed as the value-weighted local-currency market return from the Kenneth R. French Data Library less the corresponding country-level short rate from FRED.

We form forecasts at each month-end \(t\) and express forecasts at all horizons as average monthly excess returns. Let \(r^e_{m,t+j}\) denote the excess return realized in market \(m\) in the \(j\)th month after the forecast origin, with forecast horizon \(h\in\{1,3,6,12\}\). For the market--origin--horizon forecast unit \(u=(m,t,h)\), the target is

\[
u=(m,t,h),
\qquad
y_u
\equiv
y_{m,t,h}
=
\frac{1}{h}
\sum_{j=1}^{h}r^e_{m,t+j}.
\tag{2}
\]

Here, \(t\) marks the forecast origin and \(h\) the length of the future target window. Using the average excess return over the next \(h\) months keeps forecasts and realized targets on the same monthly scale across horizons. It also allows us to evaluate the outputs of both TSFM families using the same forecast-loss measure and investment rule.\footnote{The relation between point-forecast targets and evaluation loss functions is discussed by \citet{gneiting_2011}.}

Each forecast uses only the return history realized by the forecast origin. The numerical history \(X_{m,\leq t}\) for market \(m\) begins in January 1994 and expands month by month as the forecast origin advances; out-of-sample target-start months begin in January 2001. The real-time benchmark is the expanding historical-average forecast, denoted by \(\widehat y_u^{\mathrm{HA}}\), which uses only monthly returns realized by \(t\). A given forecast unit shares the same realized target, historical input, and historical-average benchmark regardless of the annual state with which it is paired.

We make matched comparisons holding the model variant \(a\) and training-information environment \(c\) fixed. Once \(u=(m,t,h)\) is defined, the market, forecast month, horizon, and numerical history determined by \(m\) and \(t\) are fixed, so we need not repeat these conditions in every subsequent equation. Replacing the annual parameter state changes only the forecasting rule applied to the same history, linking the difference between the two forecasts to parameter time.

\subsection{Annual Model Vintages}\label{tsfm-families-and-independently-trained-annual-states}

The annual parameter states come from five finance-native TSFM variants released by \citet{rahimikia_ni_wang_2025}: Chronos Tiny, Chronos Mini, Chronos Small, TimesFM 8M, and TimesFM 20M. The original study trained each annual state independently on financial numerical series. We deploy these parameters directly, without further training or model updates for our market targets. They therefore differ both from zero-shot TSFMs that directly use general-purpose pretrained weights and from models that undergo finance-specific fine-tuning starting from general-purpose weights. Each annual state is a separately formed parameter realization, rather than a model version saved successively along the same optimization path.

Chronos and TimesFM provide different forms of native forecasts, so we map their outputs to the same economic forecast target. Chronos tokenizes scaled continuous numerical observations and uses an autoregressive Transformer to generate probabilistic forecast paths \citep{ansari_etal_2024}. For each forecast unit, we generate \(S\) paths, first averaging the next \(h\) forecast steps within each path and then averaging across paths. TimesFM uses continuous patches and directly generates numerical forecasts for future steps \citep{das_kong_sen_zhou_2024}; we average its native mean-step forecasts over the forecast horizon. Both constructions yield a point forecast of average monthly excess return and can therefore be evaluated against the same \(y_u\). Panel A of Table~\ref{tab:model-design} summarizes the model families, native forecast objects, and point-forecast construction. Internet Appendix Section~\ref{sec:ia-model-forecast-construction} gives the model-specific path and quantile construction, and Section~\ref{sec:ia-parameter-registry} records the parameter-state identifiers and cutoff conventions. Model architecture and training details are given in \citet{rahimikia_ni_wang_2025} and the corresponding architecture papers.

The models' pretraining task and our forecasting task differ in frequency and level of aggregation. Annual states learn from daily firm-level excess-return series, whereas we deploy these parameters on monthly aggregate market-return histories. The models thus transfer dynamic relationships learned from daily firm-level series to monthly market-return forecasting. This bridge across frequencies and aggregation levels is the same for all annual states, so changing the annual-state year does not also change the downstream forecasting task.

Let \(v\) denote the annual state's training-data cutoff year, \(a\) the model variant, and \(c\) the training-information environment. We write the general parameter state \(\theta_\tau\) from Section~\ref{pretraining-historical-information-sets-and-temporal-exposure} more specifically as \(\theta_{a,c,v}\). For forecast unit \(u\), the forecast and its error are

\[
\widehat y_u(v)
=
f_{\theta_{a,c,v}}
\!\left(X_{m,\leq t};h\right),
\qquad
e_u(v)
=
y_u-\widehat y_u(v).
\tag{3}
\]

For given \(a\) and \(c\), we use the shorthand \(\widehat y_u(v)\) and \(e_u(v)\). Different annual states process the same \(X_{m,\leq t}\) and forecast the same \(y_u\), but form their forecasts using parameter states with different training-data cutoff years.

\subsection{Training Markets and Factor Information}\label{training-information-environments-and-exposure-channels}

The three training-information environments change the information an annual state can absorb while holding its date fixed. The US environment contains only U.S. firm-level excess returns. The Global environment adds non-U.S. firm-level excess returns to this return corpus. The Augmented (JKP-augmented) environment further adds the JKP factor series of \citet{jensen_kelly_pedersen_2023} to the Global corpus. Moving from US to Global to Augmented expands training content without changing parameter time.

The same training-information environment implies different exposure channels for different target markets. For the U.S. target, all channel assignments require the annual state's training cutoff to extend beyond the forecast origin. The US environment contains post-origin firm returns from the target market, so Panel B of Table~\ref{tab:model-design} labels this cell Direct. The Global environment contains firm returns from both the United States and other countries, yielding direct plus indirect exposure. The Augmented environment retains both channels and adds an expanded factor-information channel.

For a non-U.S. target, the US environment provides an indirect-only setting. The target market itself is excluded from training, but U.S. firm returns may still reflect contemporaneous global financial conditions. The Global environment includes both the target market and other markets in training, yielding direct plus indirect exposure; the Augmented environment adds a factor-information channel. Panel B specifies this target--environment map: comparisons across environments change training scope, comparisons across target groups change exposure channels, and timing comparisons within each cell replace only the annual state. Internet Appendix Section~\ref{sec:ia-channel-assignment} details the training corpora and the assignment of target-market, cross-market, and factor information to each comparison.

\subsection{Matched Comparisons of Model Vintages}\label{parameter-vintage-conventions-and-timing-designs}

Each annual state is indexed by the last calendar year covered by its training sample. A year-\(v\) state uses training data through the end of year \(v\) and, under our timing convention, is treated as origin-admissible from the beginning of year \(v+1\). Let \(Y(t)\) denote the calendar year of the target-start month. We define the annual-state lead relative to that year as

\[
\ell=v-Y(t).
\]

The rolling design varies the annual state around a fixed historical forecast. Lead \(-1\) corresponds to \(v=Y(t)-1\) and is the origin-aligned PIT state. Lead \(-2\) is one year older than the PIT state; its training cutoff satisfies the historical information constraint, and it serves as the stale-state benchmark. Lead \(0\) uses the annual state for the target-start year. Its training coverage extends into that year, making it a boundary-exposed state. Leads \(+1\) and \(+2\) move the training cutoff further into the post-origin period. Each alternative lead is paired with the lead \(-1\) PIT forecast on common target months. Pooled later-state evidence for leads \(0,+1,+2\) uses strict common support.

The fixed-vintage design reverses the direction of temporal variation. It holds one annual state fixed while forecast origins and target windows move across its training cutoff. We use the 2000, 2009, and 2023 states. When the entire target window precedes the state cutoff, the forecast lies in a retrospective-exposure window. A cutoff within the target window creates a boundary crossing. When the target window begins after the cutoff and the target-start year is \(v+1\), the state is the origin-aligned PIT rule. The same state becomes a stale rule only when the target-start year is later than \(v+1\). Each fixed-state forecast is paired with the origin-aligned PIT forecast for the same \(u\). The rolling design thus examines local state replacement around a fixed forecast, whereas the fixed-vintage design traces the performance of the same forecasting rule at different points in history.

\bigskip\centerline{[Insert Table~\ref{tab:model-design} about here]}\bigskip

The adjacent-vintage comparison places crossing the forecast origin and ordinary annual updating on the same one-year scale. The clean forward step moves from lead \(-2\) to lead \(-1\), with added training data strictly preceding the forecast origin. The origin-crossing step moves from lead \(-1\) to lead \(0\), with added training coverage crossing the forecast origin. Let \(N_\ell\) denote the matched predictive effect of lead \(\ell\) relative to PIT: PIT loss minus alternative-state loss, computed on the maximal matched support shared by that lead and PIT and normalized to percentage points by the historical-average MSFE on the same support. The clean forward step is \(-N_{-2}\), and the origin-crossing step is \(N_0\); positive values in both cases indicate that the newer state lowers loss. The origin-crossing contrast is

\[
\Delta_{\mathrm{cross}}
=
N_0-\bigl(-N_{-2}\bigr)
=
N_0+N_{-2}.
\tag{4}
\]

Here, \(N_0\) measures the predictive value of the origin-crossing update, and \(-N_{-2}\) measures that of the equally long clean pre-origin update. Their difference measures whether training coverage crossing the forecast origin produces an additional change in predictive value relative to ordinary annual updating. Internet Appendix Section~\ref{sec:ia-timing-designs} formalizes the matching keys, lead-specific support, pooled intersections, and target-window classifications.

Figure~\ref{fig:s33-s34-timing-designs} places the three forms of parameter-time variation on a common timeline. Panel A holds the forecast unit fixed and moves the annual state through stale, PIT, and boundary-exposed states to deeper post-origin exposure. Panel B holds the annual state fixed and moves the forecast origin and target window relative to its cutoff. The adjacent-vintage comparison uses the consecutive leads \(-2\), \(-1\), and \(0\) in Panel A to place the clean annual update and the origin-crossing update on the same scale.

\bigskip\centerline{[Insert Figure~\ref{fig:s33-s34-timing-designs} about here]}\bigskip

\subsection{Forecast and Portfolio Evaluation}\label{outcomes-and-inference}

Our empirical analysis proceeds from the information set to the investor's decision. We measure the origin-aligned model's predictive performance, the revision generated by parameter replacement, the resulting change in forecast loss, and the change in portfolio utility. Every paired comparison uses target months jointly available for the forecasts involved; pooled comparisons use the intersection across all constituent forecasts.

A meaningful parameter-time comparison first requires assessing whether the PIT forecast itself contains predictive content. This test establishes a real-time forecasting benchmark for the subsequent analysis and gives changes induced by parameter-state replacement a clear interpretation. Write

\[
\widehat y_u^{\mathrm{PIT}}
=
\widehat y_u\!\left(Y(t)-1\right),
\qquad
e_u^{\mathrm{PIT}}
=
y_u-\widehat y_u^{\mathrm{PIT}},
\]

and let \(e_u^{\mathrm{HA}}=y_u-\widehat y_u^{\mathrm{HA}}\). The PIT forecast's raw MSFE reduction relative to the expanding historical-average forecast is\footnote{Comparing forecasts through loss differentials follows the forecast-evaluation framework of \citet{diebold_mariano_1995}.}

\[
R_{\mathrm{HA}}^2
=
100
\frac{
\sum_u
\left[
\left(e_u^{\mathrm{HA}}\right)^2
-
\left(e_u^{\mathrm{PIT}}\right)^2
\right]
}{
\sum_u
\left(e_u^{\mathrm{HA}}\right)^2
}.
\tag{5}
\]

Positive values indicate that the PIT forecast has lower raw MSFE than the historical-average forecast. We also report the Clark--West adjusted loss statistic and its associated one-sided predictive-content diagnostic \citep{clark_west_2007}.

Parameter-state replacement can affect a historical backtest only if it changes model output. For an alternative state \(v\), the matched forecast revision is defined as

\[
D_u(v)
=
\widehat y_u(v)
-
\widehat y_u^{\mathrm{PIT}}.
\tag{6}
\]

A nonzero \(D_u(v)\) measures the change in the forecast generated from the same historical input by replacing the annual parameter state. We normalize the absolute forecast revision by historical-average RMSE and examine the correlation between PIT and alternative forecasts and the share of forecasts whose signs change. For Chronos, we also use the empirical Wasserstein-1 distance to compare matched predictive distributions, assessing whether parameter time also changes distributional information beyond the forecast mean.

The value of a forecast revision depends on whether it brings the forecast closer to the realized target. For annual-state assignment \(v\), matched predictive value is defined as

\[
N(v)
=
100
\frac{
\sum_u
\left[
\left(e_u^{\mathrm{PIT}}\right)^2
-
e_u(v)^2
\right]
}{
\sum_u
\left(e_u^{\mathrm{HA}}\right)^2
}.
\tag{7}
\]

Positive values indicate that the alternative state has lower squared loss than PIT; negative values indicate that the origin-aligned forecast is more accurate. In rolling analyses, a fixed lead \(\ell\) assigns an annual state to each forecast unit through \(v_\ell(u)=Y(t_u)+\ell\). In fixed-vintage analyses, \(v\) identifies a fixed calendar-year state. All subsequent tables and figures of PIT-relative timing effects follow this sign convention.

We evaluate investor value through the asset allocations implied by the two types of forecasts \citep{kandel_stambaugh_1996,granger_pesaran_2000}. The headline specification uses \(h=1\): the PIT portfolio is formed from the lead \(-1\) forecast, and the exposed portfolio uses the equal-weight average forecast from leads \(0,+1,+2\) on strict common support. Holding the mean-variance allocation rule fixed, we measure the investor effect as the difference in annualized certainty-equivalent returns:\footnote{Learning and decision objectives are linked explicitly in portfolio optimization and predict-then-optimize research \citep{ban_elkaroui_lim_2018,elmachtoub_grigas_2022}.}

\[
\Delta CER
=
CER_{\mathrm{Exposed}}
-
CER_{\mathrm{PIT}},
\]

Positive values indicate that the exposed forecast raises the certainty-equivalent return; negative values indicate that the PIT forecast is more valuable to the investor. The headline implementation sets relative risk aversion to 3, restricts the market weight to \([0,1.5]\), and estimates variance from an expanding sample of observations preceding the target month. Internet Appendix Section~\ref{sec:ia-estimands-portfolios} defines the revision statistics, design-specific loss aggregation, portfolio weights, and complete-tranche implementation for longer investment horizons.

Statistical inference follows the temporal structure of each outcome. All loss-based estimates use target months shared by the forecasts being compared. Because multi-period forecast targets overlap, Bartlett--Newey--West HAC inference uses \(h-1\) lags and is supplemented by paired circular moving-block bootstrap intervals \citep{newey_west_1987,politis_romano_1992}. International equal-market estimates first normalize within each market by its historical-average loss scale and then average equally across markets, preventing high-volatility markets from receiving mechanically greater weight in the aggregate. Cross-market distribution summaries and the medians and breadth counts of investment results are computed as specified in the corresponding exhibits. Internet Appendix Section~\ref{sec:ia-support-inference} gives the sample intersections, ratio influence functions, bootstrap procedures, multiplicity rules, and international clustered inference.
\section{Return Predictability and Forecast Revisions}\label{point-in-time-predictive-relevance-and-forecast-revision}

Dated model states produce informative return forecasts and materially different assessments of future market conditions. We first establish predictive performance relative to the historical average and then measure how annual-state replacement changes forecast levels, market direction, and predictive distributions.

\subsection{Point-in-Time Return Predictability}\label{predictive-content-of-the-origin-aligned-forecast}

The expanding historical-average forecast provides a strictly real-time benchmark for the PIT forecast. It uses only returns realized by the forecast origin and is a standard out-of-sample benchmark in equity-premium forecasting \citep{welch_goyal_2008,campbell_thompson_2008,neely_rapach_tu_zhou_2014,goyal_welch_zafirov_2024}. \citet{li_li_lyu_yu_2025} study how predictive information can improve on this benchmark. Table~\ref{tab:pit-relevance} uses the US training-information environment and the all-available sample to compare the origin-aligned forecasts of the five TSFMs with the historical-average forecast. The first line of each cell reports raw \(R^2_{\mathrm{HA}}\), with positive values indicating lower MSFE for the PIT forecast; parentheses contain one-sided Clark--West diagnostic \(p\)-values computed using a normal approximation \citep{clark_west_2007}.

\bigskip\centerline{[Insert Table~\ref{tab:pit-relevance} about here]}\bigskip

The origin-aligned models extract predictive information from U.S. and international return histories. Raw MSFE reductions are particularly pronounced for Chronos forecasts. For U.S. Chronos Tiny, \(R^2_{\mathrm{HA}}\) rises from 3.45 at \(h=1\) to 15.12 at \(h=6\) and remains at 14.31 percentage points at \(h=12\). In the international aggregate, the 12-month values for Chronos Tiny, Mini, and Small are 12.82, 12.91, and 10.60 percentage points, respectively. The reported Clark--West diagnostic has a one-sided \(p\)-value below 0.05 in 15 of 20 model--horizon cells in each target group. Internet Appendix Figure~\ref{fig:x41-pit-relevance} displays the raw and adjusted comparisons with their uncertainty intervals. Tables~\ref{tab:ia-x41-us-corpora} and~\ref{tab:ia-x41-international-corpora} extend the benchmark evidence to all three training environments and, for the international sample, report each of the 13 markets separately. Table~\ref{tab:ia-x41-balanced} repeats the exercise on the common January 2002--December 2021 window.

\subsection{Revisions to Market Return Forecasts}\label{forecast-revision-under-temporal-exposure}

\begin{figure}[p]
\centering
\includegraphics[width=0.95\textwidth]{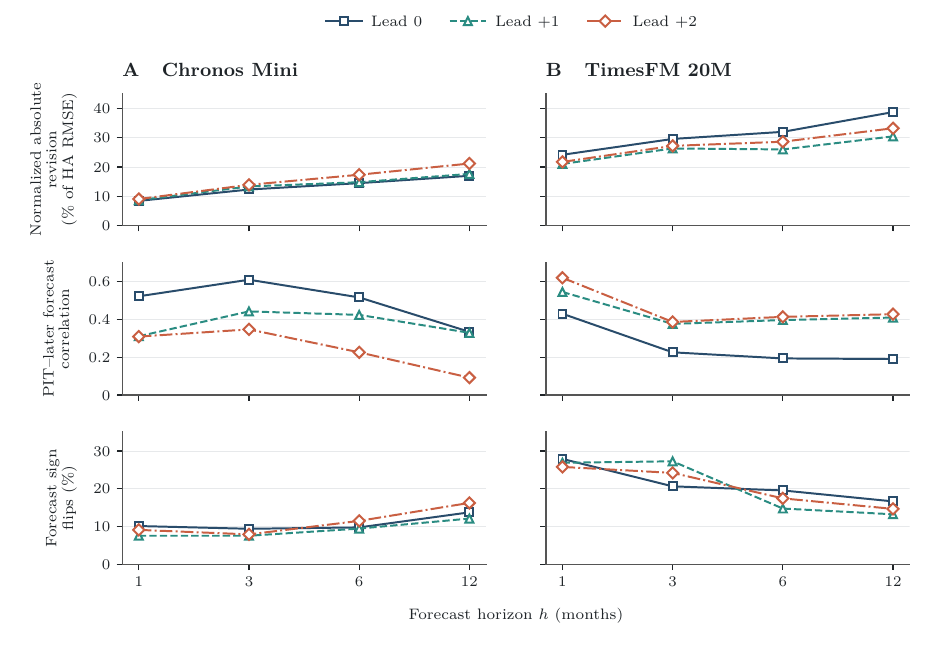}
\caption[Forecast Revisions across Rolling Annual States]{%
  \textbf{Forecast revisions across rolling annual states}.
  Forecast revisions compare later annual states with the origin-aligned point-in-time (PIT) state, $D_u(v)=\widehat y_u(v)-\widehat y_u^{\mathrm{PIT}}$. Panels A and B show Chronos Mini and TimesFM 20M. From top to bottom, the rows give mean absolute revision as a percentage of historical-average root mean squared error (RMSE) on the same paired support, the PIT--later forecast correlation, and the percentage of matched months with different forecast signs. Squares, triangles, and diamonds identify leads $0,+1,+2$, each compared with PIT lead $-1$, at horizons of 1, 3, 6, and 12 months. Historical inputs and forecast units remain fixed. The corresponding matched target-month counts are 276, 264, and 252. Estimates use the all-available U.S. equity-premium sample and the US training-information environment.
}
\label{fig:s42-forecast-revision}
\end{figure}

Replacing the annual parameter state materially revises this real-time forecast. For an alternative state \(v\), the matched revision is \(D_u(v)=\widehat y_u(v)-\widehat y_u^{\mathrm{PIT}}\), with the forecast unit and numerical history held fixed. We describe output movement along three complementary dimensions. Normalized absolute revision divides the mean absolute change by the historical-average RMSE on the same paired support. Forecast correlation measures comovement over time between PIT and later-state forecasts. The sign-change share records the fraction of months in which parameter replacement changes the predicted return direction. Figure~\ref{fig:s42-forecast-revision} focuses on the U.S. equity premium and the US training-information environment. It pairs each of the rolling leads \(0,+1,+2\) with PIT lead \(-1\) and reports results for Chronos Mini and TimesFM 20M at the four horizons.

\bigskip\centerline{[Insert Figure~\ref{fig:s42-forecast-revision} about here]}\bigskip

Forecast revisions are sizable relative to the error scale of the real-time benchmark. For Chronos Mini, normalized revisions at the one-month horizon range from 8.47\% to 9.10\% of historical-average RMSE. At the 12-month horizon, normalized revisions increase to between 17.01\% and 21.18\%. TimesFM 20M is more sensitive to annual-state replacement: its normalized revisions range from 21.05\% to 38.68\% across all its lead--horizon cells in Figure~\ref{fig:s42-forecast-revision}.

Output movement changes both the time-series content and the economic direction of forecasts. The lowest PIT--later forecast correlation in Figure~\ref{fig:s42-forecast-revision} is 0.092, for Chronos Mini at lead \(+2\) and \(h=12\). Within the same figure, the largest sign-change share is 27.9\%, for TimesFM 20M at lead \(0\) and \(h=1\); this annual-state replacement changes the forecast sign in more than one quarter of matched months. Annual states thus do more than rescale the same forecast: they can also change the model's assessment of future market direction.

Forecast revisions also occur beyond the models and rolling comparisons shown in Figure~\ref{fig:s42-forecast-revision}. Considering rolling leads \(0,+1,+2\) together with fixed 2009 and fixed 2023, median normalized revisions in the U.S. market range from 17.4\% to 21.4\% of benchmark RMSE across the five comparisons. The median sign-change shares for fixed 2009 and fixed 2023 are 26.6\% and 17.8\%, respectively. For international markets, we first take the median across 13 markets within each model--horizon cell and then the median across the 20 cells; the corresponding normalized revisions range from 12.5\% to 15.5\%. Parameter-state replacement therefore produces observable forecast movement in both rolling and fixed-vintage designs and for both U.S. and non-U.S. targets. Internet Appendix Table~\ref{tab:ia-x42-point-revision} reports normalized movement, forecast correlation, and sign changes for each model, horizon, annual-state comparison, and training environment. Figure~\ref{fig:ia-x42-revision-breadth} summarizes their medians and interquartile ranges for U.S. and international targets, showing how the revisions vary across training samples and timing designs.

Chronos predictive distributions also change with the annual state. For Chronos Mini in the U.S. equity-premium sample, the empirical Wasserstein-1 distance ranges from 25.31 to 44.47 monthly-return basis points for rolling leads \(0,+1,+2\), and from 27.99 to 64.89 basis points for the fixed 2009 and fixed 2023 comparisons. Both the HAC and circular-block-bootstrap 95\% intervals lie above zero in these cells. Internet Appendix Figure~\ref{fig:x42-chronos-distribution-movement} displays the rolling comparisons, and Table~\ref{tab:ia-x42-chronos-distribution-movement} reports all three Chronos variants under rolling and retrospective fixed-state exposure. Figure~\ref{fig:ia-x42-native-breadth} extends the distributional evidence across training environments and target markets using Chronos quantile-grid movement and TimesFM native-quantile displacement. These exhibits establish the breadth of the changes in point and distributional forecasts. The next section evaluates their consequences for forecast accuracy and investment decisions.
\section{Predictive Performance and Asset Allocation}\label{the-predictive-and-investor-value-of-parameter-time-exposure}

In the US training-information environment, later model vintages generate predominantly unfavorable revisions to market-return forecasts. The loss in predictive value appears when the training cutoff crosses the forecast origin, extends across rolling and fixed-vintage comparisons, and translates into lower median investor value under the constrained allocation rule. The analysis follows these results from forecast errors to portfolio returns.

\subsection{Forecast Accuracy at the Information Boundary}\label{crossing-the-forecast-origin}

The decline in predictive value is concentrated in the first annual update that crosses the forecast origin, rather than in the preceding pre-origin update. Each adjacent-vintage comparison advances parameter time by one year. The clean forward step moves from lead \(-2\) to origin-aligned lead \(-1\), with added training information strictly before the forecast origin. The origin-crossing step moves from lead \(-1\) to lead \(0\), with added training coverage entering the target-start year. Table~\ref{tab:origin-crossing} evaluates each step on its maximal matched support with the PIT forecast and expresses both as moves forward in parameter time. Positive values indicate that the newer annual state lowers forecast loss.

\bigskip\centerline{[Insert Table~\ref{tab:origin-crossing} about here]}\bigskip

The mean predictive effect of the pre-origin forward step is \(+0.72\) historical-average MSFE percentage points in the United States and \(+0.85\) points across the 13 non-U.S. markets. The update that brings training coverage into the target-start year lowers matched predictive value by \(5.65\) and \(5.71\) points, respectively. The decline is therefore concentrated in the annual-state replacement that first crosses the forecast origin.

Relative to the pre-origin step, the origin-crossing contrast is \(-6.37\) percentage points in the United States and \(-6.56\) points internationally. All 13 non-U.S. market-level differences are negative, and the international contrast has a market-clustered \(t\)-statistic of \(-7.08\). Moving the cutoff across the origin is thus associated with a larger deterioration in average predictive performance than the preceding one-year update.

\subsection{Forecast Accuracy with Rolling Model Vintages}\label{the-rolling-path-beyond-the-forecast-origin}

The rolling evidence shows that later annual states can lose incremental value relative to the origin-aligned rule even while retaining predictive content relative to the historical average. Within each model--horizon cell, Table~\ref{tab:x43-lead-performance} reevaluates the PIT forecast and leads \(0,+1,+2\) on the same 252 target-start months and reports each forecast's raw out-of-sample \(R^2_{\mathrm{HA}}\) relative to the expanding historical-average forecast. Common support makes the rows within a cell directly comparable. The one-sided Clark--West \(p\)-values in parentheses report the adjusted-loss diagnostic relative to the historical-average benchmark.

\bigskip\centerline{[Insert Table~\ref{tab:x43-lead-performance} about here]}\bigskip

Later annual states retain absolute predictive content in many cells, especially within the Chronos family, but their raw performance is usually weaker than PIT. At \(h=6\), for example, Chronos Tiny's \(R^2_{\mathrm{HA}}\) falls from 17.44 for PIT to 12.77, 4.86, and 6.91 for leads \(0,+1,+2\), respectively. For TimesFM 20M, it falls from 11.69 to \(-3.79\), \(-22.62\), and \(-8.25\). Across the 20 model--horizon cells, raw \(R^2_{\mathrm{HA}}\) is below PIT in 17, 17, and 19 cells for leads \(0,+1,+2\), respectively. A later-state forecast can therefore continue to outperform the historical average while losing incremental value relative to the origin-aligned forecasting rule.

\begin{landscape}
\begin{figure}[p]
\centering
\includegraphics[width=0.95\linewidth]{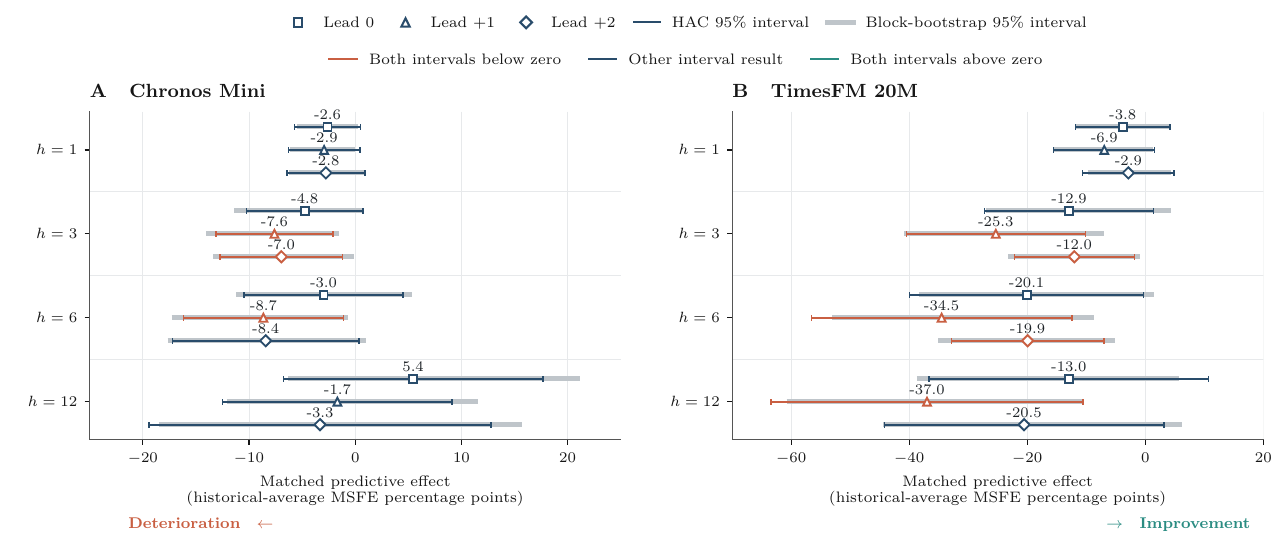}
\caption[Matched Predictive Effects of Rolling Annual States]{%
  \textbf{Matched predictive effects of rolling annual states}.
  Matched predictive effects are PIT mean squared forecast error (MSFE) minus later-state MSFE, scaled by historical-average MSFE on the same paired support and expressed in percentage points. Positive values favor the later state over the origin-aligned point-in-time (PIT) forecast. Panels A and B show Chronos Mini and TimesFM 20M on model-specific horizontal scales. Squares, triangles, and diamonds identify leads $0,+1,+2$. Thin lines give 95\% Bartlett--Newey--West HAC intervals with $h-1$ lags; thick light-gray lines give 95\% percentile intervals from 2,000 circular moving-block bootstrap replications with block length $\max(h,6)$. Terracotta indicates that both intervals lie below zero, teal that both lie above zero, and navy all other cases. Each lead uses its maximal matched support with PIT. The horizon $h$ is in months. Estimates use the all-available U.S. equity-premium sample and the US training-information environment.
}\label{fig:x43-matched-effects}

\end{figure}
\end{landscape}

Figure~\ref{fig:x43-matched-effects} expresses the same comparison as direct PIT-relative effects and reports estimation uncertainty for Chronos Mini and TimesFM 20M. Each of the three exposed leads is evaluated on its maximal matched support with the PIT forecast.

\bigskip\centerline{[Insert Figure~\ref{fig:x43-matched-effects} about here]}\bigskip

Matched effects are predominantly negative. Of the 24 lead--model--horizon estimates in the figure, 8 have both types of confidence intervals below zero, and none have both types above zero. Pooling leads \(0,+1,+2\) on their strict common support yields negative effects in 18 of the 20 model--horizon cells, with a median of \(-5.94\) historical-average MSFE percentage points. Each lead-specific estimate uses its pairwise support with the PIT forecast, whereas the pooled estimate uses strict common support for the three exposed leads and PIT.

Internet Appendix Table~\ref{tab:x43-pooled-effects} reports the full pooled model--horizon estimates. Table~\ref{tab:ia-x43-complete-inference} gives lead-specific coefficients, standard errors, one- and two-sided tests, and both interval procedures for all five models, including the pre-origin comparison. Figure~\ref{fig:ia-x43-all-model-effects} displays the complete rolling effect grid.

International markets show the same overall ranking. In the US training-information environment, pooled rolling effects are negative in all 20 model--horizon cells after taking the median across the 13 markets. Internet Appendix Tables~\ref{tab:ia-x43-us-corpora} and~\ref{tab:ia-x43-international-corpora} report the U.S. and equal-market rolling forecasts under every training environment. Table~\ref{tab:ia-x43-balanced} repeats the rolling comparison on the 240-month January 2002--December 2021 window. Figures~\ref{fig:ia-x47-chronos-profiles} and~\ref{fig:ia-x47-timesfm-profiles} place forecast movement alongside matched predictive value on this common support, including the stale lead. Together, these exhibits connect the size of the revision to the loss in predictive value within each model family.

\subsection{Forecast Accuracy with Fixed Model Vintages}\label{a-common-parameter-state-across-historical-origins}

The fixed-vintage evidence reaches the same conclusion while holding the forecasting rule fixed. Table~\ref{tab:x44-fixed-effects} reports PIT-relative matched effects for the 2000, 2009, and 2023 annual states across five models and four horizons. Each cell pairs the fixed-state forecast with the origin-aligned PIT forecast on the same target months. The full deployment sample covers retrospective exposure, boundary crossing, and post-cutoff use of the fixed rule.

\bigskip\centerline{[Insert Table~\ref{tab:x44-fixed-effects} about here]}\bigskip

Negative effects dominate the fixed-state grid. Of the 60 state--model--horizon estimates, 52 are negative, with an overall median of \(-9.29\) historical-average MSFE percentage points. The median effects for the 2000, 2009, and 2023 states are \(-15.10\), \(-3.99\), and \(-8.69\), respectively. This sequence first improves and then deteriorates, without producing a recency ranking as the annual-state year advances. Among the negative estimates, 23 are significant at the 5\% level; no positive estimate reaches the same threshold.

Internet Appendix Table~\ref{tab:ia-x44-us-inference} reports fixed-state confidence intervals and multiplicity-adjusted tests of improvement. Tables~\ref{tab:ia-x44-us-corpora} and~\ref{tab:ia-x44-international-corpora} extend the fixed-state comparisons to Global and Augmented training, and Table~\ref{tab:ia-x44-balanced} evaluates U.S. and international performance over the common 240-month window.

\begin{landscape}
\begin{figure}[p]
\centering
\includegraphics[width=0.95\linewidth]{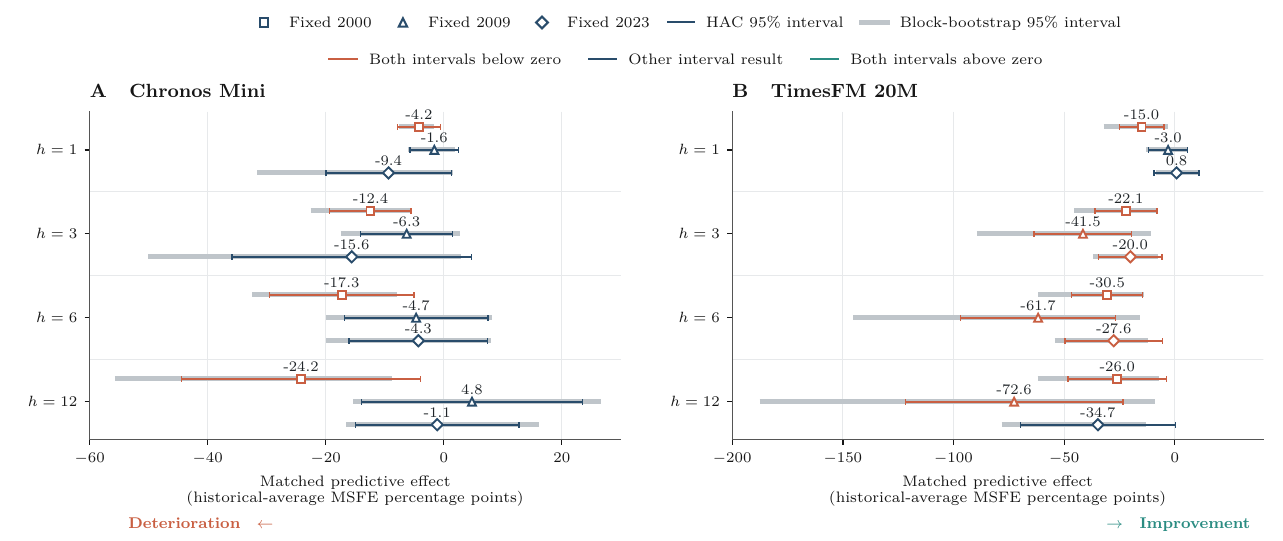}
\caption[Matched Predictive Effects of Fixed Annual States]{%
  \textbf{Matched predictive effects of fixed annual states}.
  Fixed-state effects are the reduction in mean squared forecast error (MSFE) relative to origin-aligned point-in-time (PIT) forecasts, scaled by historical-average MSFE on the same paired support and expressed in percentage points. Positive values favor the fixed state. Panels A and B show Chronos Mini and TimesFM 20M on model-specific horizontal scales. Squares, triangles, and diamonds identify the 2000, 2009, and 2023 states. Thin lines give 95\% Bartlett--Newey--West HAC intervals with $h-1$ lags; thick light-gray lines give 95\% percentile intervals from 2,000 circular moving-block bootstrap replications with block length $\max(h,6)$. Terracotta indicates that both intervals lie below zero, teal that both lie above zero, and navy all other cases. Each state uses its full available matched deployment support with PIT. The horizon $h$ is in months. Estimates use the all-available U.S. equity-premium sample and the US training-information environment.
}\label{fig:x44-fixed-effects}

\end{figure}
\end{landscape}

Figure~\ref{fig:x44-fixed-effects} compares effect magnitudes and estimation uncertainty for Chronos Mini and TimesFM 20M across the 2000, 2009, and 2023 annual states and four forecast horizons.

\bigskip\centerline{[Insert Figure~\ref{fig:x44-fixed-effects} about here]}\bigskip

The fixed-state comparisons reveal how a forecasting rule's value varies with its position in historical time. Internet Appendix Figure~\ref{fig:ia-x44-all-model-alignment} groups effects by the distance between the fixed annual state and the origin-aligned PIT year. Table~\ref{tab:ia-x44-timing-states} separates retrospective exposure, boundary crossing, and post-cutoff deployment. For the 2009 state, the median effect is \(-7.37\) in the retrospective-exposure window and \(-1.53\) in the post-cutoff window. The fixed-rule results therefore link investment-relevant forecast accuracy to the timing of the target window and the training sample.

\subsection{Predictive Performance across Timing Designs}\label{persistence-across-timing-designs}

The rolling and fixed-vintage designs provide complementary evidence on the performance of later states. The rolling design replaces the state around each forecast origin; the fixed-vintage design applies one state across different origins. Both pair the alternative forecast with PIT for the same forecast unit. Table~\ref{tab:cross-design} places the pooled rolling effect alongside the fixed-2009 and fixed-2023 retrospective effects. Three positive effects define a common gain, and three negative effects define a common loss. Opposite fixed-state signs define a fixed-vintage mixed pattern; a common fixed-state sign opposite to the rolling sign defines a protocol disagreement.

\bigskip\centerline{[Insert Table~\ref{tab:cross-design} about here]}\bigskip

Common losses dominate under U.S. training. Twelve of the 20 U.S. model--horizon cells have negative effects in all three designs, and none have three positive effects. The international evidence in Internet Appendix Table~\ref{tab:ia-x45-international-patterns} yields 167 common losses and 15 common gains across 260 market--model--horizon cells. The concentration of losses across two different timing designs reinforces the finding that substituting later states generally reduces predictive value in the reference training environment.

\subsection{The Economic Value of Market Timing}\label{investor-value}

The constrained allocation exercise translates these forecast revisions into lower median investor value. Investors adjust market positions according to forecast signs and magnitudes, so the economic comparison evaluates the portfolio returns generated by each forecasting rule \citep{kandel_stambaugh_1996,granger_pesaran_2000}. Recent learning-based portfolio methods likewise evaluate models through investment objectives \citep{simon_weibels_zimmermann_2026,guijarro_pelger_zanotti_2025,feng_jiang_li_song_wang_2026}.

Table~\ref{tab:investor-value} applies the one-month allocation rule in Section~\ref{outcomes-and-inference} to matched PIT and rolling-exposed forecasts.\footnote{Related portfolio evaluations study the economic value of return and volatility forecasts \citep{marquering_verbeek_2004} and the role of constraints in portfolio estimation \citep{demiguel_garlappi_nogales_uppal_2009}.}

\bigskip\centerline{[Insert Table~\ref{tab:investor-value} about here]}\bigskip

The paired median \(\Delta CER\) across the five U.S. models is \(-1.77\) percentage points per year. Internationally, the median across 65 market--model cells is \(-2.14\) points, and the cross-market medians for all five models are negative. Fifty-six of the 65 international point estimates are below zero. Internet Appendix Figure~\ref{fig:ia-x46-market-distribution} displays the individual U.S. estimates and the international market distributions, linking the summary CER differences to their breadth across markets.

The constrained strategy produces negative median CER differences at each of the three reported risk-aversion levels, \(\gamma\in\{1,3,5\}\), in both target groups (Internet Appendix Table~\ref{tab:ia-x46-gamma-grid}, Panel A). The complete-tranche portfolios in Table~\ref{tab:ia-x46-longer-horizons} extend the comparison to investment horizons of three, six, and twelve months. Their U.S. CER differences are negative in 14 of 15 model--horizon cells, and the international cross-market medians are negative in all 15 cells. The realized-return samples contain 250, 247, and 241 months, respectively, after all staggered tranches become active.

The predictive and portfolio comparisons give a consistent economic interpretation of the reference specification. Later model states change expected-return signals and the market positions formed from them, while the origin-aligned forecasts deliver lower forecast loss and higher median utility under the constrained investment rule. The next section explains these outcomes through the training information and the error-correcting content of forecast revisions.
\section{Training Information and Predictive Value}\label{training-information-and-revision-to-value-accounting}

Training information shapes both the size of a forecast revision and its value for return prediction. We first compare the training samples underlying the origin-aligned forecasts, then examine how international returns and factor information change the effects of later-state substitution. An exact loss decomposition connects those changes to the errors remaining in the PIT forecast.

\subsection{Point-in-Time Predictability across Training Samples}\label{the-point-in-time-benchmark-across-training-environments}

The US training environment provides the strongest origin-aligned benchmark in most model--horizon cells. It yields the highest raw \(R_{\mathrm{HA}}^2\) in 16 of 20 U.S. cells and 17 of 20 cells in the equal-market international aggregate. Internet Appendix Tables~\ref{tab:ia-x41-us-corpora} and~\ref{tab:ia-x41-international-corpora} report these comparisons by model, horizon, and training sample. Figure~\ref{fig:x52-absolute-training-performance} displays the corresponding PIT-to-historical-average MSFE ratios for Chronos Mini and TimesFM 20M, making the differences in baseline forecast quality visible on a common relative-loss scale.

We use US training as the reference environment and measure how Global and Augmented training change the later-minus-PIT effect within each environment. Baseline performance measures the quality of the origin-aligned rule; the matched effect measures the incremental value of substituting another annual state. Keeping both quantities explicit connects training-sample comparisons to the investor's choice of forecasting rule.

\subsection{Training Content and Look-Ahead Effects}\label{timing-effects-across-targettraining-environments}

\begin{figure}[p]
\centering
\includegraphics[width=0.95\textwidth]{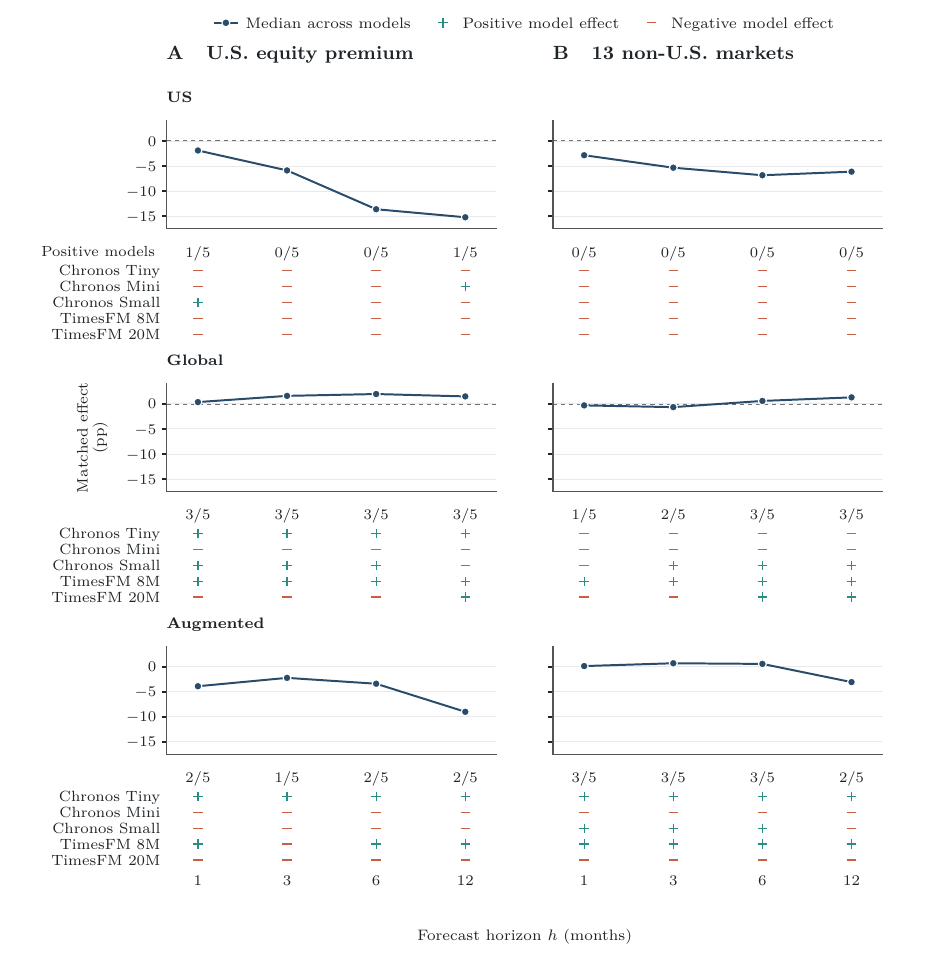}
\caption[Matched Predictive Effects across Training-Information Environments]{%
  \textbf{Matched predictive effects across training-information environments}.
  Horizon profiles summarize pooled rolling effects across the US, Global, and Augmented training-information environments. Columns A and B correspond to the U.S. equity premium and the 13 non-U.S. markets; rows correspond to the three environments. Pooled effects equally average the PIT-relative loss advantages of leads $0,+1,+2$ on strict common support, where PIT denotes the origin-aligned point-in-time state. At each horizon $h=1,3,6,12$ months, the upper profile connects the median effect across the five model variants. International effects first take the median across markets within each model--horizon cell, then across models at that horizon. The lower matrices show each model's effect sign: a teal $+$ denotes a positive effect and a terracotta $-$ a negative effect. Counts indicate positive model effects out of five. Values are historical-average mean squared forecast error percentage points; positive values favor the later-state comparison. All six profiles share a vertical scale.
}
\label{fig:training-environments}
\end{figure}

Training content materially changes the distribution of outcomes from parameter-state replacement across the six target--training cells. Following the target--training map in Table~\ref{tab:model-design} Panel B, Figure~\ref{fig:training-environments} shows the horizon profiles for the U.S. target and the 13 non-U.S. targets in the US, Global, and Augmented environments. The upper profile in each panel connects the median pooled rolling effect across the five models at \(h=1,3,6,12\). The lower matrix shows the direction of each model's effect at each horizon and reports the number of models with positive effects at that horizon. For the international results, we first take the median across the 13 markets within each model--horizon cell and then summarize across the five models at the same horizon.

\bigskip\centerline{[Insert Figure~\ref{fig:training-environments} about here]}\bigskip

For the U.S. target, global return information improves the median matched predictive effect, while factor augmentation changes that improvement. In the US training-information environment, the pooled rolling median effect is \(-5.94\) historical-average MSFE percentage points, with only 2 of the 20 model--horizon cells positive. The Global environment retains direct exposure to the target market while adding indirect information from other markets. The median effect rises to \(+1.37\), and the number of positive cells increases to 12. When the Augmented environment adds JKP factor information, the median effect falls back to \(-3.64\), with only 7 positive cells. The change from US to Global training improves the balance of revisions across the model--horizon grid, while the Augmented results reflect a different combination of return and factor information.

The role of target-market coverage is clearer for non-U.S. targets. The US environment provides indirect-only exposure in these markets, with a pooled median effect of \(-5.19\) and none of the 20 model--horizon cells positive. The Global environment includes the target market itself in training, expanding the exposure channel to direct plus indirect exposure. The median effect moves to \(-0.16\), and the number of positive cells increases to 9. The Augmented environment has a median effect of \(+0.35\), with 11 positive cells. The Global environment adds both target-market information and broader cross-market information. Relative to the indirect-only US environment, this expansion of training information shifts the matched effects markedly toward zero. Parameter timing determines whether an annual state is exposed; training content determines what forecast revision the state can produce; and the alignment of the revision with the target ultimately determines its value.

\subsection{Forecast Revisions and Error Correction}\label{loss-accounting-for-forecast-revisions}

Matched predictive value admits an exact decomposition into the error-correcting component of a revision and its movement cost. Let \(e_u^{\mathrm{PIT}}\) denote the error of the origin-aligned forecast and \(D_u(v)\) the forecast revision from annual state \(v\) relative to PIT. Since the alternative-state error satisfies \(e_u(v)=e_u^{\mathrm{PIT}}-D_u(v)\), the squared-loss difference between PIT and the alternative state is

\[
\left(e_u^{\mathrm{PIT}}\right)^2-e_u(v)^2
=
2e_u^{\mathrm{PIT}}D_u(v)-D_u(v)^2.
\tag{8}
\]

The first term measures whether the revision moves in the direction of the PIT error, while the second penalizes the distance the forecast moves. This identity distinguishes a change in the forecast from an improvement in it: a revision generates a benefit only when it corrects the original error, whereas every revision incurs a quadratic movement cost regardless of its direction.

We normalize both components by historical-average MSFE on the same matched support:

\[
A(v)
=
100\frac{
2\overline{e_u^{\mathrm{PIT}}D_u(v)}
}{
\overline{(e_u^{\mathrm{HA}})^2}
},
\qquad
P(v)
=
100\frac{
\overline{D_u(v)^2}
}{
\overline{(e_u^{\mathrm{HA}})^2}
},
\qquad
N(v)=A(v)-P(v).
\tag{9}
\]

The alignment benefit \(A(v)\) increases when the revision points toward the realized target; the movement penalty \(P(v)\) depends only on the squared magnitude of the revision; and net matched predictive value \(N(v)\) equals the matched predictive effect used in Section~\ref{the-predictive-and-investor-value-of-parameter-time-exposure}. A larger revision increases both the potential error-correction benefit and the movement penalty. Their balance determines the predictive value of the alternative annual state.

\begin{figure}[p]
\centering
\includegraphics[width=0.95\textwidth]{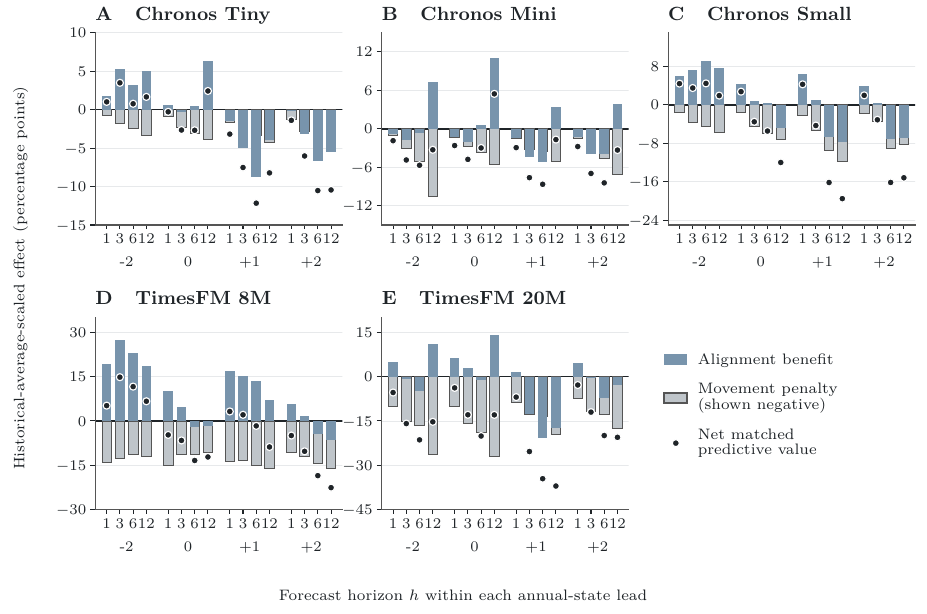}
\caption[Squared-Loss Decomposition of Forecast Revisions]{%
  \textbf{Squared-loss decomposition of forecast revisions}.
  Matched predictive value is decomposed into alignment benefit and movement penalty for five model variants: Chronos Tiny, Mini, and Small, and TimesFM 8M and 20M, in Panels A--E. Within each panel, groups identify leads $-2,0,+1,+2$, with horizons $h=1,3,6,12$ months within each group. For alternative-minus-PIT revision $D$ and origin-aligned point-in-time (PIT) error $e^{\mathrm{PIT}}$, colored bars give $A=100\,2\overline{e^{\mathrm{PIT}}D}/\overline{(e^{\mathrm{HA}})^2}$; light-gray bars give $-P=-100\,\overline{D^2}/\overline{(e^{\mathrm{HA}})^2}$; and black circles give $N=A-P$. Horizon labels appear on the horizontal axis. All averages within a comparison use the same matched support. Values are historical-average mean squared forecast error percentage points; positive $N$ means lower loss than PIT. Lead $-2$ replaces PIT with the stale state, reversing the pre-origin forward step in Table 3. Model panels have separate vertical scales. Estimates use the all-available U.S. equity-premium sample and the US training-information environment.
}
\label{fig:revision-accounting}
\end{figure}

\bigskip\centerline{[Insert Figure~\ref{fig:revision-accounting} about here]}\bigskip

Figure~\ref{fig:revision-accounting} applies this accounting to all five model variants at annual-state leads \(-2,0,+1,+2\) and \(h=1,3,6,12\). Lead \(-2\) follows the PIT-relative replacement direction used in the accounting analysis: the stale state replaces PIT. This direction is the reverse of the clean pre-origin update in Table~\ref{tab:origin-crossing}, which replaces lead \(-2\) with the origin-aligned PIT state at lead \(-1\).

Across models, the movement penalties of exposed annual states generally exceed their alignment benefits. TimesFM revisions are particularly large at medium and long horizons, with correspondingly pronounced penalties. The Chronos components are more concentrated, and their balance likewise generally favors the origin-aligned forecast. Internet Appendix Figure~\ref{fig:x51-fixed-loss-accounting} gives the decomposition over the full matched deployment periods of the 2000, 2009, and 2023 states. Tables~\ref{tab:ia-x51-us-loss-accounting} and~\ref{tab:ia-x51-international-loss-accounting} report rolling and retrospective fixed-state components, respectively with U.S. uncertainty intervals and international medians, interquartile ranges, and positive-market counts. These companion exhibits separate error correction from revision magnitude at the model--horizon level.

\subsection{Break-Even Condition for Predictive Gains}\label{alignment-efficiency-and-the-break-even-condition}

Scale-free accounting gives the break-even condition that a forecast revision must satisfy. For a nonzero matched revision, let \(e=e^{\mathrm{PIT}}\), let \(D\) denote the revision, and define relative revision scale and alignment efficiency as

\[
r
=
\sqrt{
\frac{\mathbb E[D^2]}
{\mathbb E[e^2]}
},
\qquad
\kappa
=
\frac{
\mathbb E[eD]
}{
\sqrt{\mathbb E[e^2]\mathbb E[D^2]}
}.
\]

Let \(\Delta L\equiv e^2-(e-D)^2\). Substituting these two quantities into equation (8) gives

\[
\frac{\mathbb E[\Delta L]}{\mathbb E[e^2]}
=
2\kappa r-r^2,
\qquad
\mathbb E[\Delta L]>0
\iff
\kappa>\frac r2.
\tag{10}
\]

\(r\) measures the size of the revision relative to the error remaining in the PIT forecast, while \(\kappa\) measures its alignment efficiency with that error per unit of revision. A larger revision raises the alignment required to break even: the model must move the forecast toward the realized target more reliably to offset the quadratic cost of departing from the PIT forecast.

\bigskip\centerline{[Insert Table~\ref{tab:alignment-efficiency} about here]}\bigskip

Rolling-exposed and fixed-deployment revisions are sizable but achieve little error correction. In the U.S. rolling and fixed comparisons, \(r\) is \(0.273\) and \(0.348\), respectively, with corresponding break-even thresholds of \(0.137\) and \(0.174\); actual \(\kappa\) is \(-0.014\) and \(0.006\). In the international rolling and fixed comparisons, \(r\) is \(0.312\) and \(0.373\), respectively. The required \(\kappa\) is \(0.156\) and \(0.186\), but the actual values are only \(0.024\) and \(0.041\). These revisions are thus approximately 27\% to 37\% of the error remaining in the PIT forecast, yet their realized alignment falls well short of the level needed to offset this movement. Only 12\% to 17\% of the corresponding comparison cells yield positive net matched predictive value.

The pre-origin comparison places the exposed-state alignment shortfall in context. Table~\ref{tab:alignment-efficiency} replaces PIT with the stale lead \(-2\) forecast, following the PIT-relative direction of the accounting exercise. The pre-origin forward step in Section~\ref{crossing-the-forecast-origin} runs in the opposite direction, from lead \(-2\) to lead \(-1\). For the PIT-to-stale revision, \(\kappa\) is \(0.136\) in the United States and \(0.168\) internationally, close to the respective break-even thresholds of \(0.133\) and \(0.154\); net-gain shares are 60\% and 53\%. This comparison illustrates the role of alignment: revisions of similar scale approach break-even when they correct a larger share of the remaining forecast error.

The training-information environment affects this gap by changing both movement and alignment. For the U.S. target in the rolling design, Global increases normalized movement by approximately 2.87 to 6.86 benchmark-RMSE percentage points relative to US, while raising net matched predictive value by 2.50 to 27.27 historical-average MSFE percentage points. In this summary, added alignment exceeds the added penalty for all five models. The component balance in the Augmented environment is more mixed: additional factor information sometimes improves alignment and sometimes mainly increases revision scale. For non-U.S. targets, Global adds target-market information to the indirect-only US environment, improving revision alignment on average without generally crossing the break-even threshold. Internet Appendix Table~\ref{tab:x52-training-environments} reports the U.S. component differences for Global and Augmented training relative to US training under rolling and retrospective fixed-state comparisons. Figure~\ref{fig:ia-x52-international-corpus-shifts} summarizes these differences across the 13 non-U.S. markets using medians, interquartile ranges, and full ranges. Tables~\ref{tab:ia-x52-us-corpus-components} and~\ref{tab:ia-x52-international-corpus-components} provide the full model--horizon components within each training environment, preserving the distinction between baseline component levels and differences across training samples.

These results connect three empirical dimensions of parameter-time look-ahead. Chronology determines whether a set of parameters is exposed relative to a forecast origin; training content determines how that parameter state revises the same forecast; and the realized alignment of the revision with the error remaining in the PIT forecast determines whether the movement creates predictive value. Training coverage that crosses the forecast origin is therefore an information-set fact, rather than a sufficient statistic for backtest gains.

\section{Conclusion}\label{conclusion}

The information set of a financial forecast includes the data used to train its forecasting rule. Pretrained models make this parameter information set especially important because the same parameters can be applied repeatedly to historical investment decisions. Our matched-vintage design follows the consequences of changing those parameters from return forecasts to prediction errors and portfolio outcomes.

Annual financial foundation models provide observable training cutoffs and repeatable numerical forecasts. Origin-aligned states extract predictive information from historical returns. Substituting later states materially changes forecast levels, directions, and predictive distributions.In the reference U.S.-training environment, those revisions predominantly lower predictive accuracy across U.S. and international markets. Compared with an equally long pre-origin update, the update that crosses the forecast origin lowers matched predictive value. Under the constrained market-timing rule, they also lower median certainty-equivalent returns. Cross-market training information changes these outcomes by changing the balance between error correction and revision magnitude. Temporal exposure alone does not determine whether a pretrained model improves historical predictive accuracy or investor value.

The economic interpretation follows directly from the forecast-error accounting. A revision improves prediction when its alignment with the remaining error is sufficient to offset the squared movement in the forecast. Training dates determine the historical information constraint; the revisions generated by the trained rule determine how that information changes the forecast; and realized returns and portfolio decisions determine its predictive and investor value. Temporal overlap identifies whether a model has encountered
future information. The forecast revisions induced by that exposure, together with their matched predictive and economic consequences, determine whether the historical backtest exhibits favorable look-ahead bias.

\clearpage
\singlespacing

\clearpage
\singlespacing

\clearpage
\definecolor{ink}{HTML}{1F2428}
\definecolor{midgray}{HTML}{6C7278}
\definecolor{lightgray}{HTML}{BFC5CA}
\definecolor{rulegray}{HTML}{6C7278}

\begin{figure}[p]
\centering
\resizebox{\textwidth}{!}{%

\begin{minipage}{7.35in}
\centering

\begin{tikzpicture}[x=1cm,y=1cm,font=\fontsize{8.29}{9.55}\selectfont,>=Latex]
  \path[use as bounding box] (0,0) rectangle (18.5,4.25);

  \node[anchor=west,font=\fontsize{10.44}{12}\selectfont\bfseries] at (0,4.02) {A\quad Rolling design};
  \node[anchor=east,font=\fontsize{8.29}{9.55}\selectfont\itshape] at (18.5,4.02)
    {Hold the forecast unit fixed; move the annual state $v=Y(t)+\ell$};
  \draw[ink,line width=0.62pt] (0,3.73) -- (18.5,3.73);

  \draw[decorate,decoration={brace,amplitude=4pt,raise=2pt},ink,line width=0.45pt]
    (8.62,3.05) -- (15.52,3.05);
  \node[anchor=south,font=\fontsize{8.29}{9.55}\selectfont] at (12.07,3.23)
    {Later-state pool: $\ell=0,+1,+2$, each versus PIT};

  \draw[->,ink,line width=0.74pt] (1.25,2.45) -- (17.45,2.45);

  \foreach \x/\lead in {2.30/$-2$,5.55/$-1$,8.80/$0$,12.05/$+1$,15.30/$+2$}{
    \node[anchor=south,font=\fontsize{8.29}{9.55}\selectfont] at (\x,2.71) {$\ell=\lead$};
  }

  \node[circle,draw=midgray,line width=0.75pt,minimum size=7.4pt,inner sep=0pt,fill=white] at (2.30,2.45) {};
  \node[diamond,draw=ink,fill=ink,line width=0.74pt,minimum size=8.5pt,inner sep=0pt] at (5.55,2.45) {};
  \node[rectangle,draw=ink,line width=0.74pt,minimum size=8.2pt,inner sep=0pt,pattern=north east lines,pattern color=midgray] at (8.80,2.45) {};
  \node[circle,draw=ink,line width=0.62pt,minimum size=7.4pt,inner sep=0pt,fill=midgray] at (12.05,2.45) {};
  \node[circle,draw=ink,line width=0.62pt,minimum size=7.4pt,inner sep=0pt,fill=midgray] at (15.30,2.45) {};

  \node[align=center,anchor=north,text width=2.55cm,font=\fontsize{8.29}{9.55}\selectfont] at (2.30,2.10) {Stale but admissible};
  \node[align=center,anchor=north,text width=2.65cm,font=\fontsize{8.29}{9.55}\selectfont\bfseries] at (5.55,2.10) {Origin-aligned PIT};
  \node[align=center,anchor=north,text width=2.55cm,font=\fontsize{8.29}{9.55}\selectfont] at (8.80,2.10) {Boundary-exposed};
  \node[align=center,anchor=north,text width=2.55cm,font=\fontsize{8.29}{9.55}\selectfont] at (12.05,2.10) {Post-origin exposed};
  \node[align=center,anchor=north,text width=2.55cm,font=\fontsize{8.29}{9.55}\selectfont] at (15.30,2.10) {Post-origin exposed};

  \draw[->,rulegray,line width=0.74pt] (2.48,1.15) -- (5.37,1.15);
  \node[align=center,anchor=north,text width=2.80cm,font=\fontsize{8.29}{9.55}\selectfont] at (3.93,0.98)
    {Clean pre-origin\\step};
  \draw[->,ink,line width=0.74pt] (5.73,1.15) -- (8.62,1.15);
  \node[align=center,anchor=north,text width=2.80cm,font=\fontsize{8.29}{9.55}\selectfont] at (7.18,0.98)
    {Origin-crossing\\step};
\end{tikzpicture}

\vspace{9pt}

\begin{tikzpicture}[x=1cm,y=1cm,font=\fontsize{8.29}{9.55}\selectfont,>=Latex]
  \path[use as bounding box] (0,0) rectangle (18.5,4.75);

  \node[anchor=west,font=\fontsize{10.44}{12}\selectfont\bfseries] at (0,4.52) {B\quad Fixed-vintage design};
  \node[anchor=east,font=\fontsize{8.29}{9.55}\selectfont\itshape] at (18.5,4.52)
    {Hold the annual state fixed; move the target window $W_u$ across $\tau_v$};
  \draw[ink,line width=0.62pt] (0,4.23) -- (18.5,4.23);

  \fill[midgray] (0.28,3.76) rectangle (0.61,3.98);
  \draw[ink,line width=0.35pt] (0.28,3.76) rectangle (0.61,3.98);
  \node[anchor=west,font=\fontsize{8.29}{9.55}\selectfont] at (0.73,3.87) {Retrospective exposure};

  \fill[pattern=north east lines,pattern color=midgray] (4.29,3.76) rectangle (4.62,3.98);
  \draw[ink,line width=0.35pt] (4.29,3.76) rectangle (4.62,3.98);
  \node[anchor=west,font=\fontsize{8.29}{9.55}\selectfont] at (4.74,3.87) {Boundary crossing};

  \fill[lightgray] (7.76,3.76) rectangle (8.09,3.98);
  \draw[ink,line width=0.35pt] (7.76,3.76) rectangle (8.09,3.98);
  \node[anchor=west,font=\fontsize{8.29}{9.55}\selectfont] at (8.21,3.87) {Origin-aligned PIT};

  \fill[white] (11.18,3.76) rectangle (11.51,3.98);
  \draw[ink,line width=0.45pt] (11.18,3.76) rectangle (11.51,3.98);
  \node[anchor=west,font=\fontsize{8.29}{9.55}\selectfont] at (11.63,3.87) {Stale deployment};

  \node[diamond,draw=ink,fill=ink,line width=0.45pt,minimum size=6.8pt,inner sep=0pt] at (15.35,3.87) {};
  \node[anchor=west,font=\fontsize{8.29}{9.55}\selectfont] at (15.58,3.87) {Training cutoff $\tau_v$};

  \foreach \yy/\cut/\state in {2.95/5.25/2000,2.10/9.65/2009,1.25/14.35/2023}{
    \node[anchor=east,font=\fontsize{8.29}{9.55}\selectfont] at (2.62,\yy) {State $v=\state$};
    \fill[midgray] (3.00,{\yy-0.18}) rectangle ({\cut-0.34},{\yy+0.18});
    \fill[pattern=north east lines,pattern color=midgray] ({\cut-0.34},{\yy-0.18}) rectangle ({\cut+0.34},{\yy+0.18});
    \fill[lightgray] ({\cut+0.34},{\yy-0.18}) rectangle ({\cut+1.44},{\yy+0.18});
    \fill[white] ({\cut+1.44},{\yy-0.18}) rectangle (17.25,{\yy+0.18});
    \draw[ink,line width=0.45pt] (3.00,{\yy-0.18}) rectangle (17.25,{\yy+0.18});
    \draw[ink,line width=0.45pt] ({\cut-0.34},{\yy-0.18}) -- ({\cut-0.34},{\yy+0.18});
    \draw[ink,line width=0.45pt] ({\cut+0.34},{\yy-0.18}) -- ({\cut+0.34},{\yy+0.18});
    \draw[ink,line width=0.35pt] ({\cut+1.44},{\yy-0.18}) -- ({\cut+1.44},{\yy+0.18});
    \node[diamond,draw=ink,fill=ink,line width=0.45pt,minimum size=6.8pt,inner sep=0pt] at (\cut,\yy) {};
  }

  \draw[->,ink,line width=0.62pt] (3.00,0.48) -- (17.25,0.48);
  \node[anchor=south west,font=\fontsize{8.29}{9.55}\selectfont] at (3.00,0.60) {Earlier target windows};
  \node[anchor=south east,font=\fontsize{8.29}{9.55}\selectfont] at (17.25,0.60) {Later target windows};
\end{tikzpicture}

\end{minipage}

}
\caption[Annual-State Timing in Rolling and Fixed-Vintage Designs]{%
  \textbf{Annual-state timing in rolling and fixed-vintage designs}.
  Panel A varies the annual state while holding the model, training environment, forecast unit $u=(m,t,h)$, numerical history, and inference protocol fixed. State year $v=Y(t)+\ell$, where $Y(t)$ is the target-start year. Lead $-1$ is origin-aligned point-in-time (PIT), lead $-2$ is admissible but stale, and leads $0,+1,+2$ have post-origin training coverage. Arrows mark the pre-origin transition from lead $-2$ to lead $-1$ and the origin-crossing transition from lead $-1$ to lead $0$; the brace groups the three later leads for the pooled comparison. Panel B holds the annual state fixed and moves the complete target window $W_u=[s_u,b_u]$ across the end-of-year training cutoff $\tau_v$. Shading distinguishes retrospective exposure, boundary crossing, PIT deployment in target-start year $v+1$, and stale deployment thereafter. Diamonds mark $\tau_v$. Each alternative forecast is paired with PIT for the same forecast unit and numerical history.
}
\label{fig:s33-s34-timing-designs}

\end{figure}
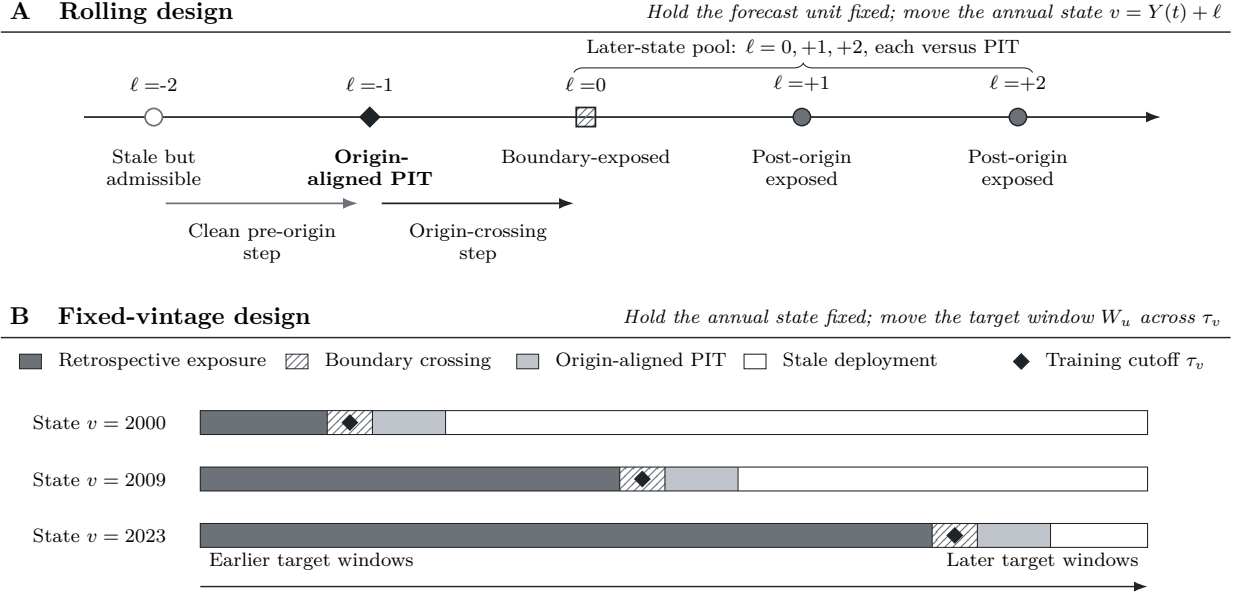
\clearpage
\clearpage
\clearpage
\clearpage
\clearpage
\clearpage
\begingroup
\makeatletter
\setlength{\@fptop}{0pt}
\makeatother
\begingroup
\setlength{\tabcolsep}{4.2pt}
\renewcommand{\arraystretch}{1.10}
\newcolumntype{L}[1]{>{\RaggedRight\arraybackslash}p{#1}}
\newcolumntype{Y}{>{\RaggedRight\arraybackslash}X}

\begin{table}[p]
\centering
\begin{threeparttable}
\caption[TSFM Forecast Construction, Training Information, and Annual-State Timing]{%
  \textbf{TSFM forecast construction, training information, and annual-state timing}
}
\label{tab:model-design}

\small

\begin{tabularx}{\textwidth}{@{}L{0.78in}L{1.02in}YL{2.36in}@{}}
\toprule
\multicolumn{4}{@{}l}{\textit{Panel A. Model families and point-forecast construction}} \\
\midrule
\textbf{Family} &
\textbf{Evaluated variants} &
\textbf{Native forecast object} &
\textbf{Point forecast used in this paper} \\
\midrule
Chronos &
Tiny, Mini, Small &
Probabilistic forecast paths &
Generate $S$ paths; average the $h$ steps within each path, then average across paths \\

TimesFM &
8M, 20M &
Native future-step mean and quantile forecasts &
Average the native means over the $h$-step prediction horizon \\
\end{tabularx}

\vspace{6pt}

\begin{tabularx}{\textwidth}{@{}L{0.90in}YL{1.47in}L{1.70in}@{}}
\toprule
\multicolumn{4}{@{}l}{\textit{Panel B. Training information and exposure channels}} \\
\midrule
\textbf{Environment} &
\textbf{Documented training content} &
\textbf{U.S. target} &
\textbf{Non-U.S. target market} \\
\midrule
US &
U.S. firm-level excess returns &
Direct &
Indirect only \\

Global &
U.S. and non-U.S. firm-level excess returns &
Direct + indirect &
Direct + indirect \\

Augmented &
Global returns plus JKP factor series &
Direct + indirect; expanded factor channel &
Direct + indirect; expanded factor channel \\
\end{tabularx}

\vspace{6pt}

\begin{tabularx}{\textwidth}{@{}L{0.66in}L{1.48in}L{1.88in}Y@{}}
\toprule
\multicolumn{4}{@{}l}{\textit{Panel C. Annual-state timing conventions}} \\
\midrule
\textbf{Design} &
\textbf{Annual states} &
\textbf{Training coverage and real-time availability} &
\textbf{Role in the design} \\
\midrule
Rolling &
$v=Y(t)+\ell$, $\ell\in\{-2,-1,0,+1,+2\}$ &
Expanding sample through year $v$; available from the beginning of year $v+1$ &
$-2$: stale but admissible; $-1$: origin-aligned PIT; $0$: boundary-exposed; $+1,+2$: post-origin exposed \\

Fixed &
$v\in\{2000,2009,2023\}$, held fixed across historical origins &
Expanding sample through the fixed year; available from the beginning of the following year &
Cutoff after the full target window: retrospective exposure; cutoff inside the window: boundary crossing; target window after the cutoff: origin-aligned PIT in year $v+1$, then stale deployment \\
\bottomrule
\end{tabularx}

\begin{tablenotes}[flushleft]
\footnotesize
\item \textit{Notes:} Panel A lists the two time-series foundation model (TSFM) families, five variants, and procedures used to obtain point forecasts of average monthly excess returns. The annual states were released by \citet{rahimikia_ni_wang_2025} and independently trained on expanding financial samples ending in the stated years, using the architectures of \citet{ansari_etal_2024} and \citet{das_kong_sen_zhou_2024}. The return training data are daily firm-level excess returns; deployment uses monthly market histories without target-specific parameter updating. Panel B gives the US, Global, and Augmented training-information environments and their exposure channels. JKP refers to the factor series of \citet{jensen_kelly_pedersen_2023}. Direct exposure refers to post-origin information from the target market; indirect exposure refers to post-origin information from related markets, securities, factors, or common economic conditions. These labels apply only when training coverage extends beyond the forecast origin. Panel C gives the timing conventions: $v$ is the training-data cutoff year, $Y(t)$ is the target-start year, and $\ell=v-Y(t)$. A year-$v$ state is treated as admissible from the beginning of year $v+1$, and PIT has lead $-1$. Fixed-state timing depends on the cutoff's position relative to the complete $h$-month target window.
\end{tablenotes}

\end{threeparttable}
\end{table}

\endgroup
\clearpage
\begin{table}[p]
\centering
\begin{threeparttable}
\caption[Point-in-Time Return Forecasting Performance]{%
  \textbf{Point-in-time return forecasting performance}
}
\label{tab:pit-relevance}
\renewcommand{\arraystretch}{1.10}
\small
\begin{tabular*}{\textwidth}{@{\extracolsep{\fill}}lcccc@{\hspace{1.15em}}cccc@{}}
\toprule
& \multicolumn{4}{c}{United States}
& \multicolumn{4}{c}{\makecell[c]{International markets\\[-0.4ex]\footnotesize Equal-weight aggregate}} \\
\cmidrule(lr){2-5} \cmidrule(lr){6-9}
Model & $h=1$ & $h=3$ & $h=6$ & $h=12$
      & $h=1$ & $h=3$ & $h=6$ & $h=12$ \\
\midrule
Chronos Tiny & \makecell[c]{\ensuremath{3.45^{***}}\\[-0.35ex]{\scriptsize \ensuremath{(0.005)}}} & \makecell[c]{\ensuremath{9.26^{***}}\\[-0.35ex]{\scriptsize \ensuremath{(0.002)}}} & \makecell[c]{\ensuremath{15.12^{***}}\\[-0.35ex]{\scriptsize \ensuremath{(0.002)}}} & \makecell[c]{\ensuremath{14.31^{**}}\\[-0.35ex]{\scriptsize \ensuremath{(0.040)}}} & \makecell[c]{\ensuremath{1.77^{***}}\\[-0.35ex]{\scriptsize \ensuremath{(0.003)}}} & \makecell[c]{\ensuremath{4.55^{***}}\\[-0.35ex]{\scriptsize \ensuremath{(0.003)}}} & \makecell[c]{\ensuremath{9.08^{***}}\\[-0.35ex]{\scriptsize \ensuremath{(0.002)}}} & \makecell[c]{\ensuremath{12.82^{**}}\\[-0.35ex]{\scriptsize \ensuremath{(0.017)}}} \\
Chronos Mini & \makecell[c]{\ensuremath{3.70^{***}}\\[-0.35ex]{\scriptsize \ensuremath{(0.002)}}} & \makecell[c]{\ensuremath{10.32^{***}}\\[-0.35ex]{\scriptsize \ensuremath{(<0.001)}}} & \makecell[c]{\ensuremath{13.61^{***}}\\[-0.35ex]{\scriptsize \ensuremath{(0.002)}}} & \makecell[c]{\ensuremath{9.95^{*}}\\[-0.35ex]{\scriptsize \ensuremath{(0.078)}}} & \makecell[c]{\ensuremath{1.76^{***}}\\[-0.35ex]{\scriptsize \ensuremath{(0.005)}}} & \makecell[c]{\ensuremath{5.56^{***}}\\[-0.35ex]{\scriptsize \ensuremath{(<0.001)}}} & \makecell[c]{\ensuremath{9.18^{***}}\\[-0.35ex]{\scriptsize \ensuremath{(0.001)}}} & \makecell[c]{\ensuremath{12.91^{**}}\\[-0.35ex]{\scriptsize \ensuremath{(0.018)}}} \\
Chronos Small & \makecell[c]{\ensuremath{0.29}\\[-0.35ex]{\scriptsize \ensuremath{(0.159)}}} & \makecell[c]{\ensuremath{6.00^{**}}\\[-0.35ex]{\scriptsize \ensuremath{(0.014)}}} & \makecell[c]{\ensuremath{10.53^{***}}\\[-0.35ex]{\scriptsize \ensuremath{(0.008)}}} & \makecell[c]{\ensuremath{13.71^{**}}\\[-0.35ex]{\scriptsize \ensuremath{(0.029)}}} & \makecell[c]{\ensuremath{-0.99}\\[-0.35ex]{\scriptsize \ensuremath{(0.130)}}} & \makecell[c]{\ensuremath{0.60^{**}}\\[-0.35ex]{\scriptsize \ensuremath{(0.020)}}} & \makecell[c]{\ensuremath{6.38^{***}}\\[-0.35ex]{\scriptsize \ensuremath{(0.003)}}} & \makecell[c]{\ensuremath{10.60^{**}}\\[-0.35ex]{\scriptsize \ensuremath{(0.017)}}} \\
\addlinespace[0.35em]
TimesFM 8M & \makecell[c]{\ensuremath{-9.83}\\[-0.35ex]{\scriptsize \ensuremath{(0.443)}}} & \makecell[c]{\ensuremath{-5.65}\\[-0.35ex]{\scriptsize \ensuremath{(0.296)}}} & \makecell[c]{\ensuremath{2.15^{**}}\\[-0.35ex]{\scriptsize \ensuremath{(0.043)}}} & \makecell[c]{\ensuremath{9.19^{***}}\\[-0.35ex]{\scriptsize \ensuremath{(0.005)}}} & \makecell[c]{\ensuremath{-25.25}\\[-0.35ex]{\scriptsize \ensuremath{(0.682)}}} & \makecell[c]{\ensuremath{-12.23}\\[-0.35ex]{\scriptsize \ensuremath{(0.137)}}} & \makecell[c]{\ensuremath{-0.71^{**}}\\[-0.35ex]{\scriptsize \ensuremath{(0.011)}}} & \makecell[c]{\ensuremath{5.06^{***}}\\[-0.35ex]{\scriptsize \ensuremath{(0.004)}}} \\
TimesFM 20M & \makecell[c]{\ensuremath{-7.55}\\[-0.35ex]{\scriptsize \ensuremath{(0.313)}}} & \makecell[c]{\ensuremath{1.38^{**}}\\[-0.35ex]{\scriptsize \ensuremath{(0.020)}}} & \makecell[c]{\ensuremath{9.69^{**}}\\[-0.35ex]{\scriptsize \ensuremath{(0.011)}}} & \makecell[c]{\ensuremath{7.46^{**}}\\[-0.35ex]{\scriptsize \ensuremath{(0.011)}}} & \makecell[c]{\ensuremath{-14.53}\\[-0.35ex]{\scriptsize \ensuremath{(0.459)}}} & \makecell[c]{\ensuremath{-10.01^{**}}\\[-0.35ex]{\scriptsize \ensuremath{(0.018)}}} & \makecell[c]{\ensuremath{-6.55^{**}}\\[-0.35ex]{\scriptsize \ensuremath{(0.044)}}} & \makecell[c]{\ensuremath{-31.11}\\[-0.35ex]{\scriptsize \ensuremath{(0.317)}}} \\
\midrule
Target-start months & 288 & 286 & 283 & 277 & 288 & 288 & 288 & 288 \\
\bottomrule
\end{tabular*}

\begin{tablenotes}[flushleft]
\footnotesize
\item \textit{Notes:} Raw out-of-sample $R^2_{\mathrm{HA}}$ measures the percentage-point reduction in mean squared forecast error (MSFE) of origin-aligned point-in-time (PIT) forecasts relative to the expanding historical average. Positive values indicate lower loss for PIT. Parentheses contain one-sided Clark--West diagnostic $p$-values computed using a normal approximation to the statistic based on the nesting-adjusted loss differential, which can be significant even when raw $R^2_{\mathrm{HA}}$ is negative. International estimates first normalize each market's monthly raw and adjusted loss differentials by its historical-average MSFE, then average the 13 markets equally over common target-start months. Bartlett--Newey--West HAC inference uses $h-1$ lags, where $h$ is the forecast horizon in months. $^{*}$, $^{**}$, and $^{***}$ indicate significance at the 10\%, 5\%, and 1\% levels. Estimates use PIT lead $-1$, the all-available sample, and the US training-information environment. The final row gives sample sizes.
\end{tablenotes}

\end{threeparttable}
\end{table}
\clearpage
\begin{table}[p]
\centering
\begin{threeparttable}
\caption[Predictive Effects of Pre-Origin and Origin-Crossing Annual Updates]{%
  \textbf{Predictive effects of pre-origin and origin-crossing annual updates}
}
\label{tab:origin-crossing}
\renewcommand{\arraystretch}{1.10}
\small
\setlength{\tabcolsep}{3pt}
\begin{tabularx}{\textwidth}{@{}>{\raggedright\arraybackslash}X*{2}{>{\centering\arraybackslash}p{0.145\textwidth}}@{\hspace{0.75em}}*{2}{>{\centering\arraybackslash}p{0.145\textwidth}}@{}}
\toprule
\multicolumn{5}{@{}l}{\textit{Panel A. One-year annual-state updates}} \\
\midrule
& \multicolumn{2}{c}{United States} & \multicolumn{2}{c}{13 non-U.S. markets} \\
\cmidrule(lr){2-3} \cmidrule(lr){4-5}
& \multicolumn{1}{c}{\makecell[c]{\footnotesize Clean pre-origin\\[-0.35ex]\footnotesize lead $-1$ vs. $-2$}}
& \multicolumn{1}{c}{\makecell[c]{\footnotesize Origin crossing\\[-0.35ex]\footnotesize lead $0$ vs. $-1$}}
& \multicolumn{1}{c}{\makecell[c]{\footnotesize Clean pre-origin\\[-0.35ex]\footnotesize lead $-1$ vs. $-2$}}
& \multicolumn{1}{c}{\makecell[c]{\footnotesize Origin crossing\\[-0.35ex]\footnotesize lead $0$ vs. $-1$}} \\
\midrule
Mean effect (pp)              & $0.72$ & $-5.65$ & $0.85$ & $-5.71$ \\
Median effect (pp)            & $-1.32$ & $-4.28$ & $-0.88$ & $-3.81$ \\
Standard error               & $2.01$  & $1.44$  & $0.81$  & $0.56$ \\
$t$-statistic                & $0.36$ & $-3.92$ & $1.05$ & $-10.15$ \\
$p$-value (two-sided)        & $0.724$  & $0.001$  & $0.294$  & $<0.001$ \\
\addlinespace[0.35em]
Cells with interval below zero & 2  & 1  & 36  & 54 \\
Cells with interval above zero & 2  & 0  & 32  & 2  \\
\addlinespace[0.35em]
Cells                        & 20 & 20 & 260 & 260 \\
\midrule
\multicolumn{5}{@{}l}{\textit{Panel B. Origin-crossing minus pre-origin updates}} \\
\midrule
& \multicolumn{2}{c}{United States} & \multicolumn{2}{c}{13 non-U.S. markets} \\
\cmidrule(lr){2-3} \cmidrule(lr){4-5}
Difference (pp)                        & \multicolumn{2}{c}{$-6.37$} & \multicolumn{2}{c}{$-6.56$} \\
$t$, cells treated as independent      & \multicolumn{2}{c}{$-2.23$} & \multicolumn{2}{c}{$-6.60$} \\
$t$, clustered by market               & \multicolumn{2}{c}{n/a}     & \multicolumn{2}{c}{$-7.08$} \\
$t$, clustered by architecture         & \multicolumn{2}{c}{$-1.21$} & \multicolumn{2}{c}{$-1.40$} \\
Markets with negative difference       & \multicolumn{2}{c}{1 of 1}  & \multicolumn{2}{c}{13 of 13} \\
Architectures with negative difference & \multicolumn{2}{c}{3 of 5}  & \multicolumn{2}{c}{4 of 5} \\
\bottomrule
\end{tabularx}
\begin{tablenotes}[flushleft]
\footnotesize
\item \textit{Notes:} The pre-origin update moves from lead $-2$ to the origin-aligned point-in-time (PIT) state at lead $-1$; the origin-crossing update moves from lead $-1$ to lead $0$. Panel A summarizes the two one-year updates, and Panel B reports the origin-crossing effect minus the pre-origin effect, calculated cell by cell. Each update uses its maximal matched support with PIT and is scaled by the historical-average mean squared forecast error on that support. Effects are in percentage points; positive values mean that the newer state lowers forecast loss. U.S. summaries are across model--horizon cells, and international summaries are across market--model--horizon cells in the 13 non-U.S. markets. The above- and below-zero counts report cells whose 95\% HAC confidence intervals lie entirely above and below zero, respectively. The $t$-statistics use cell-level, market-clustered, or architecture-clustered inference, as indicated. Estimates use the all-available sample and the US training-information environment.
\end{tablenotes}
\end{threeparttable}
\end{table}
\clearpage
\begin{table}[!htbp]
\centering
\captionsetup{font=tabletitle}
\caption[Return Forecasting Performance across Rolling Annual States]{%
  \textbf{Return forecasting performance across rolling annual states}
}\label{tab:x43-lead-performance}
\begin{threeparttable}
\footnotesize
\renewcommand\cellset{\def\arraystretch{0.85}\setlength{\extrarowheight}{0pt}\nomakegapedcells}
\setlength{\tabcolsep}{5.5pt}
\renewcommand{\arraystretch}{0.88}
\begin{tabular*}{\textwidth}{@{\extracolsep{\fill}}llrrrr@{}}
\toprule
Model & Annual-state lead & $h=1$ & $h=3$ & $h=6$ & $h=12$ \\
\midrule
\multicolumn{6}{@{}l}{\textit{Panel A. Chronos Tiny}} \\[-1pt]
Chronos Tiny & \textbf{PIT ($-1$)} & \makecell[r]{3.74$^{***}$\\{\scriptsize (0.007)}} & \makecell[r]{11.56$^{***}$\\{\scriptsize (0.001)}} & \makecell[r]{17.44$^{***}$\\{\scriptsize (0.002)}} & \makecell[r]{16.74$^{**}$\\{\scriptsize (0.032)}} \\
 & Lead $0$ & \makecell[r]{3.19$^{**}$\\{\scriptsize (0.017)}} & \makecell[r]{7.05$^{**}$\\{\scriptsize (0.013)}} & \makecell[r]{12.77$^{***}$\\{\scriptsize (0.007)}} & \makecell[r]{17.41$^{**}$\\{\scriptsize (0.029)}} \\
 & Lead $+1$ & \makecell[r]{1.83$^{*}$\\{\scriptsize (0.087)}} & \makecell[r]{4.48$^{*}$\\{\scriptsize (0.055)}} & \makecell[r]{4.86$^{*}$\\{\scriptsize (0.063)}} & \makecell[r]{8.50$^{*}$\\{\scriptsize (0.054)}} \\
 & Lead $+2$ & \makecell[r]{2.34$^{**}$\\{\scriptsize (0.038)}} & \makecell[r]{5.53$^{**}$\\{\scriptsize (0.017)}} & \makecell[r]{6.91$^{**}$\\{\scriptsize (0.017)}} & \makecell[r]{6.30$^{*}$\\{\scriptsize (0.094)}} \\
\addlinespace[0.5pt]
\multicolumn{6}{@{}l}{\textit{Panel B. Chronos Mini}} \\[-1pt]
Chronos Mini & \textbf{PIT ($-1$)} & \makecell[r]{3.70$^{***}$\\{\scriptsize (0.003)}} & \makecell[r]{10.75$^{***}$\\{\scriptsize ($<$0.001)}} & \makecell[r]{13.86$^{***}$\\{\scriptsize (0.003)}} & \makecell[r]{10.18$^{*}$\\{\scriptsize (0.086)}} \\
 & Lead $0$ & \makecell[r]{1.44$^{*}$\\{\scriptsize (0.079)}} & \makecell[r]{8.21$^{***}$\\{\scriptsize (0.001)}} & \makecell[r]{13.37$^{***}$\\{\scriptsize (0.001)}} & \makecell[r]{17.63$^{**}$\\{\scriptsize (0.022)}} \\
 & Lead $+1$ & \makecell[r]{0.74\\{\scriptsize (0.189)}} & \makecell[r]{3.91$^{**}$\\{\scriptsize (0.023)}} & \makecell[r]{5.52$^{**}$\\{\scriptsize (0.011)}} & \makecell[r]{8.51$^{*}$\\{\scriptsize (0.068)}} \\
 & Lead $+2$ & \makecell[r]{0.93\\{\scriptsize (0.170)}} & \makecell[r]{3.79$^{**}$\\{\scriptsize (0.025)}} & \makecell[r]{5.42$^{***}$\\{\scriptsize (0.007)}} & \makecell[r]{6.86$^{**}$\\{\scriptsize (0.047)}} \\
\addlinespace[0.5pt]
\multicolumn{6}{@{}l}{\textit{Panel C. Chronos Small}} \\[-1pt]
Chronos Small & \textbf{PIT ($-1$)} & \makecell[r]{1.81$^{*}$\\{\scriptsize (0.098)}} & \makecell[r]{9.30$^{***}$\\{\scriptsize (0.007)}} & \makecell[r]{13.54$^{***}$\\{\scriptsize (0.005)}} & \makecell[r]{13.87$^{**}$\\{\scriptsize (0.035)}} \\
 & Lead $0$ & \makecell[r]{3.63$^{**}$\\{\scriptsize (0.011)}} & \makecell[r]{5.58$^{**}$\\{\scriptsize (0.012)}} & \makecell[r]{8.40$^{**}$\\{\scriptsize (0.013)}} & \makecell[r]{4.69\\{\scriptsize (0.132)}} \\
 & Lead $+1$ & \makecell[r]{3.60$^{***}$\\{\scriptsize (0.004)}} & \makecell[r]{2.87$^{**}$\\{\scriptsize (0.038)}} & \makecell[r]{-6.30\\{\scriptsize (0.449)}} & \makecell[r]{-7.48\\{\scriptsize (0.394)}} \\
 & Lead $+2$ & \makecell[r]{3.76$^{**}$\\{\scriptsize (0.014)}} & \makecell[r]{6.17$^{**}$\\{\scriptsize (0.016)}} & \makecell[r]{-2.61\\{\scriptsize (0.282)}} & \makecell[r]{-1.31\\{\scriptsize (0.238)}} \\
\addlinespace[0.5pt]
\multicolumn{6}{@{}l}{\textit{Panel D. TimesFM 8M}} \\[-1pt]
TimesFM 8M & \textbf{PIT ($-1$)} & \makecell[r]{-11.88\\{\scriptsize (0.559)}} & \makecell[r]{-2.98\\{\scriptsize (0.203)}} & \makecell[r]{6.91$^{**}$\\{\scriptsize (0.011)}} & \makecell[r]{14.28$^{***}$\\{\scriptsize (0.001)}} \\
 & Lead $0$ & \makecell[r]{-17.43\\{\scriptsize (0.860)}} & \makecell[r]{-12.49\\{\scriptsize (0.716)}} & \makecell[r]{-11.96\\{\scriptsize (0.714)}} & \makecell[r]{-1.67$^{*}$\\{\scriptsize (0.098)}} \\
 & Lead $+1$ & \makecell[r]{-7.07\\{\scriptsize (0.144)}} & \makecell[r]{-2.13$^{*}$\\{\scriptsize (0.072)}} & \makecell[r]{3.36$^{*}$\\{\scriptsize (0.054)}} & \makecell[r]{3.14$^{**}$\\{\scriptsize (0.049)}} \\
 & Lead $+2$ & \makecell[r]{-16.82\\{\scriptsize (0.873)}} & \makecell[r]{-13.25\\{\scriptsize (0.594)}} & \makecell[r]{-11.58\\{\scriptsize (0.521)}} & \makecell[r]{-8.26\\{\scriptsize (0.324)}} \\
\addlinespace[0.5pt]
\multicolumn{6}{@{}l}{\textit{Panel E. TimesFM 20M}} \\[-1pt]
TimesFM 20M & \textbf{PIT ($-1$)} & \makecell[r]{-10.19\\{\scriptsize (0.512)}} & \makecell[r]{2.67$^{**}$\\{\scriptsize (0.025)}} & \makecell[r]{11.69$^{**}$\\{\scriptsize (0.011)}} & \makecell[r]{7.03$^{**}$\\{\scriptsize (0.014)}} \\
 & Lead $0$ & \makecell[r]{-13.62\\{\scriptsize (0.807)}} & \makecell[r]{-8.30\\{\scriptsize (0.350)}} & \makecell[r]{-3.79\\{\scriptsize (0.202)}} & \makecell[r]{2.38$^{*}$\\{\scriptsize (0.081)}} \\
 & Lead $+1$ & \makecell[r]{-14.45\\{\scriptsize (0.849)}} & \makecell[r]{-19.74\\{\scriptsize (0.840)}} & \makecell[r]{-22.62\\{\scriptsize (0.764)}} & \makecell[r]{-29.38\\{\scriptsize (0.820)}} \\
 & Lead $+2$ & \makecell[r]{-13.05\\{\scriptsize (0.694)}} & \makecell[r]{-9.35\\{\scriptsize (0.487)}} & \makecell[r]{-8.25\\{\scriptsize (0.334)}} & \makecell[r]{-13.51\\{\scriptsize (0.424)}} \\
\midrule
\multicolumn{2}{@{}l}{Joint target-start months} & 252 & 252 & 252 & 252 \\
\bottomrule
\end{tabular*}

\begin{tablenotes}[flushleft]
\footnotesize
\item \textit{Notes:} Raw out-of-sample $R^2_{\mathrm{HA}}$ compares the origin-aligned point-in-time (PIT) state at lead $-1$ and later states at leads $0,+1,+2$ with the expanding historical-average forecast. Values are percentage-point reductions in mean squared forecast error (MSFE) relative to historical-average MSFE; positive values favor the model forecast over the historical average. Within each model--horizon cell, all four forecasts use the same 252 target-start months. The three later states have training coverage beyond the forecast origin. Parentheses contain one-sided Clark--West diagnostic $p$-values against the historical average, computed using a normal approximation to the statistic based on the nesting-adjusted loss differential and Bartlett--Newey--West HAC inference with $h-1$ lags. This adjusted test can be significant when raw $R^2_{\mathrm{HA}}$ is negative. $^{*}$, $^{**}$, and $^{***}$ indicate significance at 10\%, 5\%, and 1\%. The horizon $h$ is in months. Estimates use the U.S. equity premium and the US training-information environment.
\end{tablenotes}

\end{threeparttable}
\end{table}
\clearpage
\begin{table}[!htbp]
\centering
\begin{threeparttable}
\caption[Predictive Effects of Fixed Annual Parameter States]{%
  \textbf{Predictive effects of fixed annual parameter states}
}\label{tab:x44-fixed-effects}
\footnotesize
\setlength{\tabcolsep}{0pt}
\renewcommand{\arraystretch}{0.96}
\begin{tabular*}{\textwidth}{@{\extracolsep{\fill}}lrrrr@{}}
\toprule
Fixed annual state & $h=1$ & $h=3$ & $h=6$ & $h=12$ \\
\midrule
\multicolumn{5}{@{}l}{\textit{Panel A. Chronos Tiny}} \\
Fixed 2000 & \makecell[r]{-5.33$^{**}$\\{\scriptsize (0.028)}} & \makecell[r]{-14.32$^{**}$\\{\scriptsize (0.020)}} & \makecell[r]{-25.96$^{**}$\\{\scriptsize (0.016)}} & \makecell[r]{-32.77$^{**}$\\{\scriptsize (0.025)}} \\
Fixed 2009 & \makecell[r]{-3.68\\{\scriptsize (0.110)}} & \makecell[r]{-9.63$^{**}$\\{\scriptsize (0.038)}} & \makecell[r]{-9.90\\{\scriptsize (0.143)}} & \makecell[r]{4.66\\{\scriptsize (0.685)}} \\
Fixed 2023 & \makecell[r]{-3.08\\{\scriptsize (0.274)}} & \makecell[r]{-8.43$^{***}$\\{\scriptsize (0.007)}} & \makecell[r]{-8.96$^{**}$\\{\scriptsize (0.010)}} & \makecell[r]{-5.21\\{\scriptsize (0.310)}} \\
\addlinespace[1pt]
\multicolumn{5}{@{}l}{\textit{Panel B. Chronos Mini}} \\
Fixed 2000 & \makecell[r]{-4.22$^{**}$\\{\scriptsize (0.024)}} & \makecell[r]{-12.43$^{***}$\\{\scriptsize ($<$0.001)}} & \makecell[r]{-17.30$^{***}$\\{\scriptsize (0.006)}} & \makecell[r]{-24.21$^{**}$\\{\scriptsize (0.019)}} \\
Fixed 2009 & \makecell[r]{-1.60\\{\scriptsize (0.448)}} & \makecell[r]{-6.29\\{\scriptsize (0.115)}} & \makecell[r]{-4.69\\{\scriptsize (0.451)}} & \makecell[r]{4.78\\{\scriptsize (0.617)}} \\
Fixed 2023 & \makecell[r]{-9.36$^{*}$\\{\scriptsize (0.084)}} & \makecell[r]{-15.60\\{\scriptsize (0.132)}} & \makecell[r]{-4.31\\{\scriptsize (0.473)}} & \makecell[r]{-1.12\\{\scriptsize (0.874)}} \\
\addlinespace[1pt]
\multicolumn{5}{@{}l}{\textit{Panel C. Chronos Small}} \\
Fixed 2000 & \makecell[r]{0.46\\{\scriptsize (0.804)}} & \makecell[r]{-5.64\\{\scriptsize (0.152)}} & \makecell[r]{-10.80\\{\scriptsize (0.141)}} & \makecell[r]{-17.64\\{\scriptsize (0.140)}} \\
Fixed 2009 & \makecell[r]{1.57\\{\scriptsize (0.504)}} & \makecell[r]{0.93\\{\scriptsize (0.792)}} & \makecell[r]{-0.70\\{\scriptsize (0.815)}} & \makecell[r]{-4.30\\{\scriptsize (0.350)}} \\
Fixed 2023 & \makecell[r]{-11.01$^{*}$\\{\scriptsize (0.053)}} & \makecell[r]{-19.25$^{**}$\\{\scriptsize (0.022)}} & \makecell[r]{-15.92$^{**}$\\{\scriptsize (0.032)}} & \makecell[r]{-32.30$^{*}$\\{\scriptsize (0.078)}} \\
\addlinespace[1pt]
\multicolumn{5}{@{}l}{\textit{Panel D. TimesFM 8M}} \\
Fixed 2000 & \makecell[r]{-11.33$^{**}$\\{\scriptsize (0.015)}} & \makecell[r]{-9.22\\{\scriptsize (0.238)}} & \makecell[r]{-15.21\\{\scriptsize (0.188)}} & \makecell[r]{-23.82\\{\scriptsize (0.111)}} \\
Fixed 2009 & \makecell[r]{-2.38\\{\scriptsize (0.581)}} & \makecell[r]{-1.94\\{\scriptsize (0.746)}} & \makecell[r]{-6.92\\{\scriptsize (0.364)}} & \makecell[r]{-10.38\\{\scriptsize (0.111)}} \\
Fixed 2023 & \makecell[r]{9.62$^{*}$\\{\scriptsize (0.078)}} & \makecell[r]{1.78\\{\scriptsize (0.786)}} & \makecell[r]{-1.78\\{\scriptsize (0.763)}} & \makecell[r]{-7.77\\{\scriptsize (0.212)}} \\
\addlinespace[1pt]
\multicolumn{5}{@{}l}{\textit{Panel E. TimesFM 20M}} \\
Fixed 2000 & \makecell[r]{-14.99$^{***}$\\{\scriptsize (0.004)}} & \makecell[r]{-22.07$^{***}$\\{\scriptsize (0.002)}} & \makecell[r]{-30.53$^{***}$\\{\scriptsize ($<$0.001)}} & \makecell[r]{-26.00$^{**}$\\{\scriptsize (0.022)}} \\
Fixed 2009 & \makecell[r]{-3.01\\{\scriptsize (0.499)}} & \makecell[r]{-41.48$^{***}$\\{\scriptsize ($<$0.001)}} & \makecell[r]{-61.75$^{***}$\\{\scriptsize ($<$0.001)}} & \makecell[r]{-72.59$^{***}$\\{\scriptsize (0.004)}} \\
Fixed 2023 & \makecell[r]{0.82\\{\scriptsize (0.875)}} & \makecell[r]{-20.05$^{***}$\\{\scriptsize (0.006)}} & \makecell[r]{-27.60$^{**}$\\{\scriptsize (0.014)}} & \makecell[r]{-34.74$^{*}$\\{\scriptsize (0.053)}} \\
\midrule
Target-start months & 288 & 286 & 283 & 277 \\
\bottomrule
\end{tabular*}

\begin{tablenotes}[flushleft]
\footnotesize
\item \textit{Notes:} Matched predictive effects compare the fixed annual states of 2000, 2009, and 2023 with origin-aligned point-in-time (PIT) forecasts. Each effect is PIT mean squared forecast error (MSFE) minus fixed-state MSFE, divided by historical-average MSFE on the same target months and expressed in percentage points. Positive values favor the fixed state; negative values favor PIT. Each state uses its full available pairwise support, which across states includes retrospective-exposure, boundary-crossing, origin-aligned, and stale periods. Parentheses contain two-sided Bartlett--Newey--West HAC $p$-values with $h-1$ lags, where $h$ is the horizon in months. $^{*}$, $^{**}$, and $^{***}$ indicate significance at 10\%, 5\%, and 1\%. Estimates use the all-available U.S. equity-premium sample and the US training-information environment. Matched target-start counts appear in the final row.
\end{tablenotes}

\end{threeparttable}
\end{table}
\clearpage
\begin{table}[p]
\centering
\begin{threeparttable}
\caption[Predictive Effects across Rolling and Fixed-Vintage Designs]{%
  \textbf{Predictive effects across rolling and fixed-vintage designs}
}
\label{tab:cross-design}
\small
\renewcommand{\arraystretch}{1.10}
\setlength{\tabcolsep}{3.5pt}
\begin{tabular*}{\textwidth}{@{\extracolsep{\fill}}lcrrrl@{}}
\toprule
Model & $h$ & \multicolumn{3}{c}{Matched predictive effect relative to PIT} & Cross-design \\
\cmidrule(lr){3-5}\cmidrule(l){6-6}
& & Rolling pooled & \shortstack{Fixed 2009\\(retrospective)} & \shortstack{Fixed 2023\\(retrospective)} & Pattern \\
\midrule
\multirow{4}{*}{Chronos Tiny} & 1 & -1.29 & -6.05 & -3.14 & Common loss \\
 & 3 & \textbf{-5.88} & -16.15 & \textbf{-8.82} & Common loss \\
 & 6 & \textbf{-9.26} & -18.97 & \textbf{-9.64} & Common loss \\
 & 12 & -6.01 & -8.69 & -7.04 & Common loss \\
\addlinespace[3pt]
\multirow{4}{*}{Chronos Mini} & 1 & -2.66 & -5.66 & -9.52 & Common loss \\
 & 3 & \textbf{-5.45} & \textbf{-11.61} & -15.85 & Common loss \\
 & 6 & -5.75 & -11.95 & -4.11 & Common loss \\
 & 12 & 0.82 & -9.68 & -0.41 & Protocol disagreement \\
\addlinespace[3pt]
\multirow{4}{*}{Chronos Small} & 1 & 1.86 & 3.90 & -11.24 & Fixed-vintage mixed \\
 & 3 & -4.42 & 0.95 & \textbf{-19.31} & Fixed-vintage mixed \\
 & 6 & -13.72 & -0.56 & \textbf{-15.67} & Common loss \\
 & 12 & -15.24 & 0.67 & -30.80 & Fixed-vintage mixed \\
\addlinespace[3pt]
\multirow{4}{*}{TimesFM 8M} & 1 & -1.89 & 7.85 & 9.81 & Protocol disagreement \\
 & 3 & -6.31 & 7.90 & 0.44 & Protocol disagreement \\
 & 6 & -13.63 & 0.34 & -4.71 & Fixed-vintage mixed \\
 & 12 & \textbf{-16.55} & -9.38 & -10.71 & Common loss \\
\addlinespace[3pt]
\multirow{4}{*}{TimesFM 20M} & 1 & -3.52 & -1.43 & 0.83 & Fixed-vintage mixed \\
 & 3 & \textbf{-15.13} & \textbf{-56.98} & \textbf{-18.96} & Common loss \\
 & 6 & \textbf{-23.24} & \textbf{-78.69} & \textbf{-23.46} & Common loss \\
 & 12 & -20.53 & \textbf{-106.24} & -27.43 & Common loss \\
\midrule
\multicolumn{6}{@{}l@{}}{\footnotesize \textit{Pattern totals:}\quad Common gain: 0\quad Common loss: 12\quad Fixed-vintage mixed: 5\quad Protocol disagreement: 3} \\
\bottomrule
\end{tabular*}

\begin{tablenotes}[flushleft]
\footnotesize
\item \textit{Notes:} The three columns compare pooled rolling effects with the fixed-2009 and fixed-2023 retrospective-exposure effects for the U.S. equity premium. All effects are relative to origin-aligned point-in-time (PIT) forecasts and expressed in historical-average mean squared forecast error percentage points; positive values favor the alternative state. Rolling effects equally average the loss advantages of leads $0,+1,+2$ after restricting the three leads and PIT to their strict common support. Each fixed state uses its retrospective-exposure window. Common gain and common loss require three positive or three negative point estimates, respectively. Fixed-vintage mixed denotes opposite signs for the two fixed-state estimates; protocol disagreement denotes two fixed-state estimates with a common sign opposite to the rolling estimate. Bold estimates have both 95\% Bartlett--Newey--West HAC and circular moving-block bootstrap intervals excluding zero in the same direction. The horizon $h$ is in months. Estimates use the all-available sample and the US training-information environment.
\end{tablenotes}

\end{threeparttable}
\end{table}
\clearpage
\begin{table}[p]
\centering
\begin{threeparttable}
\caption[Portfolio Performance under Point-in-Time and Post-Origin Forecasts]{%
  \textbf{Portfolio performance under point-in-time and post-origin forecasts}
}
\label{tab:investor-value}
\small
\renewcommand{\arraystretch}{1.10}
\setlength{\tabcolsep}{3.8pt}
\begin{tabular*}{\textwidth}{@{\extracolsep{\fill}}lrrrcrr@{}}
\toprule
\multicolumn{7}{@{}l}{\textit{Panel A. United States}} \\
\midrule
Model & \multicolumn{4}{c}{Annualized certainty-equivalent return} & \multicolumn{2}{c}{Annualized Sharpe ratio} \\
\cmidrule(lr){2-5}\cmidrule(l){6-7}
& PIT & \shortstack{Rolling\\exposed} & $\Delta CER$ & 95\% interval & PIT & \shortstack{Rolling\\exposed} \\
\midrule
Chronos Tiny & 6.44 & 4.67 & -1.77 & [-4.18, 0.34] & 0.632 & 0.529 \\
Chronos Mini & 6.34 & 3.57 & -2.78 & [-6.50, 0.52] & 0.630 & 0.469 \\
Chronos Small & 3.30 & 6.46 & 3.16 & [-0.57, 7.50] & 0.450 & 0.639 \\
TimesFM 8M & -0.25 & -3.06 & -2.81 & [-6.95, 1.30] & 0.237 & 0.096 \\
TimesFM 20M & -1.13 & -0.44 & 0.68 & [-2.89, 4.62] & 0.216 & 0.260 \\
\addlinespace[2pt]
\textit{Column median} & 3.30 & 3.57 & -1.77 & -- & 0.450 & 0.469 \\
\multicolumn{7}{@{}l@{}}{\footnotesize Positive $\Delta CER$: 2/5\quad 95\% CI below 0: 0/5\quad 95\% CI above 0: 0/5} \\
\midrule
\multicolumn{7}{@{}l}{\textit{Panel B. International markets}} \\
\midrule
Model & \multicolumn{3}{c}{Median annualized CER} & \multicolumn{3}{c}{Breadth across 13 markets} \\
\cmidrule(lr){2-4}\cmidrule(l){5-7}
& PIT & \shortstack{Rolling\\exposed} & $\Delta CER$ & $\Delta CER>0$ & \shortstack{95\% CI\\below 0} & \shortstack{95\% CI\\above 0} \\
\midrule
Chronos Tiny & 2.20 & 1.02 & -1.35 & 2/13 & 2/13 & 0/13 \\
Chronos Mini & 1.76 & -0.12 & -2.17 & 0/13 & 3/13 & 0/13 \\
Chronos Small & 2.20 & 0.65 & -0.44 & 5/13 & 2/13 & 0/13 \\
TimesFM 8M & -3.21 & -5.97 & -2.87 & 2/13 & 2/13 & 0/13 \\
TimesFM 20M & -1.98 & -6.17 & -3.54 & 0/13 & 1/13 & 0/13 \\
\addlinespace[2pt]
\textit{All 65 cells} & 0.70 & -0.99 & -2.14 & 9/65 & 10/65 & 0/65 \\
\bottomrule
\end{tabular*}

\begin{tablenotes}[flushleft]
\footnotesize
\item \textit{Notes:} Portfolio performance is measured by annualized certainty-equivalent returns (CERs), annualized Sharpe ratios, and paired CER differences between origin-aligned point-in-time (PIT) and rolling post-origin forecasts. Panel A gives the five U.S. model estimates. Panel B gives model-specific medians and interval counts across 13 non-U.S. markets, followed by the summary for all 65 market--model cells. PIT uses lead $-1$; the post-origin forecast equally averages leads $0,+1,+2$ on strict common support. Both portfolios use the same mean--variance rule with risk aversion $\gamma=3$, market weights in $[0,1.5]$, and variance estimated from returns realized before the target month. CER and $\Delta CER=CER_{\mathrm{Exposed}}-CER_{\mathrm{PIT}}$ are in annualized percentage points; positive differences favor the post-origin forecast. CER levels are summarized by column, whereas median differences are calculated from paired comparisons. Brackets contain 95\% percentile intervals from 1,000 paired circular moving-block bootstrap replications with block length six. Bold differences have intervals excluding zero. Panel B counts markets with intervals entirely below or above zero. Results use the one-month horizon, the US training-information environment, and 252 target-start months from January 2001 through December 2021.
\end{tablenotes}

\end{threeparttable}
\end{table}
\clearpage
\begin{table}[p]
\centering
\begin{threeparttable}
\caption[Forecast Revision Scale and Alignment Efficiency]{%
  \textbf{Forecast revision scale and alignment efficiency}
}
\label{tab:alignment-efficiency}
\renewcommand{\arraystretch}{1.10}
\small
\begin{tabular*}{\textwidth}{@{\extracolsep{\fill}}lccc@{}}
\toprule
& \makecell[c]{PIT-relative pre-origin\\[-0.4ex]\footnotesize PIT vs.\ stale lead $-2$}
& \makecell[c]{Exposed, rolling\\[-0.4ex]\footnotesize leads $0,+1,+2$}
& \makecell[c]{Fixed-state deployment\\[-0.4ex]\footnotesize full support: 2000, 2009, 2023} \\
\midrule
\multicolumn{4}{@{}l}{\textit{Panel A. United States (descriptive)}} \\
\midrule
Alignment efficiency $\kappa$, mean   & $0.136$ & $-0.014$ & $0.006$ \\
Alignment efficiency $\kappa$, median & $0.127$ & $-0.014$ & $0.005$ \\
Revision scale $r$                    & $0.266$ & $0.273$  & $0.348$ \\
Break-even threshold $r/2$            & $0.133$ & $0.137$  & $0.174$ \\
Margin $\kappa - r/2$                 & $0.004$ & $-0.150$ & $-0.168$ \\
Share of cells with net gain          & $0.60$  & $0.12$   & $0.13$ \\
Cells                                 & 20      & 60       & 60 \\
\midrule
\multicolumn{4}{@{}l}{\textit{Panel B. 13 non-U.S. markets}} \\
\midrule
Alignment efficiency $\kappa$, mean   & $0.168$ & $0.024$  & $0.041$ \\
\quad standard error                  & $(0.010)$ & $(0.016)$ & $(0.012)$ \\
Alignment efficiency $\kappa$, median & $0.158$ & $0.024$  & $0.034$ \\
Revision scale $r$                    & $0.308$ & $0.312$  & $0.373$ \\
Break-even threshold $r/2$            & $0.154$ & $0.156$  & $0.186$ \\
Margin $\kappa - r/2$                 & $0.014$ & $-0.132$ & $-0.145$ \\
\quad market-clustered $t$            & $(1.74)$ & $(-10.0)$ & $(-14.7)$ \\
Share of cells with net gain          & $0.53$  & $0.17$   & $0.17$ \\
Cells                                 & 260     & 780      & 780 \\
\bottomrule
\end{tabular*}
\begin{tablenotes}[flushleft]
\footnotesize
\item \textit{Notes:} Revision scale and alignment are calculated relative to the remaining point-in-time (PIT) forecast error. For revision $D$ and PIT error $e$, $r=\sqrt{E[D^2]/E[e^2]}$ and $\kappa=E[eD]/\sqrt{E[e^2]E[D^2]}$. A positive cell-level margin $\kappa-r/2$ indicates lower expected squared loss. The pre-origin column replaces PIT with stale lead $-2$; the rolling-exposed column compares PIT with leads $0,+1,+2$. The fixed-state column uses the full matched deployment support of states 2000, 2009, and 2023, including retrospective-exposure, boundary-crossing, origin-aligned, and stale periods. Rows report means of cell-level $\kappa$, $r$, thresholds, and margins, with medians and net-gain shares shown separately. Panel A cells are model--horizon--comparison combinations; Panel B adds the market dimension. In Panel B, parentheses contain market-clustered standard errors for mean $\kappa$ and $t$-statistics for mean margins. Rows are rounded independently. Estimates use the all-available sample and the US training-information environment.
\end{tablenotes}
\end{threeparttable}
\end{table}
\clearpage
\endgroup

\clearpage
\doublespacing
\begingroup
\makeatletter
\renewcommand{\normalsize}{\@setfontsize\normalsize{11}{13.2}}
\makeatother
\normalsize
\appendix
\renewcommand{\thesection}{IA.\arabic{section}}
\renewcommand{\thesubsection}{\thesection.\arabic{subsection}}
\pagenumbering{arabic}
\renewcommand*{\thepage}{Appendix-\arabic{page}}
\setcounter{page}{1}

\begin{center}
	{\Large\bfseries Internet Appendix\par}
\end{center}

\counterwithin{table}{section}
\counterwithin{figure}{section}
\renewcommand{\thetable}{\thesection.\arabic{table}}
\renewcommand{\thefigure}{\thesection.\arabic{figure}}

\section{Forecast Construction and Empirical Methods}\label{sec:ia-model-provenance}

This section documents the forecast construction, training samples, and dated model states used in main-text Section~\ref{markets-annual-tsfm-states-and-matched-design}.

\subsection{From Model Outputs to Return Forecasts}\label{sec:ia-model-forecast-construction}

Chronos maps mean-scaled continuous observations into discrete tokens and processes the resulting sequence with a T5-based autoregressive Transformer \citep{ansari_etal_2024}. For each forecast unit and annual parameter state, the model generates \(S\) simulated paths, where \(S\) is the number of paths used in the corresponding forecast-generation setting. Let \(\widehat r^{(v,s)}_{u,j}\) denote forecast step \(j\) from path \(s\) under annual parameter state \(v\). The point forecast used in the empirical analysis is

\[
\widehat y_{u,\mathrm{Chronos}}^{(v)}
=
\frac{1}{S}\sum_{s=1}^{S}
\left(
\frac{1}{h}\sum_{j=1}^{h}
\widehat r^{(v,s)}_{u,j}
\right).
\]

The same construction is applied to the Tiny, Mini, and Small variants. Each variant has a maximum context length of 512 steps and a native forecast horizon of 64 steps \citep{ansari_etal_2024}.

TimesFM maps continuous observations into nonoverlapping patches and processes the patch sequence with a causal decoder-only Transformer \citep{das_kong_sen_zhou_2024}. Let \(\widehat\mu^{(v)}_{u,j}\) denote the native mean forecast at step \(j\). The empirical point forecast is

\[
\widehat y_{u,\mathrm{TimesFM}}^{(v)}
=
\frac{1}{h}\sum_{j=1}^{h}
\widehat\mu^{(v)}_{u,j}.
\]

The financial 8M and 20M variants are the scaled-down TimesFM 2 models of \citet{rahimikia_ni_wang_2025} and share this forecast construction. The released configurations used here specify a 512-step context and a 128-step output horizon. Native quantile outputs are retained as distributional forecast objects, while the mean-step path supplies the point forecast used in matched-loss comparisons. The empirical horizons of 1, 3, 6, and 12 months lie within the native forecast horizon of both model families, and the monthly histories used in the analysis lie within their common context capacity.

\subsection{Training Markets and Information Channels}\label{sec:ia-channel-assignment}

The released annual parameter states are trained separately on expanding financial samples through each state year \citep{rahimikia_ni_wang_2025}. The US environment contains U.S. firm-level excess-return series; Global adds firm-level excess returns from non-U.S. markets; Augmented adds JKP factor series to the Global return sample \citep{jensen_kelly_pedersen_2023}. These factors span investment, value, low risk, quality, seasonality, momentum, profitability, and related firm characteristics, supplying an additional predictor-based information channel.

Exposure channels are assigned relative to the prediction target. For the U.S. equity-premium target, US supplies direct target-market information, while Global and Augmented combine that direct channel with information from other markets and factors. For a selected non-U.S. market, US supplies indirect-only information because the target market is absent from its training sample. Global and Augmented include the target market and related external information and therefore supply direct-plus-indirect exposure. This target-specific assignment is applied uniformly across architectures and annual parameter states.

\subsection{Model Vintages and Training Cutoffs}\label{sec:ia-parameter-registry}

Each parameter state is identified by the tuple \((a,c,v)\) and linked to a model repository identifier and an immutable model revision. The registry uses the same architecture and training-information-environment labels across rolling and fixed designs, so forecasts can be matched without changing the model family or training-information environment.

For training environment \(c\), the year-\(v\) parameter state is estimated on the expanding corpus

\[
\mathcal D_{c,v}
=
\left\{z\in\mathcal D_c:
\operatorname{date}(z)\leq \operatorname{December}(v)
\right\},
\]
and is assigned an empirical origin-admissibility date at the beginning of year \(v+1\). If \(Y(t)\) denotes the calendar year containing the target-start month, origin-relative status is

\[
\operatorname{status}\!\left(v,Y(t)\right)
=
\begin{cases}
\text{stale}, & v<Y(t)-1,\\
\text{origin-aligned PIT}, & v=Y(t)-1,\\
\text{boundary-exposed}, & v=Y(t),\\
\text{post-origin exposed}, & v>Y(t).
\end{cases}
\]

The rolling design applies this year-based origin-status rule to \(v=Y(t)+\ell\), with \(\ell\in\{-2,-1,0,+1,+2\}\). The fixed design applies the annual states \(v\in\{2000,2009,2023\}\) to every historical origin and classifies their timing from the exact target-window boundaries defined in Section~\ref{sec:ia-timing-designs}. The two conventions express the same annual-state--target ordering at the temporal resolution required by each design.

\subsection{Matched Forecasts and Timing Classifications}\label{sec:ia-timing-designs}

Forecasts are matched by exact equality of the market or panel identifier, target-start date, target-end date, horizon, architecture, training-information environment, and numerical-history identity. These keys operationalize the canonical forecast unit \(u\). Within each matched unit, an alternative parameter state is linked to the unique origin-aligned PIT state \(\theta_{a,c,Y(t_u)-1}\).

For the rolling design and \(\ell\in\{-2,0,+1,+2\}\), define the lead-specific pairwise support as

\[
\mathcal P_{\ell}
=
\left\{
u:
\widehat y_u^{(Y(t_u)+\ell)}
\text{ and }
\widehat y_u^{(Y(t_u)-1)}
\text{ are jointly available}
\right\}.
\]

Every lead-specific comparison is computed on \(\mathcal P_{\ell}\). The pooled later-state comparison uses

\[
\mathcal P_{\mathrm{pool}}
=
\mathcal P_{0}
\cap
\mathcal P_{+1}
\cap
\mathcal P_{+2},
\]

which holds the forecast units fixed across the boundary-exposed and post-origin annual states.

For a fixed annual state \(v\in\{2000,2009,2023\}\), define the fixed-to-PIT pairwise support as

\[
\mathcal F_v
=
\left\{
u:
\widehat y_u^{(v)}
\text{ and }
\widehat y_u^{(Y(t_u)-1)}
\text{ are jointly available}
\right\}.
\]

Each fixed-to-PIT comparison uses \(\mathcal F_v\). Direct comparisons among common fixed states use the triple-common support

\[
\mathcal F_{\mathrm{triple}}
=
\mathcal F_{2000}
\cap
\mathcal F_{2009}
\cap
\mathcal F_{2023}.
\]

For the fixed design, let \(W_u=[s_u,b_u]\) be the complete target window and let \(\tau_v\) be December 31 of annual-state year \(v\). The target-window timing class is

\[
q(u,v)
=
\begin{cases}
\text{retrospective exposure}, & b_u\leq \tau_v,\\
\text{boundary crossing}, & s_u\leq \tau_v<b_u,\\
\text{post-cutoff deployment}, & s_u>\tau_v.
\end{cases}
\]

The classification is applied separately at each horizon because \(b_u\) moves with \(h\). The post-cutoff class combines the first origin-aligned PIT deployment year with subsequent stale deployments. Lead-specific and fixed-to-PIT statistics use their pairwise supports, while pooled lead and cross-state statistics use the corresponding common-support intersections above.

\subsection{Forecast Evaluation and Portfolio Construction}\label{sec:ia-estimands-portfolios}

For any matched support \(\mathcal S\), write

\[
\overline z_{\mathcal S}
=
\frac{1}{|\mathcal S|}
\sum_{u\in\mathcal S}z_u.
\]

All normalized revision and matched-loss quantities use the same support-specific historical-average scale. This convention preserves the economic units of the target while expressing forecast movement and predictive value relative to the real-time benchmark.

\subsubsection*{Measures of Forecast Revision}\label{revision-summaries}

For alternative annual state \(v\), normalized mean absolute forecast movement is

\[
M_{\mathcal S}^{(v)}
=
\frac{
\overline{|D_u(v)|}_{\mathcal S}
}{
\sqrt{\overline{(e_u^{\mathrm{HA}})^2}_{\mathcal S}}
}.
\]

Exhibits stated in percentage units report \(100M_{\mathcal S}^{(v)}\).

Forecast comovement is the Pearson correlation

\[
\rho_{\mathcal S}^{(v)}
=
\operatorname{Corr}_{u\in\mathcal S}
\left(
\widehat y_u^{\mathrm{PIT}},
\widehat y_u(v)
\right),
\]
and the sign-flip share is

\[
\operatorname{Flip}_{\mathcal S}^{(v)}
=
\frac{1}{|\mathcal S|}
\sum_{u\in\mathcal S}
\mathbf 1
\left\{
\mathbf 1(\widehat y_u^{\mathrm{PIT}}<0)
\neq
\mathbf 1(\widehat y_u(v)<0)
\right\}.
\]

For Chronos, let \(F_{u,\mathrm{PIT}}\) and \(F_{u,v}\) be the empirical distributions of horizon-averaged predictive paths for the matched PIT and alternative forecasts. Their model-native movement is

\[
W_{1,u}^{(v)}
=
\int_0^1
\left|
F_{u,v}^{-1}(p)
-
F_{u,\mathrm{PIT}}^{-1}(p)
\right|
\,dp.
\]

The reported cell-level Wasserstein statistic is the mean of \(W_{1,u}^{(v)}\) over the relevant matched support. It captures movement in the location, dispersion, and shape of the predictive path distribution.

For TimesFM, model-native movement is computed from the forecast quantile vector \(\mathcal Q=\{0.10,0.50,0.90\}\):

\[
Q_{1,u}^{(v)}
=
\frac{1}{3}
\sum_{p\in\mathcal Q}
\left|
\widehat q_{u,p}^{(v)}
-
\widehat q_{u,p}^{\mathrm{PIT}}
\right|,
\]
with interval-width revision

\[
\Delta IW_u^{(v)}
=
\left(
\widehat q_{u,0.90}^{(v)}
-
\widehat q_{u,0.10}^{(v)}
\right)
-
\left(
\widehat q_{u,0.90}^{\mathrm{PIT}}
-
\widehat q_{u,0.10}^{\mathrm{PIT}}
\right).
\]

Chronos path-distribution distances and TimesFM quantile-vector distances retain their model-family-specific scales.

\subsubsection*{Aggregating Predictive Effects across Backtests}\label{design-specific-predictive-value-aggregation}

For rolling lead \(\ell\), the lead-specific effect is \(N_{\mathcal P_{\ell}}^{(\ell)}\). The pooled later-state effect first places PIT and all three later leads on \(\mathcal P_{\mathrm{pool}}\), then averages the three unit-level PIT-relative loss advantages:

\[
N_{\mathrm{pool}}
=
100
\frac{
\displaystyle
\sum_{u\in\mathcal P_{\mathrm{pool}}}
\frac{1}{3}
\sum_{\ell\in\{0,+1,+2\}}
\left[
(e_u^{\mathrm{PIT}})^2-(e_u^{(\ell)})^2
\right]
}{
\displaystyle
\sum_{u\in\mathcal P_{\mathrm{pool}}}
(e_u^{\mathrm{HA}})^2
}.
\]

For fixed annual state \(v\), the pairwise effect is \(N_{\mathcal F_v}^{(v)}\). The signed exposure depth measures the annual state's distance from the origin-aligned PIT year:

\[
g_u^{(v)}
=
v-[Y(t_u)-1].
\]

It is combined with the exact target-window class \(q(u,v)\) from Section~\ref{sec:ia-timing-designs}. The alignment profile uses the integer bins \(\{\leq-16,-15{:}{-}6,-5{:}{-}2,-1{:}0,1{:}5,6{:}15,\geq16\}\). Within each market--architecture--training-environment--horizon cell, unit-level PIT-relative loss advantages are scaled by that cell's historical-average MSFE and then averaged within signed-depth bins.

\subsubsection*{Market-Timing Signals and Portfolio Returns}\label{investor-forecasts-and-portfolio-returns}

The investor sample uses the US training-information environment and the strict joint support of rolling leads \(0,+1,+2\). For each market, architecture, horizon, and target month, the matched source contains one common PIT forecast, historical-average forecast, and realized target. In this subsection, \(t\) indexes the target month, whose forecast origin is month-end \(t-1\). The exposed forecast is the equal-weight mean

\[
\widehat r_{t,\mathrm{Exposed}}
=
\frac{1}{3}
\left(
\widehat r_t^{(0)}
+
\widehat r_t^{(+1)}
+
\widehat r_t^{(+2)}
\right).
\]

Let \(\mathcal I_{m,t-1}\) contain market \(m\)'s realized monthly excess returns strictly before target month \(t\). The real-time risk estimate is the sample variance

\[
\widehat\sigma_{m,t}^2
=
\operatorname{Var}
\left(
r_{m,\tau}^{e}:\tau<t
\right),
\]

formed from at least 12 prior observations. PIT, Exposed, and historical-average strategies use this common variance estimate. For risk aversion \(\gamma\), the raw risky-asset position is \(\widehat r_{t,s}/(\gamma\widehat\sigma_{m,t}^2)\); the headline constrained strategy projects it onto \([0,1.5]\).

At horizon \(h\), each forecast-origin tranche invests one-\(h\)th of wealth for \(h\) months. Once all tranches are active, the aggregate position and portfolio excess return are

\[
\overline w_{t,s}^{(h)}
=
\frac{1}{h}
\sum_{j=0}^{h-1}w_{t-j,s},
\qquad
r_{p,t,s}^{(h)}
=
\overline w_{t,s}^{(h)}r_{m,t}^{e}.
\]

The common January 2001--December 2021 target-month grid contains 252 observations per market--architecture--horizon cell. Complete-tranche portfolio returns number 252, 250, 247, and 241 months for horizons of 1, 3, 6, and 12 months, respectively. The one-month, \(\gamma=3\), constrained-weight specification is the headline investor endpoint.

For a monthly portfolio-return sequence, annualized certainty-equivalent return and Sharpe ratio are

\[
CER_s
=
1200
\left[
\overline r_{p,s}
-
\frac{\gamma}{2}
\widehat{\operatorname{Var}}(r_{p,s})
\right],
\qquad
SR_s
=
\sqrt{12}
\frac{
\overline r_{p,s}
}{
\widehat\sigma(r_{p,s})
}.
\]

The investor-value contrast is \(\Delta CER=CER_{\mathrm{Exposed}}-CER_{\mathrm{PIT}}\). The risk-aversion grid uses \(\gamma\in\{1,3,5\}\) with constrained and unconstrained positions at the one-month horizon; the longer-horizon implementation uses \(\gamma=3\) and constrained positions. Section 3.5 and Section~\ref{sec:ia-support-inference} define the corresponding sampling and inferential procedures.

\subsection{Evaluation Samples and Statistical Inference}\label{sec:ia-support-inference}

\subsubsection*{Matched Return Samples}\label{sample-and-support-definitions}

For estimand \(j\), the all-available support is
\[
\mathcal S_j^{\mathrm{AA}}
=
\left\{
u:
\text{every forecast, benchmark, and outcome required by }j
\text{ is available for }u
\right\}.
\]

The common-calendar counterpart is
\[
\mathcal S_j^{\mathrm{CC}}
=
\mathcal S_j^{\mathrm{AA}}
\cap
\left\{
u:
\text{January 2002}
\leq s_u\leq
\text{December 2021}
\right\},
\]
where \(s_u\) is the target-start month. The PIT-versus-HA comparison uses the intersection of the PIT, historical-average, and realized-target series. Rolling and fixed comparisons use \(\mathcal P_{\ell}\), \(\mathcal P_{\mathrm{pool}}\), \(\mathcal F_v\), and \(\mathcal F_{\mathrm{triple}}\) as defined in Section~\ref{sec:ia-timing-designs}. The investor support is \(\mathcal P_{\mathrm{pool}}\) restricted to the US training-information environment and target months from January 2001 through December 2021, as defined in Section~\ref{sec:ia-estimands-portfolios}.

Support is formed separately by empirical environment, market, architecture, training-information environment, horizon, and timing comparison. The U.S. series and the 13 international market series enter separate result families.

\subsubsection*{Inference for Normalized Forecast Loss}\label{ratio-influence-function-and-time-series-inference}

Let

\[
\widehat\theta
=
100\frac{\overline d}{\overline b},
\qquad
\widehat\psi_t
=
\frac{100}{\overline b}
\left(
d_t
-
\frac{\widehat\theta}{100}b_t
\right),
\]
where \(d_t\) is the relevant monthly loss advantage and \(b_t=(e_t^{\mathrm{HA}})^2\). For a target-month series of length \(T\), define

\[
\widehat\Gamma_k
=
\frac{1}{T}
\sum_{t=k+1}^{T}
\left(
\widehat\psi_t-\overline{\widehat\psi}
\right)
\left(
\widehat\psi_{t-k}-\overline{\widehat\psi}
\right).
\]

The Bartlett--Newey--West standard error is

\[
\widehat{\operatorname{se}}_{\mathrm{HAC}}(\widehat\theta)
=
\left[
\frac{1}{T}
\left\{
\widehat\Gamma_0
+
2
\sum_{k=1}^{h-1}
\left(1-\frac{k}{h}\right)
\widehat\Gamma_k
\right\}
\right]^{1/2}.
\]

The 95\% HAC interval is \(\widehat\theta\pm1.959964\,\widehat{\operatorname{se}}_{\mathrm{HAC}}(\widehat\theta)\). Predictive estimands also use 2,000 circular moving-block replications with block length \(L=\max(h,6)\). Every replication forms target-month blocks with wraparound and recomputes \(100\overline d^{*}/\overline b^{*}\); the 2.5th and 97.5th percentiles provide the bootstrap interval.

\subsubsection*{Inference for Predictive and Investor Value}\label{endpoint-specific-procedures}

For the PIT-versus-HA comparison, the raw monthly loss advantage is

\[
d_t^{\mathrm{raw}}
=
(e_t^{\mathrm{HA}})^2
-
(e_t^{\mathrm{PIT}})^2.
\]

The Clark--West adjusted loss advantage is

\[
d_t^{\mathrm{CW}}
=
(e_t^{\mathrm{HA}})^2
-
\left[
(e_t^{\mathrm{PIT}})^2
-
\left(
\widehat y_t^{\mathrm{PIT}}
-
\widehat y_t^{\mathrm{HA}}
\right)^2
\right].
\]

The one-sided Clark--West diagnostic uses a normal approximation with the Bartlett--Newey--West standard error of \(\overline d^{\mathrm{CW}}\) and lag \(h-1\). Forecast-revision and model-native movement statistics are aggregated on their exact matched supports. Rolling effects, fixed effects, and loss-accounting components use their target-month sequences with the same horizon-specific HAC and moving-block settings. Alignment, movement penalty, and net loss improvement use common resampled target months in each loss-accounting draw.

Fixed timing-state cells are designated for formal HAC and block-bootstrap intervals when their support contains at least 24 target months; every timing-state cell retains its point estimate and exact \(N\). Cellwise predictive p-values are supplemented by Benjamini--Hochberg q-values within protocol-defined families \citep{benjamini_hochberg_1995}. Rolling market-level families are defined by market, training-information environment, and sample. For fixed-vintage results, each BH family holds the market, training-information environment, sample definition, and result family fixed and pools the model, fixed-vintage, and horizon cells. Equal-market panel families are defined by training-information environment and sample.

Investor inference uses 1,000 paired circular moving-block replications with block length six. Each draw applies common month indices to the PIT, Exposed, and historical-average portfolio-return sequences and recomputes their CER levels and contrasts. The 2.5th and 97.5th percentiles of the resulting \(\Delta CER\) distribution form its 95\% interval.

\subsubsection*{International Aggregation and Cross-Design Comparisons}\label{international-aggregation-and-cross-design-classification}

For market \(m\), monthly loss advantages are normalized by that market's historical-average MSFE:

\[
z_{m,t}
=
100
\frac{
d_{m,t}
}{
\overline{(e_{m,t}^{\mathrm{HA}})^2}
}.
\]

Market-level effects are computed before the cross-market distribution is summarized. The equal-market pooled coefficient is

\[
\widehat\beta^{\mathrm{F13}}
=
\frac{1}{13}
\sum_{m=1}^{13}
\overline z_m.
\]

Its panel representation assigns each market total weight \(1/13\). Inference reports market-clustered, target-month-clustered, and two-way-clustered standard errors \citep{cameron_gelbach_miller_2011}. The Webb six-point wild-cluster bootstrap applies 9,999 replications to the 13 market-level effects \citep{webb_2023}. Distributional summaries report the median, interquartile range, positive-market count, and the number of markets whose HAC and block-bootstrap intervals lie on the same side of zero.

For architecture \(a\) and horizon \(h\), define the cross-design vector

\[
\mathbf C_{a,h}
=
\left(
N_{\mathrm{pool},a,h}^{\mathrm{R}},
N_{2009,a,h}^{\mathrm{F}},
N_{2023,a,h}^{\mathrm{F}}
\right).
\]

Three positive entries define a common gain and three negative entries define a common loss. Opposite signs for the two fixed effects define a fixed-vintage mixed pattern. A common sign for the two fixed effects combined with the opposite rolling sign defines a protocol disagreement. Joint interval support is recorded when both the HAC and circular moving-block intervals for an effect lie on the same side of zero.

\clearpage
\section{Point-in-Time Return Predictability}\label{sec:ia-pit-results}

The following figure and tables extend the benchmark evidence in main-text Sections~\ref{predictive-content-of-the-origin-aligned-forecast} and~\ref{the-point-in-time-benchmark-across-training-environments} across training samples, individual markets, and the common evaluation window.
\begin{figure}[!htbp]
\centering
\includegraphics[width=0.95\textwidth]{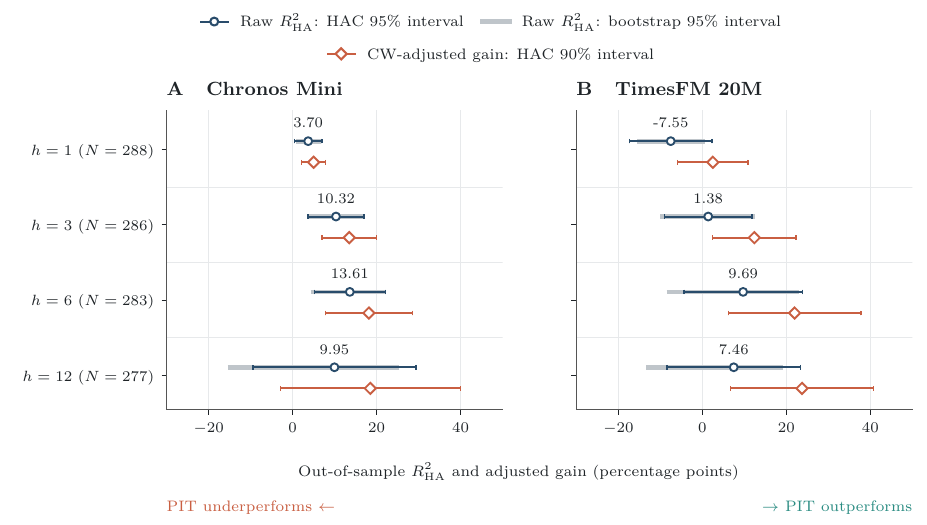}
\caption[Point-in-time forecasting performance and Clark--West-adjusted gains]{%
  \textbf{Point-in-time forecasting performance and Clark--West-adjusted gains.}
  Origin-aligned point-in-time (PIT) forecasts are compared with the expanding historical average for Chronos Mini (Panel A) and TimesFM 20M (Panel B) at horizons $h=1,3,6,12$ months. Navy circles give raw out-of-sample $R^2_{\mathrm{HA}}$, the percentage-point reduction in mean squared forecast error (MSFE). Thin navy lines give 95\% Bartlett--Newey--West HAC intervals; thick light-gray lines give 95\% percentile intervals from 2,000 circular moving-block bootstrap replications with block length $\max(h,6)$. Terracotta diamonds give $100\overline d_{\mathrm{CW}}/\overline{(e^{\mathrm{HA}})^2}$ with two-sided 90\% HAC intervals, where $d_{\mathrm{CW},t}=(e_t^{\mathrm{HA}})^2-(e_t^{\mathrm{PIT}})^2+(\widehat y_t^{\mathrm{PIT}}-\widehat y_t^{\mathrm{HA}})^2$. A 90\% adjusted interval entirely above zero corresponds to rejection by the one-sided 5\% Clark--West test. HAC inference uses $h-1$ lags. Estimates use PIT lead $-1$, the all-available U.S. equity-premium sample, and the US training-information environment. Matched target-month counts accompany the horizon labels.
}
\label{fig:x41-pit-relevance}

\end{figure}
\FloatBarrier

\begin{table}[!htbp]
\centering
\begin{threeparttable}
\caption[U.S. point-in-time forecasting performance by training environment]{%
  \textbf{U.S. point-in-time forecasting performance by training environment.}
}
\label{tab:ia-x41-us-corpora}
\renewcommand{\arraystretch}{1.08}
\small
\begin{tabular*}{\textwidth}{@{\extracolsep{\fill}}lcccc@{}}
\toprule
Model & $h=1$ & $h=3$ & $h=6$ & $h=12$ \\
\midrule
\multicolumn{5}{@{}l}{\textit{Panel A: US training-information environment}} \\
\addlinespace[0.15em]
Chronos Tiny & \makecell[c]{\ensuremath{3.45^{***}}\\[-0.35ex]{\scriptsize \ensuremath{(0.005)}}} & \makecell[c]{\ensuremath{9.26^{***}}\\[-0.35ex]{\scriptsize \ensuremath{(0.002)}}} & \makecell[c]{\ensuremath{15.12^{***}}\\[-0.35ex]{\scriptsize \ensuremath{(0.002)}}} & \makecell[c]{\ensuremath{14.31^{**}}\\[-0.35ex]{\scriptsize \ensuremath{(0.040)}}} \\
Chronos Mini & \makecell[c]{\ensuremath{3.70^{***}}\\[-0.35ex]{\scriptsize \ensuremath{(0.002)}}} & \makecell[c]{\ensuremath{10.32^{***}}\\[-0.35ex]{\scriptsize \ensuremath{(<0.001)}}} & \makecell[c]{\ensuremath{13.61^{***}}\\[-0.35ex]{\scriptsize \ensuremath{(0.002)}}} & \makecell[c]{\ensuremath{9.95^{*}}\\[-0.35ex]{\scriptsize \ensuremath{(0.078)}}} \\
Chronos Small & \makecell[c]{\ensuremath{0.29}\\[-0.35ex]{\scriptsize \ensuremath{(0.159)}}} & \makecell[c]{\ensuremath{6.00^{**}}\\[-0.35ex]{\scriptsize \ensuremath{(0.014)}}} & \makecell[c]{\ensuremath{10.53^{***}}\\[-0.35ex]{\scriptsize \ensuremath{(0.008)}}} & \makecell[c]{\ensuremath{13.71^{**}}\\[-0.35ex]{\scriptsize \ensuremath{(0.029)}}} \\
\addlinespace[0.25em]
TimesFM 8M & \makecell[c]{\ensuremath{-9.83}\\[-0.35ex]{\scriptsize \ensuremath{(0.443)}}} & \makecell[c]{\ensuremath{-5.65}\\[-0.35ex]{\scriptsize \ensuremath{(0.296)}}} & \makecell[c]{\ensuremath{2.15^{**}}\\[-0.35ex]{\scriptsize \ensuremath{(0.043)}}} & \makecell[c]{\ensuremath{9.19^{***}}\\[-0.35ex]{\scriptsize \ensuremath{(0.005)}}} \\
TimesFM 20M & \makecell[c]{\ensuremath{-7.55}\\[-0.35ex]{\scriptsize \ensuremath{(0.313)}}} & \makecell[c]{\ensuremath{1.38^{**}}\\[-0.35ex]{\scriptsize \ensuremath{(0.020)}}} & \makecell[c]{\ensuremath{9.69^{**}}\\[-0.35ex]{\scriptsize \ensuremath{(0.011)}}} & \makecell[c]{\ensuremath{7.46^{**}}\\[-0.35ex]{\scriptsize \ensuremath{(0.011)}}} \\
\addlinespace[0.45em]
\midrule
\addlinespace[0.25em]
\multicolumn{5}{@{}l}{\textit{Panel B: Global training-information environment}} \\
\addlinespace[0.15em]
Chronos Tiny & \makecell[c]{\ensuremath{-0.44}\\[-0.35ex]{\scriptsize \ensuremath{(0.155)}}} & \makecell[c]{\ensuremath{-0.17^{*}}\\[-0.35ex]{\scriptsize \ensuremath{(0.098)}}} & \makecell[c]{\ensuremath{2.59^{*}}\\[-0.35ex]{\scriptsize \ensuremath{(0.071)}}} & \makecell[c]{\ensuremath{4.27^{*}}\\[-0.35ex]{\scriptsize \ensuremath{(0.077)}}} \\
Chronos Mini & \makecell[c]{\ensuremath{-1.65}\\[-0.35ex]{\scriptsize \ensuremath{(0.163)}}} & \makecell[c]{\ensuremath{-0.93^{*}}\\[-0.35ex]{\scriptsize \ensuremath{(0.063)}}} & \makecell[c]{\ensuremath{-0.41^{*}}\\[-0.35ex]{\scriptsize \ensuremath{(0.056)}}} & \makecell[c]{\ensuremath{2.97^{**}}\\[-0.35ex]{\scriptsize \ensuremath{(0.019)}}} \\
Chronos Small & \makecell[c]{\ensuremath{-1.90}\\[-0.35ex]{\scriptsize \ensuremath{(0.152)}}} & \makecell[c]{\ensuremath{-9.00}\\[-0.35ex]{\scriptsize \ensuremath{(0.219)}}} & \makecell[c]{\ensuremath{-4.71^{*}}\\[-0.35ex]{\scriptsize \ensuremath{(0.089)}}} & \makecell[c]{\ensuremath{5.83^{**}}\\[-0.35ex]{\scriptsize \ensuremath{(0.045)}}} \\
\addlinespace[0.25em]
TimesFM 8M & \makecell[c]{\ensuremath{-20.31}\\[-0.35ex]{\scriptsize \ensuremath{(0.812)}}} & \makecell[c]{\ensuremath{-50.49}\\[-0.35ex]{\scriptsize \ensuremath{(0.684)}}} & \makecell[c]{\ensuremath{-72.48}\\[-0.35ex]{\scriptsize \ensuremath{(0.712)}}} & \makecell[c]{\ensuremath{-38.79}\\[-0.35ex]{\scriptsize \ensuremath{(0.417)}}} \\
TimesFM 20M & \makecell[c]{\ensuremath{-9.00}\\[-0.35ex]{\scriptsize \ensuremath{(0.123)}}} & \makecell[c]{\ensuremath{2.66^{**}}\\[-0.35ex]{\scriptsize \ensuremath{(0.011)}}} & \makecell[c]{\ensuremath{-2.45}\\[-0.35ex]{\scriptsize \ensuremath{(0.121)}}} & \makecell[c]{\ensuremath{-26.85}\\[-0.35ex]{\scriptsize \ensuremath{(0.508)}}} \\
\addlinespace[0.45em]
\midrule
\addlinespace[0.25em]
\multicolumn{5}{@{}l}{\textit{Panel C: Augmented training-information environment}} \\
\addlinespace[0.15em]
Chronos Tiny & \makecell[c]{\ensuremath{1.36^{*}}\\[-0.35ex]{\scriptsize \ensuremath{(0.085)}}} & \makecell[c]{\ensuremath{2.25^{**}}\\[-0.35ex]{\scriptsize \ensuremath{(0.033)}}} & \makecell[c]{\ensuremath{4.74^{**}}\\[-0.35ex]{\scriptsize \ensuremath{(0.029)}}} & \makecell[c]{\ensuremath{8.64^{**}}\\[-0.35ex]{\scriptsize \ensuremath{(0.023)}}} \\
Chronos Mini & \makecell[c]{\ensuremath{0.90^{*}}\\[-0.35ex]{\scriptsize \ensuremath{(0.056)}}} & \makecell[c]{\ensuremath{3.35^{**}}\\[-0.35ex]{\scriptsize \ensuremath{(0.014)}}} & \makecell[c]{\ensuremath{8.01^{***}}\\[-0.35ex]{\scriptsize \ensuremath{(0.005)}}} & \makecell[c]{\ensuremath{10.95^{***}}\\[-0.35ex]{\scriptsize \ensuremath{(0.009)}}} \\
Chronos Small & \makecell[c]{\ensuremath{-0.54}\\[-0.35ex]{\scriptsize \ensuremath{(0.123)}}} & \makecell[c]{\ensuremath{0.28^{*}}\\[-0.35ex]{\scriptsize \ensuremath{(0.065)}}} & \makecell[c]{\ensuremath{7.49^{***}}\\[-0.35ex]{\scriptsize \ensuremath{(0.008)}}} & \makecell[c]{\ensuremath{15.87^{***}}\\[-0.35ex]{\scriptsize \ensuremath{(0.003)}}} \\
\addlinespace[0.25em]
TimesFM 8M & \makecell[c]{\ensuremath{-11.31}\\[-0.35ex]{\scriptsize \ensuremath{(0.800)}}} & \makecell[c]{\ensuremath{-2.66}\\[-0.35ex]{\scriptsize \ensuremath{(0.279)}}} & \makecell[c]{\ensuremath{-2.90}\\[-0.35ex]{\scriptsize \ensuremath{(0.224)}}} & \makecell[c]{\ensuremath{-3.34}\\[-0.35ex]{\scriptsize \ensuremath{(0.164)}}} \\
TimesFM 20M & \makecell[c]{\ensuremath{-68336.63}\\[-0.35ex]{\scriptsize \ensuremath{(0.892)}}} & \makecell[c]{\ensuremath{-46263.12}\\[-0.35ex]{\scriptsize \ensuremath{(0.955)}}} & \makecell[c]{\ensuremath{-23078.85}\\[-0.35ex]{\scriptsize \ensuremath{(0.903)}}} & \makecell[c]{\ensuremath{-3070.73}\\[-0.35ex]{\scriptsize \ensuremath{(0.844)}}} \\
\bottomrule
\end{tabular*}

\par\vspace{6pt}\nointerlineskip
\exhibitnotes{%
  U.S. point-in-time (PIT) forecasting performance is reported under the US, Global, and Augmented training-information environments in Panels A, B, and C. Entries are raw out-of-sample $R^2_{\mathrm{HA}}$, the percentage-point reduction in mean squared forecast error relative to the expanding historical average. Parentheses contain one-sided Clark--West $p$-values. Because the test uses an adjusted loss differential, significance can accompany a negative raw value. $^{*}$, $^{**}$, and $^{***}$ indicate significance at 10\%, 5\%, and 1\%. Estimates use PIT lead $-1$ and the all-available U.S. sample. Bartlett--Newey--West HAC inference uses $h-1$ lags, where $h$ is the horizon in months.
}
\end{threeparttable}
\end{table}

\clearpage
\begingroup
\begin{landscape}
\setlength{\tabcolsep}{6.0pt}
\fontsize{11}{13.6}\selectfont
\setlength{\tabcolsep}{4.00pt}
\setlength{\linewidth}{547.33pt}
\begin{longtable}{@{}lllcccc@{}}
\caption[International point-in-time forecasting performance by market and training environment]{%
  \textbf{International point-in-time forecasting performance by market and training environment.}
}
\label{tab:ia-x41-international-corpora}\\
\multicolumn{7}{@{}p{\linewidth}@{}}{\exhibitnotes{%
  International point-in-time (PIT) forecasting performance is reported for the US, Global, and Augmented training-information environments. Each block begins with the equal-market aggregate and then gives estimates for the 13 individual non-U.S. markets. Entries are raw out-of-sample $R^2_{\mathrm{HA}}$ in percentage points, measuring the reduction in mean squared forecast error (MSFE) relative to the expanding historical average. Parentheses contain one-sided Clark--West $p$-values based on adjusted loss differentials. For the aggregate, each market's monthly raw and adjusted loss differentials are first normalized by its historical-average MSFE and then averaged equally over common target months. $^{*}$, $^{**}$, and $^{***}$ indicate Clark--West significance at 10\%, 5\%, and 1\%. Estimates use PIT lead $-1$ and the all-available sample. Bartlett--Newey--West HAC inference uses $h-1$ lags, where $h$ is in months.
}}\\
\noalign{\vskip 6pt}
\toprule
Training environment & Market & Model & $h=1$ & $h=3$ & $h=6$ & $h=12$ \\
\midrule
\endfirsthead
\multicolumn{7}{@{}l}{\tablename\ \thetable\ continued}\\[0.25em]
\toprule
Training environment & Market & Model & $h=1$ & $h=3$ & $h=6$ & $h=12$ \\
\midrule
\endhead
\midrule
\multicolumn{7}{r}{Continued on next page}\\
\endfoot
\bottomrule
\endlastfoot
US & Equal-weight aggregate & Chronos Tiny & \makecell[c]{\ensuremath{1.77^{***}}\\[-0.35ex]{\scriptsize \ensuremath{(0.003)}}} & \makecell[c]{\ensuremath{4.55^{***}}\\[-0.35ex]{\scriptsize \ensuremath{(0.003)}}} & \makecell[c]{\ensuremath{9.08^{***}}\\[-0.35ex]{\scriptsize \ensuremath{(0.002)}}} & \makecell[c]{\ensuremath{12.82^{**}}\\[-0.35ex]{\scriptsize \ensuremath{(0.017)}}} \\
US & Equal-weight aggregate & Chronos Mini & \makecell[c]{\ensuremath{1.76^{***}}\\[-0.35ex]{\scriptsize \ensuremath{(0.005)}}} & \makecell[c]{\ensuremath{5.56^{***}}\\[-0.35ex]{\scriptsize \ensuremath{(<0.001)}}} & \makecell[c]{\ensuremath{9.18^{***}}\\[-0.35ex]{\scriptsize \ensuremath{(0.001)}}} & \makecell[c]{\ensuremath{12.91^{**}}\\[-0.35ex]{\scriptsize \ensuremath{(0.018)}}} \\
US & Equal-weight aggregate & Chronos Small & \makecell[c]{\ensuremath{-0.99}\\[-0.35ex]{\scriptsize \ensuremath{(0.130)}}} & \makecell[c]{\ensuremath{0.60^{**}}\\[-0.35ex]{\scriptsize \ensuremath{(0.020)}}} & \makecell[c]{\ensuremath{6.38^{***}}\\[-0.35ex]{\scriptsize \ensuremath{(0.003)}}} & \makecell[c]{\ensuremath{10.60^{**}}\\[-0.35ex]{\scriptsize \ensuremath{(0.017)}}} \\
\addlinespace[0.18em]
US & Equal-weight aggregate & TimesFM 8M & \makecell[c]{\ensuremath{-25.25}\\[-0.35ex]{\scriptsize \ensuremath{(0.682)}}} & \makecell[c]{\ensuremath{-12.23}\\[-0.35ex]{\scriptsize \ensuremath{(0.137)}}} & \makecell[c]{\ensuremath{-0.71^{**}}\\[-0.35ex]{\scriptsize \ensuremath{(0.011)}}} & \makecell[c]{\ensuremath{5.06^{***}}\\[-0.35ex]{\scriptsize \ensuremath{(0.004)}}} \\
US & Equal-weight aggregate & TimesFM 20M & \makecell[c]{\ensuremath{-14.53}\\[-0.35ex]{\scriptsize \ensuremath{(0.459)}}} & \makecell[c]{\ensuremath{-10.01^{**}}\\[-0.35ex]{\scriptsize \ensuremath{(0.018)}}} & \makecell[c]{\ensuremath{-6.55^{**}}\\[-0.35ex]{\scriptsize \ensuremath{(0.044)}}} & \makecell[c]{\ensuremath{-31.11}\\[-0.35ex]{\scriptsize \ensuremath{(0.317)}}} \\
\end{longtable}
\end{landscape}

\endgroup
\clearpage
\begin{table}[!htbp]
\centering
\begin{threeparttable}
\caption[Point-in-time forecasting performance in the balanced sample]{%
  \textbf{Point-in-time forecasting performance in the balanced sample.}
}
\label{tab:ia-x41-balanced}
\renewcommand{\arraystretch}{1.10}
\small
\begin{tabular*}{\textwidth}{@{\extracolsep{\fill}}lcccc@{\hspace{1.15em}}cccc@{}}
\toprule
& \multicolumn{4}{c}{United States}
& \multicolumn{4}{c}{\makecell[c]{International markets\\[-0.4ex]\footnotesize Equal-weight aggregate}} \\
\cmidrule(lr){2-5} \cmidrule(lr){6-9}
Model & $h=1$ & $h=3$ & $h=6$ & $h=12$
      & $h=1$ & $h=3$ & $h=6$ & $h=12$ \\
\midrule
Chronos Tiny & \makecell[c]{\ensuremath{2.89^{**}}\\[-0.35ex]{\scriptsize \ensuremath{(0.019)}}} & \makecell[c]{\ensuremath{9.16^{***}}\\[-0.35ex]{\scriptsize \ensuremath{(0.006)}}} & \makecell[c]{\ensuremath{13.73^{***}}\\[-0.35ex]{\scriptsize \ensuremath{(0.007)}}} & \makecell[c]{\ensuremath{7.99^{*}}\\[-0.35ex]{\scriptsize \ensuremath{(0.053)}}} & \makecell[c]{\ensuremath{0.90^{**}}\\[-0.35ex]{\scriptsize \ensuremath{(0.023)}}} & \makecell[c]{\ensuremath{3.12^{**}}\\[-0.35ex]{\scriptsize \ensuremath{(0.011)}}} & \makecell[c]{\ensuremath{6.73^{***}}\\[-0.35ex]{\scriptsize \ensuremath{(0.005)}}} & \makecell[c]{\ensuremath{7.92^{**}}\\[-0.35ex]{\scriptsize \ensuremath{(0.028)}}} \\
Chronos Mini & \makecell[c]{\ensuremath{3.08^{***}}\\[-0.35ex]{\scriptsize \ensuremath{(0.009)}}} & \makecell[c]{\ensuremath{9.01^{***}}\\[-0.35ex]{\scriptsize \ensuremath{(0.004)}}} & \makecell[c]{\ensuremath{10.77^{**}}\\[-0.35ex]{\scriptsize \ensuremath{(0.012)}}} & \makecell[c]{\ensuremath{3.02}\\[-0.35ex]{\scriptsize \ensuremath{(0.189)}}} & \makecell[c]{\ensuremath{0.93^{**}}\\[-0.35ex]{\scriptsize \ensuremath{(0.027)}}} & \makecell[c]{\ensuremath{4.26^{***}}\\[-0.35ex]{\scriptsize \ensuremath{(0.003)}}} & \makecell[c]{\ensuremath{6.84^{***}}\\[-0.35ex]{\scriptsize \ensuremath{(0.004)}}} & \makecell[c]{\ensuremath{8.03^{**}}\\[-0.35ex]{\scriptsize \ensuremath{(0.033)}}} \\
\bottomrule
\end{tabular*}

\par\vspace{6pt}\nointerlineskip
\exhibitnotes{%
  The balanced January 2002--December 2021 sample contains 240 target-start months at each horizon. Entries are raw out-of-sample $R^2_{\mathrm{HA}}$ for origin-aligned point-in-time (PIT) forecasts, in percentage points, with one-sided Clark--West $p$-values in parentheses. Positive raw values indicate lower mean squared forecast error (MSFE) than the expanding historical average. International estimates normalize each market's monthly loss differential by its historical-average MSFE, then average the 13 markets equally over common target months. International Clark--West tests use the monthly average of normalized adjusted loss differentials. Bartlett--Newey--West HAC inference uses $h-1$ lags, where $h$ is in months. $^{*}$, $^{**}$, and $^{***}$ indicate Clark--West significance at 10\%, 5\%, and 1\%. Estimates use PIT lead $-1$ and the US training-information environment.
}
\end{threeparttable}
\end{table}

\clearpage

\section{Revisions to Return Forecasts and Predictive Distributions}\label{sec:ia-revision-results}

This section supports main-text Section~\ref{forecast-revision-under-temporal-exposure} with the complete point-forecast, distributional, and model-native revision comparisons.
\begin{figure}[H]
\centering
\includegraphics[width=0.95\linewidth]{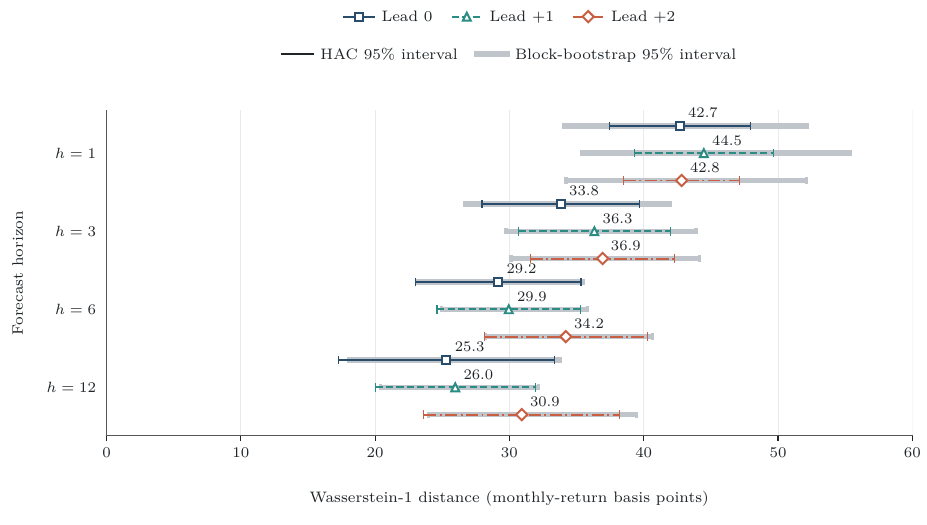}
\caption[Wasserstein distances between Chronos predictive distributions]{%
  \textbf{Wasserstein distances between Chronos predictive distributions.}
  Empirical Wasserstein-1 distances compare Chronos Mini's horizon-averaged predictive-path distributions under origin-aligned point-in-time (PIT) and later annual states. Distances are averaged over matched target months and expressed in monthly-return basis points; zero indicates identical empirical distributions. Squares, triangles, and diamonds identify leads $0,+1,+2$, each relative to PIT lead $-1$, at horizons $h=1,3,6,12$ months. Thin lines give 95\% Bartlett--Newey--West HAC intervals with $h-1$ lags; thick light-gray lines give 95\% percentile intervals from 2,000 circular moving-block bootstrap replications with block length $\max(h,6)$. Matched target-month counts are 276, 264, and 252 for the three leads, respectively. Comparisons use the all-available U.S. equity-premium sample and the US training-information environment.
}
\label{fig:x42-chronos-distribution-movement}

\end{figure}

\clearpage
\begin{figure}[H]
\centering
\includegraphics[width=0.95\linewidth]{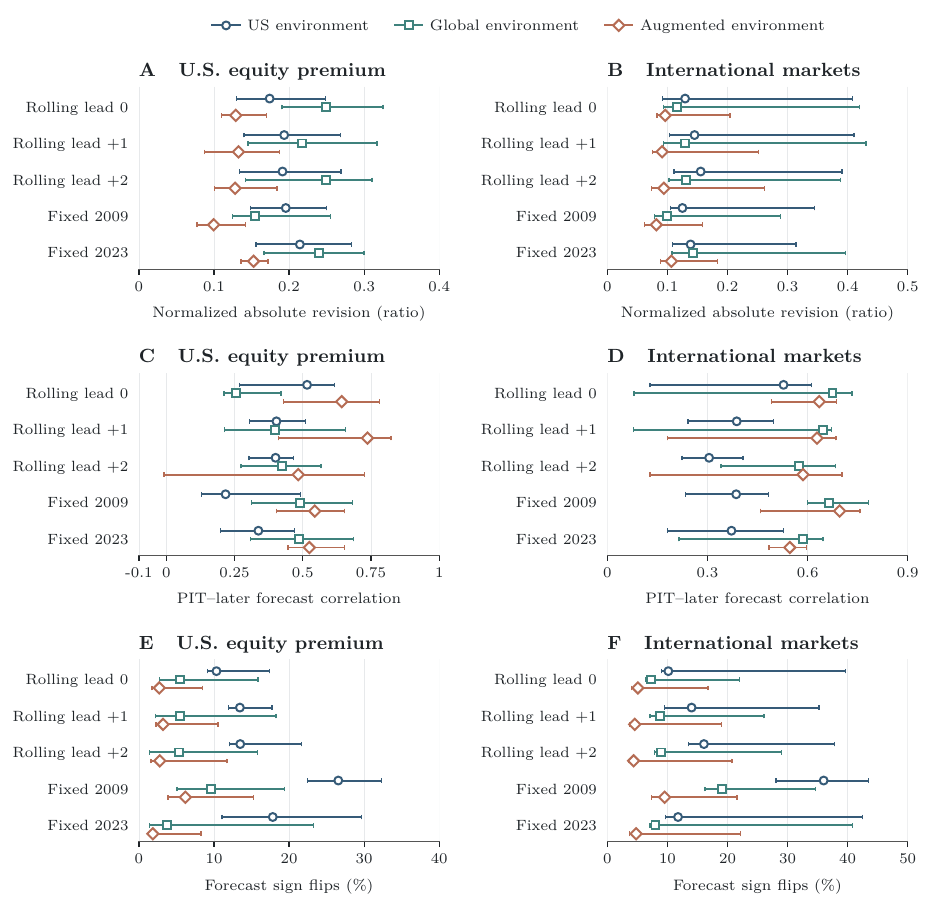}
\caption[Point-forecast revision summaries across designs and training environments]{%
  \textbf{Point-forecast revision summaries across designs and training environments.}
  Point-forecast revisions are summarized across five TSFM variants and horizons of 1, 3, 6, and 12 months. The left column (Panels A, C, and E) shows the U.S. equity premium, and the right column (Panels B, D, and F) shows the 13 non-U.S. markets. The rows give mean absolute revision divided by historical-average RMSE, PIT--alternative forecast correlation, and the percentage of forecast sign changes. Each panel uses its own horizontal scale. Normalized revision is a ratio. Markers give medians across 20 model--horizon cells, and horizontal bars give interquartile ranges. International statistics first take the median across markets within each model--horizon cell. Colors identify the US, Global, and Augmented training environments. Rolling comparisons use leads $0,+1,+2$ relative to origin-aligned point-in-time (PIT) lead $-1$; fixed comparisons use the retrospective-exposure windows of states 2009 and 2023. Estimates use the all-available sample.
}
\label{fig:ia-x42-revision-breadth}

\end{figure}

\clearpage
\begin{figure}[H]
\centering
\includegraphics[width=0.95\linewidth]{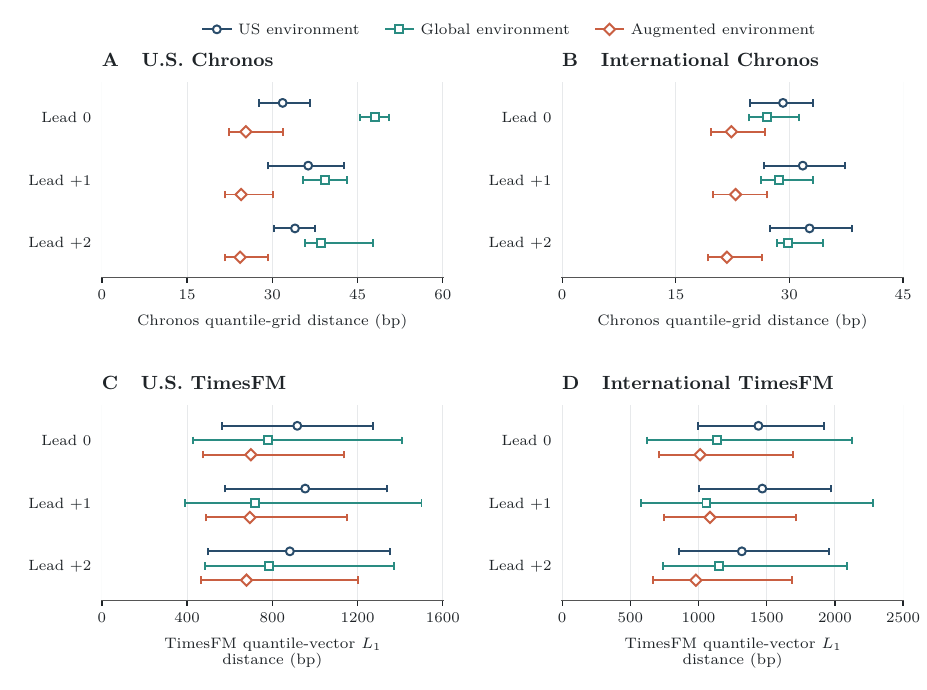}
\caption[Model-native forecast revisions across training environments]{%
  \textbf{Model-native forecast revisions across training environments.}
  Model-native forecast movement is measured for rolling leads $0,+1,+2$ relative to origin-aligned point-in-time (PIT) forecasts under the US, Global, and Augmented training-information environments. Panels A and B give Chronos predictive quantile-grid movement for the U.S. equity premium and the 13 non-U.S. markets. Panels C and D give mean $L_1$ displacement of the native TimesFM quantile vector for the same target groups. Markers show medians across model--horizon cells within each family, and horizontal bars show interquartile ranges. International statistics first take the median across the 13 markets within each model--horizon cell. Distances are in monthly-return basis points. The two families retain their respective native metrics; each panel uses its own horizontal scale.
}
\label{fig:ia-x42-native-breadth}

\end{figure}

\FloatBarrier
\clearpage
\begin{landscape}
\begingroup
\fontsize{11}{13.6}\selectfont
\setlength{\tabcolsep}{4.00pt}
\setlength{\linewidth}{524.50pt}
\begin{longtable}{@{}llccccccccccc@{}}
\caption[Point-forecast revisions by annual state and training environment]{%
  \textbf{Point-forecast revisions by annual state and training environment.}
}
\label{tab:ia-x42-point-revision}\\
\multicolumn{13}{@{}p{\linewidth}@{}}{\exhibitnotes{%
  Point-forecast revisions are summarized for five TSFM variants under the US, Global, and Augmented training-information environments. Panel A gives U.S. equity-premium estimates; Panel B gives the median of each statistic calculated separately across the 13 non-U.S. markets. For $D=\widehat y^{\mathrm{alternative}}-\widehat y^{\mathrm{PIT}}$, normalized distance is $\overline{|D|}/\mathrm{RMSE}_{\mathrm{HA}}$ on the same paired support, reported as a ratio. Correlation is the Pearson correlation between forecasts, and sign flips are the percentage of matched forecasts with different signs. Rolling comparisons use leads $0,+1,+2$ relative to origin-aligned point-in-time (PIT) lead $-1$; fixed comparisons use the retrospective-exposure windows of states 2009 and 2023. The horizon $h$ is in months. Support counts target months in Panel A and markets in Panel B. Estimates use the all-available sample.
}}\\
\noalign{\vskip 6pt}
\toprule
& & & \multicolumn{3}{c}{US environment} & \multicolumn{3}{c}{Global environment} & \multicolumn{3}{c}{Augmented environment} & \\
Comparison & Model & $h$ & Norm. & Corr. & Flip & Norm. & Corr. & Flip & Norm. & Corr. & Flip & Support \\
& & & dist. & & (\%) & dist. & & (\%) & dist. & & (\%) & \\
\cmidrule(lr){4-6}\cmidrule(lr){7-9}\cmidrule(lr){10-12}
\midrule
\endfirsthead
\multicolumn{13}{c}{\tablename\ \thetable\ -- continued}\\
\toprule
& & & \multicolumn{3}{c}{US environment} & \multicolumn{3}{c}{Global environment} & \multicolumn{3}{c}{Augmented environment} & \\
Comparison & Model & $h$ & Norm. & Corr. & Flip & Norm. & Corr. & Flip & Norm. & Corr. & Flip & Support \\
& & & dist. & & (\%) & dist. & & (\%) & dist. & & (\%) & \\
\cmidrule(lr){4-6}\cmidrule(lr){7-9}\cmidrule(lr){10-12}
\midrule
\endhead
\midrule
\multicolumn{13}{r}{Continued on next page}\\
\endfoot
\bottomrule
\endlastfoot
\multicolumn{13}{@{}l}{\textit{Panel A: U.S. Equity Premium}} \\
\addlinespace[2pt]
lead 0 & Chronos Tiny & 1 & 0.068 & 0.681 & 8.3 & 0.088 & 0.608 & 1.4 & 0.078 & 0.801 & 2.5 & 276 \\*
 &  & 3 & 0.110 & 0.714 & 7.6 & 0.161 & 0.550 & 2.9 & 0.101 & 0.827 & 1.8 & 276 \\*
 &  & 6 & 0.132 & 0.699 & 8.3 & 0.206 & 0.429 & 3.3 & 0.116 & 0.807 & 2.9 & 276 \\*
 &  & 12 & 0.153 & 0.638 & 7.6 & 0.261 & 0.312 & 4.3 & 0.131 & 0.773 & 3.3 & 276 \\
\addlinespace[1pt]
 & Chronos Mini & 1 & 0.085 & 0.523 & 10.1 & 0.111 & 0.418 & 0.7 & 0.091 & 0.765 & 1.1 & 276 \\*
 &  & 3 & 0.123 & 0.611 & 9.4 & 0.171 & 0.341 & 0.4 & 0.111 & 0.830 & 1.4 & 276 \\*
 &  & 6 & 0.145 & 0.517 & 9.8 & 0.197 & 0.205 & 1.1 & 0.113 & 0.827 & 1.4 & 276 \\
\end{longtable}

\endgroup
\end{landscape}
\clearpage

\begingroup
\begingroup
\small
\fontsize{10}{12}\selectfont
\setlength{\tabcolsep}{4.00pt}
\setlength{\linewidth}{373.01pt}
\begin{longtable}{@{}llccccc@{}}
\caption[Chronos predictive-distribution distances across annual states]{%
  \textbf{Chronos predictive-distribution distances across annual states.}
}\label{tab:ia-x42-chronos-distribution-movement}\\
\multicolumn{7}{@{}p{\linewidth}@{}}{\exhibitnotes{%
  Empirical Wasserstein-1 distances measure changes in the horizon-averaged predictive-path distributions of Chronos. Panel A compares rolling leads $0,+1,+2$ with origin-aligned point-in-time (PIT) lead $-1$; Panel B compares fixed states 2009 and 2023 with PIT within their retrospective-exposure windows. Entries are mean distances across matched target months, in monthly-return basis points. The horizon $h$ is in months, and $N$ counts matched target months. The 95\% HAC intervals use Bartlett--Newey--West estimation with $h-1$ lags. The 95\% bootstrap intervals use 2,000 circular moving-block replications with block length $\max(h,6)$ and percentile endpoints. Estimates use the all-available U.S. equity-premium sample and the US training-information environment.
}}\\
\noalign{\vskip 6pt}
\toprule
Model & \shortstack{Annual\\state} & $h$ & Mean $W_1$ & \shortstack{HAC 95\%\\interval} & \shortstack{Block-bootstrap\\95\% interval} & $N$ \\
 & & (months) & (bp) & & & \\
\midrule
\endfirsthead
\multicolumn{7}{@{}l}{\tablename\ \thetable\ (continued)}\\
\addlinespace[2pt]
\toprule
Model & \shortstack{Annual\\state} & $h$ & Mean $W_1$ & \shortstack{HAC 95\%\\interval} & \shortstack{Block-bootstrap\\95\% interval} & $N$ \\
 & & (months) & (bp) & & & \\
\midrule
\endhead
\midrule
\multicolumn{7}{r}{Continued on next page}\\
\endfoot
\bottomrule
\endlastfoot
\multicolumn{7}{@{}l}{\textit{Panel A: Rolling Annual States}} \\
\addlinespace[2pt]
Chronos Tiny & Lead 0 & 1 & 32.57 & [29.09, 36.04] & [26.49, 39.07] & 276 \\*
 &  & 3 & 28.90 & [24.15, 33.65] & [22.95, 35.44] & 276 \\*
 &  & 6 & 25.53 & [20.19, 30.87] & [20.40, 31.02] & 276 \\*
 &  & 12 & 21.70 & [16.30, 27.10] & [16.47, 27.14] & 276 \\
\addlinespace[2pt]
Chronos Tiny & Lead +1 & 1 & 38.97 & [33.75, 44.19] & [29.02, 50.73] & 264 \\*
 &  & 3 & 31.99 & [26.96, 37.02] & [25.01, 38.96] & 264 \\*
 &  & 6 & 27.67 & [21.87, 33.46] & [22.05, 33.54] & 264 \\
\end{longtable}
\endgroup

\endgroup

\clearpage

\section{Forecast Accuracy with Rolling Model Vintages}\label{sec:ia-rolling-results}

These exhibits extend main-text Section~\ref{the-rolling-path-beyond-the-forecast-origin} across model variants, training environments, and matched sample definitions.
\FloatBarrier

Figures~\ref{fig:ia-x47-chronos-profiles} and \ref{fig:ia-x47-timesfm-profiles} report the complete variant-level lead profiles underlying the main-text size-matched comparison. Each row holds the architecture variant fixed and traces the four forecast horizons across the stale, boundary, and post-origin states. The left panels measure forecast movement, while the right panels measure the predictive value created by that movement.

\begin{figure}[p]
\centering
\includegraphics[width=0.95\textwidth]{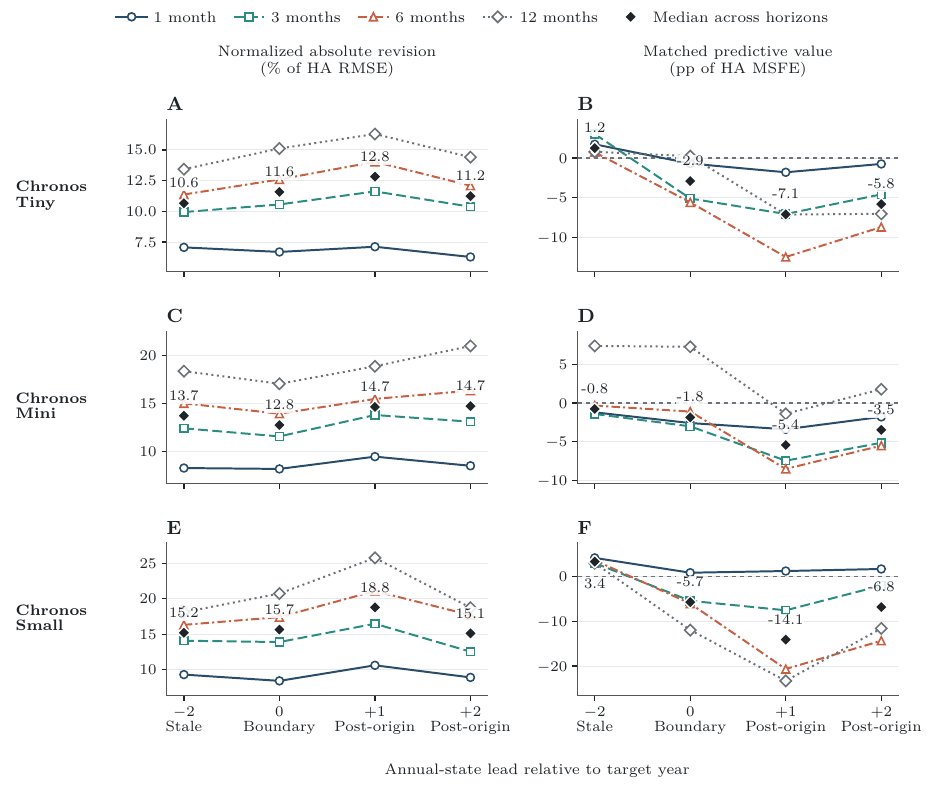}
\caption[Chronos forecast revisions and matched predictive effects]{%
  \textbf{Chronos forecast revisions and matched predictive effects.}
  Chronos Tiny, Mini, and Small occupy the three rows. Left panels show mean absolute forecast revision as a percentage of historical-average RMSE; right panels show PIT mean squared forecast error (MSFE) minus alternative-state MSFE, scaled by historical-average MSFE and expressed in percentage points. Lines distinguish horizons $h=1,3,6,12$ months, and black diamonds mark the median across horizons at each lead. All estimates use the same 240 target-start months from January 2002 through December 2021, jointly available for origin-aligned point-in-time (PIT) forecasts and leads $-2,0,+1,+2$. Positive matched value favors the alternative state. Lead $-2$ replaces PIT with the stale state, reversing the pre-origin forward step in Table 3. Panels have separate vertical scales; zero in the right panels indicates equal forecast loss. Estimates use the U.S. equity premium and the US training-information environment.
}
\label{fig:ia-x47-chronos-profiles}

\end{figure}

\clearpage
\begin{figure}[p]
\centering
\includegraphics[width=0.95\textwidth]{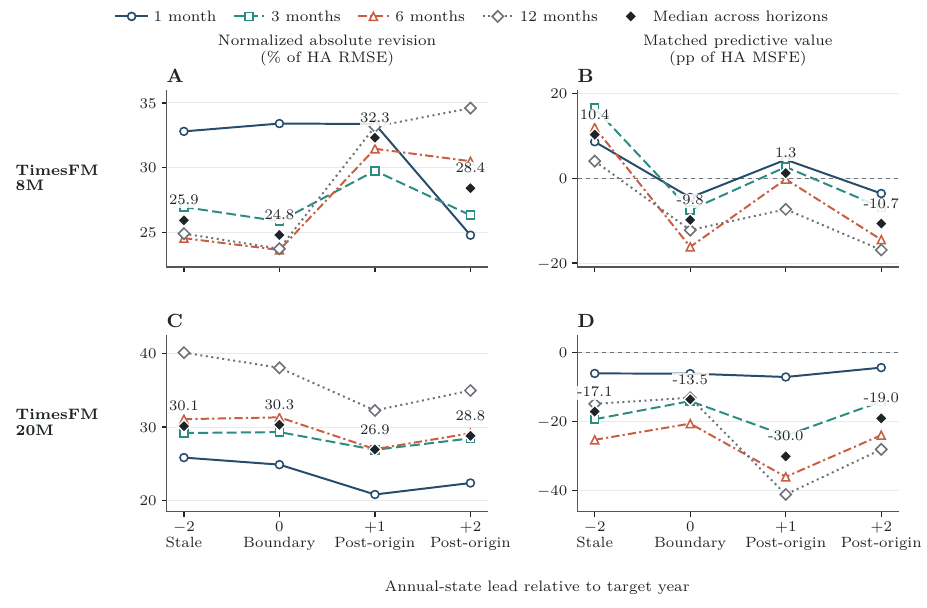}
\caption[TimesFM forecast revisions and matched predictive effects]{%
  \textbf{TimesFM forecast revisions and matched predictive effects.}
  TimesFM 8M and 20M occupy the two rows. Left panels show mean absolute forecast revision as a percentage of historical-average RMSE; right panels show PIT mean squared forecast error (MSFE) minus alternative-state MSFE, scaled by historical-average MSFE and expressed in percentage points. Lines distinguish horizons $h=1,3,6,12$ months, and black diamonds mark the median across horizons at each lead. All estimates use the same 240 target-start months from January 2002 through December 2021, jointly available for origin-aligned point-in-time (PIT) forecasts and leads $-2,0,+1,+2$. Positive matched value favors the alternative state. Lead $-2$ replaces PIT with the stale state, reversing the pre-origin forward step in Table 3. Panels have separate vertical scales; zero in the right panels indicates equal forecast loss. Estimates use the U.S. equity premium and the US training-information environment.
}
\label{fig:ia-x47-timesfm-profiles}

\end{figure}

\clearpage
\begin{figure}[!htbp]
\centering
\includegraphics[width=0.95\textwidth]{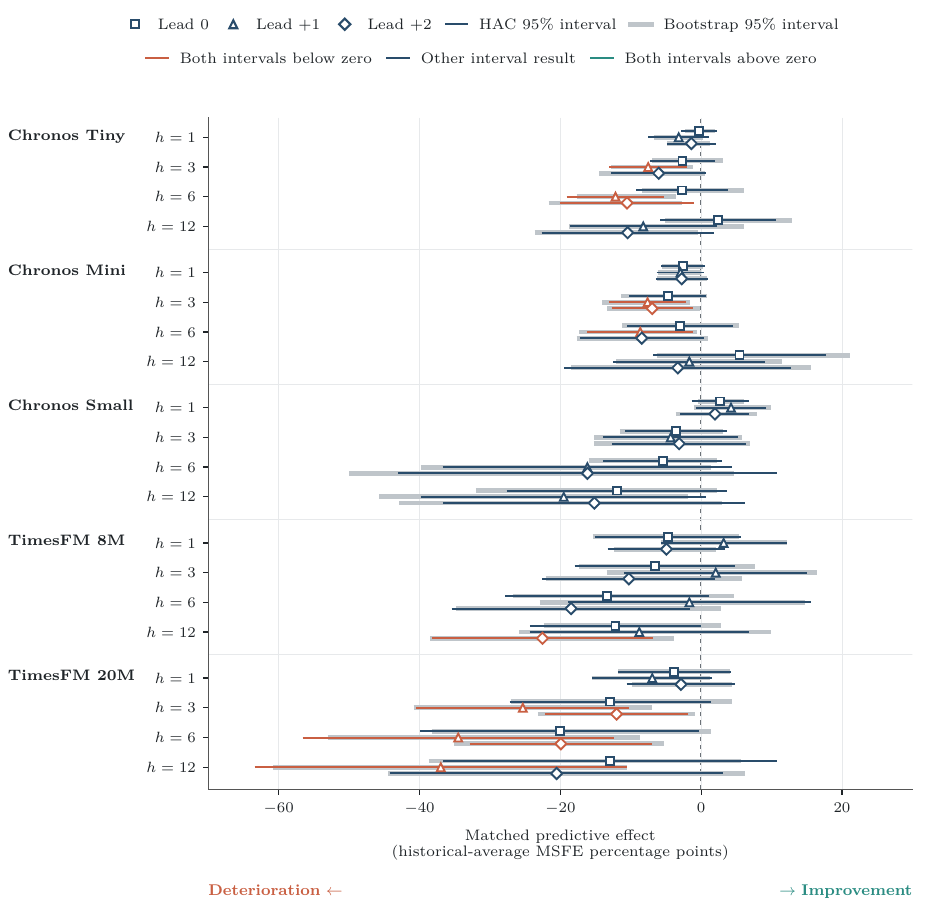}
\caption[Rolling matched predictive effects across all model variants]{%
  \textbf{Rolling matched predictive effects across all model variants.}
  Rolling matched predictive effects are shown for all five TSFM variants at horizons $h=1,3,6,12$ months. Within each model--horizon row, squares, triangles, and diamonds identify leads $0,+1,+2$ relative to origin-aligned point-in-time (PIT) lead $-1$. Effects equal PIT mean squared forecast error (MSFE) minus alternative-state MSFE, divided by historical-average MSFE on the same paired support and expressed in percentage points. Positive values favor the alternative state. Thin colored lines give 95\% Bartlett--Newey--West HAC intervals with $h-1$ lags; thick light-gray lines give 95\% percentile intervals from 2,000 circular moving-block bootstrap replications with block length $\max(h,6)$. Each lead uses its maximal matched support with PIT. Estimates use the all-available U.S. equity-premium sample and the US training-information environment.
}\label{fig:ia-x43-all-model-effects}

\end{figure}

\clearpage
\begin{table}[!htbp]
\centering
\caption[Pooled predictive effects of rolling annual states]{%
  \textbf{Pooled predictive effects of rolling annual states.}
}\label{tab:x43-pooled-effects}
\begin{threeparttable}
\small
\setlength{\tabcolsep}{6pt}
\renewcommand{\arraystretch}{1.20}
\begin{tabular*}{\textwidth}{@{\extracolsep{\fill}}lrrrr@{}}
\toprule
Model & $h=1$ & $h=3$ & $h=6$ & $h=12$ \\
\midrule
Chronos Tiny & \makecell[r]{-1.29\\{\scriptsize (0.335)}} & \makecell[r]{-5.88$^{***}$\\{\scriptsize (0.009)}} & \makecell[r]{-9.26$^{***}$\\{\scriptsize ($<$0.001)}} & \makecell[r]{-6.01\\{\scriptsize (0.148)}} \\
Chronos Mini & \makecell[r]{-2.66$^{*}$\\{\scriptsize (0.063)}} & \makecell[r]{-5.45$^{**}$\\{\scriptsize (0.023)}} & \makecell[r]{-5.75\\{\scriptsize (0.114)}} & \makecell[r]{0.82\\{\scriptsize (0.900)}} \\
Chronos Small & \makecell[r]{1.86\\{\scriptsize (0.394)}} & \makecell[r]{-4.42\\{\scriptsize (0.259)}} & \makecell[r]{-13.72\\{\scriptsize (0.112)}} & \makecell[r]{-15.24$^{*}$\\{\scriptsize (0.053)}} \\
\addlinespace[2pt]
TimesFM 8M & \makecell[r]{-1.89\\{\scriptsize (0.654)}} & \makecell[r]{-6.31\\{\scriptsize (0.229)}} & \makecell[r]{-13.63$^{*}$\\{\scriptsize (0.056)}} & \makecell[r]{-16.55$^{***}$\\{\scriptsize (0.006)}} \\
TimesFM 20M & \makecell[r]{-3.52\\{\scriptsize (0.322)}} & \makecell[r]{-15.13$^{***}$\\{\scriptsize (0.010)}} & \makecell[r]{-23.24$^{***}$\\{\scriptsize (0.003)}} & \makecell[r]{-20.53$^{**}$\\{\scriptsize (0.031)}} \\
\midrule
Joint target-start months & 252 & 252 & 252 & 252 \\
\bottomrule
\end{tabular*}

\par\vspace{6pt}\nointerlineskip
\exhibitnotes{%
  Pooled rolling effects equally average the PIT-relative squared-loss advantages of leads $0,+1,+2$. Within each model--horizon cell, the three later leads and origin-aligned point-in-time (PIT) lead $-1$ use strict common target-start support. The average loss advantage is divided by historical-average mean squared forecast error (MSFE) on that support and expressed in percentage points; positive values favor the later states. Parentheses contain two-sided Bartlett--Newey--West HAC $p$-values with $h-1$ lags, where $h$ is in months. $^{*}$, $^{**}$, and $^{***}$ indicate significance at 10\%, 5\%, and 1\%. Estimates use the all-available U.S. equity-premium sample and the US training-information environment. The final row gives 252 joint target-start months at every horizon.
}
\end{threeparttable}
\end{table}

\clearpage

\renewcommand{\arraystretch}{0.96}
\setlength{\tabcolsep}{5pt}
\small
\begin{longtable}{@{\extracolsep{\fill}}lllrrrr@{}}
\caption[U.S. rolling forecasting performance by training environment]{%
  \textbf{U.S. rolling forecasting performance by training environment.}
}\label{tab:ia-x43-us-corpora}\\
\multicolumn{7}{@{}p{\linewidth}@{}}{\exhibitnotes{%
  U.S. raw out-of-sample $R^2_{\mathrm{HA}}$ is reported under the US, Global, and Augmented training-information environments. Within each model--environment--horizon cell, origin-aligned point-in-time (PIT) lead $-1$ and leads $0,+1,+2$ share the same joint target-month support. Values are percentage-point reductions in mean squared forecast error relative to the expanding historical average. Parentheses contain one-sided Clark--West $p$-values based on adjusted loss differentials, which can be significant when the raw value is negative. Bartlett--Newey--West HAC inference uses $h-1$ lags, where $h$ is in months. $^{*}$, $^{**}$, and $^{***}$ indicate Clark--West significance at 10\%, 5\%, and 1\%. Estimates use the all-available sample.
}}\\
\noalign{\vskip 6pt}
\toprule
Training environment & Model & Annual State & $h=1$ & $h=3$ & $h=6$ & $h=12$ \\
\midrule
\endfirsthead
\multicolumn{7}{l}{\textit{\tablename\ \thetable\ (continued)}}\\
\toprule
Training environment & Model & Annual State & $h=1$ & $h=3$ & $h=6$ & $h=12$ \\
\midrule
\endhead
\midrule
\multicolumn{7}{r}{\textit{Continued on next page}}\\
\endfoot
\bottomrule
\endlastfoot
\multicolumn{7}{@{}l}{\textit{Panel A: US training environment}} \\
US & Chronos Tiny & \textbf{PIT ($-1$)} & \makecell[r]{3.74$^{***}$\\{\scriptsize (0.007)}} & \makecell[r]{11.56$^{***}$\\{\scriptsize (0.001)}} & \makecell[r]{17.44$^{***}$\\{\scriptsize (0.002)}} & \makecell[r]{16.74$^{**}$\\{\scriptsize (0.032)}} \\
 &  & Lead $0$ & \makecell[r]{3.19$^{**}$\\{\scriptsize (0.017)}} & \makecell[r]{7.05$^{**}$\\{\scriptsize (0.013)}} & \makecell[r]{12.77$^{***}$\\{\scriptsize (0.007)}} & \makecell[r]{17.41$^{**}$\\{\scriptsize (0.029)}} \\
 &  & Lead $+1$ & \makecell[r]{1.83$^{*}$\\{\scriptsize (0.087)}} & \makecell[r]{4.48$^{*}$\\{\scriptsize (0.055)}} & \makecell[r]{4.86$^{*}$\\{\scriptsize (0.063)}} & \makecell[r]{8.50$^{*}$\\{\scriptsize (0.054)}} \\
 &  & Lead $+2$ & \makecell[r]{2.34$^{**}$\\{\scriptsize (0.038)}} & \makecell[r]{5.53$^{**}$\\{\scriptsize (0.017)}} & \makecell[r]{6.91$^{**}$\\{\scriptsize (0.017)}} & \makecell[r]{6.30$^{*}$\\{\scriptsize (0.094)}} \\
 & Chronos Mini & \textbf{PIT ($-1$)} & \makecell[r]{3.70$^{***}$\\{\scriptsize (0.003)}} & \makecell[r]{10.75$^{***}$\\{\scriptsize ($<$0.001)}} & \makecell[r]{13.86$^{***}$\\{\scriptsize (0.003)}} & \makecell[r]{10.18$^{*}$\\{\scriptsize (0.086)}} \\
 &  & Lead $0$ & \makecell[r]{1.44$^{*}$\\{\scriptsize (0.079)}} & \makecell[r]{8.21$^{***}$\\{\scriptsize (0.001)}} & \makecell[r]{13.37$^{***}$\\{\scriptsize (0.001)}} & \makecell[r]{17.63$^{**}$\\{\scriptsize (0.022)}} \\
 &  & Lead $+1$ & \makecell[r]{0.74\\{\scriptsize (0.189)}} & \makecell[r]{3.91$^{**}$\\{\scriptsize (0.023)}} & \makecell[r]{5.52$^{**}$\\{\scriptsize (0.011)}} & \makecell[r]{8.51$^{*}$\\{\scriptsize (0.068)}} \\
 &  & Lead $+2$ & \makecell[r]{0.93\\{\scriptsize (0.170)}} & \makecell[r]{3.79$^{**}$\\{\scriptsize (0.025)}} & \makecell[r]{5.42$^{***}$\\{\scriptsize (0.007)}} & \makecell[r]{6.86$^{**}$\\{\scriptsize (0.047)}} \\
 & Chronos Small & \textbf{PIT ($-1$)} & \makecell[r]{1.81$^{*}$\\{\scriptsize (0.098)}} & \makecell[r]{9.30$^{***}$\\{\scriptsize (0.007)}} & \makecell[r]{13.54$^{***}$\\{\scriptsize (0.005)}} & \makecell[r]{13.87$^{**}$\\{\scriptsize (0.035)}} \\
 &  & Lead $0$ & \makecell[r]{3.63$^{**}$\\{\scriptsize (0.011)}} & \makecell[r]{5.58$^{**}$\\{\scriptsize (0.012)}} & \makecell[r]{8.40$^{**}$\\{\scriptsize (0.013)}} & \makecell[r]{4.69\\{\scriptsize (0.132)}} \\
 &  & Lead $+1$ & \makecell[r]{3.60$^{***}$\\{\scriptsize (0.004)}} & \makecell[r]{2.87$^{**}$\\{\scriptsize (0.038)}} & \makecell[r]{-6.30\\{\scriptsize (0.449)}} & \makecell[r]{-7.48\\{\scriptsize (0.394)}} \\
 &  & Lead $+2$ & \makecell[r]{3.76$^{**}$\\{\scriptsize (0.014)}} & \makecell[r]{6.17$^{**}$\\{\scriptsize (0.016)}} & \makecell[r]{-2.61\\{\scriptsize (0.282)}} & \makecell[r]{-1.31\\{\scriptsize (0.238)}} \\
 & TimesFM 8M & \textbf{PIT ($-1$)} & \makecell[r]{-11.88\\{\scriptsize (0.559)}} & \makecell[r]{-2.98\\{\scriptsize (0.203)}} & \makecell[r]{6.91$^{**}$\\{\scriptsize (0.011)}} & \makecell[r]{14.28$^{***}$\\{\scriptsize (0.001)}} \\
 &  & Lead $0$ & \makecell[r]{-17.43\\{\scriptsize (0.860)}} & \makecell[r]{-12.49\\{\scriptsize (0.716)}} & \makecell[r]{-11.96\\{\scriptsize (0.714)}} & \makecell[r]{-1.67$^{*}$\\{\scriptsize (0.098)}} \\
 &  & Lead $+1$ & \makecell[r]{-7.07\\{\scriptsize (0.144)}} & \makecell[r]{-2.13$^{*}$\\{\scriptsize (0.072)}} & \makecell[r]{3.36$^{*}$\\{\scriptsize (0.054)}} & \makecell[r]{3.14$^{**}$\\{\scriptsize (0.049)}} \\
 &  & Lead $+2$ & \makecell[r]{-16.82\\{\scriptsize (0.873)}} & \makecell[r]{-13.25\\{\scriptsize (0.594)}} & \makecell[r]{-11.58\\{\scriptsize (0.521)}} & \makecell[r]{-8.26\\{\scriptsize (0.324)}} \\
 & TimesFM 20M & \textbf{PIT ($-1$)} & \makecell[r]{-10.19\\{\scriptsize (0.512)}} & \makecell[r]{2.67$^{**}$\\{\scriptsize (0.025)}} & \makecell[r]{11.69$^{**}$\\{\scriptsize (0.011)}} & \makecell[r]{7.03$^{**}$\\{\scriptsize (0.014)}} \\
 &  & Lead $0$ & \makecell[r]{-13.62\\{\scriptsize (0.807)}} & \makecell[r]{-8.30\\{\scriptsize (0.350)}} & \makecell[r]{-3.79\\{\scriptsize (0.202)}} & \makecell[r]{2.38$^{*}$\\{\scriptsize (0.081)}} \\
 &  & Lead $+1$ & \makecell[r]{-14.45\\{\scriptsize (0.849)}} & \makecell[r]{-19.74\\{\scriptsize (0.840)}} & \makecell[r]{-22.62\\{\scriptsize (0.764)}} & \makecell[r]{-29.38\\{\scriptsize (0.820)}} \\
 &  & Lead $+2$ & \makecell[r]{-13.05\\{\scriptsize (0.694)}} & \makecell[r]{-9.35\\{\scriptsize (0.487)}} & \makecell[r]{-8.25\\{\scriptsize (0.334)}} & \makecell[r]{-13.51\\{\scriptsize (0.424)}} \\
\addlinespace[2pt]
\multicolumn{7}{@{}l}{\textit{Panel B. Global training-information environment}} \\*
Global & Chronos Tiny & \textbf{PIT ($-1$)} & \makecell[r]{0.34\\{\scriptsize (0.148)}} & \makecell[r]{1.37\\{\scriptsize (0.122)}} & \makecell[r]{2.40\\{\scriptsize (0.117)}} & \makecell[r]{2.82\\{\scriptsize (0.118)}} \\
 &  & Lead $0$ & \makecell[r]{-1.06\\{\scriptsize (0.260)}} & \makecell[r]{-3.53\\{\scriptsize (0.250)}} & \makecell[r]{-5.34\\{\scriptsize (0.237)}} & \makecell[r]{-9.39\\{\scriptsize (0.215)}} \\
 &  & Lead $+1$ & \makecell[r]{1.17\\{\scriptsize (0.143)}} & \makecell[r]{6.45$^{**}$\\{\scriptsize (0.017)}} & \makecell[r]{9.02$^{***}$\\{\scriptsize (0.005)}} & \makecell[r]{9.55$^{**}$\\{\scriptsize (0.019)}} \\
 &  & Lead $+2$ & \makecell[r]{1.97$^{*}$\\{\scriptsize (0.088)}} & \makecell[r]{6.00$^{**}$\\{\scriptsize (0.024)}} & \makecell[r]{9.42$^{***}$\\{\scriptsize (0.004)}} & \makecell[r]{12.78$^{**}$\\{\scriptsize (0.020)}} \\
 & Chronos Mini & \textbf{PIT ($-1$)} & \makecell[r]{-1.72\\{\scriptsize (0.182)}} & \makecell[r]{2.79$^{**}$\\{\scriptsize (0.046)}} & \makecell[r]{2.06$^{*}$\\{\scriptsize (0.052)}} & \makecell[r]{3.17$^{**}$\\{\scriptsize (0.022)}} \\
 &  & Lead $0$ & \makecell[r]{-3.69\\{\scriptsize (0.386)}} & \makecell[r]{-2.27\\{\scriptsize (0.224)}} & \makecell[r]{4.15$^{*}$\\{\scriptsize (0.062)}} & \makecell[r]{4.44$^{*}$\\{\scriptsize (0.051)}} \\
 &  & Lead $+1$ & \makecell[r]{-2.74\\{\scriptsize (0.300)}} & \makecell[r]{-7.40\\{\scriptsize (0.364)}} & \makecell[r]{-8.32\\{\scriptsize (0.265)}} & \makecell[r]{-10.69\\{\scriptsize (0.203)}} \\
 &  & Lead $+2$ & \makecell[r]{0.28\\{\scriptsize (0.169)}} & \makecell[r]{3.79$^{*}$\\{\scriptsize (0.097)}} & \makecell[r]{6.92$^{*}$\\{\scriptsize (0.073)}} & \makecell[r]{7.24$^{**}$\\{\scriptsize (0.037)}} \\
 & Chronos Small & \textbf{PIT ($-1$)} & \makecell[r]{-2.48\\{\scriptsize (0.214)}} & \makecell[r]{-8.37\\{\scriptsize (0.249)}} & \makecell[r]{-4.86\\{\scriptsize (0.110)}} & \makecell[r]{4.84$^{*}$\\{\scriptsize (0.061)}} \\
 &  & Lead $0$ & \makecell[r]{0.15\\{\scriptsize (0.122)}} & \makecell[r]{2.51$^{*}$\\{\scriptsize (0.076)}} & \makecell[r]{7.34$^{**}$\\{\scriptsize (0.010)}} & \makecell[r]{9.39$^{**}$\\{\scriptsize (0.012)}} \\
 &  & Lead $+1$ & \makecell[r]{-1.50\\{\scriptsize (0.242)}} & \makecell[r]{-2.36\\{\scriptsize (0.180)}} & \makecell[r]{-4.78\\{\scriptsize (0.234)}} & \makecell[r]{0.62$^{*}$\\{\scriptsize (0.064)}} \\
 &  & Lead $+2$ & \makecell[r]{-2.32\\{\scriptsize (0.344)}} & \makecell[r]{3.61$^{*}$\\{\scriptsize (0.059)}} & \makecell[r]{4.69$^{*}$\\{\scriptsize (0.050)}} & \makecell[r]{3.38$^{*}$\\{\scriptsize (0.070)}} \\
 & TimesFM 8M & \textbf{PIT ($-1$)} & \makecell[r]{-16.79\\{\scriptsize (0.698)}} & \makecell[r]{-49.07\\{\scriptsize (0.624)}} & \makecell[r]{-70.33\\{\scriptsize (0.621)}} & \makecell[r]{-34.49\\{\scriptsize (0.355)}} \\
 &  & Lead $0$ & \makecell[r]{-7.26\\{\scriptsize (0.187)}} & \makecell[r]{-22.92\\{\scriptsize (0.221)}} & \makecell[r]{-37.48\\{\scriptsize (0.513)}} & \makecell[r]{-18.46\\{\scriptsize (0.316)}} \\
 &  & Lead $+1$ & \makecell[r]{-13.51\\{\scriptsize (0.369)}} & \makecell[r]{-24.44\\{\scriptsize (0.168)}} & \makecell[r]{-25.47$^{**}$\\{\scriptsize (0.045)}} & \makecell[r]{-20.19\\{\scriptsize (0.111)}} \\
 &  & Lead $+2$ & \makecell[r]{-16.53\\{\scriptsize (0.750)}} & \makecell[r]{-34.94\\{\scriptsize (0.629)}} & \makecell[r]{-48.45\\{\scriptsize (0.624)}} & \makecell[r]{-26.28\\{\scriptsize (0.304)}} \\
 & TimesFM 20M & \textbf{PIT ($-1$)} & \makecell[r]{-11.22\\{\scriptsize (0.148)}} & \makecell[r]{4.43$^{***}$\\{\scriptsize (0.008)}} & \makecell[r]{1.89$^{*}$\\{\scriptsize (0.070)}} & \makecell[r]{-21.84\\{\scriptsize (0.400)}} \\
 &  & Lead $0$ & \makecell[r]{-9.33$^{**}$\\{\scriptsize (0.049)}} & \makecell[r]{5.26$^{***}$\\{\scriptsize ($<$0.001)}} & \makecell[r]{13.47$^{***}$\\{\scriptsize ($<$0.001)}} & \makecell[r]{3.26$^{**}$\\{\scriptsize (0.020)}} \\
 &  & Lead $+1$ & \makecell[r]{-24.47\\{\scriptsize (0.887)}} & \makecell[r]{-23.11\\{\scriptsize (0.918)}} & \makecell[r]{-3.58\\{\scriptsize (0.215)}} & \makecell[r]{14.39$^{*}$\\{\scriptsize (0.064)}} \\
 &  & Lead $+2$ & \makecell[r]{-16.79\\{\scriptsize (0.293)}} & \makecell[r]{-14.01\\{\scriptsize (0.340)}} & \makecell[r]{-18.85\\{\scriptsize (0.746)}} & \makecell[r]{-27.27\\{\scriptsize (0.516)}} \\
\addlinespace[2pt]
\multicolumn{7}{@{}l}{\textit{Panel C. Augmented training-information environment}} \\*
Augmented & Chronos Tiny & \textbf{PIT ($-1$)} & \makecell[r]{1.72\\{\scriptsize (0.132)}} & \makecell[r]{3.61$^{*}$\\{\scriptsize (0.055)}} & \makecell[r]{5.02$^{*}$\\{\scriptsize (0.054)}} & \makecell[r]{7.61$^{**}$\\{\scriptsize (0.045)}} \\
 &  & Lead $0$ & \makecell[r]{3.65$^{**}$\\{\scriptsize (0.030)}} & \makecell[r]{9.31$^{***}$\\{\scriptsize (0.004)}} & \makecell[r]{11.16$^{***}$\\{\scriptsize (0.006)}} & \makecell[r]{13.33$^{**}$\\{\scriptsize (0.016)}} \\
 &  & Lead $+1$ & \makecell[r]{2.07$^{*}$\\{\scriptsize (0.095)}} & \makecell[r]{4.23$^{**}$\\{\scriptsize (0.035)}} & \makecell[r]{8.46$^{***}$\\{\scriptsize (0.006)}} & \makecell[r]{9.39$^{*}$\\{\scriptsize (0.056)}} \\
 &  & Lead $+2$ & \makecell[r]{2.45$^{*}$\\{\scriptsize (0.071)}} & \makecell[r]{7.14$^{***}$\\{\scriptsize (0.007)}} & \makecell[r]{10.25$^{***}$\\{\scriptsize (0.006)}} & \makecell[r]{12.87$^{**}$\\{\scriptsize (0.018)}} \\
 & Chronos Mini & \textbf{PIT ($-1$)} & \makecell[r]{2.91$^{*}$\\{\scriptsize (0.052)}} & \makecell[r]{7.88$^{**}$\\{\scriptsize (0.011)}} & \makecell[r]{11.01$^{***}$\\{\scriptsize (0.004)}} & \makecell[r]{11.88$^{**}$\\{\scriptsize (0.014)}} \\
 &  & Lead $0$ & \makecell[r]{1.70\\{\scriptsize (0.118)}} & \makecell[r]{5.63$^{**}$\\{\scriptsize (0.033)}} & \makecell[r]{8.44$^{**}$\\{\scriptsize (0.016)}} & \makecell[r]{10.07$^{**}$\\{\scriptsize (0.034)}} \\
 &  & Lead $+1$ & \makecell[r]{3.22$^{*}$\\{\scriptsize (0.072)}} & \makecell[r]{7.96$^{***}$\\{\scriptsize (0.007)}} & \makecell[r]{11.61$^{***}$\\{\scriptsize (0.001)}} & \makecell[r]{11.92$^{**}$\\{\scriptsize (0.010)}} \\
 &  & Lead $+2$ & \makecell[r]{-12.66\\{\scriptsize (0.754)}} & \makecell[r]{-2.01\\{\scriptsize (0.211)}} & \makecell[r]{-0.68\\{\scriptsize (0.119)}} & \makecell[r]{-13.33\\{\scriptsize (0.414)}} \\
 & Chronos Small & \textbf{PIT ($-1$)} & \makecell[r]{1.63\\{\scriptsize (0.122)}} & \makecell[r]{4.69$^{*}$\\{\scriptsize (0.072)}} & \makecell[r]{10.08$^{**}$\\{\scriptsize (0.013)}} & \makecell[r]{16.03$^{***}$\\{\scriptsize (0.006)}} \\
 &  & Lead $0$ & \makecell[r]{-13.76\\{\scriptsize (0.816)}} & \makecell[r]{-6.16\\{\scriptsize (0.513)}} & \makecell[r]{-1.61\\{\scriptsize (0.217)}} & \makecell[r]{-15.30\\{\scriptsize (0.462)}} \\
 &  & Lead $+1$ & \makecell[r]{5.08$^{**}$\\{\scriptsize (0.011)}} & \makecell[r]{8.92$^{***}$\\{\scriptsize (0.006)}} & \makecell[r]{10.32$^{***}$\\{\scriptsize (0.006)}} & \makecell[r]{12.79$^{**}$\\{\scriptsize (0.017)}} \\
 &  & Lead $+2$ & \makecell[r]{1.91$^{*}$\\{\scriptsize (0.059)}} & \makecell[r]{4.65$^{*}$\\{\scriptsize (0.053)}} & \makecell[r]{11.36$^{***}$\\{\scriptsize (0.009)}} & \makecell[r]{13.04$^{**}$\\{\scriptsize (0.030)}} \\
 & TimesFM 8M & \textbf{PIT ($-1$)} & \makecell[r]{-11.35\\{\scriptsize (0.718)}} & \makecell[r]{-0.85\\{\scriptsize (0.181)}} & \makecell[r]{-1.80\\{\scriptsize (0.187)}} & \makecell[r]{-2.66\\{\scriptsize (0.179)}} \\
 &  & Lead $0$ & \makecell[r]{-5.52\\{\scriptsize (0.664)}} & \makecell[r]{0.76\\{\scriptsize (0.114)}} & \makecell[r]{1.18$^{*}$\\{\scriptsize (0.097)}} & \makecell[r]{5.49$^{*}$\\{\scriptsize (0.054)}} \\
 &  & Lead $+1$ & \makecell[r]{0.41\\{\scriptsize (0.121)}} & \makecell[r]{-1.16\\{\scriptsize (0.243)}} & \makecell[r]{9.63$^{**}$\\{\scriptsize (0.044)}} & \makecell[r]{5.88\\{\scriptsize (0.139)}} \\
 &  & Lead $+2$ & \makecell[r]{-7.79\\{\scriptsize (0.723)}} & \makecell[r]{-2.86\\{\scriptsize (0.431)}} & \makecell[r]{1.91$^{*}$\\{\scriptsize (0.079)}} & \makecell[r]{3.00\\{\scriptsize (0.157)}} \\
 & TimesFM 20M & \textbf{PIT ($-1$)} & \makecell[r]{-10.07\\{\scriptsize (0.974)}} & \makecell[r]{-4.05\\{\scriptsize (0.497)}} & \makecell[r]{-3.15\\{\scriptsize (0.387)}} & \makecell[r]{-6.84\\{\scriptsize (0.542)}} \\
 &  & Lead $0$ & \makecell[r]{-3.05\\{\scriptsize (0.632)}} & \makecell[r]{1.69\\{\scriptsize (0.120)}} & \makecell[r]{7.19$^{**}$\\{\scriptsize (0.032)}} & \makecell[r]{5.25$^{**}$\\{\scriptsize (0.028)}} \\
 &  & Lead $+1$ & \makecell[r]{-0.54\\{\scriptsize (0.272)}} & \makecell[r]{-1.00\\{\scriptsize (0.337)}} & \makecell[r]{5.34$^{**}$\\{\scriptsize (0.022)}} & \makecell[r]{7.43$^{**}$\\{\scriptsize (0.018)}} \\
 &  & Lead $+2$ & \makecell[r]{-95519.24\\{\scriptsize (0.960)}} & \makecell[r]{-79159.27\\{\scriptsize (0.936)}} & \makecell[r]{-38714.49\\{\scriptsize (0.780)}} & \makecell[r]{-19536.90\\{\scriptsize (0.243)}} \\
\end{longtable}

\normalsize
\renewcommand{\arraystretch}{1.0}

\clearpage
\renewcommand{\arraystretch}{0.96}
\small
\begin{longtable}{@{\extracolsep{\fill}}lllrrrr@{}}
\caption[International rolling forecasting performance by training environment]{%
  \textbf{International rolling forecasting performance by training environment.}
}\label{tab:ia-x43-international-corpora}\\
\multicolumn{7}{@{}p{\linewidth}@{}}{\exhibitnotes{%
  Equal-market raw out-of-sample $R^2_{\mathrm{HA}}$ summarizes the 13 non-U.S. markets under the US, Global, and Augmented training-information environments. Origin-aligned point-in-time (PIT) lead $-1$ and leads $0,+1,+2$ share the same joint support within each comparison. Each market's monthly raw and adjusted loss differentials are normalized by its historical-average mean squared forecast error (MSFE), then averaged equally across markets over common target months. Raw $R^2_{\mathrm{HA}}$ is in percentage points; positive values indicate lower loss than the historical average. Parentheses contain one-sided Clark--West $p$-values based on adjusted loss differentials. Bartlett--Newey--West HAC inference uses $h-1$ lags, where $h$ is in months. $^{*}$, $^{**}$, and $^{***}$ indicate Clark--West significance at 10\%, 5\%, and 1\%. Estimates use the all-available international sample.
}}\\
\noalign{\vskip 6pt}
\toprule
Training environment & Model & Annual State & $h=1$ & $h=3$ & $h=6$ & $h=12$ \\
\midrule
\endfirsthead
\multicolumn{7}{l}{\textit{\tablename\ \thetable\ (continued)}}\\
\toprule
Training environment & Model & Annual State & $h=1$ & $h=3$ & $h=6$ & $h=12$ \\
\midrule
\endhead
\midrule
\multicolumn{7}{r}{\textit{Continued on next page}}\\
\endfoot
\bottomrule
\endlastfoot
\multicolumn{7}{@{}l}{\textit{Panel A: US training environment}} \\
US & Chronos Tiny & \textbf{PIT ($-1$)} & \makecell[r]{1.81$^{***}$\\{\scriptsize (0.005)}} & \makecell[r]{5.11$^{***}$\\{\scriptsize (0.003)}} & \makecell[r]{9.69$^{***}$\\{\scriptsize (0.002)}} & \makecell[r]{13.58$^{**}$\\{\scriptsize (0.017)}} \\
 &  & Lead $0$ & \makecell[r]{0.65$^{**}$\\{\scriptsize (0.046)}} & \makecell[r]{2.03$^{**}$\\{\scriptsize (0.044)}} & \makecell[r]{4.91$^{**}$\\{\scriptsize (0.024)}} & \makecell[r]{9.48$^{**}$\\{\scriptsize (0.027)}} \\
 &  & Lead $+1$ & \makecell[r]{-0.24\\{\scriptsize (0.176)}} & \makecell[r]{0.67$^{*}$\\{\scriptsize (0.082)}} & \makecell[r]{2.59$^{**}$\\{\scriptsize (0.038)}} & \makecell[r]{7.15$^{**}$\\{\scriptsize (0.037)}} \\
 &  & Lead $+2$ & \makecell[r]{-0.82\\{\scriptsize (0.104)}} & \makecell[r]{2.10$^{***}$\\{\scriptsize (0.005)}} & \makecell[r]{4.54$^{***}$\\{\scriptsize (0.003)}} & \makecell[r]{5.56$^{**}$\\{\scriptsize (0.025)}} \\
 & Chronos Mini & \textbf{PIT ($-1$)} & \makecell[r]{1.83$^{***}$\\{\scriptsize (0.007)}} & \makecell[r]{5.83$^{***}$\\{\scriptsize ($<$0.001)}} & \makecell[r]{9.57$^{***}$\\{\scriptsize (0.002)}} & \makecell[r]{13.67$^{**}$\\{\scriptsize (0.018)}} \\
 &  & Lead $0$ & \makecell[r]{-1.24\\{\scriptsize (0.271)}} & \makecell[r]{1.89$^{**}$\\{\scriptsize (0.032)}} & \makecell[r]{5.50$^{***}$\\{\scriptsize (0.008)}} & \makecell[r]{10.23$^{**}$\\{\scriptsize (0.023)}} \\
 &  & Lead $+1$ & \makecell[r]{-1.17\\{\scriptsize (0.382)}} & \makecell[r]{0.54$^{*}$\\{\scriptsize (0.069)}} & \makecell[r]{3.59$^{**}$\\{\scriptsize (0.017)}} & \makecell[r]{7.29$^{*}$\\{\scriptsize (0.050)}} \\
 &  & Lead $+2$ & \makecell[r]{-1.56\\{\scriptsize (0.533)}} & \makecell[r]{0.27$^{**}$\\{\scriptsize (0.049)}} & \makecell[r]{2.71$^{***}$\\{\scriptsize (0.006)}} & \makecell[r]{6.15$^{**}$\\{\scriptsize (0.036)}} \\
 & Chronos Small & \textbf{PIT ($-1$)} & \makecell[r]{-0.90\\{\scriptsize (0.130)}} & \makecell[r]{1.11$^{**}$\\{\scriptsize (0.020)}} & \makecell[r]{7.11$^{***}$\\{\scriptsize (0.003)}} & \makecell[r]{11.46$^{**}$\\{\scriptsize (0.017)}} \\
 &  & Lead $0$ & \makecell[r]{-0.97\\{\scriptsize (0.164)}} & \makecell[r]{0.19$^{*}$\\{\scriptsize (0.066)}} & \makecell[r]{3.96$^{**}$\\{\scriptsize (0.017)}} & \makecell[r]{5.74$^{*}$\\{\scriptsize (0.052)}} \\
 &  & Lead $+1$ & \makecell[r]{0.72$^{**}$\\{\scriptsize (0.013)}} & \makecell[r]{-0.94\\{\scriptsize (0.106)}} & \makecell[r]{-0.29$^{*}$\\{\scriptsize (0.078)}} & \makecell[r]{4.50$^{*}$\\{\scriptsize (0.069)}} \\
 &  & Lead $+2$ & \makecell[r]{-0.11$^{**}$\\{\scriptsize (0.038)}} & \makecell[r]{1.25$^{***}$\\{\scriptsize (0.002)}} & \makecell[r]{0.68$^{**}$\\{\scriptsize (0.024)}} & \makecell[r]{1.45$^{*}$\\{\scriptsize (0.073)}} \\
 & TimesFM 8M & \textbf{PIT ($-1$)} & \makecell[r]{-25.94\\{\scriptsize (0.826)}} & \makecell[r]{-10.07\\{\scriptsize (0.177)}} & \makecell[r]{1.53$^{**}$\\{\scriptsize (0.011)}} & \makecell[r]{7.61$^{***}$\\{\scriptsize (0.002)}} \\
 &  & Lead $0$ & \makecell[r]{-34.90\\{\scriptsize (0.987)}} & \makecell[r]{-27.53\\{\scriptsize (0.963)}} & \makecell[r]{-24.43\\{\scriptsize (0.920)}} & \makecell[r]{-14.07\\{\scriptsize (0.324)}} \\
 &  & Lead $+1$ & \makecell[r]{-23.87\\{\scriptsize (0.405)}} & \makecell[r]{-16.49\\{\scriptsize (0.336)}} & \makecell[r]{-13.16\\{\scriptsize (0.336)}} & \makecell[r]{-22.79\\{\scriptsize (0.679)}} \\
 &  & Lead $+2$ & \makecell[r]{-33.67\\{\scriptsize (0.995)}} & \makecell[r]{-27.05\\{\scriptsize (0.841)}} & \makecell[r]{-26.02\\{\scriptsize (0.831)}} & \makecell[r]{-30.83\\{\scriptsize (0.884)}} \\
 & TimesFM 20M & \textbf{PIT ($-1$)} & \makecell[r]{-16.42\\{\scriptsize (0.672)}} & \makecell[r]{-7.43$^{**}$\\{\scriptsize (0.025)}} & \makecell[r]{-4.01$^{*}$\\{\scriptsize (0.057)}} & \makecell[r]{-28.19\\{\scriptsize (0.332)}} \\
 &  & Lead $0$ & \makecell[r]{-16.93\\{\scriptsize (0.539)}} & \makecell[r]{-15.13\\{\scriptsize (0.157)}} & \makecell[r]{-12.84\\{\scriptsize (0.148)}} & \makecell[r]{-11.49$^{*}$\\{\scriptsize (0.095)}} \\
 &  & Lead $+1$ & \makecell[r]{-22.99\\{\scriptsize (0.979)}} & \makecell[r]{-35.46\\{\scriptsize (0.957)}} & \makecell[r]{-43.28\\{\scriptsize (0.908)}} & \makecell[r]{-55.72\\{\scriptsize (0.926)}} \\
 &  & Lead $+2$ & \makecell[r]{-25.42\\{\scriptsize (0.988)}} & \makecell[r]{-27.19\\{\scriptsize (0.752)}} & \makecell[r]{-26.02\\{\scriptsize (0.470)}} & \makecell[r]{-38.02\\{\scriptsize (0.534)}} \\
\addlinespace[2pt]
\multicolumn{7}{@{}l}{\textit{Panel B. Global training-information environment}} \\*
Global & Chronos Tiny & \textbf{PIT ($-1$)} & \makecell[r]{0.03$^{**}$\\{\scriptsize (0.034)}} & \makecell[r]{1.52$^{**}$\\{\scriptsize (0.013)}} & \makecell[r]{4.36$^{***}$\\{\scriptsize (0.006)}} & \makecell[r]{8.27$^{**}$\\{\scriptsize (0.015)}} \\
 &  & Lead $0$ & \makecell[r]{0.47$^{**}$\\{\scriptsize (0.019)}} & \makecell[r]{2.28$^{***}$\\{\scriptsize (0.007)}} & \makecell[r]{4.64$^{***}$\\{\scriptsize (0.005)}} & \makecell[r]{7.57$^{**}$\\{\scriptsize (0.011)}} \\
 &  & Lead $+1$ & \makecell[r]{-0.76\\{\scriptsize (0.168)}} & \makecell[r]{1.24$^{**}$\\{\scriptsize (0.026)}} & \makecell[r]{4.03$^{***}$\\{\scriptsize (0.008)}} & \makecell[r]{7.28$^{**}$\\{\scriptsize (0.016)}} \\
 &  & Lead $+2$ & \makecell[r]{-0.09$^{*}$\\{\scriptsize (0.075)}} & \makecell[r]{0.14$^{*}$\\{\scriptsize (0.053)}} & \makecell[r]{1.78$^{**}$\\{\scriptsize (0.023)}} & \makecell[r]{4.28$^{**}$\\{\scriptsize (0.031)}} \\
 & Chronos Mini & \textbf{PIT ($-1$)} & \makecell[r]{0.27$^{**}$\\{\scriptsize (0.036)}} & \makecell[r]{2.23$^{**}$\\{\scriptsize (0.013)}} & \makecell[r]{5.35$^{***}$\\{\scriptsize (0.005)}} & \makecell[r]{9.03$^{**}$\\{\scriptsize (0.011)}} \\
 &  & Lead $0$ & \makecell[r]{-0.98\\{\scriptsize (0.172)}} & \makecell[r]{0.15$^{*}$\\{\scriptsize (0.058)}} & \makecell[r]{2.76$^{**}$\\{\scriptsize (0.021)}} & \makecell[r]{3.89$^{**}$\\{\scriptsize (0.040)}} \\
 &  & Lead $+1$ & \makecell[r]{0.44$^{***}$\\{\scriptsize (0.010)}} & \makecell[r]{2.14$^{***}$\\{\scriptsize (0.004)}} & \makecell[r]{3.53$^{***}$\\{\scriptsize (0.006)}} & \makecell[r]{4.33$^{**}$\\{\scriptsize (0.039)}} \\
 &  & Lead $+2$ & \makecell[r]{0.27$^{**}$\\{\scriptsize (0.032)}} & \makecell[r]{2.47$^{***}$\\{\scriptsize (0.002)}} & \makecell[r]{4.60$^{***}$\\{\scriptsize (0.003)}} & \makecell[r]{6.31$^{**}$\\{\scriptsize (0.017)}} \\
 & Chronos Small & \textbf{PIT ($-1$)} & \makecell[r]{-0.03$^{**}$\\{\scriptsize (0.025)}} & \makecell[r]{0.72$^{**}$\\{\scriptsize (0.017)}} & \makecell[r]{2.44$^{**}$\\{\scriptsize (0.015)}} & \makecell[r]{4.17$^{**}$\\{\scriptsize (0.034)}} \\
 &  & Lead $0$ & \makecell[r]{0.07$^{**}$\\{\scriptsize (0.025)}} & \makecell[r]{2.23$^{***}$\\{\scriptsize (0.005)}} & \makecell[r]{5.63$^{***}$\\{\scriptsize (0.002)}} & \makecell[r]{7.74$^{**}$\\{\scriptsize (0.015)}} \\
 &  & Lead $+1$ & \makecell[r]{-0.02$^{**}$\\{\scriptsize (0.026)}} & \makecell[r]{0.84$^{**}$\\{\scriptsize (0.026)}} & \makecell[r]{2.85$^{**}$\\{\scriptsize (0.010)}} & \makecell[r]{4.66$^{**}$\\{\scriptsize (0.030)}} \\
 &  & Lead $+2$ & \makecell[r]{0.08$^{**}$\\{\scriptsize (0.026)}} & \makecell[r]{1.04$^{**}$\\{\scriptsize (0.012)}} & \makecell[r]{3.48$^{***}$\\{\scriptsize (0.006)}} & \makecell[r]{6.58$^{**}$\\{\scriptsize (0.014)}} \\
 & TimesFM 8M & \textbf{PIT ($-1$)} & \makecell[r]{-41.14\\{\scriptsize (0.878)}} & \makecell[r]{-114.80\\{\scriptsize (0.801)}} & \makecell[r]{-162.56\\{\scriptsize (0.803)}} & \makecell[r]{-90.54\\{\scriptsize (0.606)}} \\
 &  & Lead $0$ & \makecell[r]{-29.59\\{\scriptsize (0.569)}} & \makecell[r]{-84.06\\{\scriptsize (0.576)}} & \makecell[r]{-124.57\\{\scriptsize (0.740)}} & \makecell[r]{-69.37\\{\scriptsize (0.488)}} \\
 &  & Lead $+1$ & \makecell[r]{-32.66\\{\scriptsize (0.162)}} & \makecell[r]{-70.20\\{\scriptsize (0.105)}} & \makecell[r]{-84.92$^{**}$\\{\scriptsize (0.045)}} & \makecell[r]{-61.11\\{\scriptsize (0.162)}} \\
 &  & Lead $+2$ & \makecell[r]{-44.94\\{\scriptsize (0.864)}} & \makecell[r]{-98.39\\{\scriptsize (0.811)}} & \makecell[r]{-125.36\\{\scriptsize (0.719)}} & \makecell[r]{-66.73\\{\scriptsize (0.462)}} \\
 & TimesFM 20M & \textbf{PIT ($-1$)} & \makecell[r]{-26.12\\{\scriptsize (0.194)}} & \makecell[r]{-14.37$^{*}$\\{\scriptsize (0.077)}} & \makecell[r]{-15.81\\{\scriptsize (0.275)}} & \makecell[r]{-58.59\\{\scriptsize (0.665)}} \\
 &  & Lead $0$ & \makecell[r]{-14.34$^{***}$\\{\scriptsize (0.006)}} & \makecell[r]{3.19$^{***}$\\{\scriptsize ($<$0.001)}} & \makecell[r]{2.39$^{**}$\\{\scriptsize (0.013)}} & \makecell[r]{-25.43\\{\scriptsize (0.220)}} \\
 &  & Lead $+1$ & \makecell[r]{-43.05\\{\scriptsize (0.992)}} & \makecell[r]{-27.94\\{\scriptsize (0.900)}} & \makecell[r]{-8.76\\{\scriptsize (0.135)}} & \makecell[r]{-9.16\\{\scriptsize (0.101)}} \\
 &  & Lead $+2$ & \makecell[r]{-33.10\\{\scriptsize (0.616)}} & \makecell[r]{-28.99\\{\scriptsize (0.780)}} & \makecell[r]{-32.86\\{\scriptsize (0.926)}} & \makecell[r]{-47.20\\{\scriptsize (0.460)}} \\
\addlinespace[2pt]
\multicolumn{7}{@{}l}{\textit{Panel C. Augmented training-information environment}} \\*
Augmented & Chronos Tiny & \textbf{PIT ($-1$)} & \makecell[r]{0.31$^{*}$\\{\scriptsize (0.064)}} & \makecell[r]{1.43$^{**}$\\{\scriptsize (0.010)}} & \makecell[r]{2.73$^{***}$\\{\scriptsize (0.007)}} & \makecell[r]{3.92$^{**}$\\{\scriptsize (0.030)}} \\
 &  & Lead $0$ & \makecell[r]{-2.30\\{\scriptsize (0.271)}} & \makecell[r]{1.13$^{***}$\\{\scriptsize (0.006)}} & \makecell[r]{3.84$^{***}$\\{\scriptsize (0.005)}} & \makecell[r]{5.40$^{**}$\\{\scriptsize (0.029)}} \\
 &  & Lead $+1$ & \makecell[r]{-0.47$^{*}$\\{\scriptsize (0.068)}} & \makecell[r]{0.91$^{***}$\\{\scriptsize (0.007)}} & \makecell[r]{4.14$^{***}$\\{\scriptsize (0.005)}} & \makecell[r]{6.77$^{**}$\\{\scriptsize (0.028)}} \\
 &  & Lead $+2$ & \makecell[r]{-0.16$^{*}$\\{\scriptsize (0.077)}} & \makecell[r]{2.26$^{***}$\\{\scriptsize (0.002)}} & \makecell[r]{5.08$^{***}$\\{\scriptsize (0.001)}} & \makecell[r]{6.90$^{**}$\\{\scriptsize (0.017)}} \\
 & Chronos Mini & \textbf{PIT ($-1$)} & \makecell[r]{0.68$^{**}$\\{\scriptsize (0.037)}} & \makecell[r]{2.35$^{***}$\\{\scriptsize (0.006)}} & \makecell[r]{5.05$^{***}$\\{\scriptsize (0.002)}} & \makecell[r]{8.46$^{**}$\\{\scriptsize (0.013)}} \\
 &  & Lead $0$ & \makecell[r]{-2.67\\{\scriptsize (0.375)}} & \makecell[r]{-1.44$^{**}$\\{\scriptsize (0.048)}} & \makecell[r]{0.94$^{**}$\\{\scriptsize (0.015)}} & \makecell[r]{3.80$^{**}$\\{\scriptsize (0.041)}} \\
 &  & Lead $+1$ & \makecell[r]{-0.25\\{\scriptsize (0.143)}} & \makecell[r]{1.68$^{***}$\\{\scriptsize (0.004)}} & \makecell[r]{4.03$^{***}$\\{\scriptsize (0.003)}} & \makecell[r]{5.81$^{**}$\\{\scriptsize (0.029)}} \\
 &  & Lead $+2$ & \makecell[r]{-2.81\\{\scriptsize (0.440)}} & \makecell[r]{-2.21$^{*}$\\{\scriptsize (0.077)}} & \makecell[r]{1.70$^{***}$\\{\scriptsize (0.007)}} & \makecell[r]{-1.78$^{*}$\\{\scriptsize (0.087)}} \\
 & Chronos Small & \textbf{PIT ($-1$)} & \makecell[r]{-0.73\\{\scriptsize (0.235)}} & \makecell[r]{0.08$^{*}$\\{\scriptsize (0.075)}} & \makecell[r]{3.38$^{***}$\\{\scriptsize (0.008)}} & \makecell[r]{7.89$^{**}$\\{\scriptsize (0.011)}} \\
 &  & Lead $0$ & \makecell[r]{-1.90\\{\scriptsize (0.220)}} & \makecell[r]{0.93$^{***}$\\{\scriptsize (0.007)}} & \makecell[r]{3.29$^{***}$\\{\scriptsize (0.004)}} & \makecell[r]{2.66$^{**}$\\{\scriptsize (0.040)}} \\
 &  & Lead $+1$ & \makecell[r]{0.26$^{**}$\\{\scriptsize (0.022)}} & \makecell[r]{2.09$^{***}$\\{\scriptsize (0.002)}} & \makecell[r]{4.30$^{***}$\\{\scriptsize (0.002)}} & \makecell[r]{3.99$^{**}$\\{\scriptsize (0.032)}} \\
 &  & Lead $+2$ & \makecell[r]{-0.37\\{\scriptsize (0.130)}} & \makecell[r]{0.61$^{*}$\\{\scriptsize (0.052)}} & \makecell[r]{3.60$^{**}$\\{\scriptsize (0.010)}} & \makecell[r]{5.67$^{**}$\\{\scriptsize (0.033)}} \\
 & TimesFM 8M & \textbf{PIT ($-1$)} & \makecell[r]{-14.78\\{\scriptsize (0.928)}} & \makecell[r]{-5.34\\{\scriptsize (0.452)}} & \makecell[r]{-8.43\\{\scriptsize (0.411)}} & \makecell[r]{-9.79\\{\scriptsize (0.302)}} \\
 &  & Lead $0$ & \makecell[r]{-9.79\\{\scriptsize (0.945)}} & \makecell[r]{-1.82\\{\scriptsize (0.148)}} & \makecell[r]{-6.46\\{\scriptsize (0.543)}} & \makecell[r]{-3.93\\{\scriptsize (0.175)}} \\
 &  & Lead $+1$ & \makecell[r]{1.22$^{**}$\\{\scriptsize (0.033)}} & \makecell[r]{-0.28\\{\scriptsize (0.121)}} & \makecell[r]{7.30\\{\scriptsize (0.126)}} & \makecell[r]{-3.49\\{\scriptsize (0.279)}} \\
 &  & Lead $+2$ & \makecell[r]{-11.15\\{\scriptsize (0.752)}} & \makecell[r]{-3.14\\{\scriptsize (0.326)}} & \makecell[r]{-1.14$^{*}$\\{\scriptsize (0.097)}} & \makecell[r]{-0.54\\{\scriptsize (0.172)}} \\
 & TimesFM 20M & \textbf{PIT ($-1$)} & \makecell[r]{-6.72\\{\scriptsize (0.943)}} & \makecell[r]{-5.53\\{\scriptsize (0.714)}} & \makecell[r]{-7.14\\{\scriptsize (0.721)}} & \makecell[r]{-11.57\\{\scriptsize (0.820)}} \\
 &  & Lead $0$ & \makecell[r]{-3.60\\{\scriptsize (0.714)}} & \makecell[r]{0.13$^{*}$\\{\scriptsize (0.087)}} & \makecell[r]{3.91$^{*}$\\{\scriptsize (0.058)}} & \makecell[r]{0.84\\{\scriptsize (0.131)}} \\
 &  & Lead $+1$ & \makecell[r]{-0.79\\{\scriptsize (0.113)}} & \makecell[r]{1.38$^{**}$\\{\scriptsize (0.035)}} & \makecell[r]{5.10$^{**}$\\{\scriptsize (0.010)}} & \makecell[r]{7.12$^{**}$\\{\scriptsize (0.025)}} \\
 &  & Lead $+2$ & \makecell[r]{-217206.19\\{\scriptsize (0.936)}} & \makecell[r]{-153983.36\\{\scriptsize (0.870)}} & \makecell[r]{-74795.84\\{\scriptsize (0.623)}} & \makecell[r]{-35416.62\\{\scriptsize (0.171)}} \\
\end{longtable}

\normalsize
\renewcommand{\arraystretch}{1.0}

\setlength{\tabcolsep}{6pt}
\clearpage
\begin{landscape}
\begingroup
\small
\renewcommand{\arraystretch}{0.98}
\fontsize{10}{12}\selectfont
\setlength{\tabcolsep}{4.00pt}
\setlength{\linewidth}{512.59pt}
\begin{longtable}{@{}lccccccccccc@{}}
\caption[Forecast-cell inference for rolling annual-state effects]{%
  \textbf{Forecast-cell inference for rolling annual-state effects.}
}\label{tab:ia-x43-complete-inference}\\
\multicolumn{12}{@{}p{\linewidth}@{}}{\exhibitnotes{%
  Forecast-cell fixed-effect coefficients compare rolling annual states with origin-aligned point-in-time (PIT) lead $-1$. Coefficients are in historical-average mean squared forecast error percentage points; positive values favor the alternative state. Panel A gives leads $0,+1,+2$, and Panel B gives the PIT-to-stale comparison at lead $-2$. Columns $p_D$ and $p_I$ contain one-sided HAC $p$-values for deterioration and improvement; $p_2$ contains the two-sided HAC $p$-value. HAC standard errors and 95\% intervals use Bartlett--Newey--West estimation with $h-1$ lags. Bootstrap 95\% intervals and two-sided $p$-values use 2,000 circular moving-block replications with block length $\max(h,6)$. The horizon $h$ is in months, and $N$ counts matched target months. Each row uses maximal paired support with PIT in the all-available U.S. equity-premium sample under the US training-information environment.
}}\\
\noalign{\vskip 6pt}
\toprule
Model & Lead & $h$ & $\widehat\beta$ & HAC SE & HAC 95\% CI & $p_D$ & $p_I$ & $p_2$ & Boot. 95\% CI & Boot. $p_2$ & $N$ \\
\midrule
\endfirsthead
\multicolumn{12}{l}{\textit{\tablename\ \thetable\ (continued)}}\\
\toprule
Model & Lead & $h$ & $\widehat\beta$ & HAC SE & HAC 95\% CI & $p_D$ & $p_I$ & $p_2$ & Boot. 95\% CI & Boot. $p_2$ & $N$ \\
\midrule
\endhead
\midrule
\multicolumn{12}{r}{\textit{Continued on next page}}\\
\endfoot
\bottomrule
\endlastfoot
\multicolumn{12}{@{}l}{\textit{Panel A: Exposed annual states}} \\
Chronos Tiny & 0 & 1 & -0.29 & 1.29 & [-2.83, 2.24] & 0.410 & 0.590 & 0.820 & [-2.27, 1.92] & 0.757 & 276 \\
 &  & 3 & -2.66 & 2.37 & [-7.30, 1.98] & 0.131 & 0.869 & 0.261 & [-6.96, 3.16] & 0.243 & 276 \\
 &  & 6 & -2.69 & 3.33 & [-9.22, 3.84] & 0.209 & 0.791 & 0.419 & [-8.45, 6.15] & 0.371 & 276 \\
 &  & 12 & 2.42 & 4.20 & [-5.82, 10.65] & 0.717 & 0.283 & 0.565 & [-5.15, 12.93] & 0.618 & 276 \\
\addlinespace[1pt]
 & +1 & 1 & -3.19 & 2.20 & [-7.51, 1.13] & 0.074 & 0.926 & 0.148 & [-6.74, 0.20] & 0.072 & 264 \\
 &  & 3 & -7.53 & 2.83 & [-13.07, -1.99] & 0.004 & 0.996 & 0.008 & [-12.78, -1.11] & 0.005 & 264 \\
 &  & 6 & -12.16 & 3.54 & [-19.11, -5.22] & $<$0.001 & 1.000 & $<$0.001 & [-17.68, -3.59] & $<$0.001 & 264 \\
 &  & 12 & -8.23 & 5.34 & [-18.70, 2.24] & 0.062 & 0.938 & 0.123 & [-18.75, 6.06] & 0.128 & 264 \\
\addlinespace[1pt]
 & +2 & 1 & -1.40 & 1.77 & [-4.87, 2.07] & 0.215 & 0.785 & 0.429 & [-4.80, 1.30] & 0.364 & 252 \\
 &  & 3 & -6.03 & 3.45 & [-12.79, 0.73] & 0.040 & 0.960 & 0.080 & [-14.45, 0.51] & 0.156 & 252 \\
 &  & 6 & -10.53 & 4.87 & [-20.08, -0.98] & 0.015 & 0.985 & 0.031 & [-21.66, -2.78] & 0.062 & 252 \\
 &  & 12 & -10.45 & 6.24 & [-22.67, 1.78] & 0.047 & 0.953 & 0.094 & [-23.64, -0.40] & 0.127 & 252 \\
\addlinespace[1pt]
\addlinespace[2pt]
Chronos Mini & 0 & 1 & -2.61 & 1.57 & [-5.70, 0.47] & 0.049 & 0.951 & 0.097 & [-5.52, 0.23] & 0.071 & 276 \\
 &  & 3 & -4.77 & 2.80 & [-10.24, 0.71] & 0.044 & 0.956 & 0.088 & [-11.40, 0.78] & 0.152 & 276 \\
 &  & 6 & -2.99 & 3.82 & [-10.48, 4.50] & 0.217 & 0.783 & 0.434 & [-11.23, 5.36] & 0.458 & 276 \\
 &  & 12 & 5.45 & 6.24 & [-6.78, 17.67] & 0.809 & 0.191 & 0.382 & [-6.31, 21.21] & 0.440 & 276 \\
\addlinespace[1pt]
 & +1 & 1 & -2.93 & 1.72 & [-6.31, 0.44] & 0.044 & 0.956 & 0.088 & [-6.17, -0.07] & 0.079 & 264 \\
 &  & 3 & -7.62 & 2.81 & [-13.12, -2.11] & 0.003 & 0.997 & 0.007 & [-14.04, -1.52] & 0.027 & 264 \\
 &  & 6 & -8.65 & 3.85 & [-16.19, -1.11] & 0.012 & 0.988 & 0.025 & [-17.29, -0.64] & 0.048 & 264 \\
 &  & 12 & -1.68 & 5.51 & [-12.49, 9.12] & 0.380 & 0.620 & 0.760 & [-12.08, 11.52] & 0.764 & 264 \\
\addlinespace[1pt]
 & +2 & 1 & -2.77 & 1.88 & [-6.44, 0.91] & 0.070 & 0.930 & 0.140 & [-6.22, 0.83] & 0.120 & 252 \\
 &  & 3 & -6.96 & 2.95 & [-12.74, -1.19] & 0.009 & 0.991 & 0.018 & [-13.36, -0.16] & 0.031 & 252 \\
 &  & 6 & -8.43 & 4.48 & [-17.22, 0.35] & 0.030 & 0.970 & 0.060 & [-17.68, 1.00] & 0.065 & 252 \\
 &  & 12 & -3.32 & 8.22 & [-19.42, 12.79] & 0.343 & 0.657 & 0.686 & [-18.45, 15.67] & 0.662 & 252 \\
\addlinespace[1pt]
\addlinespace[2pt]
Chronos Small & 0 & 1 & 2.72 & 2.08 & [-1.35, 6.79] & 0.905 & 0.095 & 0.190 & [-0.46, 6.04] & 0.100 & 276 \\
 &  & 3 & -3.54 & 3.71 & [-10.81, 3.72] & 0.170 & 0.830 & 0.339 & [-11.53, 3.15] & 0.370 & 276 \\
 &  & 6 & -5.46 & 4.29 & [-13.88, 2.96] & 0.102 & 0.898 & 0.204 & [-16.00, 2.21] & 0.270 & 276 \\
 &  & 12 & -12.00 & 7.99 & [-27.65, 3.65] & 0.066 & 0.934 & 0.133 & [-31.99, 2.20] & 0.195 & 276 \\
\addlinespace[1pt]
 & +1 & 1 & 4.24 & 2.56 & [-0.77, 9.26] & 0.951 & 0.049 & 0.097 & [-1.00, 9.91] & 0.135 & 264 \\
 &  & 3 & -4.33 & 4.91 & [-13.95, 5.29] & 0.189 & 0.811 & 0.378 & [-15.16, 5.82] & 0.408 & 264 \\
 &  & 6 & -16.17 & 10.49 & [-36.73, 4.40] & 0.062 & 0.938 & 0.123 & [-39.83, 1.37] & 0.162 & 264 \\
 &  & 12 & -19.52 & 10.31 & [-39.73, 0.69] & 0.029 & 0.971 & 0.058 & [-45.75, -1.82] & 0.133 & 264 \\
\addlinespace[1pt]
 & +2 & 1 & 1.95 & 2.50 & [-2.94, 6.85] & 0.783 & 0.217 & 0.434 & [-3.63, 7.87] & 0.496 & 252 \\
 &  & 3 & -3.13 & 4.87 & [-12.68, 6.42] & 0.261 & 0.739 & 0.521 & [-15.24, 6.90] & 0.654 & 252 \\
 &  & 6 & -16.16 & 13.74 & [-43.08, 10.77] & 0.120 & 0.880 & 0.240 & [-49.96, 4.62] & 0.288 & 252 \\
 &  & 12 & -15.18 & 10.95 & [-36.64, 6.28] & 0.083 & 0.917 & 0.166 & [-42.86, 2.96] & 0.224 & 252 \\
\addlinespace[1pt]
\addlinespace[2pt]
TimesFM 8M & 0 & 1 & -4.72 & 5.31 & [-15.12, 5.68] & 0.187 & 0.813 & 0.374 & [-15.43, 5.37] & 0.375 & 276 \\
 &  & 3 & -6.59 & 5.81 & [-17.98, 4.80] & 0.128 & 0.872 & 0.257 & [-17.42, 7.68] & 0.235 & 276 \\
 &  & 6 & -13.36 & 7.39 & [-27.84, 1.11] & 0.035 & 0.965 & 0.070 & [-26.76, 4.65] & 0.052 & 276 \\
 &  & 12 & -12.17 & 6.21 & [-24.35, 0.00] & 0.025 & 0.975 & 0.050 & [-22.38, 2.87] & 0.025 & 276 \\
\addlinespace[1pt]
 & +1 & 1 & 3.20 & 4.56 & [-5.75, 12.14] & 0.758 & 0.242 & 0.484 & [-5.74, 12.22] & 0.494 & 264 \\
 &  & 3 & 2.07 & 6.65 & [-10.96, 15.10] & 0.622 & 0.378 & 0.756 & [-13.44, 16.49] & 0.743 & 264 \\
 &  & 6 & -1.68 & 8.81 & [-18.95, 15.60] & 0.425 & 0.575 & 0.849 & [-22.95, 14.70] & 0.870 & 264 \\
 &  & 12 & -8.79 & 7.94 & [-24.36, 6.77] & 0.134 & 0.866 & 0.268 & [-25.82, 9.93] & 0.300 & 264 \\
\addlinespace[1pt]
 & +2 & 1 & -4.94 & 4.27 & [-13.30, 3.43] & 0.124 & 0.876 & 0.247 & [-12.32, 2.06] & 0.178 & 252 \\
 &  & 3 & -10.27 & 6.27 & [-22.56, 2.02] & 0.051 & 0.949 & 0.101 & [-22.10, 5.82] & 0.105 & 252 \\
 &  & 6 & -18.48 & 8.65 & [-35.45, -1.52] & 0.016 & 0.984 & 0.033 & [-34.80, 2.88] & 0.028 & 252 \\
 &  & 12 & -22.54 & 8.01 & [-38.24, -6.84] & 0.002 & 0.998 & 0.005 & [-38.51, -3.88] & 0.006 & 252 \\
\addlinespace[1pt]
\addlinespace[2pt]
TimesFM 20M & 0 & 1 & -3.83 & 4.08 & [-11.82, 4.17] & 0.174 & 0.826 & 0.348 & [-11.76, 4.08] & 0.313 & 276 \\
 &  & 3 & -12.92 & 7.28 & [-27.20, 1.36] & 0.038 & 0.962 & 0.076 & [-27.04, 4.42] & 0.071 & 276 \\
 &  & 6 & -20.10 & 10.13 & [-39.95, -0.25] & 0.024 & 0.976 & 0.047 & [-38.29, 1.43] & 0.031 & 276 \\
 &  & 12 & -12.96 & 12.08 & [-36.64, 10.73] & 0.142 & 0.858 & 0.284 & [-38.71, 5.69] & 0.252 & 276 \\
\addlinespace[1pt]
 & +1 & 1 & -6.95 & 4.36 & [-15.50, 1.60] & 0.056 & 0.944 & 0.111 & [-15.56, 1.23] & 0.113 & 264 \\
 &  & 3 & -25.34 & 7.73 & [-40.49, -10.18] & $<$0.001 & 0.999 & 0.001 & [-40.85, -6.97] & 0.002 & 264 \\
 &  & 6 & -34.51 & 11.28 & [-56.62, -12.41] & 0.001 & 0.999 & 0.002 & [-53.04, -8.71] & $<$0.001 & 264 \\
 &  & 12 & -36.99 & 13.50 & [-63.45, -10.53] & 0.003 & 0.997 & 0.006 & [-60.79, -10.52] & $<$0.001 & 264 \\
\addlinespace[1pt]
 & +2 & 1 & -2.87 & 3.94 & [-10.60, 4.86] & 0.234 & 0.766 & 0.467 & [-9.78, 4.36] & 0.388 & 252 \\
 &  & 3 & -12.02 & 5.20 & [-22.21, -1.84] & 0.010 & 0.990 & 0.021 & [-23.19, -0.85] & 0.036 & 252 \\
 &  & 6 & -19.94 & 6.61 & [-32.89, -6.98] & 0.001 & 0.999 & 0.003 & [-35.09, -5.21] & 0.021 & 252 \\
 &  & 12 & -20.54 & 12.10 & [-44.24, 3.17] & 0.045 & 0.955 & 0.090 & [-44.45, 6.19] & 0.090 & 252 \\
\addlinespace[1pt]
\addlinespace[2pt]
\addlinespace[5pt]
\multicolumn{12}{@{}l}{\textit{Panel B: Stale annual-state comparison}} \\
Chronos Tiny & -2 & 1 & 1.00 & 1.17 & [-1.30, 3.30] & 0.803 & 0.197 & 0.394 & [-1.78, 3.57] & 0.393 & 276 \\
 &  & 3 & 3.48 & 2.53 & [-1.48, 8.44] & 0.915 & 0.085 & 0.170 & [-2.52, 8.60] & 0.203 & 274 \\
 &  & 6 & 0.76 & 3.11 & [-5.35, 6.86] & 0.596 & 0.404 & 0.808 & [-6.70, 6.61] & 0.813 & 271 \\
 &  & 12 & 1.64 & 4.32 & [-6.82, 10.11] & 0.648 & 0.352 & 0.703 & [-9.78, 10.79] & 0.733 & 265 \\
\addlinespace[1pt]
\addlinespace[2pt]
Chronos Mini & -2 & 1 & -1.85 & 1.42 & [-4.62, 0.92] & 0.095 & 0.905 & 0.191 & [-4.56, 0.63] & 0.181 & 276 \\
 &  & 3 & -4.86 & 3.54 & [-11.79, 2.07] & 0.085 & 0.915 & 0.169 & [-13.63, 2.17] & 0.260 & 274 \\
 &  & 6 & -5.69 & 4.52 & [-14.55, 3.18] & 0.104 & 0.896 & 0.209 & [-16.33, 2.52] & 0.262 & 271 \\
 &  & 12 & -3.27 & 8.12 & [-19.18, 12.64] & 0.343 & 0.657 & 0.687 & [-24.39, 14.08] & 0.679 & 265 \\
\addlinespace[1pt]
\addlinespace[2pt]
Chronos Small & -2 & 1 & 4.40 & 1.81 & [0.86, 7.94] & 0.993 & 0.007 & 0.015 & [1.14, 8.10] & 0.019 & 276 \\
 &  & 3 & 3.48 & 3.86 & [-4.09, 11.05] & 0.816 & 0.184 & 0.368 & [-6.38, 12.55] & 0.399 & 274 \\
 &  & 6 & 4.44 & 4.70 & [-4.78, 13.66] & 0.827 & 0.173 & 0.345 & [-7.53, 13.40] & 0.336 & 271 \\
 &  & 12 & 1.91 & 5.06 & [-8.00, 11.81] & 0.647 & 0.353 & 0.706 & [-11.21, 10.50] & 0.688 & 265 \\
\addlinespace[1pt]
\addlinespace[2pt]
TimesFM 8M & -2 & 1 & 5.17 & 5.06 & [-4.74, 15.08] & 0.847 & 0.153 & 0.307 & [-7.26, 16.73] & 0.371 & 276 \\
 &  & 3 & 14.78 & 5.96 & [3.10, 26.47] & 0.993 & 0.007 & 0.013 & [0.23, 26.33] & 0.011 & 274 \\
 &  & 6 & 11.60 & 6.56 & [-1.26, 24.46] & 0.961 & 0.039 & 0.077 & [-3.55, 24.15] & 0.068 & 271 \\
 &  & 12 & 6.64 & 6.18 & [-5.47, 18.75] & 0.859 & 0.141 & 0.282 & [-7.25, 21.73] & 0.313 & 265 \\
\addlinespace[1pt]
\addlinespace[2pt]
TimesFM 20M & -2 & 1 & -5.40 & 3.74 & [-12.72, 1.92] & 0.074 & 0.926 & 0.148 & [-12.45, 1.84] & 0.126 & 276 \\
 &  & 3 & -15.89 & 6.58 & [-28.79, -3.00] & 0.008 & 0.992 & 0.016 & [-29.55, -1.31] & 0.027 & 274 \\
 &  & 6 & -21.40 & 7.56 & [-36.21, -6.59] & 0.002 & 0.998 & 0.005 & [-38.33, -4.93] & 0.020 & 271 \\
 &  & 12 & -15.30 & 9.85 & [-34.61, 4.01] & 0.060 & 0.940 & 0.120 & [-42.65, 4.14] & 0.202 & 265 \\
\end{longtable}
\normalsize
\renewcommand{\arraystretch}{1.0}

\endgroup
\end{landscape}
\clearpage
\begin{table}[!htbp]
\centering
\caption[Rolling forecasting performance in the balanced sample]{%
  \textbf{Rolling forecasting performance in the balanced sample.}
}\label{tab:ia-x43-balanced}
\begin{threeparttable}
\scriptsize
\setlength{\tabcolsep}{4.2pt}
\renewcommand{\arraystretch}{1.12}
\begin{tabular*}{\textwidth}{@{\extracolsep{\fill}}llrrrr@{}}
\toprule
Model & Annual State & $h=1$ & $h=3$ & $h=6$ & $h=12$ \\
\midrule
\multicolumn{6}{@{}l}{\textit{Panel A: Absolute performance relative to the historical average}}\\[-1pt]
Chronos Mini & \textbf{PIT ($-1$)} & \makecell[r]{3.08$^{***}$\\{\scriptsize (0.009)}} & \makecell[r]{9.01$^{***}$\\{\scriptsize (0.004)}} & \makecell[r]{10.77$^{**}$\\{\scriptsize (0.012)}} & \makecell[r]{3.02\\{\scriptsize (0.189)}} \\
 & Lead $0$ & \makecell[r]{0.49\\{\scriptsize (0.196)}} & \makecell[r]{6.00$^{***}$\\{\scriptsize (0.006)}} & \makecell[r]{9.66$^{***}$\\{\scriptsize (0.003)}} & \makecell[r]{10.26$^{**}$\\{\scriptsize (0.044)}} \\
 & Lead $+1$ & \makecell[r]{-0.31\\{\scriptsize (0.363)}} & \makecell[r]{1.58$^{*}$\\{\scriptsize (0.095)}} & \makecell[r]{2.28$^{**}$\\{\scriptsize (0.040)}} & \makecell[r]{1.63\\{\scriptsize (0.186)}} \\
 & Lead $+2$ & \makecell[r]{1.29\\{\scriptsize (0.145)}} & \makecell[r]{3.89$^{**}$\\{\scriptsize (0.031)}} & \makecell[r]{5.28$^{**}$\\{\scriptsize (0.010)}} & \makecell[r]{4.78$^{*}$\\{\scriptsize (0.084)}} \\
\addlinespace[2pt]
TimesFM 20M & \textbf{PIT ($-1$)} & \makecell[r]{-7.31\\{\scriptsize (0.428)}} & \makecell[r]{3.32$^{**}$\\{\scriptsize (0.028)}} & \makecell[r]{13.34$^{***}$\\{\scriptsize (0.009)}} & \makecell[r]{8.04$^{**}$\\{\scriptsize (0.015)}} \\
 & Lead $0$ & \makecell[r]{-13.38\\{\scriptsize (0.778)}} & \makecell[r]{-10.70\\{\scriptsize (0.463)}} & \makecell[r]{-7.20\\{\scriptsize (0.304)}} & \makecell[r]{-5.01\\{\scriptsize (0.177)}} \\
 & Lead $+1$ & \makecell[r]{-14.36\\{\scriptsize (0.824)}} & \makecell[r]{-20.73\\{\scriptsize (0.826)}} & \makecell[r]{-22.62\\{\scriptsize (0.735)}} & \makecell[r]{-33.04\\{\scriptsize (0.810)}} \\
 & Lead $+2$ & \makecell[r]{-11.64\\{\scriptsize (0.640)}} & \makecell[r]{-10.83\\{\scriptsize (0.547)}} & \makecell[r]{-10.55\\{\scriptsize (0.426)}} & \makecell[r]{-19.98\\{\scriptsize (0.584)}} \\
\addlinespace[2pt]
\midrule
\multicolumn{6}{@{}l}{\textit{Panel B: Pooled later-state effects relative to PIT}}\\[-1pt]
Chronos Tiny & Pooled $0,+1,+2$ & \makecell[r]{-1.09\\{\scriptsize (0.444)}} & \makecell[r]{-5.58$^{**}$\\{\scriptsize (0.020)}} & \makecell[r]{-8.91$^{***}$\\{\scriptsize (0.002)}} & \makecell[r]{-4.64\\{\scriptsize (0.287)}} \\
Chronos Mini & Pooled $0,+1,+2$ & \makecell[r]{-2.59$^{*}$\\{\scriptsize (0.092)}} & \makecell[r]{-5.19$^{**}$\\{\scriptsize (0.050)}} & \makecell[r]{-5.03\\{\scriptsize (0.196)}} & \makecell[r]{2.54\\{\scriptsize (0.727)}} \\
Chronos Small & Pooled $0,+1,+2$ & \makecell[r]{1.32\\{\scriptsize (0.568)}} & \makecell[r]{-4.96\\{\scriptsize (0.249)}} & \makecell[r]{-13.68\\{\scriptsize (0.140)}} & \makecell[r]{-15.63$^{*}$\\{\scriptsize (0.085)}} \\
\addlinespace[1pt]
TimesFM 8M & Pooled $0,+1,+2$ & \makecell[r]{-1.22\\{\scriptsize (0.788)}} & \makecell[r]{-3.86\\{\scriptsize (0.486)}} & \makecell[r]{-10.29\\{\scriptsize (0.153)}} & \makecell[r]{-12.15$^{**}$\\{\scriptsize (0.045)}} \\
TimesFM 20M & Pooled $0,+1,+2$ & \makecell[r]{-5.82$^{*}$\\{\scriptsize (0.099)}} & \makecell[r]{-17.41$^{***}$\\{\scriptsize (0.004)}} & \makecell[r]{-26.80$^{***}$\\{\scriptsize ($<$0.001)}} & \makecell[r]{-27.38$^{***}$\\{\scriptsize ($<$0.001)}} \\
\bottomrule
\end{tabular*}

\par\vspace{6pt}\nointerlineskip
\exhibitnotes{%
  The balanced January 2002--December 2021 U.S. sample is evaluated under the US training-information environment. Panel A gives raw out-of-sample $R^2_{\mathrm{HA}}$ for Chronos Mini and TimesFM 20M, with one-sided Clark--West $p$-values in parentheses. PIT lead $-1$ and leads $0,+1,+2$ share joint support. Panel B gives pooled PIT-relative effects for all five models, averaging the three later leads' squared-loss advantages on common support, with two-sided HAC $p$-values in parentheses. Both panels use historical-average MSFE percentage points, but Panel A compares each forecast with the historical average and Panel B compares later states with origin-aligned point-in-time (PIT) forecasts. Bartlett--Newey--West HAC inference uses $h-1$ lags, where $h$ is in months. $^{*}$, $^{**}$, and $^{***}$ indicate 10\%, 5\%, and 1\% significance for the test specified in each panel.
}
\end{threeparttable}
\end{table}


\clearpage
\section{Limitations}\label{sec:limitations}

The evidence covers five model variants from two numerical forecasting families, trained on daily firm-level data and applied to monthly aggregate equity returns. The findings do not establish performance across all pretrained architectures or asset classes. Extending this evaluation to language-model forecasts requires specifying prompts, retrieved context, sampling procedures, and the mapping from text outputs to return forecasts or investment decisions.

\clearpage

\section{Forecast Accuracy with Fixed Model Vintages}\label{sec:ia-fixed-results}

This section provides the inference, training-sample, target-window, and calendar-alignment evidence underlying main-text Section~\ref{a-common-parameter-state-across-historical-origins}.
\FloatBarrier

\begin{figure}[H]
\centering
\includegraphics[width=0.95\linewidth]{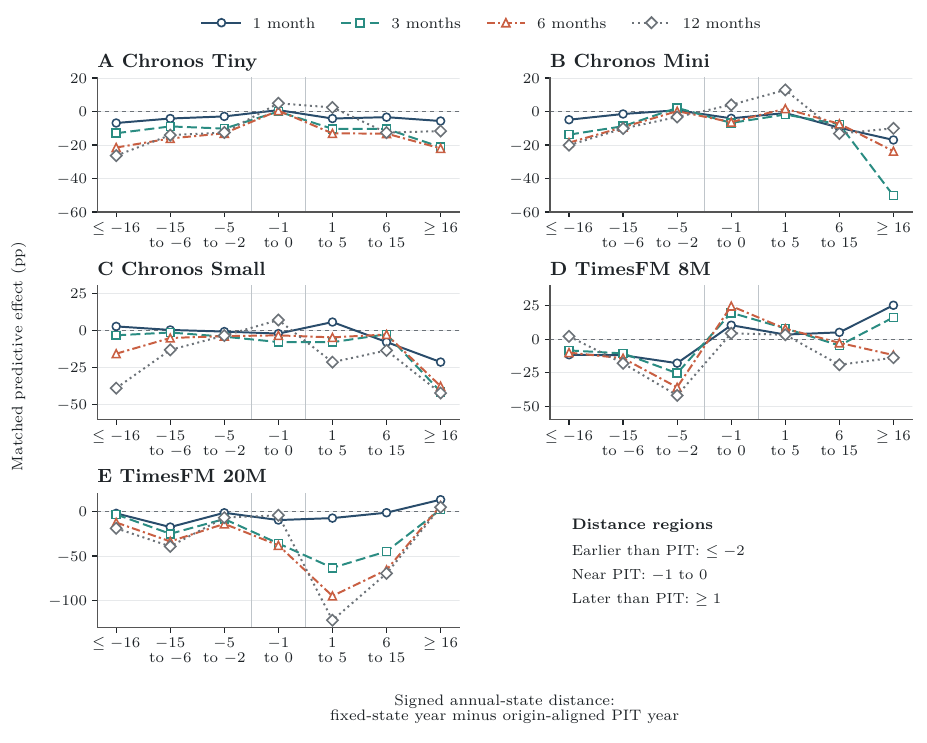}
\caption[Fixed-state predictive effects by distance from the PIT year]{%
  \textbf{Fixed-state predictive effects by distance from the PIT year.}
  Fixed-state matched effects are grouped by distance from the origin-aligned point-in-time (PIT) year. For state year $v$ and target-start year $Y(t_u)$, $g_u^{(v)}=v-[Y(t_u)-1]$: negative values place the state before PIT, zero identifies PIT, and positive values place it after PIT. Within each model--horizon cell, monthly PIT-minus-fixed-state squared-loss advantages are normalized by the cell's historical-average MSFE and averaged within the displayed integer-distance bins. Effects are in percentage points; positive effects favor the fixed state. Panels cover all five model variants on separate vertical scales. Lines distinguish horizons $h=1,3,6,12$ months. Estimates use the all-available U.S. equity-premium sample and the US training-information environment.
}\label{fig:ia-x44-all-model-alignment}

\end{figure}

\clearpage
\begingroup
\clearpage
\fontsize{11}{13.6}\selectfont
\setlength{\tabcolsep}{4.00pt}
\setlength{\linewidth}{425.61pt}
\begin{longtable}{@{}lcccccccc@{}}
\caption[Inference for U.S. fixed annual-state effects]{%
  \textbf{Inference for U.S. fixed annual-state effects.}
}\label{tab:ia-x44-us-inference}\\
\multicolumn{9}{@{}p{\linewidth}@{}}{\exhibitnotes{%
  Fixed-state matched predictive effects compare annual states 2000, 2009, and 2023 with origin-aligned point-in-time (PIT) forecasts. Effects equal PIT mean squared forecast error (MSFE) minus fixed-state MSFE, divided by historical-average MSFE on the same support and expressed in percentage points. Positive values favor the fixed state. Each state uses its full available matched deployment support, and $N$ counts target months. HAC 95\% intervals and two-sided $p$-values use Bartlett--Newey--West estimation with $h-1$ lags, where $h$ is in months. Bootstrap 95\% intervals use 2,000 circular moving-block replications with block length $\max(h,6)$. Benjamini--Hochberg $q$-values adjust one-sided HAC tests of improvement across the 60 displayed model--vintage--horizon cells. Estimates use the all-available U.S. equity-premium sample and the US training-information environment.
}}\\
\noalign{\vskip 6pt}
\toprule
Model & Fixed & $h$ & $\widehat\Delta$ & $N$ & HAC 95\% CI & HAC $p_2$ & Boot. 95\% CI & BH $q$ \\
\midrule
\endfirsthead
\multicolumn{9}{l}{\textit{\tablename\ \thetable\ (continued)}}\\
\toprule
Model & Fixed & $h$ & $\widehat\Delta$ & $N$ & HAC 95\% CI & HAC $p_2$ & Boot. 95\% CI & BH $q$ \\
\midrule
\endhead
\midrule\multicolumn{9}{r}{\textit{Continued on next page}}\\\endfoot
\bottomrule\endlastfoot
Chronos Tiny & 2000 & 1 & -5.33 & 288 & [-10.08, -0.58] & 0.028 & [-9.95, -1.58] & 1.000 \\
 &  & 3 & -14.32 & 286 & [-26.39, -2.25] & 0.020 & [-29.15, -2.65] & 1.000 \\
 &  & 6 & -25.96 & 283 & [-47.03, -4.89] & 0.016 & [-52.00, -10.56] & 1.000 \\
 &  & 12 & -32.77 & 277 & [-61.42, -4.11] & 0.025 & [-68.37, -13.36] & 1.000 \\
\addlinespace[1pt]
 & 2009 & 1 & -3.68 & 288 & [-8.19, 0.83] & 0.110 & [-8.43, 0.38] & 1.000 \\
 &  & 3 & -9.63 & 286 & [-18.71, -0.55] & 0.038 & [-24.77, 0.55] & 1.000 \\
 &  & 6 & -9.90 & 283 & [-23.13, 3.33] & 0.143 & [-30.14, 5.06] & 1.000 \\
 &  & 12 & 4.66 & 277 & [-17.84, 27.15] & 0.685 & [-19.15, 28.88] & 1.000 \\
\addlinespace[1pt]
 & 2023 & 1 & -3.08 & 288 & [-8.59, 2.44] & 0.274 & [-8.78, 1.16] & 1.000 \\
 &  & 3 & -8.43 & 286 & [-14.60, -2.26] & 0.007 & [-19.96, -0.59] & 1.000 \\
 &  & 6 & -8.96 & 283 & [-15.81, -2.11] & 0.010 & [-22.73, -0.09] & 1.000 \\
 &  & 12 & -5.21 & 277 & [-15.26, 4.84] & 0.310 & [-19.61, 6.54] & 1.000 \\
\addlinespace[1pt]
\addlinespace[3pt]
Chronos Mini & 2000 & 1 & -4.22 & 288 & [-7.89, -0.55] & 0.024 & [-7.59, -1.59] & 1.000 \\
 &  & 3 & -12.43 & 286 & [-19.32, -5.53] & $<$0.001 & [-22.59, -5.54] & 1.000 \\
 &  & 6 & -17.30 & 283 & [-29.52, -5.07] & 0.006 & [-32.46, -7.94] & 1.000 \\
 &  & 12 & -24.21 & 277 & [-44.42, -3.99] & 0.019 & [-55.76, -8.73] & 1.000 \\
\addlinespace[1pt]
 & 2009 & 1 & -1.60 & 288 & [-5.75, 2.54] & 0.448 & [-6.09, 1.83] & 1.000 \\
 &  & 3 & -6.29 & 286 & [-14.12, 1.53] & 0.115 & [-17.44, 2.79] & 1.000 \\
 &  & 6 & -4.69 & 283 & [-16.87, 7.50] & 0.451 & [-19.98, 8.12] & 1.000 \\
 &  & 12 & 4.78 & 277 & [-13.97, 23.52] & 0.617 & [-15.34, 26.67] & 1.000 \\
\addlinespace[1pt]
 & 2023 & 1 & -9.36 & 288 & [-19.98, 1.26] & 0.084 & [-31.60, 1.52] & 1.000 \\
 &  & 3 & -15.60 & 286 & [-35.90, 4.70] & 0.132 & [-50.07, 2.84] & 1.000 \\
 &  & 6 & -4.31 & 283 & [-16.07, 7.45] & 0.473 & [-19.95, 8.09] & 1.000 \\
 &  & 12 & -1.12 & 277 & [-14.97, 12.73] & 0.874 & [-16.53, 16.13] & 1.000 \\
\addlinespace[1pt]
\addlinespace[3pt]
Chronos Small & 2000 & 1 & 0.46 & 288 & [-3.15, 4.06] & 0.804 & [-2.80, 4.23] & 1.000 \\
 &  & 3 & -5.64 & 286 & [-13.36, 2.08] & 0.152 & [-15.47, 2.60] & 1.000 \\
 &  & 6 & -10.80 & 283 & [-25.18, 3.57] & 0.141 & [-26.09, 0.58] & 1.000 \\
 &  & 12 & -17.64 & 277 & [-41.06, 5.77] & 0.140 & [-46.71, -0.49] & 1.000 \\
\addlinespace[1pt]
 & 2009 & 1 & 1.57 & 288 & [-3.03, 6.17] & 0.504 & [-2.54, 6.16] & 1.000 \\
 &  & 3 & 0.93 & 286 & [-5.94, 7.79] & 0.792 & [-6.56, 8.01] & 1.000 \\
 &  & 6 & -0.70 & 283 & [-6.59, 5.19] & 0.815 & [-7.80, 5.27] & 1.000 \\
 &  & 12 & -4.30 & 277 & [-13.31, 4.72] & 0.350 & [-16.43, 6.45] & 1.000 \\
\addlinespace[1pt]
 & 2023 & 1 & -11.01 & 288 & [-22.16, 0.14] & 0.053 & [-34.27, 2.66] & 1.000 \\
 &  & 3 & -19.25 & 286 & [-35.75, -2.76] & 0.022 & [-44.92, -2.51] & 1.000 \\
 &  & 6 & -15.92 & 283 & [-30.50, -1.34] & 0.032 & [-38.46, -1.05] & 1.000 \\
 &  & 12 & -32.30 & 277 & [-68.21, 3.61] & 0.078 & [-76.59, -4.09] & 1.000 \\
\addlinespace[1pt]
\addlinespace[3pt]
TimesFM 8M & 2000 & 1 & -11.33 & 288 & [-20.43, -2.24] & 0.015 & [-25.79, 0.46] & 1.000 \\
 &  & 3 & -9.22 & 286 & [-24.55, 6.10] & 0.238 & [-28.23, 7.72] & 1.000 \\
 &  & 6 & -15.21 & 283 & [-37.87, 7.46] & 0.188 & [-41.98, 3.03] & 1.000 \\
 &  & 12 & -23.82 & 277 & [-53.07, 5.44] & 0.111 & [-64.37, -3.18] & 1.000 \\
\addlinespace[1pt]
 & 2009 & 1 & -2.38 & 288 & [-10.84, 6.07] & 0.581 & [-12.51, 9.84] & 1.000 \\
 &  & 3 & -1.94 & 286 & [-13.64, 9.77] & 0.746 & [-14.34, 11.05] & 1.000 \\
 &  & 6 & -6.92 & 283 & [-21.86, 8.02] & 0.364 & [-23.83, 6.76] & 1.000 \\
 &  & 12 & -10.38 & 277 & [-23.14, 2.37] & 0.111 & [-28.16, -0.09] & 1.000 \\
\addlinespace[1pt]
 & 2023 & 1 & 9.62 & 288 & [-1.08, 20.32] & 0.078 & [0.09, 23.92] & 1.000 \\
 &  & 3 & 1.78 & 286 & [-11.07, 14.64] & 0.786 & [-11.53, 18.78] & 1.000 \\
 &  & 6 & -1.78 & 283 & [-13.37, 9.80] & 0.763 & [-14.76, 10.81] & 1.000 \\
 &  & 12 & -7.77 & 277 & [-19.96, 4.42] & 0.212 & [-26.85, 5.83] & 1.000 \\
\addlinespace[1pt]
\addlinespace[3pt]
TimesFM 20M & 2000 & 1 & -14.99 & 288 & [-25.08, -4.91] & 0.004 & [-32.13, -3.17] & 1.000 \\
 &  & 3 & -22.07 & 286 & [-36.08, -8.06] & 0.002 & [-45.45, -7.63] & 1.000 \\
 &  & 6 & -30.53 & 283 & [-46.61, -14.45] & $<$0.001 & [-61.80, -14.10] & 1.000 \\
 &  & 12 & -26.00 & 277 & [-48.27, -3.73] & 0.022 & [-62.03, -6.94] & 1.000 \\
\addlinespace[1pt]
 & 2009 & 1 & -3.01 & 288 & [-11.74, 5.72] & 0.499 & [-12.86, 5.92] & 1.000 \\
 &  & 3 & -41.48 & 286 & [-63.54, -19.42] & $<$0.001 & [-89.43, -10.58] & 1.000 \\
 &  & 6 & -61.75 & 283 & [-96.78, -26.72] & $<$0.001 & [-145.30, -15.66] & 1.000 \\
 &  & 12 & -72.59 & 277 & [-121.80, -23.38] & 0.004 & [-187.60, -9.08] & 1.000 \\
\addlinespace[1pt]
 & 2023 & 1 & 0.82 & 288 & [-9.34, 10.97] & 0.875 & [-8.66, 10.46] & 1.000 \\
 &  & 3 & -20.05 & 286 & [-34.34, -5.76] & 0.006 & [-36.98, -7.77] & 1.000 \\
 &  & 6 & -27.60 & 283 & [-49.63, -5.56] & 0.014 & [-54.27, -11.91] & 1.000 \\
 &  & 12 & -34.74 & 277 & [-69.86, 0.38] & 0.053 & [-78.24, -12.82] & 1.000 \\
\addlinespace[1pt]
\addlinespace[3pt]
\end{longtable}

\endgroup
\clearpage
\begingroup
\singlespacing\normalsize
\renewcommand{\arraystretch}{1}
\setlength{\tabcolsep}{4pt}
\begin{longtable}{@{}>{\raggedright\arraybackslash}p{74pt}>{\raggedleft\arraybackslash}p{30pt}*{4}{>{\raggedleft\arraybackslash}p{\dimexpr(\linewidth-144pt)/4\relax}}@{}}
\caption[U.S. fixed annual-state effects by training environment]{%
  \textbf{U.S. fixed annual-state effects by training environment.}
}\label{tab:ia-x44-us-corpora}\\
\multicolumn{6}{@{}p{\linewidth}@{}}{\exhibitnotes{%
  U.S. matched predictive effects are reported for fixed annual states 2000, 2009, and 2023 under the US, Global, and Augmented training-information environments. Each effect is PIT mean squared forecast error (MSFE) minus fixed-state MSFE, scaled by historical-average MSFE on the same paired support and expressed in percentage points. Positive values favor the fixed state over origin-aligned point-in-time (PIT) forecasts. Each state uses its full available matched deployment support. Parentheses contain two-sided Bartlett--Newey--West HAC $p$-values with $h-1$ lags, where $h$ is in months. $^{*}$, $^{**}$, and $^{***}$ indicate significance at 10\%, 5\%, and 1\%. Estimates use the all-available sample.
}}\\
\noalign{\vskip 6pt}
\toprule
Model & Fixed & $h=1$ & $h=3$ & $h=6$ & $h=12$ \\
\midrule
\endfirsthead
\multicolumn{6}{@{}l}{\textit{\tablename\ \thetable\ (continued)}}\\
\toprule
Model & Fixed & $h=1$ & $h=3$ & $h=6$ & $h=12$ \\
\midrule
\endhead
\midrule\multicolumn{6}{r}{\textit{Continued on next page}}\\\endfoot
\bottomrule\endlastfoot
\multicolumn{6}{@{}l}{\textbf{US training environment}}\\*
\addlinespace[3pt]
Chronos Tiny & 2000 & \makecell[r]{-5.33$^{**}$\\{\scriptsize (0.028)}} & \makecell[r]{-14.32$^{**}$\\{\scriptsize (0.020)}} & \makecell[r]{-25.96$^{**}$\\{\scriptsize (0.016)}} & \makecell[r]{-32.77$^{**}$\\{\scriptsize (0.025)}} \\*
 & 2009 & \makecell[r]{-3.68\\{\scriptsize (0.110)}} & \makecell[r]{-9.63$^{**}$\\{\scriptsize (0.038)}} & \makecell[r]{-9.90\\{\scriptsize (0.143)}} & \makecell[r]{4.66\\{\scriptsize (0.685)}} \\*
 & 2023 & \makecell[r]{-3.08\\{\scriptsize (0.274)}} & \makecell[r]{-8.43$^{***}$\\{\scriptsize (0.007)}} & \makecell[r]{-8.96$^{**}$\\{\scriptsize (0.010)}} & \makecell[r]{-5.21\\{\scriptsize (0.310)}} \\
\addlinespace[3pt]
Chronos Mini & 2000 & \makecell[r]{-4.22$^{**}$\\{\scriptsize (0.024)}} & \makecell[r]{-12.43$^{***}$\\{\scriptsize ($<$0.001)}} & \makecell[r]{-17.30$^{***}$\\{\scriptsize (0.006)}} & \makecell[r]{-24.21$^{**}$\\{\scriptsize (0.019)}} \\*
 & 2009 & \makecell[r]{-1.60\\{\scriptsize (0.448)}} & \makecell[r]{-6.29\\{\scriptsize (0.115)}} & \makecell[r]{-4.69\\{\scriptsize (0.451)}} & \makecell[r]{4.78\\{\scriptsize (0.617)}} \\*
 & 2023 & \makecell[r]{-9.36$^{*}$\\{\scriptsize (0.084)}} & \makecell[r]{-15.60\\{\scriptsize (0.132)}} & \makecell[r]{-4.31\\{\scriptsize (0.473)}} & \makecell[r]{-1.12\\{\scriptsize (0.874)}} \\
\addlinespace[3pt]
Chronos Small & 2000 & \makecell[r]{0.46\\{\scriptsize (0.804)}} & \makecell[r]{-5.64\\{\scriptsize (0.152)}} & \makecell[r]{-10.80\\{\scriptsize (0.141)}} & \makecell[r]{-17.64\\{\scriptsize (0.140)}} \\*
 & 2009 & \makecell[r]{1.57\\{\scriptsize (0.504)}} & \makecell[r]{0.93\\{\scriptsize (0.792)}} & \makecell[r]{-0.70\\{\scriptsize (0.815)}} & \makecell[r]{-4.30\\{\scriptsize (0.350)}} \\*
 & 2023 & \makecell[r]{-11.01$^{*}$\\{\scriptsize (0.053)}} & \makecell[r]{-19.25$^{**}$\\{\scriptsize (0.022)}} & \makecell[r]{-15.92$^{**}$\\{\scriptsize (0.032)}} & \makecell[r]{-32.30$^{*}$\\{\scriptsize (0.078)}} \\
\addlinespace[3pt]
TimesFM 8M & 2000 & \makecell[r]{-11.33$^{**}$\\{\scriptsize (0.015)}} & \makecell[r]{-9.22\\{\scriptsize (0.238)}} & \makecell[r]{-15.21\\{\scriptsize (0.188)}} & \makecell[r]{-23.82\\{\scriptsize (0.111)}} \\*
 & 2009 & \makecell[r]{-2.38\\{\scriptsize (0.581)}} & \makecell[r]{-1.94\\{\scriptsize (0.746)}} & \makecell[r]{-6.92\\{\scriptsize (0.364)}} & \makecell[r]{-10.38\\{\scriptsize (0.111)}} \\*
 & 2023 & \makecell[r]{9.62$^{*}$\\{\scriptsize (0.078)}} & \makecell[r]{1.78\\{\scriptsize (0.786)}} & \makecell[r]{-1.78\\{\scriptsize (0.763)}} & \makecell[r]{-7.77\\{\scriptsize (0.212)}} \\
\addlinespace[3pt]
TimesFM 20M & 2000 & \makecell[r]{-14.99$^{***}$\\{\scriptsize (0.004)}} & \makecell[r]{-22.07$^{***}$\\{\scriptsize (0.002)}} & \makecell[r]{-30.53$^{***}$\\{\scriptsize ($<$0.001)}} & \makecell[r]{-26.00$^{**}$\\{\scriptsize (0.022)}} \\*
 & 2009 & \makecell[r]{-3.01\\{\scriptsize (0.499)}} & \makecell[r]{-41.48$^{***}$\\{\scriptsize ($<$0.001)}} & \makecell[r]{-61.75$^{***}$\\{\scriptsize ($<$0.001)}} & \makecell[r]{-72.59$^{***}$\\{\scriptsize (0.004)}} \\*
 & 2023 & \makecell[r]{0.82\\{\scriptsize (0.875)}} & \makecell[r]{-20.05$^{***}$\\{\scriptsize (0.006)}} & \makecell[r]{-27.60$^{**}$\\{\scriptsize (0.014)}} & \makecell[r]{-34.74$^{*}$\\{\scriptsize (0.053)}} \\
\pagebreak[4]
\multicolumn{6}{@{}l}{\textbf{Global training environment}}\\*
\addlinespace[3pt]
Chronos Tiny & 2000 & \makecell[r]{-1.17\\{\scriptsize (0.786)}} & \makecell[r]{-5.94\\{\scriptsize (0.600)}} & \makecell[r]{-8.86\\{\scriptsize (0.609)}} & \makecell[r]{-13.21\\{\scriptsize (0.588)}} \\*
 & 2009 & \makecell[r]{1.14\\{\scriptsize (0.711)}} & \makecell[r]{0.79\\{\scriptsize (0.803)}} & \makecell[r]{-0.26\\{\scriptsize (0.970)}} & \makecell[r]{0.52\\{\scriptsize (0.964)}} \\*
 & 2023 & \makecell[r]{-4.61\\{\scriptsize (0.218)}} & \makecell[r]{-9.45\\{\scriptsize (0.162)}} & \makecell[r]{-0.85\\{\scriptsize (0.890)}} & \makecell[r]{8.15\\{\scriptsize (0.337)}} \\
\addlinespace[3pt]
Chronos Mini & 2000 & \makecell[r]{-2.19\\{\scriptsize (0.645)}} & \makecell[r]{-5.14\\{\scriptsize (0.536)}} & \makecell[r]{-6.33\\{\scriptsize (0.664)}} & \makecell[r]{-14.86\\{\scriptsize (0.507)}} \\*
 & 2009 & \makecell[r]{2.27\\{\scriptsize (0.573)}} & \makecell[r]{-0.48\\{\scriptsize (0.878)}} & \makecell[r]{3.75\\{\scriptsize (0.592)}} & \makecell[r]{9.73\\{\scriptsize (0.381)}} \\*
 & 2023 & \makecell[r]{-4.25\\{\scriptsize (0.170)}} & \makecell[r]{-5.99\\{\scriptsize (0.275)}} & \makecell[r]{2.75\\{\scriptsize (0.677)}} & \makecell[r]{4.95\\{\scriptsize (0.535)}} \\
\addlinespace[3pt]
Chronos Small & 2000 & \makecell[r]{-2.36\\{\scriptsize (0.722)}} & \makecell[r]{-6.32\\{\scriptsize (0.753)}} & \makecell[r]{-28.89\\{\scriptsize (0.436)}} & \makecell[r]{-74.06\\{\scriptsize (0.295)}} \\*
 & 2009 & \makecell[r]{1.48\\{\scriptsize (0.604)}} & \makecell[r]{10.25\\{\scriptsize (0.198)}} & \makecell[r]{9.31\\{\scriptsize (0.313)}} & \makecell[r]{6.10\\{\scriptsize (0.309)}} \\*
 & 2023 & \makecell[r]{-0.17\\{\scriptsize (0.948)}} & \makecell[r]{2.05\\{\scriptsize (0.809)}} & \makecell[r]{7.15\\{\scriptsize (0.441)}} & \makecell[r]{7.51\\{\scriptsize (0.385)}} \\
\addlinespace[3pt]
TimesFM 8M & 2000 & \makecell[r]{19.94$^{***}$\\{\scriptsize ($<$0.001)}} & \makecell[r]{47.68$^{**}$\\{\scriptsize (0.042)}} & \makecell[r]{67.06\\{\scriptsize (0.120)}} & \makecell[r]{26.10\\{\scriptsize (0.242)}} \\*
 & 2009 & \makecell[r]{19.16$^{***}$\\{\scriptsize (0.002)}} & \makecell[r]{49.24$^{**}$\\{\scriptsize (0.041)}} & \makecell[r]{46.74\\{\scriptsize (0.227)}} & \makecell[r]{33.19\\{\scriptsize (0.216)}} \\*
 & 2023 & \makecell[r]{-38.13$^{***}$\\{\scriptsize (0.001)}} & \makecell[r]{-70.31$^{**}$\\{\scriptsize (0.036)}} & \makecell[r]{-17.00\\{\scriptsize (0.696)}} & \makecell[r]{33.04\\{\scriptsize (0.278)}} \\
\addlinespace[3pt]
TimesFM 20M & 2000 & \makecell[r]{2.79\\{\scriptsize (0.669)}} & \makecell[r]{-11.98$^{*}$\\{\scriptsize (0.059)}} & \makecell[r]{-8.34\\{\scriptsize (0.117)}} & \makecell[r]{13.48\\{\scriptsize (0.319)}} \\*
 & 2009 & \makecell[r]{-38.46$^{***}$\\{\scriptsize ($<$0.001)}} & \makecell[r]{-33.37$^{***}$\\{\scriptsize ($<$0.001)}} & \makecell[r]{-12.14$^{*}$\\{\scriptsize (0.088)}} & \makecell[r]{-29.96\\{\scriptsize (0.255)}} \\*
 & 2023 & \makecell[r]{4.03\\{\scriptsize (0.553)}} & \makecell[r]{-4.99\\{\scriptsize (0.408)}} & \makecell[r]{-2.09\\{\scriptsize (0.667)}} & \makecell[r]{20.96\\{\scriptsize (0.167)}} \\
\pagebreak[4]
\multicolumn{6}{@{}l}{\textbf{Augmented training environment}}\\*
\addlinespace[3pt]
Chronos Tiny & 2000 & \makecell[r]{-1.65\\{\scriptsize (0.554)}} & \makecell[r]{-1.50\\{\scriptsize (0.638)}} & \makecell[r]{-1.43\\{\scriptsize (0.776)}} & \makecell[r]{-6.55\\{\scriptsize (0.455)}} \\*
 & 2009 & \makecell[r]{-0.60\\{\scriptsize (0.803)}} & \makecell[r]{0.77\\{\scriptsize (0.829)}} & \makecell[r]{3.16\\{\scriptsize (0.381)}} & \makecell[r]{3.34\\{\scriptsize (0.548)}} \\*
 & 2023 & \makecell[r]{-0.64\\{\scriptsize (0.800)}} & \makecell[r]{0.90\\{\scriptsize (0.870)}} & \makecell[r]{3.67\\{\scriptsize (0.546)}} & \makecell[r]{0.19\\{\scriptsize (0.978)}} \\
\addlinespace[3pt]
Chronos Mini & 2000 & \makecell[r]{-0.79\\{\scriptsize (0.797)}} & \makecell[r]{-2.48\\{\scriptsize (0.533)}} & \makecell[r]{-4.12\\{\scriptsize (0.406)}} & \makecell[r]{-6.18\\{\scriptsize (0.355)}} \\*
 & 2009 & \makecell[r]{0.66\\{\scriptsize (0.715)}} & \makecell[r]{-1.19\\{\scriptsize (0.674)}} & \makecell[r]{-1.13\\{\scriptsize (0.723)}} & \makecell[r]{0.22\\{\scriptsize (0.957)}} \\*
 & 2023 & \makecell[r]{-0.34\\{\scriptsize (0.927)}} & \makecell[r]{-6.67\\{\scriptsize (0.475)}} & \makecell[r]{0.95\\{\scriptsize (0.904)}} & \makecell[r]{4.02\\{\scriptsize (0.542)}} \\
\addlinespace[3pt]
Chronos Small & 2000 & \makecell[r]{-0.11\\{\scriptsize (0.972)}} & \makecell[r]{0.07\\{\scriptsize (0.989)}} & \makecell[r]{-3.93\\{\scriptsize (0.538)}} & \makecell[r]{-9.12\\{\scriptsize (0.228)}} \\*
 & 2009 & \makecell[r]{1.94\\{\scriptsize (0.407)}} & \makecell[r]{2.40\\{\scriptsize (0.550)}} & \makecell[r]{-1.99\\{\scriptsize (0.556)}} & \makecell[r]{-7.05$^{*}$\\{\scriptsize (0.082)}} \\*
 & 2023 & \makecell[r]{1.23\\{\scriptsize (0.478)}} & \makecell[r]{1.02\\{\scriptsize (0.703)}} & \makecell[r]{-1.41\\{\scriptsize (0.690)}} & \makecell[r]{-5.47\\{\scriptsize (0.314)}} \\
\addlinespace[3pt]
TimesFM 8M & 2000 & \makecell[r]{5.53\\{\scriptsize (0.190)}} & \makecell[r]{1.15\\{\scriptsize (0.661)}} & \makecell[r]{3.70\\{\scriptsize (0.438)}} & \makecell[r]{7.27\\{\scriptsize (0.113)}} \\*
 & 2009 & \makecell[r]{-24.30$^{**}$\\{\scriptsize (0.016)}} & \makecell[r]{-14.70$^{**}$\\{\scriptsize (0.022)}} & \makecell[r]{-30.71\\{\scriptsize (0.135)}} & \makecell[r]{-24.53\\{\scriptsize (0.283)}} \\*
 & 2023 & \makecell[r]{10.41$^{**}$\\{\scriptsize (0.014)}} & \makecell[r]{1.77\\{\scriptsize (0.569)}} & \makecell[r]{2.90\\{\scriptsize (0.501)}} & \makecell[r]{4.19\\{\scriptsize (0.347)}} \\
\addlinespace[3pt]
TimesFM 20M & 2000 & \makecell[r]{\ensuremath{6.83\times10^{4}}$^{***}$\\{\scriptsize (0.002)}} & \makecell[r]{\ensuremath{4.63\times10^{4}}$^{*}$\\{\scriptsize (0.096)}} & \makecell[r]{\ensuremath{2.31\times10^{4}}\\{\scriptsize (0.219)}} & \makecell[r]{\ensuremath{3.07\times10^{3}}\\{\scriptsize (0.337)}} \\*
 & 2009 & \makecell[r]{\ensuremath{6.83\times10^{4}}$^{***}$\\{\scriptsize (0.002)}} & \makecell[r]{\ensuremath{4.63\times10^{4}}$^{*}$\\{\scriptsize (0.095)}} & \makecell[r]{\ensuremath{2.31\times10^{4}}\\{\scriptsize (0.219)}} & \makecell[r]{\ensuremath{3.07\times10^{3}}\\{\scriptsize (0.337)}} \\*
 & 2023 & \makecell[r]{\ensuremath{-1.47\times10^{6}}$^{***}$\\{\scriptsize ($<$0.001)}} & \makecell[r]{\ensuremath{-1.19\times10^{6}}$^{***}$\\{\scriptsize ($<$0.001)}} & \makecell[r]{\ensuremath{-6.21\times10^{5}}$^{***}$\\{\scriptsize ($<$0.001)}} & \makecell[r]{\ensuremath{-3.34\times10^{5}}$^{***}$\\{\scriptsize ($<$0.001)}} \\
\end{longtable}
\endgroup
\clearpage
\begingroup
\singlespacing\normalsize
\renewcommand{\arraystretch}{1}
\setlength{\tabcolsep}{4pt}
\begin{longtable}{@{}>{\raggedright\arraybackslash}p{74pt}>{\raggedleft\arraybackslash}p{30pt}*{4}{>{\raggedleft\arraybackslash}p{\dimexpr(\linewidth-144pt)/4\relax}}@{}}
\caption[International fixed annual-state effects by training environment]{%
  \textbf{International fixed annual-state effects by training environment.}
}\label{tab:ia-x44-international-corpora}\\
\multicolumn{6}{@{}p{\linewidth}@{}}{\exhibitnotes{%
  Equal-market matched effects summarize fixed annual states 2000, 2009, and 2023 in the 13 non-U.S. markets under the US, Global, and Augmented training-information environments. Each market's monthly PIT-minus-fixed-state squared-loss differential is normalized by its historical-average mean squared forecast error (MSFE), then averaged equally across markets over common target months. Effects are in percentage points; positive values favor the fixed state over origin-aligned point-in-time (PIT) forecasts. Each state uses its full available matched deployment support. Parentheses contain two-sided Bartlett--Newey--West HAC $p$-values with $h-1$ lags, where $h$ is in months. $^{*}$, $^{**}$, and $^{***}$ indicate significance at 10\%, 5\%, and 1\%. Estimates use the all-available international sample.
}}\\
\noalign{\vskip 6pt}
\toprule
Model & Fixed & $h=1$ & $h=3$ & $h=6$ & $h=12$ \\
\midrule
\endfirsthead
\multicolumn{6}{@{}l}{\textit{\tablename\ \thetable\ (continued)}}\\
\toprule
Model & Fixed & $h=1$ & $h=3$ & $h=6$ & $h=12$ \\
\midrule
\endhead
\midrule\multicolumn{6}{r}{\textit{Continued on next page}}\\\endfoot
\bottomrule\endlastfoot
\multicolumn{6}{@{}l}{\textbf{US training environment}}\\*
\addlinespace[3pt]
Chronos Tiny & 2000 & \makecell[r]{-2.69\\{\scriptsize (0.162)}} & \makecell[r]{-6.66\\{\scriptsize (0.132)}} & \makecell[r]{-12.63$^{*}$\\{\scriptsize (0.053)}} & \makecell[r]{-19.65$^{**}$\\{\scriptsize (0.016)}} \\*
 & 2009 & \makecell[r]{-3.12$^{***}$\\{\scriptsize (0.004)}} & \makecell[r]{-5.98$^{**}$\\{\scriptsize (0.010)}} & \makecell[r]{-7.90$^{**}$\\{\scriptsize (0.013)}} & \makecell[r]{-6.04$^{*}$\\{\scriptsize (0.081)}} \\*
 & 2023 & \makecell[r]{-2.58$^{**}$\\{\scriptsize (0.021)}} & \makecell[r]{-4.13$^{**}$\\{\scriptsize (0.027)}} & \makecell[r]{-6.17$^{**}$\\{\scriptsize (0.040)}} & \makecell[r]{-5.34\\{\scriptsize (0.112)}} \\
\addlinespace[3pt]
Chronos Mini & 2000 & \makecell[r]{-1.61\\{\scriptsize (0.255)}} & \makecell[r]{-5.80$^{*}$\\{\scriptsize (0.056)}} & \makecell[r]{-9.51$^{**}$\\{\scriptsize (0.048)}} & \makecell[r]{-16.02$^{**}$\\{\scriptsize (0.014)}} \\*
 & 2009 & \makecell[r]{-2.02$^{**}$\\{\scriptsize (0.015)}} & \makecell[r]{-3.83$^{***}$\\{\scriptsize (0.002)}} & \makecell[r]{-4.28$^{***}$\\{\scriptsize (0.007)}} & \makecell[r]{-5.09$^{*}$\\{\scriptsize (0.088)}} \\*
 & 2023 & \makecell[r]{-4.49$^{**}$\\{\scriptsize (0.022)}} & \makecell[r]{-6.75$^{*}$\\{\scriptsize (0.065)}} & \makecell[r]{-3.73\\{\scriptsize (0.126)}} & \makecell[r]{-4.00\\{\scriptsize (0.201)}} \\
\addlinespace[3pt]
Chronos Small & 2000 & \makecell[r]{-0.56\\{\scriptsize (0.686)}} & \makecell[r]{-2.50\\{\scriptsize (0.262)}} & \makecell[r]{-4.59\\{\scriptsize (0.184)}} & \makecell[r]{-9.20$^{*}$\\{\scriptsize (0.075)}} \\*
 & 2009 & \makecell[r]{0.91\\{\scriptsize (0.532)}} & \makecell[r]{-0.07\\{\scriptsize (0.974)}} & \makecell[r]{-3.89$^{**}$\\{\scriptsize (0.040)}} & \makecell[r]{-6.42$^{**}$\\{\scriptsize (0.019)}} \\*
 & 2023 & \makecell[r]{-0.60\\{\scriptsize (0.625)}} & \makecell[r]{-1.81\\{\scriptsize (0.347)}} & \makecell[r]{-3.82\\{\scriptsize (0.151)}} & \makecell[r]{-3.17\\{\scriptsize (0.347)}} \\
\addlinespace[3pt]
TimesFM 8M & 2000 & \makecell[r]{-26.48$^{***}$\\{\scriptsize ($<$0.001)}} & \makecell[r]{-31.81$^{***}$\\{\scriptsize (0.002)}} & \makecell[r]{-33.89$^{**}$\\{\scriptsize (0.012)}} & \makecell[r]{-40.29$^{**}$\\{\scriptsize (0.011)}} \\*
 & 2009 & \makecell[r]{-13.36$^{**}$\\{\scriptsize (0.030)}} & \makecell[r]{-15.01$^{*}$\\{\scriptsize (0.073)}} & \makecell[r]{-20.25$^{*}$\\{\scriptsize (0.062)}} & \makecell[r]{-18.15$^{**}$\\{\scriptsize (0.025)}} \\*
 & 2023 & \makecell[r]{20.86$^{***}$\\{\scriptsize ($<$0.001)}} & \makecell[r]{-1.41\\{\scriptsize (0.807)}} & \makecell[r]{-1.00\\{\scriptsize (0.869)}} & \makecell[r]{-5.77\\{\scriptsize (0.388)}} \\
\addlinespace[3pt]
TimesFM 20M & 2000 & \makecell[r]{-16.34$^{**}$\\{\scriptsize (0.013)}} & \makecell[r]{-14.72$^{*}$\\{\scriptsize (0.057)}} & \makecell[r]{-21.61$^{***}$\\{\scriptsize (0.007)}} & \makecell[r]{5.74\\{\scriptsize (0.524)}} \\*
 & 2009 & \makecell[r]{-12.18$^{**}$\\{\scriptsize (0.019)}} & \makecell[r]{-88.51$^{***}$\\{\scriptsize ($<$0.001)}} & \makecell[r]{-121.43$^{***}$\\{\scriptsize ($<$0.001)}} & \makecell[r]{-139.25$^{***}$\\{\scriptsize (0.005)}} \\*
 & 2023 & \makecell[r]{-5.72\\{\scriptsize (0.110)}} & \makecell[r]{-49.83$^{***}$\\{\scriptsize ($<$0.001)}} & \makecell[r]{-57.99$^{***}$\\{\scriptsize ($<$0.001)}} & \makecell[r]{-47.51$^{***}$\\{\scriptsize (0.001)}} \\
\pagebreak[4]
\multicolumn{6}{@{}l}{\textbf{Global training environment}}\\*
\addlinespace[3pt]
Chronos Tiny & 2000 & \makecell[r]{0.78\\{\scriptsize (0.297)}} & \makecell[r]{0.99\\{\scriptsize (0.490)}} & \makecell[r]{0.41\\{\scriptsize (0.806)}} & \makecell[r]{-0.90\\{\scriptsize (0.659)}} \\*
 & 2009 & \makecell[r]{-1.14\\{\scriptsize (0.187)}} & \makecell[r]{-2.72$^{*}$\\{\scriptsize (0.060)}} & \makecell[r]{-3.74\\{\scriptsize (0.122)}} & \makecell[r]{-4.42\\{\scriptsize (0.254)}} \\*
 & 2023 & \makecell[r]{-0.83\\{\scriptsize (0.355)}} & \makecell[r]{-2.47$^{*}$\\{\scriptsize (0.091)}} & \makecell[r]{-2.59\\{\scriptsize (0.293)}} & \makecell[r]{-2.70\\{\scriptsize (0.422)}} \\
\addlinespace[3pt]
Chronos Mini & 2000 & \makecell[r]{-0.01\\{\scriptsize (0.996)}} & \makecell[r]{-0.69\\{\scriptsize (0.769)}} & \makecell[r]{-2.92\\{\scriptsize (0.302)}} & \makecell[r]{-5.61\\{\scriptsize (0.104)}} \\*
 & 2009 & \makecell[r]{-0.83\\{\scriptsize (0.439)}} & \makecell[r]{-2.08\\{\scriptsize (0.295)}} & \makecell[r]{-3.65\\{\scriptsize (0.209)}} & \makecell[r]{-4.72\\{\scriptsize (0.212)}} \\*
 & 2023 & \makecell[r]{-1.39\\{\scriptsize (0.113)}} & \makecell[r]{-2.10\\{\scriptsize (0.155)}} & \makecell[r]{-2.24\\{\scriptsize (0.441)}} & \makecell[r]{-2.58\\{\scriptsize (0.524)}} \\
\addlinespace[3pt]
Chronos Small & 2000 & \makecell[r]{0.52\\{\scriptsize (0.675)}} & \makecell[r]{0.48\\{\scriptsize (0.837)}} & \makecell[r]{-0.81\\{\scriptsize (0.801)}} & \makecell[r]{-3.42\\{\scriptsize (0.500)}} \\*
 & 2009 & \makecell[r]{-0.91\\{\scriptsize (0.167)}} & \makecell[r]{-1.62\\{\scriptsize (0.244)}} & \makecell[r]{-1.19\\{\scriptsize (0.537)}} & \makecell[r]{2.06\\{\scriptsize (0.466)}} \\*
 & 2023 & \makecell[r]{-1.52\\{\scriptsize (0.164)}} & \makecell[r]{-3.50\\{\scriptsize (0.189)}} & \makecell[r]{-1.72\\{\scriptsize (0.499)}} & \makecell[r]{0.47\\{\scriptsize (0.855)}} \\
\addlinespace[3pt]
TimesFM 8M & 2000 & \makecell[r]{43.28$^{***}$\\{\scriptsize ($<$0.001)}} & \makecell[r]{104.31$^{**}$\\{\scriptsize (0.022)}} & \makecell[r]{139.85$^{*}$\\{\scriptsize (0.091)}} & \makecell[r]{48.21\\{\scriptsize (0.219)}} \\*
 & 2009 & \makecell[r]{42.96$^{***}$\\{\scriptsize ($<$0.001)}} & \makecell[r]{110.36$^{**}$\\{\scriptsize (0.017)}} & \makecell[r]{76.79\\{\scriptsize (0.344)}} & \makecell[r]{68.86\\{\scriptsize (0.104)}} \\*
 & 2023 & \makecell[r]{-125.57$^{***}$\\{\scriptsize ($<$0.001)}} & \makecell[r]{-166.55$^{***}$\\{\scriptsize ($<$0.001)}} & \makecell[r]{-14.17\\{\scriptsize (0.861)}} & \makecell[r]{81.28$^{*}$\\{\scriptsize (0.072)}} \\
\addlinespace[3pt]
TimesFM 20M & 2000 & \makecell[r]{12.27$^{*}$\\{\scriptsize (0.059)}} & \makecell[r]{-10.90\\{\scriptsize (0.161)}} & \makecell[r]{-14.01\\{\scriptsize (0.117)}} & \makecell[r]{23.03\\{\scriptsize (0.250)}} \\*
 & 2009 & \makecell[r]{-89.90$^{***}$\\{\scriptsize ($<$0.001)}} & \makecell[r]{-33.25$^{***}$\\{\scriptsize (0.002)}} & \makecell[r]{-9.13\\{\scriptsize (0.281)}} & \makecell[r]{-105.78$^{***}$\\{\scriptsize (0.002)}} \\*
 & 2023 & \makecell[r]{16.98$^{***}$\\{\scriptsize (0.004)}} & \makecell[r]{0.05\\{\scriptsize (0.994)}} & \makecell[r]{-3.90\\{\scriptsize (0.618)}} & \makecell[r]{37.70$^{*}$\\{\scriptsize (0.053)}} \\
\pagebreak[4]
\multicolumn{6}{@{}l}{\textbf{Augmented training environment}}\\*
\addlinespace[3pt]
Chronos Tiny & 2000 & \makecell[r]{-0.97\\{\scriptsize (0.192)}} & \makecell[r]{-1.57$^{**}$\\{\scriptsize (0.047)}} & \makecell[r]{-2.90$^{***}$\\{\scriptsize (0.005)}} & \makecell[r]{-4.18$^{**}$\\{\scriptsize (0.013)}} \\*
 & 2009 & \makecell[r]{-0.38\\{\scriptsize (0.561)}} & \makecell[r]{0.33\\{\scriptsize (0.705)}} & \makecell[r]{0.88\\{\scriptsize (0.466)}} & \makecell[r]{2.19\\{\scriptsize (0.256)}} \\*
 & 2023 & \makecell[r]{-0.96\\{\scriptsize (0.341)}} & \makecell[r]{-0.40\\{\scriptsize (0.864)}} & \makecell[r]{2.58\\{\scriptsize (0.146)}} & \makecell[r]{4.42$^{*}$\\{\scriptsize (0.064)}} \\
\addlinespace[3pt]
Chronos Mini & 2000 & \makecell[r]{0.50\\{\scriptsize (0.723)}} & \makecell[r]{0.98\\{\scriptsize (0.589)}} & \makecell[r]{-0.36\\{\scriptsize (0.873)}} & \makecell[r]{-3.25\\{\scriptsize (0.276)}} \\*
 & 2009 & \makecell[r]{-0.27\\{\scriptsize (0.723)}} & \makecell[r]{0.06\\{\scriptsize (0.948)}} & \makecell[r]{-0.85\\{\scriptsize (0.430)}} & \makecell[r]{-2.40\\{\scriptsize (0.200)}} \\*
 & 2023 & \makecell[r]{-3.80\\{\scriptsize (0.250)}} & \makecell[r]{-6.07\\{\scriptsize (0.308)}} & \makecell[r]{-6.41\\{\scriptsize (0.289)}} & \makecell[r]{-6.43\\{\scriptsize (0.205)}} \\
\addlinespace[3pt]
Chronos Small & 2000 & \makecell[r]{0.14\\{\scriptsize (0.897)}} & \makecell[r]{-0.28\\{\scriptsize (0.865)}} & \makecell[r]{-2.37\\{\scriptsize (0.126)}} & \makecell[r]{-5.00$^{**}$\\{\scriptsize (0.014)}} \\*
 & 2009 & \makecell[r]{1.00\\{\scriptsize (0.137)}} & \makecell[r]{1.60\\{\scriptsize (0.230)}} & \makecell[r]{-0.05\\{\scriptsize (0.970)}} & \makecell[r]{-2.76$^{*}$\\{\scriptsize (0.063)}} \\*
 & 2023 & \makecell[r]{1.20$^{*}$\\{\scriptsize (0.065)}} & \makecell[r]{1.56\\{\scriptsize (0.111)}} & \makecell[r]{0.36\\{\scriptsize (0.779)}} & \makecell[r]{-1.68\\{\scriptsize (0.337)}} \\
\addlinespace[3pt]
TimesFM 8M & 2000 & \makecell[r]{5.92\\{\scriptsize (0.120)}} & \makecell[r]{3.37\\{\scriptsize (0.127)}} & \makecell[r]{9.00$^{*}$\\{\scriptsize (0.071)}} & \makecell[r]{12.43$^{**}$\\{\scriptsize (0.021)}} \\*
 & 2009 & \makecell[r]{-93.10$^{***}$\\{\scriptsize ($<$0.001)}} & \makecell[r]{-19.02$^{***}$\\{\scriptsize ($<$0.001)}} & \makecell[r]{-87.25$^{***}$\\{\scriptsize ($<$0.001)}} & \makecell[r]{-70.11$^{***}$\\{\scriptsize (0.004)}} \\*
 & 2023 & \makecell[r]{10.51$^{***}$\\{\scriptsize (0.003)}} & \makecell[r]{4.17$^{*}$\\{\scriptsize (0.057)}} & \makecell[r]{8.95$^{*}$\\{\scriptsize (0.081)}} & \makecell[r]{11.45$^{**}$\\{\scriptsize (0.031)}} \\
\addlinespace[3pt]
TimesFM 20M & 2000 & \makecell[r]{\ensuremath{1.91\times10^{5}}$^{***}$\\{\scriptsize ($<$0.001)}} & \makecell[r]{\ensuremath{1.23\times10^{5}}$^{*}$\\{\scriptsize (0.058)}} & \makecell[r]{\ensuremath{4.78\times10^{4}}\\{\scriptsize (0.195)}} & \makecell[r]{\ensuremath{4.35\times10^{3}}\\{\scriptsize (0.315)}} \\*
 & 2009 & \makecell[r]{\ensuremath{1.91\times10^{5}}$^{***}$\\{\scriptsize ($<$0.001)}} & \makecell[r]{\ensuremath{1.23\times10^{5}}$^{*}$\\{\scriptsize (0.058)}} & \makecell[r]{\ensuremath{4.78\times10^{4}}\\{\scriptsize (0.195)}} & \makecell[r]{\ensuremath{4.35\times10^{3}}\\{\scriptsize (0.315)}} \\*
 & 2023 & \makecell[r]{\ensuremath{-3.84\times10^{6}}$^{***}$\\{\scriptsize ($<$0.001)}} & \makecell[r]{\ensuremath{-2.79\times10^{6}}$^{***}$\\{\scriptsize ($<$0.001)}} & \makecell[r]{\ensuremath{-1.40\times10^{6}}$^{***}$\\{\scriptsize ($<$0.001)}} & \makecell[r]{\ensuremath{-7.08\times10^{5}}$^{***}$\\{\scriptsize ($<$0.001)}} \\
\end{longtable}
\endgroup
\clearpage
\begingroup
\fontsize{11}{13.6}\selectfont
\setlength{\tabcolsep}{4.00pt}
\setlength{\linewidth}{390.68pt}
\begin{longtable}{@{}lclccccc@{}}
\caption[Fixed annual-state effects by target-window timing]{%
  \textbf{Fixed annual-state effects by target-window timing.}
}\label{tab:ia-x44-timing-states}\\
\multicolumn{8}{@{}p{\linewidth}@{}}{\exhibitnotes{%
  Timing windows separate U.S. fixed-state deployment into retrospective exposure, boundary crossing, and post-cutoff deployment. Retrospective exposure places the complete target window on or before the training cutoff; boundary crossing places the cutoff inside it. Post-cutoff deployment combines the first origin-aligned PIT deployment year with subsequent stale years. Each row compares the fixed state with its origin-aligned point-in-time (PIT) forecast within the stated window. Effects equal PIT mean squared forecast error (MSFE) minus fixed-state MSFE, scaled by historical-average MSFE and expressed in percentage points; positive values favor the fixed state. The column $N$ counts matched target months. HAC 95\% intervals and two-sided $p$-values use Bartlett--Newey--West estimation with $h-1$ lags, where $h$ is in months. Formal inference requires at least 24 matched months; dashes indicate unreported entries. Estimates use the all-available U.S. sample and the US training-information environment.
}}\\
\noalign{\vskip 6pt}
\toprule
Model & Fixed & Timing region & $h$ & $\widehat\Delta$ & $N$ & HAC 95\% CI & HAC $p_2$ \\
\midrule
\endfirsthead
\multicolumn{8}{l}{\textit{\tablename\ \thetable\ (continued)}}\\
\toprule
Model & Fixed & Timing region & $h$ & $\widehat\Delta$ & $N$ & HAC 95\% CI & HAC $p_2$ \\
\midrule
\endhead
\midrule\multicolumn{8}{r}{\textit{Continued on next page}}\\\endfoot
\bottomrule\endlastfoot
Chronos Tiny & 2000 & Post-cutoff & 1 & -5.33 & 288 & [-10.08, -0.58] & 0.028 \\
 &  &  & 3 & -14.32 & 286 & [-26.39, -2.25] & 0.020 \\
 &  &  & 6 & -25.96 & 283 & [-47.03, -4.89] & 0.016 \\
 &  &  & 12 & -32.77 & 277 & [-61.42, -4.11] & 0.025 \\
\addlinespace[1pt]
 & 2009 & Retrospective & 1 & -6.05 & 108 & [-14.10, 2.01] & 0.141 \\
 &  &  & 3 & -16.15 & 106 & [-27.91, -4.39] & 0.007 \\
 &  &  & 6 & -18.97 & 103 & [-32.38, -5.55] & 0.006 \\
 &  &  & 12 & -8.69 & 97 & [-33.65, 16.28] & 0.495 \\
\addlinespace[1pt]
 &  & Boundary & 3 & -24.18 & 2 & -- & -- \\
 &  &  & 6 & 18.08 & 5 & -- & -- \\
 &  &  & 12 & -4.82 & 11 & -- & -- \\
\addlinespace[1pt]
 &  & Post-cutoff & 1 & -1.94 & 180 & [-6.87, 2.99] & 0.441 \\
 &  &  & 3 & -1.12 & 178 & [-13.83, 11.60] & 0.863 \\
 &  &  & 6 & 5.57 & 175 & [-17.46, 28.60] & 0.635 \\
 &  &  & 12 & 30.04 & 169 & [-5.14, 65.21] & 0.094 \\
\addlinespace[1pt]
 & 2023 & Retrospective & 1 & -3.14 & 276 & [-8.77, 2.48] & 0.273 \\
 &  &  & 3 & -8.82 & 274 & [-15.01, -2.63] & 0.005 \\
 &  &  & 6 & -9.64 & 271 & [-16.36, -2.93] & 0.005 \\
 &  &  & 12 & -7.04 & 265 & [-16.46, 2.39] & 0.143 \\
\addlinespace[1pt]
 &  & Boundary & 3 & 14.88 & 2 & -- & -- \\
 &  &  & 6 & 32.80 & 5 & -- & -- \\
 &  &  & 12 & 67.08 & 11 & -- & -- \\
\addlinespace[1pt]
 &  & Post-cutoff & 1 & 0.22 & 12 & -- & -- \\
 &  &  & 3 & 5.22 & 10 & -- & -- \\
 &  &  & 6 & 4.72 & 7 & -- & -- \\
 &  &  & 12 & 2.86 & 1 & -- & -- \\
\addlinespace[1pt]
\addlinespace[3pt]
Chronos Mini & 2000 & Post-cutoff & 1 & -4.22 & 288 & [-7.89, -0.55] & 0.024 \\
 &  &  & 3 & -12.43 & 286 & [-19.32, -5.53] & $<$0.001 \\
 &  &  & 6 & -17.30 & 283 & [-29.52, -5.07] & 0.006 \\
 &  &  & 12 & -24.21 & 277 & [-44.42, -3.99] & 0.019 \\
\addlinespace[1pt]
 & 2009 & Retrospective & 1 & -5.66 & 108 & [-13.09, 1.76] & 0.135 \\
 &  &  & 3 & -11.61 & 106 & [-21.50, -1.72] & 0.021 \\
 &  &  & 6 & -11.95 & 103 & [-26.03, 2.13] & 0.096 \\
 &  &  & 12 & -9.68 & 97 & [-29.56, 10.20] & 0.340 \\
\addlinespace[1pt]
 &  & Boundary & 3 & 43.34 & 2 & -- & -- \\
 &  &  & 6 & 21.92 & 5 & -- & -- \\
 &  &  & 12 & 13.20 & 11 & -- & -- \\
\addlinespace[1pt]
 &  & Post-cutoff & 1 & 1.38 & 180 & [-2.96, 5.72] & 0.534 \\
 &  &  & 3 & 0.56 & 178 & [-11.06, 12.18] & 0.925 \\
 &  &  & 6 & 7.58 & 175 & [-12.91, 28.07] & 0.468 \\
 &  &  & 12 & 28.42 & 169 & [2.33, 54.51] & 0.033 \\
\addlinespace[1pt]
 & 2023 & Retrospective & 1 & -9.52 & 276 & [-20.36, 1.31] & 0.085 \\
 &  &  & 3 & -15.85 & 274 & [-36.52, 4.82] & 0.133 \\
 &  &  & 6 & -4.11 & 271 & [-16.15, 7.93] & 0.504 \\
 &  &  & 12 & -0.41 & 265 & [-14.58, 13.75] & 0.954 \\
\addlinespace[1pt]
 &  & Boundary & 3 & -4.61 & 2 & -- & -- \\
 &  &  & 6 & -23.29 & 5 & -- & -- \\
 &  &  & 12 & -29.21 & 11 & -- & -- \\
\addlinespace[1pt]
 &  & Post-cutoff & 1 & -1.38 & 12 & -- & -- \\
 &  &  & 3 & 0.07 & 10 & -- & -- \\
 &  &  & 6 & 4.72 & 7 & -- & -- \\
 &  &  & 12 & 2.26 & 1 & -- & -- \\
\addlinespace[1pt]
\addlinespace[3pt]
Chronos Small & 2000 & Post-cutoff & 1 & 0.46 & 288 & [-3.15, 4.06] & 0.804 \\
 &  &  & 3 & -5.64 & 286 & [-13.36, 2.08] & 0.152 \\
 &  &  & 6 & -10.80 & 283 & [-25.18, 3.57] & 0.141 \\
 &  &  & 12 & -17.64 & 277 & [-41.06, 5.77] & 0.140 \\
\addlinespace[1pt]
 & 2009 & Retrospective & 1 & 3.90 & 108 & [-3.91, 11.71] & 0.328 \\
 &  &  & 3 & 0.95 & 106 & [-9.59, 11.50] & 0.859 \\
 &  &  & 6 & -0.56 & 103 & [-7.28, 6.16] & 0.871 \\
 &  &  & 12 & 0.67 & 97 & [-12.08, 13.43] & 0.918 \\
\addlinespace[1pt]
 &  & Boundary & 3 & -31.97 & 2 & -- & -- \\
 &  &  & 6 & -7.22 & 5 & -- & -- \\
 &  &  & 12 & -2.14 & 11 & -- & -- \\
\addlinespace[1pt]
 &  & Post-cutoff & 1 & -0.14 & 180 & [-5.64, 5.36] & 0.960 \\
 &  &  & 3 & 0.93 & 178 & [-6.91, 8.78] & 0.816 \\
 &  &  & 6 & -0.79 & 175 & [-12.16, 10.59] & 0.892 \\
 &  &  & 12 & -13.47 & 169 & [-27.67, 0.74] & 0.063 \\
\addlinespace[1pt]
 & 2023 & Retrospective & 1 & -11.24 & 276 & [-22.61, 0.14] & 0.053 \\
 &  &  & 3 & -19.31 & 274 & [-36.13, -2.49] & 0.024 \\
 &  &  & 6 & -15.67 & 271 & [-30.52, -0.83] & 0.039 \\
 &  &  & 12 & -30.80 & 265 & [-67.01, 5.41] & 0.095 \\
\addlinespace[1pt]
 &  & Boundary & 3 & -25.29 & 2 & -- & -- \\
 &  &  & 6 & -43.61 & 5 & -- & -- \\
 &  &  & 12 & -93.27 & 11 & -- & -- \\
\addlinespace[1pt]
 &  & Post-cutoff & 1 & -0.05 & 12 & -- & -- \\
 &  &  & 3 & -1.72 & 10 & -- & -- \\
 &  &  & 6 & 3.83 & 7 & -- & -- \\
 &  &  & 12 & -0.93 & 1 & -- & -- \\
\addlinespace[1pt]
\addlinespace[3pt]
TimesFM 8M & 2000 & Post-cutoff & 1 & -11.33 & 288 & [-20.43, -2.24] & 0.015 \\
 &  &  & 3 & -9.22 & 286 & [-24.55, 6.10] & 0.238 \\
 &  &  & 6 & -15.21 & 283 & [-37.87, 7.46] & 0.188 \\
 &  &  & 12 & -23.82 & 277 & [-53.07, 5.44] & 0.111 \\
\addlinespace[1pt]
 & 2009 & Retrospective & 1 & 7.85 & 108 & [-7.09, 22.79] & 0.303 \\
 &  &  & 3 & 7.90 & 106 & [-6.57, 22.36] & 0.285 \\
 &  &  & 6 & 0.34 & 103 & [-15.52, 16.19] & 0.967 \\
 &  &  & 12 & -9.38 & 97 & [-22.98, 4.22] & 0.176 \\
\addlinespace[1pt]
 &  & Boundary & 3 & -118.54 & 2 & -- & -- \\
 &  &  & 6 & -67.99 & 5 & -- & -- \\
 &  &  & 12 & -21.78 & 11 & -- & -- \\
\addlinespace[1pt]
 &  & Post-cutoff & 1 & -9.90 & 180 & [-18.78, -1.02] & 0.029 \\
 &  &  & 3 & -14.58 & 178 & [-29.86, 0.70] & 0.061 \\
 &  &  & 6 & -18.28 & 175 & [-42.06, 5.51] & 0.132 \\
 &  &  & 12 & -9.80 & 169 & [-37.83, 18.23] & 0.493 \\
\addlinespace[1pt]
 & 2023 & Retrospective & 1 & 9.81 & 276 & [-1.10, 20.73] & 0.078 \\
 &  &  & 3 & 0.44 & 274 & [-12.58, 13.47] & 0.947 \\
 &  &  & 6 & -4.71 & 271 & [-15.80, 6.37] & 0.405 \\
 &  &  & 12 & -10.71 & 265 & [-21.92, 0.50] & 0.061 \\
\addlinespace[1pt]
 &  & Boundary & 3 & 111.13 & 2 & -- & -- \\
 &  &  & 6 & 205.91 & 5 & -- & -- \\
 &  &  & 12 & 108.43 & 11 & -- & -- \\
\addlinespace[1pt]
 &  & Post-cutoff & 1 & 0.00 & 12 & -- & -- \\
 &  &  & 3 & 0.00 & 10 & -- & -- \\
 &  &  & 6 & 0.00 & 7 & -- & -- \\
 &  &  & 12 & 0.00 & 1 & -- & -- \\
\addlinespace[1pt]
\addlinespace[3pt]
TimesFM 20M & 2000 & Post-cutoff & 1 & -14.99 & 288 & [-25.08, -4.91] & 0.004 \\
 &  &  & 3 & -22.07 & 286 & [-36.08, -8.06] & 0.002 \\
 &  &  & 6 & -30.53 & 283 & [-46.61, -14.45] & $<$0.001 \\
 &  &  & 12 & -26.00 & 277 & [-48.27, -3.73] & 0.022 \\
\addlinespace[1pt]
 & 2009 & Retrospective & 1 & -1.43 & 108 & [-17.76, 14.90] & 0.864 \\
 &  &  & 3 & -56.98 & 106 & [-86.86, -27.09] & $<$0.001 \\
 &  &  & 6 & -78.69 & 103 & [-120.40, -36.97] & $<$0.001 \\
 &  &  & 12 & -106.24 & 97 & [-150.11, -62.37] & $<$0.001 \\
\addlinespace[1pt]
 &  & Boundary & 3 & -62.95 & 2 & -- & -- \\
 &  &  & 6 & 55.77 & 5 & -- & -- \\
 &  &  & 12 & 61.05 & 11 & -- & -- \\
\addlinespace[1pt]
 &  & Post-cutoff & 1 & -4.17 & 180 & [-13.42, 5.09] & 0.378 \\
 &  &  & 3 & -21.28 & 178 & [-52.88, 10.33] & 0.187 \\
 &  &  & 6 & -34.60 & 175 & [-95.68, 26.48] & 0.267 \\
 &  &  & 12 & -41.06 & 169 & [-135.07, 52.95] & 0.392 \\
\addlinespace[1pt]
 & 2023 & Retrospective & 1 & 0.83 & 276 & [-9.53, 11.20] & 0.875 \\
 &  &  & 3 & -18.96 & 274 & [-33.06, -4.85] & 0.008 \\
 &  &  & 6 & -23.46 & 271 & [-43.16, -3.76] & 0.020 \\
 &  &  & 12 & -27.43 & 265 & [-57.26, 2.40] & 0.072 \\
\addlinespace[1pt]
 &  & Boundary & 3 & -120.79 & 2 & -- & -- \\
 &  &  & 6 & -336.35 & 5 & -- & -- \\
 &  &  & 12 & -326.28 & 11 & -- & -- \\
\addlinespace[1pt]
 &  & Post-cutoff & 1 & 0.00 & 12 & -- & -- \\
 &  &  & 3 & 0.00 & 10 & -- & -- \\
 &  &  & 6 & -0.00 & 7 & -- & -- \\
 &  &  & 12 & 0.00 & 1 & -- & -- \\
\addlinespace[1pt]
\addlinespace[3pt]
\end{longtable}

\endgroup
\clearpage
\begingroup
\fontsize{11}{13.6}\selectfont
\setlength{\tabcolsep}{4.00pt}
\setlength{\linewidth}{379.05pt}
\begin{longtable}{@{}llccccc@{}}
\caption[Fixed annual-state effects in the balanced sample]{%
  \textbf{Fixed annual-state effects in the balanced sample.}
}\label{tab:ia-x44-balanced}\\
\multicolumn{7}{@{}p{\linewidth}@{}}{\exhibitnotes{%
  Fixed-state matched predictive effects are evaluated over the balanced January 2002--December 2021 sample for states 2000, 2009, and 2023. U.S. and international estimates are reported separately under the US training-information environment. Effects compare each fixed state with origin-aligned point-in-time (PIT) forecasts on matched target months and are expressed in historical-average mean squared forecast error (MSFE) percentage points; positive values favor the fixed state. International effects normalize each market's monthly PIT-relative loss differential by its historical-average MSFE and average the 13 markets equally over common target months. Parentheses contain two-sided Bartlett--Newey--West HAC $p$-values with $h-1$ lags, where $h$ is in months. $^{*}$, $^{**}$, and $^{***}$ indicate significance at 10\%, 5\%, and 1\%.
}}\\
\noalign{\vskip 6pt}
\toprule
Environment & Model & Fixed & $h=1$ & $h=3$ & $h=6$ & $h=12$ \\
\midrule
\endfirsthead
\multicolumn{7}{l}{\textit{\tablename\ \thetable\ (continued)}}\\
\toprule
Environment & Model & Fixed & $h=1$ & $h=3$ & $h=6$ & $h=12$ \\
\midrule
\endhead
\midrule\multicolumn{7}{r}{\textit{Continued on next page}}\\\endfoot
\bottomrule\endlastfoot
United States & Chronos Tiny & 2000 & \makecell[c]{-6.92$^{**}$\\{\scriptsize (0.013)}} & \makecell[c]{-18.86$^{***}$\\{\scriptsize (0.010)}} & \makecell[c]{-30.66$^{**}$\\{\scriptsize (0.016)}} & \makecell[c]{-38.40$^{**}$\\{\scriptsize (0.034)}} \\
 &  & 2009 & \makecell[c]{-2.33\\{\scriptsize (0.336)}} & \makecell[c]{-7.43\\{\scriptsize (0.126)}} & \makecell[c]{-6.27\\{\scriptsize (0.378)}} & \makecell[c]{11.23\\{\scriptsize (0.267)}} \\
 &  & 2023 & \makecell[c]{-2.15\\{\scriptsize (0.467)}} & \makecell[c]{-8.10$^{**}$\\{\scriptsize (0.020)}} & \makecell[c]{-9.20$^{**}$\\{\scriptsize (0.014)}} & \makecell[c]{-4.22\\{\scriptsize (0.404)}} \\
\addlinespace[1pt]
 & Chronos Mini & 2000 & \makecell[c]{-5.15$^{**}$\\{\scriptsize (0.013)}} & \makecell[c]{-14.03$^{***}$\\{\scriptsize ($<$0.001)}} & \makecell[c]{-18.19$^{***}$\\{\scriptsize (0.009)}} & \makecell[c]{-25.20$^{**}$\\{\scriptsize (0.032)}} \\
 &  & 2009 & \makecell[c]{-0.40\\{\scriptsize (0.868)}} & \makecell[c]{-1.32\\{\scriptsize (0.750)}} & \makecell[c]{0.68\\{\scriptsize (0.915)}} & \makecell[c]{10.37\\{\scriptsize (0.261)}} \\
 &  & 2023 & \makecell[c]{-11.26\\{\scriptsize (0.108)}} & \makecell[c]{-16.56\\{\scriptsize (0.182)}} & \makecell[c]{-2.79\\{\scriptsize (0.696)}} & \makecell[c]{0.79\\{\scriptsize (0.926)}} \\
\addlinespace[1pt]
 & Chronos Small & 2000 & \makecell[c]{-1.23\\{\scriptsize (0.533)}} & \makecell[c]{-9.61$^{**}$\\{\scriptsize (0.028)}} & \makecell[c]{-14.39$^{*}$\\{\scriptsize (0.085)}} & \makecell[c]{-18.53\\{\scriptsize (0.181)}} \\
 &  & 2009 & \makecell[c]{-0.25\\{\scriptsize (0.922)}} & \makecell[c]{-1.88\\{\scriptsize (0.641)}} & \makecell[c]{-3.03\\{\scriptsize (0.355)}} & \makecell[c]{-4.44\\{\scriptsize (0.415)}} \\
 &  & 2023 & \makecell[c]{-17.70$^{**}$\\{\scriptsize (0.011)}} & \makecell[c]{-22.87$^{**}$\\{\scriptsize (0.020)}} & \makecell[c]{-17.78$^{**}$\\{\scriptsize (0.028)}} & \makecell[c]{-32.66\\{\scriptsize (0.137)}} \\
\addlinespace[1pt]
 & TimesFM 8M & 2000 & \makecell[c]{-13.28$^{**}$\\{\scriptsize (0.025)}} & \makecell[c]{-10.98\\{\scriptsize (0.249)}} & \makecell[c]{-17.65\\{\scriptsize (0.198)}} & \makecell[c]{-29.53\\{\scriptsize (0.109)}} \\
 &  & 2009 & \makecell[c]{-2.52\\{\scriptsize (0.643)}} & \makecell[c]{-2.46\\{\scriptsize (0.731)}} & \makecell[c]{-9.01\\{\scriptsize (0.320)}} & \makecell[c]{-11.23\\{\scriptsize (0.155)}} \\
 &  & 2023 & \makecell[c]{11.53$^{*}$\\{\scriptsize (0.090)}} & \makecell[c]{-0.41\\{\scriptsize (0.959)}} & \makecell[c]{-4.94\\{\scriptsize (0.416)}} & \makecell[c]{-11.97$^{*}$\\{\scriptsize (0.053)}} \\
\addlinespace[1pt]
 & TimesFM 20M & 2000 & \makecell[c]{-19.46$^{***}$\\{\scriptsize (0.003)}} & \makecell[c]{-27.07$^{***}$\\{\scriptsize (0.002)}} & \makecell[c]{-35.90$^{***}$\\{\scriptsize ($<$0.001)}} & \makecell[c]{-29.87$^{**}$\\{\scriptsize (0.032)}} \\
 &  & 2009 & \makecell[c]{-1.15\\{\scriptsize (0.817)}} & \makecell[c]{-36.28$^{***}$\\{\scriptsize (0.004)}} & \makecell[c]{-54.23$^{***}$\\{\scriptsize (0.005)}} & \makecell[c]{-61.81$^{**}$\\{\scriptsize (0.026)}} \\
 &  & 2023 & \makecell[c]{1.89\\{\scriptsize (0.684)}} & \makecell[c]{-17.98$^{**}$\\{\scriptsize (0.020)}} & \makecell[c]{-26.31$^{**}$\\{\scriptsize (0.023)}} & \makecell[c]{-31.01$^{*}$\\{\scriptsize (0.094)}} \\
\addlinespace[1pt]
\addlinespace[5pt]
International & Chronos Tiny & 2000 & \makecell[c]{-3.44\\{\scriptsize (0.108)}} & \makecell[c]{-8.26$^{*}$\\{\scriptsize (0.094)}} & \makecell[c]{-14.13$^{**}$\\{\scriptsize (0.048)}} & \makecell[c]{-21.31$^{**}$\\{\scriptsize (0.019)}} \\
 &  & 2009 & \makecell[c]{-2.04$^{*}$\\{\scriptsize (0.060)}} & \makecell[c]{-4.67$^{*}$\\{\scriptsize (0.055)}} & \makecell[c]{-6.51$^{*}$\\{\scriptsize (0.051)}} & \makecell[c]{-4.33\\{\scriptsize (0.174)}} \\
 &  & 2023 & \makecell[c]{-2.79$^{**}$\\{\scriptsize (0.031)}} & \makecell[c]{-4.44$^{**}$\\{\scriptsize (0.031)}} & \makecell[c]{-6.35$^{*}$\\{\scriptsize (0.052)}} & \makecell[c]{-4.84\\{\scriptsize (0.175)}} \\
\addlinespace[1pt]
 & Chronos Mini & 2000 & \makecell[c]{-2.16\\{\scriptsize (0.192)}} & \makecell[c]{-7.14$^{**}$\\{\scriptsize (0.036)}} & \makecell[c]{-10.91$^{**}$\\{\scriptsize (0.039)}} & \makecell[c]{-17.60$^{**}$\\{\scriptsize (0.015)}} \\
 &  & 2009 & \makecell[c]{-1.83$^{**}$\\{\scriptsize (0.027)}} & \makecell[c]{-3.69$^{***}$\\{\scriptsize (0.001)}} & \makecell[c]{-4.08$^{**}$\\{\scriptsize (0.014)}} & \makecell[c]{-4.28\\{\scriptsize (0.154)}} \\
 &  & 2023 & \makecell[c]{-4.74$^{**}$\\{\scriptsize (0.035)}} & \makecell[c]{-7.34$^{*}$\\{\scriptsize (0.077)}} & \makecell[c]{-3.60\\{\scriptsize (0.171)}} & \makecell[c]{-3.12\\{\scriptsize (0.313)}} \\
\addlinespace[1pt]
 & Chronos Small & 2000 & \makecell[c]{-1.44\\{\scriptsize (0.360)}} & \makecell[c]{-4.01$^{*}$\\{\scriptsize (0.090)}} & \makecell[c]{-5.74\\{\scriptsize (0.120)}} & \makecell[c]{-10.21$^{*}$\\{\scriptsize (0.075)}} \\
 &  & 2009 & \makecell[c]{0.52\\{\scriptsize (0.747)}} & \makecell[c]{-0.14\\{\scriptsize (0.950)}} & \makecell[c]{-3.70$^{*}$\\{\scriptsize (0.073)}} & \makecell[c]{-7.02$^{**}$\\{\scriptsize (0.024)}} \\
 &  & 2023 & \makecell[c]{-1.20\\{\scriptsize (0.342)}} & \makecell[c]{-1.95\\{\scriptsize (0.312)}} & \makecell[c]{-3.29\\{\scriptsize (0.230)}} & \makecell[c]{-1.87\\{\scriptsize (0.563)}} \\
\addlinespace[1pt]
 & TimesFM 8M & 2000 & \makecell[c]{-31.26$^{***}$\\{\scriptsize ($<$0.001)}} & \makecell[c]{-35.86$^{***}$\\{\scriptsize (0.002)}} & \makecell[c]{-37.96$^{**}$\\{\scriptsize (0.013)}} & \makecell[c]{-45.74$^{***}$\\{\scriptsize (0.008)}} \\
 &  & 2009 & \makecell[c]{-15.38$^{**}$\\{\scriptsize (0.038)}} & \makecell[c]{-17.51$^{*}$\\{\scriptsize (0.070)}} & \makecell[c]{-22.95$^{*}$\\{\scriptsize (0.061)}} & \makecell[c]{-19.75$^{**}$\\{\scriptsize (0.028)}} \\
 &  & 2023 & \makecell[c]{23.07$^{***}$\\{\scriptsize ($<$0.001)}} & \makecell[c]{-2.62\\{\scriptsize (0.694)}} & \makecell[c]{-0.65\\{\scriptsize (0.915)}} & \makecell[c]{-4.62\\{\scriptsize (0.494)}} \\
\addlinespace[1pt]
 & TimesFM 20M & 2000 & \makecell[c]{-18.76$^{**}$\\{\scriptsize (0.017)}} & \makecell[c]{-19.63$^{**}$\\{\scriptsize (0.025)}} & \makecell[c]{-25.89$^{***}$\\{\scriptsize (0.003)}} & \makecell[c]{7.82\\{\scriptsize (0.444)}} \\
 &  & 2009 & \makecell[c]{-10.89$^{*}$\\{\scriptsize (0.072)}} & \makecell[c]{-81.83$^{***}$\\{\scriptsize ($<$0.001)}} & \makecell[c]{-108.62$^{***}$\\{\scriptsize (0.002)}} & \makecell[c]{-117.48$^{**}$\\{\scriptsize (0.027)}} \\
 &  & 2023 & \makecell[c]{-7.58$^{*}$\\{\scriptsize (0.051)}} & \makecell[c]{-54.71$^{***}$\\{\scriptsize ($<$0.001)}} & \makecell[c]{-61.05$^{***}$\\{\scriptsize ($<$0.001)}} & \makecell[c]{-47.79$^{***}$\\{\scriptsize (0.002)}} \\
\addlinespace[1pt]
\end{longtable}

\endgroup
\clearpage

\setlength{\intextsep}{6pt}
\section{Predictive Performance across Timing Designs}\label{sec:ia-cross-design-results}

The following tables extend main-text Section~\ref{persistence-across-timing-designs}, covering market-level sign patterns, training samples, exposure-depth gradients, forecast objects, and sample windows.

\begingroup
\begin{table}[H]
\centering
\begin{minipage}{394.53pt}
\centering
\caption[International predictive effects across rolling and fixed-vintage designs]{%
  \textbf{International predictive effects across rolling and fixed-vintage designs.}
}
\label{tab:ia-x45-international-patterns}
\fontsize{9}{10.3}\selectfont
\renewcommand{\arraystretch}{1.00}
\setlength{\tabcolsep}{5pt}
\fontsize{9}{10.3}\selectfont
\setlength{\tabcolsep}{4.00pt}
\begin{tabular}{@{}lcccccccc@{}}
\toprule
Model & $h$ & \multicolumn{3}{c}{Median matched effect relative to PIT} & \multicolumn{4}{c}{Markets by cross-design pattern} \\
\cmidrule(lr){3-5}\cmidrule(l){6-9}
& & Rolling & Fixed 2009 & Fixed 2023 & Gain & Loss & Fixed mixed & \shortstack{Protocol\\disagree} \\
\midrule
\multirow{4}{*}{Chronos Tiny} & 1 & -1.57 & -3.20 & -2.37 & 0 & 10 & 3 & 0 \\
 & 3 & -4.16 & -8.22 & -3.86 & 0 & 11 & 2 & 0 \\
 & 6 & -6.84 & -10.14 & -6.30 & 0 & 11 & 2 & 0 \\
 & 12 & -5.69 & -8.24 & -6.07 & 2 & 10 & 0 & 1 \\
\addlinespace[2pt]
\multirow{4}{*}{Chronos Mini} & 1 & -2.84 & -3.53 & -4.32 & 0 & 11 & 2 & 0 \\
 & 3 & -5.34 & -4.86 & -6.86 & 1 & 11 & 1 & 0 \\
 & 6 & -4.25 & -4.72 & -3.16 & 1 & 9 & 2 & 1 \\
 & 12 & -5.05 & -4.43 & -4.04 & 1 & 9 & 1 & 2 \\
\addlinespace[2pt]
\multirow{4}{*}{Chronos Small} & 1 & -0.38 & 2.89 & -0.08 & 4 & 3 & 5 & 1 \\
 & 3 & -0.12 & 3.31 & -2.21 & 3 & 4 & 5 & 1 \\
 & 6 & -4.93 & -3.11 & -3.09 & 1 & 8 & 3 & 1 \\
 & 12 & -7.39 & -7.18 & -3.10 & 0 & 6 & 5 & 2 \\
\addlinespace[2pt]
\multirow{4}{*}{TimesFM 8M} & 1 & -3.85 & -6.72 & 19.94 & 2 & 0 & 7 & 4 \\
 & 3 & -9.94 & -10.82 & 0.97 & 0 & 3 & 6 & 4 \\
 & 6 & -19.33 & -17.11 & -2.38 & 0 & 8 & 4 & 1 \\
 & 12 & -29.88 & -18.01 & -8.07 & 0 & 10 & 3 & 0 \\
\addlinespace[2pt]
\multirow{4}{*}{TimesFM 20M} & 1 & -4.53 & -9.96 & -5.23 & 0 & 9 & 3 & 1 \\
 & 3 & -18.91 & -74.88 & -53.75 & 0 & 12 & 1 & 0 \\
 & 6 & -23.28 & -87.81 & -59.30 & 0 & 13 & 0 & 0 \\
 & 12 & -6.13 & -113.47 & -47.12 & 0 & 9 & 2 & 2 \\
\midrule
\multicolumn{5}{@{}l}{\textit{Totals across 260 market--model--horizon cells}} & 15 & 167 & 57 & 21 \\
\bottomrule
\end{tabular}

\par\vspace{6pt}\nointerlineskip
\exhibitnotes{%
  The first three columns give cross-market medians of pooled rolling, fixed-2009, and fixed-2023 retrospective-exposure effects for each model--horizon row. Rolling effects equally average the loss advantages of leads $0,+1,+2$ on strict common support. All effects are relative to origin-aligned point-in-time (PIT) forecasts and expressed in historical-average MSFE percentage points; positive values favor the alternative state. Common gain and common loss require three positive or three negative point estimates. Fixed-vintage mixed denotes opposite signs for the fixed-state estimates; protocol disagreement denotes fixed-state estimates with a common sign opposite to the rolling estimate. Pattern columns count the 13 non-U.S. markets, with classification performed before cross-market summarization. The horizon $h$ is in months. Estimates use all-available target months and the US training-information environment.
}
\end{minipage}
\end{table}

\endgroup

\clearpage
\begingroup
\begin{table}[H]
\centering
\begin{minipage}{393.00pt}
\centering
\caption[Cross-design sign patterns vary across training environments and sample windows]{%
  \textbf{Cross-design sign patterns vary across training environments and sample windows.}
}
\label{tab:ia-x45-corpus-sample}
\footnotesize
\renewcommand{\arraystretch}{1.10}
\setlength{\tabcolsep}{4.2pt}
\fontsize{9}{11}\selectfont
\setlength{\tabcolsep}{4.00pt}
\begin{tabular}{@{}llccccccc@{}}
\toprule
\shortstack[l]{Training\\environment} & Target set & $N$ & \shortstack{Common\\gain} & \shortstack{Common\\loss} & \shortstack{Fixed\\mixed} & \shortstack{Protocol\\disagree} & \shortstack{Joint-\\supported\\gain} & \shortstack{Joint-\\supported\\loss} \\
\midrule
\multicolumn{9}{@{}l}{\textit{Panel A. All available}} \\
\addlinespace[2pt]
US & U.S. & 20 & 0 & 12 & 5 & 3 & 0 & 2 \\
US & International & 260 & 15 & 167 & 57 & 21 & 0 & 22 \\
\addlinespace[2pt]
Global & U.S. & 20 & 4 & 3 & 6 & 7 & 0 & 0 \\
Global & International & 260 & 25 & 84 & 97 & 54 & 0 & 0 \\
\addlinespace[2pt]
Augmented & U.S. & 20 & 6 & 3 & 8 & 3 & 0 & 0 \\
Augmented & International & 260 & 59 & 36 & 138 & 27 & 0 & 0 \\
\addlinespace[2pt]
\midrule
\multicolumn{9}{@{}l}{\textit{Panel B. Balanced 2002--2021}} \\
\addlinespace[2pt]
US & U.S. & 20 & 0 & 11 & 8 & 1 & 0 & 2 \\
US & International & 260 & 14 & 164 & 66 & 16 & 0 & 4 \\
\addlinespace[2pt]
Global & U.S. & 20 & 5 & 2 & 8 & 5 & 0 & 0 \\
Global & International & 260 & 30 & 70 & 107 & 53 & 0 & 0 \\
\addlinespace[2pt]
Augmented & U.S. & 20 & 4 & 1 & 10 & 5 & 0 & 0 \\
Augmented & International & 260 & 57 & 39 & 146 & 18 & 1 & 0 \\
\addlinespace[2pt]
\bottomrule
\end{tabular}

\par\vspace{6pt}\nointerlineskip
\exhibitnotes{%
  Cross-design sign patterns are tabulated by training-information environment, target group, and sample window. Panel A uses all-available target months; Panel B uses January 2002--December 2021. U.S. rows contain 20 model--horizon cells, and international rows contain 260 market--model--horizon cells. Each classification compares the pooled rolling effect for leads $0,+1,+2$ with fixed-2009 and fixed-2023 retrospective-exposure effects, all relative to origin-aligned point-in-time (PIT) forecasts. Common gain and common loss require three positive or three negative estimates. Fixed-vintage mixed denotes opposite signs for the fixed-state estimates; protocol disagreement denotes fixed-state estimates with a common sign opposite to the rolling estimate. Joint-supported gain or loss additionally requires both 95\% HAC and circular moving-block bootstrap intervals for each of the three effects to exclude zero on that side. Counts are reported separately for US, Global, and Augmented training and the two target groups.
}
\end{minipage}
\end{table}

\endgroup
\clearpage
\begingroup
\begin{table}[H]
\centering
\begin{minipage}{382.71pt}
\centering
\caption[Predictive-value gradients across timing designs]{%
  \textbf{Predictive-value gradients across timing designs.}
}
\label{tab:ia-x45-depth-directions}
\footnotesize
\renewcommand{\arraystretch}{1.10}
\setlength{\tabcolsep}{5pt}
\fontsize{9}{11}\selectfont
\setlength{\tabcolsep}{4.00pt}
\begin{tabular}{@{}lccccccccc@{}}
\toprule
Model & $h$ & \multicolumn{3}{c}{United States} & \multicolumn{5}{c}{International markets} \\
\cmidrule(lr){3-5}\cmidrule(l){6-10}
& & Rolling & Fixed & Direction & \shortstack{Med.\\rolling} & \shortstack{Med.\\fixed} & \shortstack{Both\\$+$} & \shortstack{Both\\$-$} & \shortstack{Protocol-\\specific} \\
\midrule
\multirow{4}{*}{Chronos Tiny} & 1 & -0.55 & -0.14 & Both negative & -0.13 & -0.16 & 2 & 6 & 5 \\
 & 3 & -1.69 & -0.27 & Both negative & 0.12 & -0.05 & 2 & 4 & 7 \\
 & 6 & -3.92 & -0.08 & Both negative & 0.03 & 0.03 & 2 & 1 & 10 \\
 & 12 & -6.43 & -0.92 & Both negative & -1.44 & -0.35 & 2 & 9 & 2 \\
\addlinespace[3pt]
\multirow{4}{*}{Chronos Mini} & 1 & -0.08 & -1.27 & Both negative & -0.47 & -0.18 & 0 & 10 & 3 \\
 & 3 & -1.10 & -1.91 & Both negative & -0.98 & -0.47 & 0 & 8 & 5 \\
 & 6 & -2.72 & 0.16 & Protocol-specific & -1.44 & -0.15 & 1 & 7 & 5 \\
 & 12 & -4.38 & 0.07 & Protocol-specific & -1.36 & -0.30 & 2 & 8 & 3 \\
\addlinespace[3pt]
\multirow{4}{*}{Chronos Small} & 1 & -0.38 & -2.13 & Both negative & -0.18 & -0.28 & 2 & 5 & 6 \\
 & 3 & 0.21 & -2.08 & Protocol-specific & 0.54 & -0.54 & 2 & 2 & 9 \\
 & 6 & -5.35 & -1.33 & Both negative & -2.43 & 0.03 & 1 & 3 & 9 \\
 & 12 & -1.59 & -1.95 & Both negative & -2.86 & 0.40 & 1 & 3 & 9 \\
\addlinespace[3pt]
\multirow{4}{*}{TimesFM 8M} & 1 & -0.11 & 1.04 & Protocol-specific & -0.58 & 2.20 & 5 & 0 & 8 \\
 & 3 & -1.84 & 0.12 & Protocol-specific & -0.73 & 0.97 & 5 & 1 & 7 \\
 & 6 & -2.56 & -0.22 & Both negative & -2.22 & 1.19 & 2 & 1 & 10 \\
 & 12 & -5.18 & 0.12 & Protocol-specific & -8.10 & 0.83 & 0 & 1 & 12 \\
\addlinespace[3pt]
\multirow{4}{*}{TimesFM 20M} & 1 & 0.48 & 0.52 & Both positive & -3.48 & 0.75 & 1 & 1 & 11 \\
 & 3 & 0.45 & 3.86 & Both positive & -3.80 & 3.30 & 2 & 1 & 10 \\
 & 6 & 0.08 & 6.19 & Both positive & -2.02 & 4.89 & 2 & 1 & 10 \\
 & 12 & -3.79 & 6.42 & Protocol-specific & -10.06 & 6.41 & 0 & 1 & 12 \\
\bottomrule
\end{tabular}

\par\vspace{6pt}\nointerlineskip
\exhibitnotes{%
  Predictive-value gradients measure how matched effects vary with exposure depth. The rolling gradient is the slope of lead-specific effects across leads $0,+1,+2$; the fixed gradient is the retrospective-exposure slope across fixed annual-state years. Positive gradients indicate increasing matched predictive value, and negative gradients decreasing value. U.S. columns give model--horizon gradients and their direction classifications. International columns give median gradients and counts of both-positive, both-negative, and protocol-specific direction patterns across the 13 markets. Rolling and fixed gradients are reported separately. The horizon $h$ is in months. Estimates use all-available target months and the US training-information environment.
}
\end{minipage}
\end{table}

\endgroup
\clearpage
\begin{table}[!htbp]
\centering
\caption[Sign and interval-support concordance across forecast objects and sample windows]{%
  \textbf{Sign and interval-support concordance across forecast objects and sample windows.}
}
\label{tab:ia-x45-object-window}
\scriptsize
\renewcommand{\arraystretch}{1.13}
\setlength{\tabcolsep}{2.5pt}
\begin{tabular*}{\textwidth}{@{\extracolsep{\fill}}llrrcrrc@{}}
\toprule
& & \multicolumn{3}{c}{United States} & \multicolumn{3}{c}{International markets} \\
\cmidrule(lr){3-5}\cmidrule(l){6-8}
Domain & Comparison & Sign (\%) & Support (\%) & $N$ & Sign (\%) & Support (\%) & $N$ \\
\midrule
\multicolumn{8}{@{}l}{\textit{Panel A. Forecast-object sign concordance}} \\
\addlinespace[2pt]
Scalar loss & Predictive mean vs. median & 79.0 & -- & 210 & 82.1 & -- & 2730 \\
Directional & Mean vs. median Macro-F1 & 72.9 & -- & 140 & 77.5 & -- & 1820 \\
Distributional & Quantile loss vs. interval score & 91.7 & -- & 84 & 92.4 & -- & 2730 \\
\midrule
\multicolumn{8}{@{}l}{\textit{Panel B. All-available versus balanced-window concordance}} \\
\addlinespace[2pt]
PIT benchmark & All available vs. balanced & 100.0 & 85.0 & 20 & 90.0 & 93.8 & 260 \\
Rolling annual-state design & All available vs. balanced & 97.5 & 92.5 & 80 & 94.5 & 91.3 & 1040 \\
Fixed-vintage design & All available vs. balanced & 90.0 & 91.7 & 60 & 95.0 & 91.4 & 780 \\
\bottomrule
\end{tabular*}

\par\vspace{6pt}\nointerlineskip
\exhibitnotes{%
  Concordance percentages are calculated across common comparison cells, with counts in the adjacent $N$ columns. Panel A compares predictive-mean and predictive-median squared-loss effects, mean- and median-based Macro-F1 changes, and average quantile-loss and 80\% interval-score advantages \citep{gneiting_raftery_2007}. Panel B compares all-available and balanced January 2002--December 2021 estimates for the PIT benchmark and the rolling and fixed-vintage designs. Sign concordance requires paired estimates to have the same sign. Interval-support concordance requires the same classification under joint HAC and circular moving-block bootstrap intervals; it refers to inferential classification, not equality of forecast samples. U.S. and 13-market results use the US training-information environment and are reported separately. Dashes in Panel A indicate unreported interval-support comparisons.
}
\end{table}

\clearpage

\section{The Economic Value of Market Timing}\label{sec:ia-investor-results}

These portfolio comparisons supplement main-text Section~\ref{investor-value}.

\begin{figure}[H]
\centering
\includegraphics[width=0.95\textwidth]{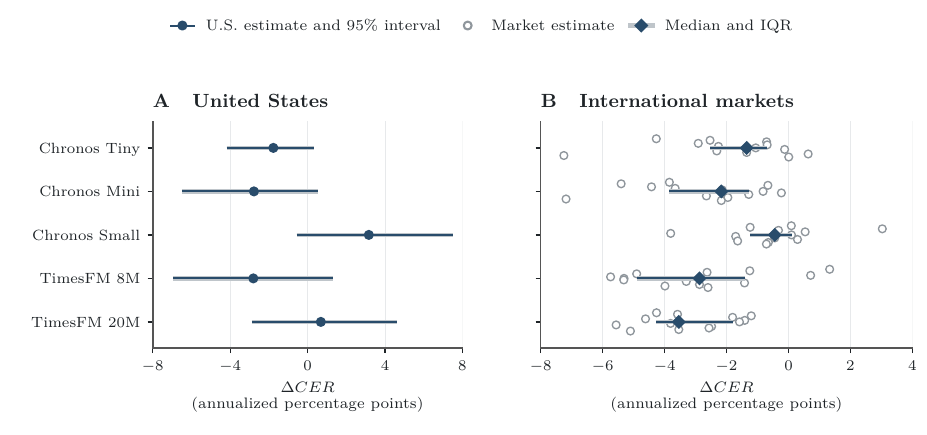}
\caption[Investor-value effects across markets]{%
  \textbf{Investor-value effects across markets.}
  Investor-value effects are $\Delta CER=CER_{\mathrm{Exposed}}-CER_{\mathrm{PIT}}$, in annualized percentage points; positive values favor post-origin forecasts. PIT uses lead $-1$, and exposed forecasts equally average leads $0,+1,+2$ on strict common support. Panel A gives the five U.S. model estimates with paired 95\% percentile intervals from 1,000 circular moving-block bootstrap replications using block length six. Panel B plots the 13 market-level estimates for each model; diamonds mark their median and horizontal bars their interquartile range. The specification uses the one-month horizon, risk aversion $\gamma=3$, market weights in $[0,1.5]$, the US training-information environment, and 252 target-start months from January 2001 through December 2021.
}
\label{fig:ia-x46-market-distribution}

\end{figure}

\clearpage
\begin{table}[!htbp]
\centering
\caption[Investor value across risk aversion and portfolio constraints]{%
  \textbf{Investor value across risk aversion and portfolio constraints.}
}
\label{tab:ia-x46-gamma-grid}
\small
\renewcommand{\arraystretch}{1.15}
\setlength{\tabcolsep}{3.3pt}
\begin{tabular*}{\textwidth}{@{\extracolsep{\fill}}crrrcrrrr@{}}
\toprule
& \multicolumn{4}{c}{United States} & \multicolumn{4}{c}{International markets} \\
\cmidrule(lr){2-5}\cmidrule(l){6-9}
$\gamma$ & Median $\Delta CER$ & $>0$ & CI $<0$ & CI $>0$ & Median $\Delta CER$ & $>0$ & CI $<0$ & CI $>0$ \\
\midrule
\multicolumn{9}{@{}l}{\textit{Panel A. Constrained weights}} \\
\addlinespace[2pt]
1 & -0.58 & 2/5 & 0/5 & 0/5 & -1.71 & 12/65 & 4/65 & 0/65 \\
3 & -1.77 & 2/5 & 0/5 & 0/5 & -2.14 & 9/65 & 10/65 & 0/65 \\
5 & -1.59 & 1/5 & 1/5 & 1/5 & -1.41 & 13/65 & 12/65 & 1/65 \\
\midrule
\multicolumn{9}{@{}l}{\textit{Panel B. Unconstrained weights}} \\
\addlinespace[2pt]
1 & 10.98 & 3/5 & 1/5 & 0/5 & -7.56 & 23/65 & 6/65 & 2/65 \\
3 & 3.66 & 3/5 & 1/5 & 1/5 & -2.52 & 23/65 & 6/65 & 2/65 \\
5 & 2.20 & 3/5 & 1/5 & 1/5 & -1.51 & 23/65 & 6/65 & 2/65 \\
\bottomrule
\end{tabular*}

\par\vspace{6pt}\nointerlineskip
\exhibitnotes{%
  Annualized certainty-equivalent-return differences, $\Delta CER=CER_{\mathrm{Exposed}}-CER_{\mathrm{PIT}}$, are reported for risk aversion $\gamma\in\{1,3,5\}$. PIT uses lead $-1$; the exposed forecast equally averages leads $0,+1,+2$ on strict common support. Panel A constrains market weights to $[0,1.5]$, whereas Panel B leaves them unconstrained. Positive differences favor exposed forecasts. U.S. summaries cover five models, and international summaries cover 65 market--model cells in 13 non-U.S. markets. Medians use paired CER differences. Counts give positive point estimates and cells with paired 95\% bootstrap intervals entirely below or above zero. Intervals use 1,000 circular moving-block replications with block length six. Results are in annualized percentage points for the one-month horizon, the US training-information environment, and 252 target-start months from January 2001 through December 2021.
}
\end{table}

\clearpage
\begin{table}[H]
\centering
\begin{minipage}{423pt}
\centering
\caption[Investor value at longer forecast horizons]{%
  \textbf{Investor value at longer forecast horizons.}
}\label{tab:ia-x46-longer-horizons}
\footnotesize
\renewcommand{\arraystretch}{1.08}
\setlength{\tabcolsep}{4pt}
\begin{tabular}{@{}>{\raggedright\arraybackslash}p{65pt}>{\centering\arraybackslash}p{16pt}>{\centering\arraybackslash}p{24pt}>{\centering\arraybackslash}p{50pt}>{\centering\arraybackslash}p{57pt}>{\centering\arraybackslash}p{52pt}>{\centering\arraybackslash}p{111pt}@{}}
\toprule
\multicolumn{7}{@{}l}{\textit{Panel A. United States}} \\
\addlinespace[2pt]
Model & $h$ & $N$ & PIT CER & Exposed CER & $\Delta CER$ & 95\% interval \\
\midrule
\multirow{3}{*}{Chronos Tiny} & 3 & 250 & 7.58 & 4.92 & -2.66 & [-5.59, -0.21] \\
 & 6 & 247 & 8.08 & 5.24 & -2.84 & [-5.69, -0.33] \\
 & 12 & 241 & 7.43 & 5.59 & -1.84 & [-3.63, -0.20] \\
\addlinespace[3pt]
\multirow{3}{*}{Chronos Mini} & 3 & 250 & 9.21 & 4.50 & -4.71 & [-8.69, -1.06] \\
 & 6 & 247 & 8.90 & 5.45 & -3.45 & [-6.98, -0.04] \\
 & 12 & 241 & 7.39 & 5.86 & -1.52 & [-3.94, 0.76] \\
\addlinespace[3pt]
\multirow{3}{*}{Chronos Small} & 3 & 250 & 6.27 & 6.42 & 0.15 & [-3.39, 3.69] \\
 & 6 & 247 & 6.60 & 5.93 & -0.67 & [-4.00, 2.71] \\
 & 12 & 241 & 6.22 & 4.91 & -1.31 & [-3.43, 0.75] \\
\addlinespace[3pt]
\multirow{3}{*}{TimesFM 8M} & 3 & 250 & 3.99 & 1.40 & -2.59 & [-6.80, 1.61] \\
 & 6 & 247 & 7.29 & 3.57 & -3.72 & [-7.65, 0.31] \\
 & 12 & 241 & 8.63 & 5.96 & -2.67 & [-6.10, 0.66] \\
\addlinespace[3pt]
\multirow{3}{*}{TimesFM 20M} & 3 & 250 & 4.26 & 1.94 & -2.32 & [-8.85, 3.39] \\
 & 6 & 247 & 7.70 & 2.86 & -4.83 & [-13.09, 0.63] \\
 & 12 & 241 & 7.58 & 4.25 & -3.33 & [-10.51, 1.70] \\
\bottomrule
\end{tabular}
\par\vspace{10pt}\nointerlineskip
\begin{tabular}{@{}>{\raggedright\arraybackslash}p{65pt}>{\centering\arraybackslash}p{16pt}>{\centering\arraybackslash}p{24pt}>{\centering\arraybackslash}p{50pt}>{\centering\arraybackslash}p{57pt}>{\centering\arraybackslash}p{52pt}*{3}{>{\centering\arraybackslash}p{31.6667pt}}@{}}
\toprule
\multicolumn{9}{@{}l}{\textit{Panel B. International markets}} \\
\addlinespace[2pt]
Model & $h$ & $N$ & \shortstack[c]{Med. PIT\\CER} & \shortstack[c]{Med. exposed\\CER} & \shortstack[c]{Med.\\$\Delta CER$} & $>0$ & CI $<0$ & CI $>0$ \\
\midrule
\multirow{3}{*}{Chronos Tiny} & 3 & 250 & 2.35 & 1.05 & -1.68 & 1/13 & 4/13 & 0/13 \\
 & 6 & 247 & 3.05 & 0.88 & -1.46 & 1/13 & 4/13 & 0/13 \\
 & 12 & 241 & 2.60 & 1.81 & -1.46 & 1/13 & 4/13 & 0/13 \\
\addlinespace[3pt]
\multirow{3}{*}{Chronos Mini} & 3 & 250 & 2.88 & 1.02 & -2.01 & 0/13 & 8/13 & 0/13 \\
 & 6 & 247 & 2.55 & 0.84 & -1.70 & 1/13 & 5/13 & 0/13 \\
 & 12 & 241 & 3.13 & 1.80 & -1.28 & 1/13 & 6/13 & 0/13 \\
\addlinespace[3pt]
\multirow{3}{*}{Chronos Small} & 3 & 250 & 1.75 & 1.70 & -0.24 & 6/13 & 2/13 & 0/13 \\
 & 6 & 247 & 2.14 & 2.10 & -1.68 & 1/13 & 0/13 & 0/13 \\
 & 12 & 241 & 2.58 & 1.76 & -1.26 & 1/13 & 3/13 & 0/13 \\
\addlinespace[3pt]
\multirow{3}{*}{TimesFM 8M} & 3 & 250 & 0.84 & -2.79 & -4.00 & 0/13 & 4/13 & 0/13 \\
 & 6 & 247 & 3.86 & -1.37 & -5.22 & 0/13 & 9/13 & 0/13 \\
 & 12 & 241 & 4.53 & 0.34 & -4.92 & 0/13 & 10/13 & 0/13 \\
\addlinespace[3pt]
\multirow{3}{*}{TimesFM 20M} & 3 & 250 & 2.46 & -1.79 & -4.95 & 0/13 & 3/13 & 0/13 \\
 & 6 & 247 & 2.58 & -0.21 & -3.73 & 0/13 & 1/13 & 0/13 \\
 & 12 & 241 & 0.77 & 0.91 & -0.48 & 6/13 & 1/13 & 0/13 \\
\bottomrule
\end{tabular}
\par\vspace{6pt}\nointerlineskip
\exhibitnotes{%
  Longer-horizon portfolios use a complete-tranche implementation at $h=3,6,12$ months. Each forecast-origin tranche invests $1/h$ of wealth for $h$ months, and monthly portfolio returns are evaluated once all $h$ staggered tranches are active. The column $N$ counts these return observations: 250, 247, and 241 for the three horizons. Panel A gives U.S. model estimates with paired 95\% bootstrap intervals; Panel B gives model-specific medians and counts across 13 non-U.S. markets. $\Delta CER=CER_{\mathrm{Exposed}}-CER_{\mathrm{PIT}}$ is in annualized percentage points, with medians calculated from paired market-level differences. Exposed forecasts equally average rolling leads $0,+1,+2$ on strict common support; PIT uses lead $-1$. Specifications use $\gamma=3$, market weights in $[0,1.5]$, and the US training-information environment. Paired intervals use 1,000 circular moving-block replications with block length six. The underlying target-start grid is January 2001--December 2021.
}
\end{minipage}
\end{table}
\clearpage

\section{Forecast Revisions and Error Correction}\label{sec:ia-accounting-results}

This section reports the complete loss-accounting evidence for main-text Sections~\ref{loss-accounting-for-forecast-revisions} and~\ref{alignment-efficiency-and-the-break-even-condition}.
\begin{figure}[H]
\centering
\includegraphics[width=0.95\linewidth]{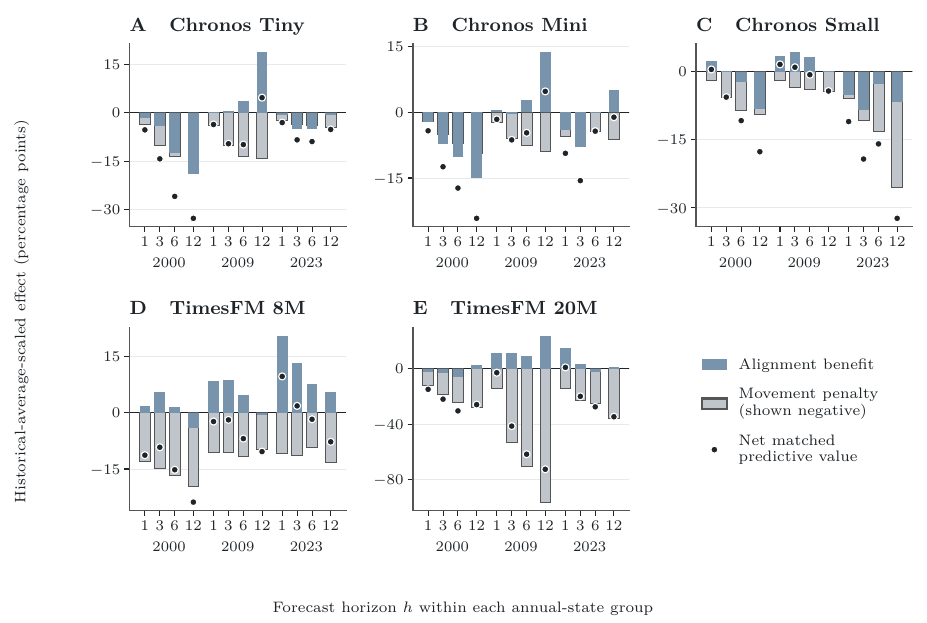}
\caption[Squared-loss decomposition for fixed annual states]{%
  \textbf{Squared-loss decomposition for fixed annual states.}
  Fixed-state matched predictive value is decomposed into alignment benefit and movement penalty relative to origin-aligned point-in-time (PIT) forecasts. Panels cover the five TSFM variants; groups identify states 2000, 2009, and 2023, with horizons $h=1,3,6,12$ months ordered within each group. For revision $D(v)=\widehat y(v)-\widehat y^{\mathrm{PIT}}$, colored bars give $A=100\,2\overline{e^{\mathrm{PIT}}D(v)}/\overline{(e^{\mathrm{HA}})^2}$; light-gray bars give $-P=-100\,\overline{D(v)^2}/\overline{(e^{\mathrm{HA}})^2}$; and black circles give $A-P$. Each state uses its full matched deployment support with PIT, and all components within a comparison use the same target months. Values are historical-average MSFE percentage points; positive net value favors the fixed state. Model panels use separate vertical scales. Estimates use the all-available U.S. equity-premium sample and the US training-information environment.
}
\label{fig:x51-fixed-loss-accounting}

\end{figure}
\FloatBarrier
\par\medskip
\clearpage
\begingroup
\makeatletter
\renewcommand{\footnotesize}{\@setfontsize\footnotesize{9}{10.3}}
\makeatother
\footnotesize
\setlength{\tabcolsep}{4pt}
\renewcommand{\arraystretch}{1.00}
\setlength{\linewidth}{424.64pt}
\begin{longtable}{@{}llccccccc@{}}
\caption[U.S. squared-loss components across annual-state comparisons]{%
  \textbf{U.S. squared-loss components across annual-state comparisons.}
}\label{tab:ia-x51-us-loss-accounting}\\
\multicolumn{9}{@{}p{\linewidth}@{}}{\exhibitnotes{%
  Squared-loss components are calculated for five TSFM variants relative to origin-aligned point-in-time (PIT) forecasts. For alternative-minus-PIT revision $D(v)$, alignment is $A=100\,2\overline{e^{\mathrm{PIT}}D(v)}/\overline{(e^{\mathrm{HA}})^2}$ and movement penalty is $P=100\,\overline{D(v)^2}/\overline{(e^{\mathrm{HA}})^2}$. Net value is $A-P$, with positive values favoring the alternative state. Components are in historical-average MSFE percentage points and use identical support within a comparison. Panel A gives rolling leads $0,+1,+2$ relative to PIT lead $-1$; Panel B gives fixed-2009 and fixed-2023 retrospective-exposure comparisons. The 95\% intervals apply to net value: HAC intervals use Bartlett--Newey--West estimation with $h-1$ lags, and bootstrap intervals use 2,000 circular moving-block replications with block length $\max(h,6)$. The final column counts matched target months; $h$ is in months. Estimates use the all-available U.S. equity-premium sample and the US training-information environment.
}}\\
\noalign{\vskip 6pt}
\toprule
Model & Comparison & $h$ & Alignment & Penalty & Net & \shortstack{HAC 95\%\\interval} & \shortstack{Bootstrap 95\%\\interval} & $N$ \\
\midrule
\endfirsthead
\multicolumn{9}{l}{\textit{\tablename\ \thetable\ (continued); US training-information environment}}\\*
\toprule
Model & Comparison & $h$ & Alignment & Penalty & Net & \shortstack{HAC 95\%\\interval} & \shortstack{Bootstrap 95\%\\interval} & $N$ \\
\midrule
\endhead
\midrule
\multicolumn{9}{r}{\textit{Continued on next page}}\\
\endfoot
\bottomrule
\endlastfoot
\multicolumn{9}{l}{\textit{Panel A: Rolling annual states}}\\*
Chronos Tiny & Lead 0 & 1 & 0.64 & 0.94 & -0.29 & [-2.83, 2.24] & [-2.38, 1.76] & 276 \\*
 & Lead 0 & 3 & -0.37 & 2.29 & -2.66 & [-7.30, 1.98] & [-7.42, 3.46] & 276 \\*
 & Lead 0 & 6 & 0.40 & 3.10 & -2.69 & [-9.22, 3.84] & [-8.73, 5.56] & 276 \\*
 & Lead 0 & 12 & 6.34 & 3.92 & 2.42 & [-5.82, 10.65] & [-5.19, 13.20] & 276 \\
Chronos Tiny & Lead +1 & 1 & -1.53 & 1.66 & -3.19 & [-7.51, 1.13] & [-6.70, 0.16] & 264 \\*
 & Lead +1 & 3 & -4.99 & 2.54 & -7.53 & [-13.07, -1.99] & [-12.83, -0.50] & 264 \\*
 & Lead +1 & 6 & -8.80 & 3.36 & -12.16 & [-19.11, -5.22] & [-17.78, -2.42] & 264 \\*
 & Lead +1 & 12 & -3.92 & 4.31 & -8.23 & [-18.70, 2.24] & [-18.78, 4.45] & 264 \\
Chronos Tiny & Lead +2 & 1 & -0.14 & 1.27 & -1.40 & [-4.87, 2.07] & [-5.10, 1.32] & 252 \\*
 & Lead +2 & 3 & -3.22 & 2.82 & -6.03 & [-12.79, 0.73] & [-14.19, 0.33] & 252 \\*
 & Lead +2 & 6 & -6.73 & 3.80 & -10.53 & [-20.08, -0.98] & [-21.23, -2.69] & 252 \\*
 & Lead +2 & 12 & -5.56 & 4.89 & -10.45 & [-22.67, 1.78] & [-24.09, -0.00] & 252 \\
\addlinespace[0pt]
Chronos Mini & Lead 0 & 1 & -1.30 & 1.31 & -2.61 & [-5.70, 0.47] & [-5.48, 0.33] & 276 \\*
 & Lead 0 & 3 & -2.02 & 2.75 & -4.77 & [-10.24, 0.71] & [-10.94, 0.71] & 276 \\*
 & Lead 0 & 6 & 0.62 & 3.60 & -2.99 & [-10.48, 4.50] & [-11.07, 4.92] & 276 \\*
 & Lead 0 & 12 & 10.98 & 5.53 & 5.45 & [-6.78, 17.67] & [-6.52, 21.02] & 276 \\
Chronos Mini & Lead +1 & 1 & -1.47 & 1.46 & -2.93 & [-6.31, 0.44] & [-6.15, 0.04] & 264 \\*
 & Lead +1 & 3 & -4.45 & 3.16 & -7.62 & [-13.12, -2.11] & [-13.96, -1.39] & 264 \\*
 & Lead +1 & 6 & -5.19 & 3.46 & -8.65 & [-16.19, -1.11] & [-16.88, -1.03] & 264 \\*
 & Lead +1 & 12 & 3.38 & 5.06 & -1.68 & [-12.49, 9.12] & [-12.76, 11.72] & 264 \\
Chronos Mini & Lead +2 & 1 & -1.30 & 1.46 & -2.77 & [-6.44, 0.91] & [-6.28, 0.64] & 252 \\*
 & Lead +2 & 3 & -3.87 & 3.09 & -6.96 & [-12.74, -1.19] & [-13.39, -0.21] & 252 \\*
 & Lead +2 & 6 & -3.88 & 4.55 & -8.43 & [-17.22, 0.35] & [-16.88, 2.07] & 252 \\*
 & Lead +2 & 12 & 3.80 & 7.11 & -3.32 & [-19.42, 12.79] & [-17.79, 14.97] & 252 \\
\addlinespace[0pt]
Chronos Small & Lead 0 & 1 & 4.38 & 1.65 & 2.72 & [-1.35, 6.79] & [-0.49, 5.78] & 276 \\*
 & Lead 0 & 3 & 0.87 & 4.41 & -3.54 & [-10.81, 3.72] & [-11.19, 3.15] & 276 \\*
 & Lead 0 & 6 & 0.44 & 5.90 & -5.46 & [-13.88, 2.96] & [-16.00, 2.20] & 276 \\*
 & Lead 0 & 12 & -4.84 & 7.15 & -12.00 & [-27.65, 3.65] & [-31.28, 2.05] & 276 \\
Chronos Small & Lead +1 & 1 & 6.39 & 2.14 & 4.24 & [-0.77, 9.26] & [-1.29, 9.34] & 264 \\*
 & Lead +1 & 3 & 1.01 & 5.33 & -4.33 & [-13.95, 5.29] & [-15.69, 4.92] & 264 \\*
 & Lead +1 & 6 & -6.71 & 9.45 & -16.17 & [-36.73, 4.40] & [-40.52, 1.08] & 264 \\*
 & Lead +1 & 12 & -7.76 & 11.76 & -19.52 & [-39.73, 0.69] & [-46.52, -1.97] & 264 \\
Chronos Small & Lead +2 & 1 & 3.83 & 1.88 & 1.95 & [-2.94, 6.85] & [-3.46, 7.68] & 252 \\*
 & Lead +2 & 3 & 0.32 & 3.45 & -3.13 & [-12.68, 6.42] & [-14.77, 6.91] & 252 \\*
 & Lead +2 & 6 & -7.10 & 9.06 & -16.16 & [-43.08, 10.77] & [-48.36, 5.22] & 252 \\*
 & Lead +2 & 12 & -6.94 & 8.24 & -15.18 & [-36.64, 6.28] & [-40.52, 1.75] & 252 \\
\addlinespace[0pt]
TimesFM 8M & Lead 0 & 1 & 10.18 & 14.91 & -4.72 & [-15.12, 5.68] & [-15.00, 4.40] & 276 \\*
 & Lead 0 & 3 & 4.73 & 11.32 & -6.59 & [-17.98, 4.80] & [-17.79, 7.24] & 276 \\*
 & Lead 0 & 6 & -2.09 & 11.27 & -13.36 & [-27.84, 1.11] & [-27.40, 4.50] & 276 \\*
 & Lead 0 & 12 & -1.72 & 10.45 & -12.17 & [-24.35, 0.00] & [-22.79, 3.56] & 276 \\
TimesFM 8M & Lead +1 & 1 & 16.87 & 13.67 & 3.20 & [-5.75, 12.14] & [-5.59, 12.18] & 264 \\*
 & Lead +1 & 3 & 15.25 & 13.18 & 2.07 & [-10.96, 15.10] & [-14.41, 16.09] & 264 \\*
 & Lead +1 & 6 & 13.37 & 15.04 & -1.68 & [-18.95, 15.60] & [-22.26, 14.22] & 264 \\*
 & Lead +1 & 12 & 7.19 & 15.98 & -8.79 & [-24.36, 6.77] & [-25.62, 9.79] & 264 \\
TimesFM 8M & Lead +2 & 1 & 5.61 & 10.54 & -4.94 & [-13.30, 3.43] & [-12.35, 2.45] & 252 \\*
 & Lead +2 & 3 & 1.66 & 11.93 & -10.27 & [-22.56, 2.02] & [-21.67, 6.02] & 252 \\*
 & Lead +2 & 6 & -4.25 & 14.23 & -18.48 & [-35.45, -1.52] & [-34.81, 2.71] & 252 \\*
 & Lead +2 & 12 & -6.52 & 16.02 & -22.54 & [-38.24, -6.84] & [-38.07, -2.60] & 252 \\
\addlinespace[0pt]
TimesFM 20M & Lead 0 & 1 & 6.20 & 10.02 & -3.83 & [-11.82, 4.17] & [-11.83, 3.86] & 276 \\*
 & Lead 0 & 3 & 2.88 & 15.80 & -12.92 & [-27.20, 1.36] & [-27.62, 4.02] & 276 \\*
 & Lead 0 & 6 & -1.31 & 18.79 & -20.10 & [-39.95, -0.25] & [-38.41, 2.58] & 276 \\*
 & Lead 0 & 12 & 14.10 & 27.06 & -12.96 & [-36.64, 10.73] & [-39.19, 7.01] & 276 \\
TimesFM 20M & Lead +1 & 1 & 1.61 & 8.56 & -6.95 & [-15.50, 1.60] & [-15.75, 0.96] & 264 \\*
 & Lead +1 & 3 & -12.57 & 12.77 & -25.34 & [-40.49, -10.18] & [-42.26, -6.90] & 264 \\*
 & Lead +1 & 6 & -20.93 & 13.58 & -34.51 & [-56.62, -12.41] & [-53.28, -9.32] & 264 \\*
 & Lead +1 & 12 & -17.54 & 19.45 & -36.99 & [-63.45, -10.53] & [-60.06, -9.78] & 264 \\
TimesFM 20M & Lead +2 & 1 & 4.61 & 7.48 & -2.87 & [-10.60, 4.86] & [-9.83, 4.70] & 252 \\*
 & Lead +2 & 3 & -0.35 & 11.67 & -12.02 & [-22.21, -1.84] & [-23.04, -1.08] & 252 \\*
 & Lead +2 & 6 & -7.10 & 12.84 & -19.94 & [-32.89, -6.98] & [-34.66, -5.70] & 252 \\*
 & Lead +2 & 12 & -3.00 & 17.53 & -20.54 & [-44.24, 3.17] & [-43.95, 4.92] & 252 \\
\addlinespace[4pt]
\multicolumn{9}{l}{\textit{Panel B: Fixed annual states - retrospective-exposure windows}}\\*
Chronos Tiny & Fixed 2009 & 1 & -1.42 & 4.62 & -6.05 & [-14.10, 2.01] & [-12.51, 2.26] & 108 \\*
 & Fixed 2009 & 3 & -7.99 & 8.16 & -16.15 & [-27.91, -4.39] & [-28.83, -0.80] & 106 \\*
 & Fixed 2009 & 6 & -10.20 & 8.77 & -18.97 & [-32.38, -5.55] & [-34.06, -1.19] & 103 \\*
 & Fixed 2009 & 12 & 1.19 & 9.87 & -8.69 & [-33.65, 16.28] & [-35.00, 42.89] & 97 \\
Chronos Tiny & Fixed 2023 & 1 & -0.64 & 2.51 & -3.14 & [-8.77, 2.48] & [-7.19, 1.54] & 276 \\*
 & Fixed 2023 & 3 & -5.29 & 3.53 & -8.82 & [-15.01, -2.63] & [-14.38, -0.89] & 274 \\*
 & Fixed 2023 & 6 & -5.58 & 4.07 & -9.64 & [-16.36, -2.93] & [-15.30, -0.91] & 271 \\*
 & Fixed 2023 & 12 & -2.52 & 4.52 & -7.04 & [-16.46, 2.39] & [-15.99, 5.26] & 265 \\
\addlinespace[0pt]
Chronos Mini & Fixed 2009 & 1 & -2.68 & 2.98 & -5.66 & [-13.09, 1.76] & [-11.95, 1.87] & 108 \\*
 & Fixed 2009 & 3 & -7.61 & 4.00 & -11.61 & [-21.50, -1.72] & [-24.14, -0.19] & 106 \\*
 & Fixed 2009 & 6 & -7.51 & 4.43 & -11.95 & [-26.03, 2.13] & [-30.19, 2.77] & 103 \\*
 & Fixed 2009 & 12 & -4.46 & 5.22 & -9.68 & [-29.56, 10.20] & [-34.07, 21.02] & 97 \\
Chronos Mini & Fixed 2023 & 1 & -4.00 & 5.52 & -9.52 & [-20.36, 1.31] & [-25.98, 1.94] & 276 \\*
 & Fixed 2023 & 3 & -7.94 & 7.91 & -15.85 & [-36.52, 4.82] & [-36.13, 4.40] & 274 \\*
 & Fixed 2023 & 6 & 0.36 & 4.46 & -4.11 & [-16.15, 7.93] & [-14.02, 10.83] & 271 \\*
 & Fixed 2023 & 12 & 5.97 & 6.38 & -0.41 & [-14.58, 13.75] & [-16.71, 14.60] & 265 \\
\addlinespace[0pt]
Chronos Small & Fixed 2009 & 1 & 6.15 & 2.25 & 3.90 & [-3.91, 11.71] & [-3.93, 10.70] & 108 \\*
 & Fixed 2009 & 3 & 4.38 & 3.43 & 0.95 & [-9.59, 11.50] & [-10.46, 10.21] & 106 \\*
 & Fixed 2009 & 6 & 3.04 & 3.60 & -0.56 & [-7.28, 6.16] & [-8.78, 6.82] & 103 \\*
 & Fixed 2009 & 12 & 5.12 & 4.45 & 0.67 & [-12.08, 13.43] & [-10.78, 31.49] & 97 \\
Chronos Small & Fixed 2023 & 1 & -5.19 & 6.05 & -11.24 & [-22.61, 0.14] & [-28.68, 3.41] & 276 \\*
 & Fixed 2023 & 3 & -8.45 & 10.86 & -19.31 & [-36.13, -2.49] & [-35.05, -3.73] & 274 \\*
 & Fixed 2023 & 6 & -2.25 & 13.43 & -15.67 & [-30.52, -0.83] & [-32.94, -0.84] & 271 \\*
 & Fixed 2023 & 12 & -5.13 & 25.67 & -30.80 & [-67.01, 5.41] & [-81.53, -3.56] & 265 \\
\addlinespace[0pt]
TimesFM 8M & Fixed 2009 & 1 & 18.86 & 11.01 & 7.85 & [-7.09, 22.79] & [-12.19, 22.85] & 108 \\*
 & Fixed 2009 & 3 & 15.12 & 7.23 & 7.90 & [-6.57, 22.36] & [-13.05, 19.94] & 106 \\*
 & Fixed 2009 & 6 & 7.20 & 6.86 & 0.34 & [-15.52, 16.19] & [-24.76, 12.36] & 103 \\*
 & Fixed 2009 & 12 & -3.93 & 5.45 & -9.38 & [-22.98, 4.22] & [-34.65, 0.07] & 97 \\
TimesFM 8M & Fixed 2023 & 1 & 20.87 & 11.06 & 9.81 & [-1.10, 20.73] & [-0.00, 21.02] & 276 \\*
 & Fixed 2023 & 3 & 11.84 & 11.40 & 0.44 & [-12.58, 13.47] & [-14.37, 12.92] & 274 \\*
 & Fixed 2023 & 6 & 3.73 & 8.44 & -4.71 & [-15.80, 6.37] & [-17.89, 5.93] & 271 \\*
 & Fixed 2023 & 12 & 1.70 & 12.41 & -10.71 & [-21.92, 0.50] & [-24.89, 1.00] & 265 \\
\addlinespace[0pt]
TimesFM 20M & Fixed 2009 & 1 & 15.86 & 17.29 & -1.43 & [-17.76, 14.90] & [-17.62, 15.49] & 108 \\*
 & Fixed 2009 & 3 & -9.43 & 47.54 & -56.98 & [-86.86, -27.09] & [-94.11, -18.84] & 106 \\*
 & Fixed 2009 & 6 & -19.19 & 59.50 & -78.69 & [-120.40, -36.97] & [-136.57, -35.74] & 103 \\*
 & Fixed 2009 & 12 & -17.25 & 89.00 & -106.24 & [-150.11, -62.37] & [-176.64, -42.94] & 97 \\
TimesFM 20M & Fixed 2023 & 1 & 15.38 & 14.55 & 0.83 & [-9.53, 11.20] & [-8.91, 10.05] & 276 \\*
 & Fixed 2023 & 3 & 4.45 & 23.40 & -18.96 & [-33.06, -4.85] & [-38.45, -7.27] & 274 \\*
 & Fixed 2023 & 6 & 0.03 & 23.49 & -23.46 & [-43.16, -3.76] & [-49.15, -8.33] & 271 \\*
 & Fixed 2023 & 12 & 4.70 & 32.13 & -27.43 & [-57.26, 2.40] & [-72.44, -6.28] & 265 \\
\end{longtable}

\endgroup


\clearpage
\begingroup
\makeatletter
\renewcommand{\footnotesize}{\@setfontsize\footnotesize{9}{10.3}}
\makeatother
\footnotesize
\setlength{\tabcolsep}{4pt}
\renewcommand{\arraystretch}{1.00}
\setlength{\linewidth}{369.31pt}
\begin{longtable}{@{}llcccccc@{}}
\caption[International squared-loss component distributions]{%
  \textbf{International squared-loss component distributions.}
}\label{tab:ia-x51-international-loss-accounting}\\[-10pt]
\multicolumn{8}{@{}p{\linewidth}@{}}{\exhibitnotes{%
  Cross-market medians summarize squared-loss components for the 13 non-U.S. markets. Alignment benefit, movement penalty, and net matched value are first calculated within each market relative to origin-aligned point-in-time (PIT) forecasts on identical matched target months. Each component is summarized separately; median net value is the median of market-level alignment minus penalty, not the difference between component medians. Net IQR spans the 25th--75th percentiles of market-level net effects, and Positive markets counts positive net effects. Panel A gives rolling leads $0,+1,+2$ relative to PIT lead $-1$; Panel B gives fixed-2009 and fixed-2023 retrospective-exposure comparisons. Values are in historical-average MSFE percentage points, and $h$ is in months. Estimates use the all-available sample and the US training-information environment.
}}\\
\noalign{\vskip 4pt}
\toprule
Model & Comparison & $h$ & \shortstack{Median\\alignment} & \shortstack{Median\\penalty} & \shortstack{Median\\net} & Net IQR & \shortstack{Positive\\markets} \\
\midrule
\endfirsthead
\multicolumn{8}{l}{\textit{\tablename\ \thetable\ (continued); US training-information environment}}\\*
\toprule
Model & Comparison & $h$ & \shortstack{Median\\alignment} & \shortstack{Median\\penalty} & \shortstack{Median\\net} & Net IQR & \shortstack{Positive\\markets} \\
\midrule
\endhead
\midrule
\multicolumn{8}{r}{\textit{Continued on next page}}\\
\endfoot
\bottomrule
\endlastfoot
\multicolumn{8}{l}{\textit{Panel A: Rolling annual states}}\\*
Chronos Tiny & Lead 0 & 1 & -0.45 & 0.74 & -0.99 & [-2.03, -0.52] & 1/13 \\*
 & Lead 0 & 3 & -1.71 & 1.16 & -2.78 & [-3.75, -1.20] & 0/13 \\*
 & Lead 0 & 6 & -2.79 & 1.42 & -4.00 & [-5.86, -2.27] & 1/13 \\*
 & Lead 0 & 12 & -1.48 & 1.81 & -2.82 & [-6.19, -1.19] & 2/13 \\
Chronos Tiny & Lead +1 & 1 & -1.88 & 0.71 & -2.48 & [-3.10, -1.33] & 2/13 \\*
 & Lead +1 & 3 & -3.01 & 1.64 & -4.44 & [-5.15, -3.85] & 2/13 \\*
 & Lead +1 & 6 & -5.06 & 2.11 & -7.54 & [-9.08, -3.73] & 2/13 \\*
 & Lead +1 & 12 & -2.32 & 2.28 & -6.03 & [-11.12, -1.16] & 2/13 \\
Chronos Tiny & Lead +2 & 1 & -1.00 & 0.97 & -1.95 & [-2.55, -0.41] & 3/13 \\*
 & Lead +2 & 3 & -1.37 & 1.95 & -2.72 & [-5.26, -1.90] & 2/13 \\*
 & Lead +2 & 6 & -3.33 & 2.49 & -5.78 & [-6.77, -2.36] & 1/13 \\*
 & Lead +2 & 12 & -3.92 & 2.68 & -7.03 & [-8.98, -2.22] & 2/13 \\
\addlinespace[2pt]
Chronos Mini & Lead 0 & 1 & -1.49 & 1.03 & -2.47 & [-4.20, -0.99] & 1/13 \\*
 & Lead 0 & 3 & -1.77 & 1.67 & -3.73 & [-5.75, -2.09] & 2/13 \\*
 & Lead 0 & 6 & -1.44 & 2.06 & -3.49 & [-6.17, -2.45] & 1/13 \\*
 & Lead 0 & 12 & -0.86 & 2.71 & -2.67 & [-4.94, -0.85] & 3/13 \\
Chronos Mini & Lead +1 & 1 & -1.87 & 0.85 & -2.58 & [-4.27, -1.18] & 0/13 \\*
 & Lead +1 & 3 & -2.72 & 1.89 & -4.88 & [-5.32, -2.84] & 1/13 \\*
 & Lead +1 & 6 & -3.18 & 2.51 & -5.37 & [-7.10, -0.81] & 2/13 \\*
 & Lead +1 & 12 & -3.27 & 3.02 & -6.24 & [-9.65, 0.28] & 4/13 \\
Chronos Mini & Lead +2 & 1 & -1.23 & 1.49 & -2.75 & [-4.28, -1.86] & 0/13 \\*
 & Lead +2 & 3 & -3.51 & 2.36 & -6.58 & [-7.49, -1.93] & 0/13 \\*
 & Lead +2 & 6 & -3.31 & 3.31 & -6.41 & [-8.47, -2.19] & 0/13 \\*
 & Lead +2 & 12 & -3.43 & 4.68 & -5.92 & [-9.22, -3.78] & 1/13 \\
\addlinespace[2pt]
Chronos Small & Lead 0 & 1 & 1.00 & 1.32 & -0.49 & [-1.87, 0.44] & 6/13 \\*
 & Lead 0 & 3 & 2.04 & 3.14 & -0.20 & [-3.63, 3.18] & 6/13 \\*
 & Lead 0 & 6 & 0.01 & 3.77 & -3.59 & [-4.99, -1.28] & 3/13 \\*
 & Lead 0 & 12 & -0.44 & 4.96 & -6.60 & [-8.74, -3.46] & 2/13 \\
Chronos Small & Lead +1 & 1 & 2.50 & 2.00 & 0.32 & [-0.71, 2.96] & 7/13 \\*
 & Lead +1 & 3 & 0.89 & 4.23 & -1.69 & [-6.45, 0.36] & 5/13 \\*
 & Lead +1 & 6 & -1.65 & 5.94 & -5.55 & [-8.41, -1.00] & 3/13 \\*
 & Lead +1 & 12 & -3.96 & 6.51 & -6.00 & [-14.61, 1.93] & 4/13 \\
Chronos Small & Lead +2 & 1 & 2.62 & 2.06 & 0.47 & [-0.84, 1.95] & 8/13 \\*
 & Lead +2 & 3 & 6.19 & 4.07 & 1.41 & [-1.84, 4.78] & 8/13 \\*
 & Lead +2 & 6 & 0.80 & 5.89 & -3.36 & [-12.92, -0.34] & 3/13 \\*
 & Lead +2 & 12 & -5.71 & 7.11 & -10.14 & [-14.46, -5.23] & 2/13 \\
\addlinespace[2pt]
TimesFM 8M & Lead 0 & 1 & 23.69 & 34.53 & -6.13 & [-12.19, -2.76] & 3/13 \\*
 & Lead 0 & 3 & 7.39 & 27.76 & -15.68 & [-21.78, -9.56] & 0/13 \\*
 & Lead 0 & 6 & 0.52 & 28.00 & -24.22 & [-27.58, -14.83] & 0/13 \\*
 & Lead 0 & 12 & 1.99 & 25.13 & -22.67 & [-25.81, -14.79] & 0/13 \\
TimesFM 8M & Lead +1 & 1 & 29.30 & 24.67 & 1.51 & [-2.48, 5.71] & 9/13 \\*
 & Lead +1 & 3 & 16.66 & 25.30 & -3.56 & [-12.98, 0.56] & 4/13 \\*
 & Lead +1 & 6 & 12.50 & 33.29 & -9.98 & [-21.36, -5.03] & 0/13 \\*
 & Lead +1 & 12 & 2.54 & 35.33 & -28.29 & [-36.18, -19.26] & 0/13 \\
TimesFM 8M & Lead +2 & 1 & 12.09 & 19.41 & -8.55 & [-11.49, -5.90] & 2/13 \\*
 & Lead +2 & 3 & 6.12 & 22.57 & -13.81 & [-21.84, -11.53] & 0/13 \\*
 & Lead +2 & 6 & 0.49 & 28.27 & -25.62 & [-37.64, -17.76] & 0/13 \\*
 & Lead +2 & 12 & -6.00 & 33.88 & -36.55 & [-48.98, -26.13] & 0/13 \\
\addlinespace[2pt]
TimesFM 20M & Lead 0 & 1 & 18.62 & 21.84 & -2.92 & [-3.94, 1.31] & 6/13 \\*
 & Lead 0 & 3 & 23.12 & 35.70 & -8.58 & [-14.73, -5.92] & 0/13 \\*
 & Lead 0 & 6 & 29.72 & 41.97 & -14.35 & [-17.50, -7.65] & 0/13 \\*
 & Lead 0 & 12 & 71.49 & 61.45 & 8.81 & [3.00, 15.39] & 11/13 \\
TimesFM 20M & Lead +1 & 1 & 11.42 & 18.74 & -7.57 & [-9.24, -2.32] & 0/13 \\*
 & Lead +1 & 3 & -0.80 & 31.21 & -25.44 & [-31.17, -21.77] & 0/13 \\*
 & Lead +1 & 6 & -5.73 & 33.42 & -35.93 & [-42.60, -31.53] & 0/13 \\*
 & Lead +1 & 12 & 20.18 & 44.73 & -28.93 & [-34.15, -21.55] & 0/13 \\
TimesFM 20M & Lead +2 & 1 & 9.25 & 19.91 & -8.61 & [-12.94, -5.09] & 0/13 \\*
 & Lead +2 & 3 & 10.74 & 35.01 & -15.43 & [-25.27, -12.28] & 0/13 \\*
 & Lead +2 & 6 & 9.78 & 35.56 & -16.08 & [-25.78, -12.69] & 0/13 \\*
 & Lead +2 & 12 & 34.68 & 49.33 & -6.35 & [-13.32, 0.18] & 4/13 \\
\addlinespace[4pt]
\multicolumn{8}{l}{\textit{Panel B: Fixed annual states - retrospective-exposure windows}}\\*
Chronos Tiny & Fixed 2009 & 1 & -2.01 & 1.26 & -3.20 & [-5.18, -2.41] & 2/13 \\*
 & Fixed 2009 & 3 & -5.25 & 2.06 & -8.22 & [-9.71, -2.25] & 1/13 \\*
 & Fixed 2009 & 6 & -7.95 & 2.05 & -10.14 & [-13.37, -3.66] & 1/13 \\*
 & Fixed 2009 & 12 & -4.35 & 1.88 & -8.24 & [-8.51, -0.86] & 3/13 \\
Chronos Tiny & Fixed 2023 & 1 & -1.10 & 1.66 & -2.37 & [-4.02, -0.74] & 1/13 \\*
 & Fixed 2023 & 3 & -1.09 & 2.39 & -3.86 & [-5.69, -1.77] & 1/13 \\*
 & Fixed 2023 & 6 & -2.51 & 2.66 & -6.30 & [-9.05, -3.83] & 1/13 \\*
 & Fixed 2023 & 12 & -3.44 & 2.53 & -6.07 & [-8.97, -2.74] & 3/13 \\
\addlinespace[2pt]
Chronos Mini & Fixed 2009 & 1 & -2.46 & 1.24 & -3.53 & [-4.49, -1.54] & 1/13 \\*
 & Fixed 2009 & 3 & -3.80 & 1.27 & -4.86 & [-5.71, -2.74] & 2/13 \\*
 & Fixed 2009 & 6 & -3.44 & 1.28 & -4.72 & [-6.10, -2.63] & 2/13 \\*
 & Fixed 2009 & 12 & -3.22 & 1.87 & -4.43 & [-8.44, -2.60] & 2/13 \\
Chronos Mini & Fixed 2023 & 1 & -3.17 & 1.68 & -4.32 & [-6.60, -1.98] & 1/13 \\*
 & Fixed 2023 & 3 & -3.52 & 2.39 & -6.86 & [-9.88, -3.09] & 1/13 \\*
 & Fixed 2023 & 6 & -1.92 & 2.68 & -3.16 & [-7.77, 0.62] & 4/13 \\*
 & Fixed 2023 & 12 & -1.49 & 3.61 & -4.04 & [-6.48, -0.42] & 3/13 \\
\addlinespace[2pt]
Chronos Small & Fixed 2009 & 1 & 4.13 & 2.32 & 2.89 & [-0.59, 5.22] & 8/13 \\*
 & Fixed 2009 & 3 & 6.76 & 3.47 & 3.31 & [-2.52, 5.60] & 9/13 \\*
 & Fixed 2009 & 6 & 0.25 & 3.63 & -3.11 & [-4.39, -0.58] & 3/13 \\*
 & Fixed 2009 & 12 & -3.09 & 3.35 & -7.18 & [-8.15, -0.06] & 3/13 \\
Chronos Small & Fixed 2023 & 1 & 0.97 & 1.63 & -0.08 & [-1.98, 0.14] & 5/13 \\*
 & Fixed 2023 & 3 & 2.50 & 4.10 & -2.21 & [-3.48, 0.46] & 4/13 \\*
 & Fixed 2023 & 6 & 1.96 & 4.76 & -3.09 & [-5.85, 0.04] & 4/13 \\*
 & Fixed 2023 & 12 & 1.69 & 4.97 & -3.10 & [-6.26, 0.85] & 4/13 \\
\addlinespace[2pt]
TimesFM 8M & Fixed 2009 & 1 & 15.20 & 22.20 & -6.72 & [-12.95, 2.22] & 6/13 \\*
 & Fixed 2009 & 3 & 4.49 & 16.98 & -10.82 & [-18.11, 1.85] & 4/13 \\*
 & Fixed 2009 & 6 & -1.78 & 14.59 & -17.11 & [-21.24, -6.64] & 2/13 \\*
 & Fixed 2009 & 12 & -10.60 & 9.06 & -18.01 & [-26.96, -14.97] & 0/13 \\
TimesFM 8M & Fixed 2023 & 1 & 39.41 & 18.33 & 19.94 & [16.87, 25.36] & 13/13 \\*
 & Fixed 2023 & 3 & 12.01 & 15.72 & 0.97 & [0.25, 4.23] & 10/13 \\*
 & Fixed 2023 & 6 & 7.94 & 12.70 & -2.38 & [-7.79, 0.34] & 4/13 \\*
 & Fixed 2023 & 12 & 5.17 & 15.18 & -8.07 & [-13.68, -2.99] & 3/13 \\
\addlinespace[2pt]
TimesFM 20M & Fixed 2009 & 1 & 14.66 & 27.94 & -9.96 & [-19.48, -2.50] & 1/13 \\*
 & Fixed 2009 & 3 & 10.43 & 94.11 & -74.88 & [-92.83, -55.93] & 1/13 \\*
 & Fixed 2009 & 6 & 12.16 & 114.12 & -87.81 & [-118.03, -76.38] & 0/13 \\*
 & Fixed 2009 & 12 & 43.11 & 160.88 & -113.47 & [-121.57, -86.19] & 1/13 \\
TimesFM 20M & Fixed 2023 & 1 & 19.98 & 25.55 & -5.23 & [-9.45, -3.16] & 2/13 \\*
 & Fixed 2023 & 3 & 12.45 & 65.91 & -53.75 & [-70.80, -32.09] & 0/13 \\*
 & Fixed 2023 & 6 & 12.78 & 67.57 & -59.30 & [-76.41, -29.49] & 0/13 \\*
 & Fixed 2023 & 12 & 32.13 & 75.36 & -47.12 & [-61.85, -23.76] & 1/13 \\
\end{longtable}

\endgroup

\clearpage

\section{Training Information and Predictive Value}\label{sec:ia-training-results}

These exhibits extend main-text Section~\ref{training-information-and-revision-to-value-accounting} by reporting component levels and changes across training samples.
\begin{figure}[H]
\centering
\includegraphics[width=0.95\linewidth]{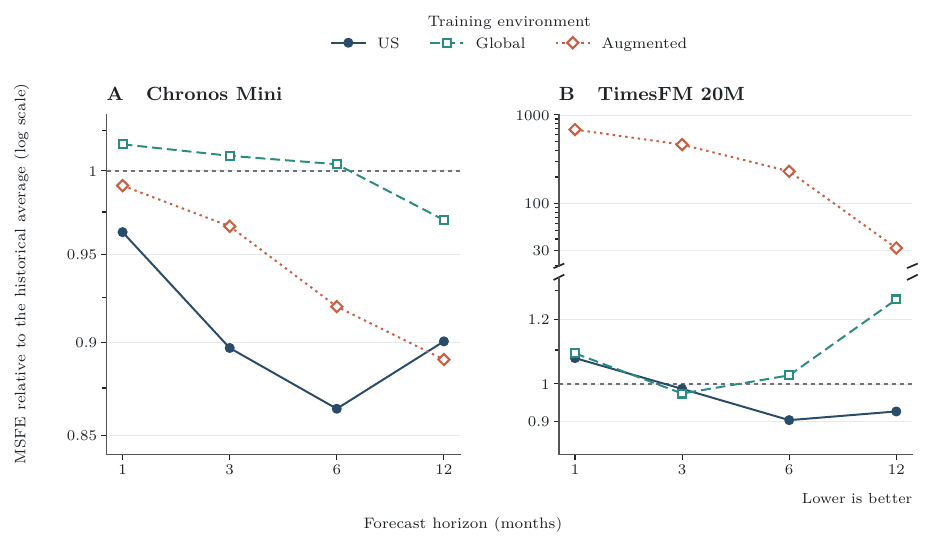}
\caption[Point-in-time relative MSFE across training environments]{%
  \textbf{Point-in-time relative MSFE across training environments.}
  Relative PIT mean squared forecast error is $Q_{\mathrm{PIT}}=\mathrm{MSFE}_{\mathrm{TSFM}}/\mathrm{MSFE}_{\mathrm{HA}}=1-R^2_{\mathrm{HA}}/100$, where HA is the expanding historical-average forecast. Values below one favor the origin-aligned point-in-time (PIT) forecast. Panels A and B show Chronos Mini and TimesFM 20M under the US, Global, and Augmented training-information environments at horizons $h=1,3,6,12$ months. Panel A has a logarithmic vertical scale; Panel B has a broken logarithmic vertical axis with separate lower and upper segments. Estimates use the all-available U.S. equity-premium sample. Table IA.2.1 gives the corresponding raw $R^2_{\mathrm{HA}}$ estimates and one-sided Clark--West inference for all five models.
}
\label{fig:x52-absolute-training-performance}

\end{figure}
\FloatBarrier

\clearpage
\begin{figure}[!htbp]
\centering
\includegraphics[width=0.95\textwidth]{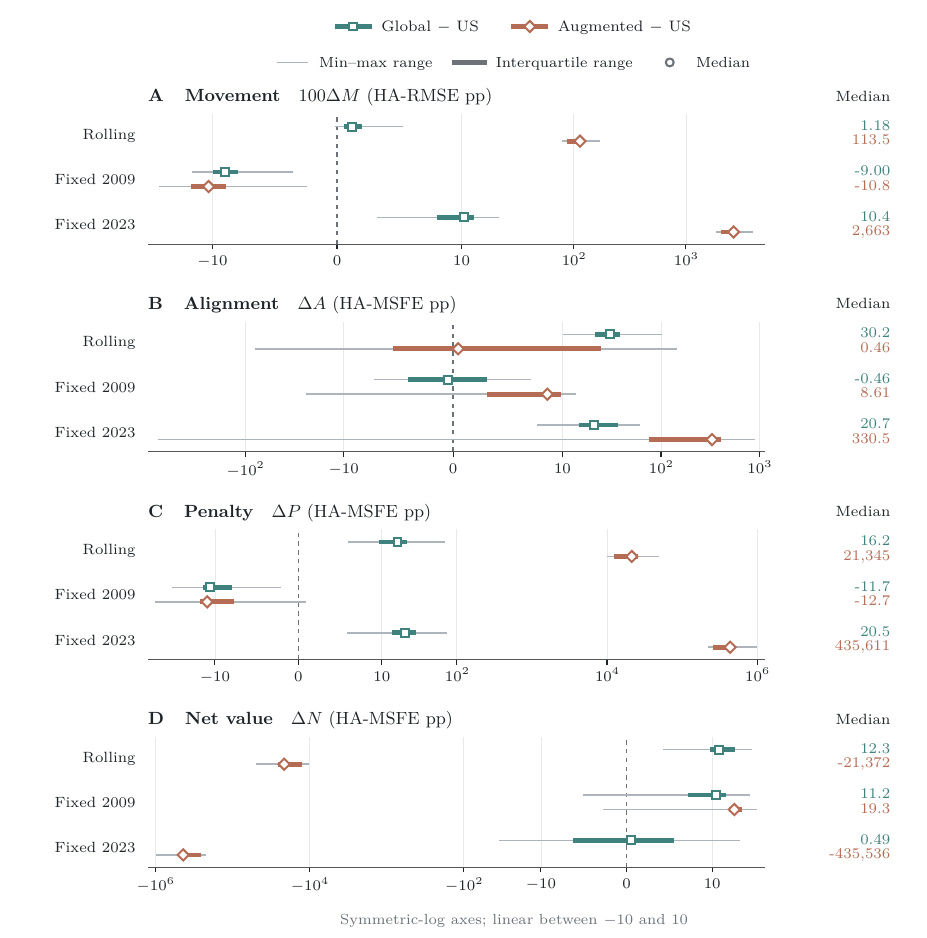}
\caption[International revision-component differences across training environments]{%
  \textbf{International revision-component differences across training environments.}
  International component changes compare Global and Augmented training with the US environment: each contrast is $X^{(c)}-X^{(\mathrm{US})}$. Panels A--D show movement, alignment, penalty, and net-value differences. Movement is in historical-average RMSE percentage points; the other components are in historical-average MSFE percentage points. Within each of the 13 non-U.S. markets, differences are equal-cell averages across the five model variants and the relevant horizons and annual-state comparisons. Thin gray lines span the minimum and maximum market averages; thick colored segments show interquartile ranges. Hollow squares and diamonds mark medians for Global--US and Augmented--US, respectively; the right column reports these medians. Rolling summaries average leads $0,+1,+2$; fixed summaries use the retrospective-exposure windows of states 2009 and 2023. Each panel has its own symmetric-log horizontal scale, with a linear region from $-10$ to $10$. Zero indicates no component change relative to US training.
}
\label{fig:ia-x52-international-corpus-shifts}

\end{figure}

\clearpage
\begin{table}[H]
\centering
\caption[U.S. revision-component differences across training environments]{%
  \textbf{U.S. revision-component differences across training environments.}
}
\label{tab:x52-training-environments}
\fontsize{10}{11.5}\selectfont
\setlength{\tabcolsep}{5.0pt}
\renewcommand{\arraystretch}{1.00}
\begin{threeparttable}
\begin{tabular*}{\linewidth}{@{\extracolsep{\fill}}llrrrr@{}}
\toprule
Model & Training contrast & $100\Delta M$ & $\Delta A$ & $\Delta P$ & $\Delta N$ \\
\midrule
\multicolumn{6}{l}{\textit{Panel A: Rolling post-origin annual states}}\\*
Chronos Tiny & Global $-$ US & 3.11 & 8.59 & 3.12 & 5.47 \\*
 & Augmented $-$ US & -1.49 & 6.65 & -0.93 & 7.58 \\
Chronos Mini & Global $-$ US & 2.87 & 11.05 & 8.54 & 2.50 \\*
 & Augmented $-$ US & -4.16 & -1.16 & 1.03 & -2.18 \\
Chronos Small & Global $-$ US & 4.31 & 15.99 & 5.63 & 10.36 \\*
 & Augmented $-$ US & -4.90 & -0.47 & -1.17 & 0.70 \\
TimesFM 8M & Global $-$ US & 3.92 & 60.22 & 32.95 & 27.27 \\*
 & Augmented $-$ US & -11.54 & 6.88 & -6.58 & 13.46 \\
TimesFM 20M & Global $-$ US & 6.86 & 25.67 & 9.74 & 15.94 \\*
 & Augmented $-$ US & 428.95 & 51.57 & $5.16\!\times\!10^{4}$ & $-5.15\!\times\!10^{4}$ \\
\addlinespace[3pt]
\multicolumn{6}{l}{\textit{Panel B: Fixed 2009 retrospective-exposure comparison}}\\*
Chronos Tiny & Global $-$ US & -10.88 & 7.89 & -3.22 & 11.11 \\*
 & Augmented $-$ US & -12.98 & 14.93 & -6.09 & 21.02 \\
Chronos Mini & Global $-$ US & -0.31 & 14.92 & 3.81 & 11.10 \\*
 & Augmented $-$ US & -6.74 & 5.44 & -3.07 & 8.51 \\
Chronos Small & Global $-$ US & 7.49 & 17.91 & 11.03 & 6.88 \\*
 & Augmented $-$ US & -6.03 & -4.02 & -2.02 & -2.01 \\
TimesFM 8M & Global $-$ US & -6.16 & -8.92 & -4.25 & -4.67 \\*
 & Augmented $-$ US & 15.92 & 20.76 & 14.45 & 6.31 \\
TimesFM 20M & Global $-$ US & -17.72 & 21.70 & -18.08 & 39.78 \\*
 & Augmented $-$ US & -52.22 & 22.30 & -49.24 & 71.55 \\
\addlinespace[3pt]
\multicolumn{6}{l}{\textit{Panel C: Fixed 2023 retrospective-exposure comparison}}\\*
Chronos Tiny & Global $-$ US & 1.31 & 6.46 & 1.17 & 5.29 \\*
 & Augmented $-$ US & 0.02 & 10.40 & 1.51 & 8.89 \\
Chronos Mini & Global $-$ US & 3.12 & 9.79 & 3.11 & 6.68 \\*
 & Augmented $-$ US & 0.71 & 10.32 & 2.66 & 7.65 \\
Chronos Small & Global $-$ US & -4.72 & 18.79 & -4.81 & 23.61 \\*
 & Augmented $-$ US & -12.00 & 7.17 & -11.04 & 18.21 \\
TimesFM 8M & Global $-$ US & 58.20 & 59.44 & 84.57 & -25.13 \\*
 & Augmented $-$ US & -10.64 & 0.38 & -5.86 & 6.24 \\
TimesFM 20M & Global $-$ US & -8.16 & 14.01 & -6.90 & 20.90 \\*
 & Augmented $-$ US & $8.51\!\times\!10^{3}$ & $1.35\!\times\!10^{3}$ & $9.11\!\times\!10^{5}$ & $-9.09\!\times\!10^{5}$ \\
\bottomrule
\end{tabular*}

\par\vspace{6pt}\nointerlineskip
\exhibitnotes{%
  Training-environment contrasts measure changes in U.S. forecast-revision components relative to US training. For component $X$ and $c\in\{\mathrm{Global},\mathrm{Augmented}\}$, $\Delta_cX=X^{(c)}-X^{(\mathrm{US})}$. Movement is $M=\overline{|D|}/\mathrm{RMSE}_{\mathrm{HA}}$, with differences reported as $100\Delta M$. Alignment $A$, penalty $P$, and net matched value $N=A-P$ are in historical-average MSFE percentage points, so $\Delta N=\Delta A-\Delta P$. Panel A equally averages each model's differences across rolling leads $0,+1,+2$ and horizons $h=1,3,6,12$ months. Panels B and C average over the four horizons for fixed states 2009 and 2023 within their retrospective-exposure windows. Every environment contrast uses identical matched target-month support. Estimates use the all-available U.S. equity-premium sample.
}
\end{threeparttable}
\end{table}
\clearpage
\begingroup
\makeatletter
\renewcommand{\footnotesize}{\@setfontsize\footnotesize{9}{10.3}}
\makeatother
\footnotesize
\setlength{\tabcolsep}{4.0pt}
\renewcommand{\arraystretch}{1.00}
\setlength{\linewidth}{426.54pt}
\begin{longtable}{@{}lcccccccc@{}}
\caption[U.S. revision components by training environment]{%
  \textbf{U.S. revision components by training environment.}
}\label{tab:ia-x52-us-corpus-components}\\
\multicolumn{9}{@{}p{\linewidth}@{}}{\exhibitnotes{%
  Forecast-revision components are reported separately for the US, Global, and Augmented training-information environments. Within each environment, the alternative state is compared with its origin-aligned point-in-time (PIT) forecast. Movement is $100M=100\overline{|D|}/\mathrm{RMSE}_{\mathrm{HA}}$; alignment benefit, movement penalty, and net value $A-P$ are in historical-average MSFE percentage points, with identical support within each comparison. Positive net value favors the alternative state. Rolling rows use leads $0,+1,+2$; fixed rows use the retrospective-exposure windows of states 2009 and 2023. The 95\% intervals apply to net value. HAC intervals use Bartlett--Newey--West estimation with $h-1$ lags; bootstrap intervals use 2,000 circular moving-block replications with block length $\max(h,6)$. The final column counts matched target months, and $h$ is in months. Estimates use the all-available U.S. equity-premium sample.
}}\\
\noalign{\vskip 6pt}
\toprule
Comparison & $h$ & $100M$ & Alignment & Penalty & Net & \shortstack{HAC 95\%\\interval} & \shortstack{Bootstrap 95\%\\interval} & $N$ \\
\midrule
\endfirsthead
\multicolumn{9}{l}{\textit{\tablename\ \thetable\ (continued)}}\\
\toprule
Comparison & $h$ & $100M$ & Alignment & Penalty & Net & \shortstack{HAC 95\%\\interval} & \shortstack{Bootstrap 95\%\\interval} & $N$ \\
\midrule
\endhead
\midrule
\multicolumn{9}{r}{\textit{Continued on next page}}\\
\endfoot
\bottomrule
\endlastfoot
\multicolumn{9}{@{}l}{\textit{US training-information environment: Chronos Tiny}}\\*
Lead 0 & 1 & 6.78 & 0.64 & 0.94 & -0.29 & [-2.83, 2.24] & [-2.38, 1.76] & 276 \\*
Lead 0 & 3 & 11.03 & -0.37 & 2.29 & -2.66 & [-7.30, 1.98] & [-7.42, 3.46] & 276 \\*
Lead 0 & 6 & 13.22 & 0.40 & 3.10 & -2.69 & [-9.22, 3.84] & [-8.73, 5.56] & 276 \\*
Lead 0 & 12 & 15.27 & 6.34 & 3.92 & 2.42 & [-5.82, 10.65] & [-5.19, 13.20] & 276 \\
Lead +1 & 1 & 8.33 & -1.53 & 1.66 & -3.19 & [-7.51, 1.13] & [-6.70, 0.16] & 264 \\*
Lead +1 & 3 & 12.23 & -4.99 & 2.54 & -7.53 & [-13.07, -1.99] & [-12.83, -0.50] & 264 \\*
Lead +1 & 6 & 14.19 & -8.80 & 3.36 & -12.16 & [-19.11, -5.22] & [-17.78, -2.42] & 264 \\
\end{longtable}

\endgroup

\clearpage
\begingroup
\makeatletter
\renewcommand{\footnotesize}{\@setfontsize\footnotesize{9}{10.3}}
\makeatother
\footnotesize
\setlength{\tabcolsep}{4.0pt}
\renewcommand{\arraystretch}{1.00}
\setlength{\linewidth}{408.19pt}
\begin{longtable}{@{}lccccccc@{}}
\caption[International revision components by training environment]{%
  \textbf{International revision components by training environment.}
}\label{tab:ia-x52-international-corpus-components}\\
\multicolumn{8}{@{}p{\linewidth}@{}}{\exhibitnotes{%
  International revision components are summarized separately for the US, Global, and Augmented training-information environments. Within each of the 13 non-U.S. markets, the alternative state and its origin-aligned point-in-time (PIT) forecast use identical matched target months. Movement is $100M=100\overline{|D|}/\mathrm{RMSE}_{\mathrm{HA}}$; alignment benefit, movement penalty, and net value $A-P$ are in historical-average MSFE percentage points. Entries are cross-market medians of the individual components. Net IQR spans the 25th--75th percentiles of market-level net effects, and Positive markets counts markets with positive net value. Net summaries are calculated from market-level alignment minus penalty. Rolling rows use leads $0,+1,+2$; fixed rows use the retrospective-exposure windows of states 2009 and 2023. The horizon $h$ is in months. Estimates use the all-available international sample.
}}\\
\noalign{\vskip 6pt}
\toprule
Comparison & $h$ & \shortstack{Median\\$100M$} & \shortstack{Median\\alignment} & \shortstack{Median\\penalty} & \shortstack{Median\\net} & Net IQR & \shortstack{Positive\\markets} \\
\midrule
\endfirsthead
\multicolumn{8}{l}{\textit{\tablename\ \thetable\ (continued)}}\\
\toprule
Comparison & $h$ & \shortstack{Median\\$100M$} & \shortstack{Median\\alignment} & \shortstack{Median\\penalty} & \shortstack{Median\\net} & Net IQR & \shortstack{Positive\\markets} \\
\midrule
\endhead
\midrule
\multicolumn{8}{r}{\textit{Continued on next page}}\\
\endfoot
\bottomrule
\endlastfoot
\multicolumn{8}{@{}l}{\textit{US training-information environment: Chronos Tiny}}\\*
Lead 0 & 1 & 5.75 & -0.45 & 0.74 & -0.99 & [-2.03, -0.52] & 1/13 \\*
Lead 0 & 3 & 7.84 & -1.71 & 1.16 & -2.78 & [-3.75, -1.20] & 0/13 \\*
Lead 0 & 6 & 8.92 & -2.79 & 1.42 & -4.00 & [-5.86, -2.27] & 1/13 \\*
Lead 0 & 12 & 10.10 & -1.48 & 1.81 & -2.82 & [-6.19, -1.19] & 2/13 \\
Lead +1 & 1 & 6.22 & -1.88 & 0.71 & -2.48 & [-3.10, -1.33] & 2/13 \\*
Lead +1 & 3 & 8.86 & -3.01 & 1.64 & -4.44 & [-5.15, -3.85] & 2/13 \\*
Lead +1 & 6 & 10.32 & -5.06 & 2.11 & -7.54 & [-9.08, -3.73] & 2/13 \\
\end{longtable}

\endgroup

\clearpage



\endgroup 
\end{document}